\documentclass[12pt]{report}
\usepackage{epsfig}
\usepackage{epsf}
\usepackage[cp866]{inputenc}
\usepackage[english,russian]{babel}

\begin{document}
\fontsize{14pt}{21pt}
\begin{rm}
\thispagestyle{empty}

\begin{center}
ФГУП ГНЦ РФ
\end{center}

\begin{center}
ИНСТИТУТ ТЕОРЕТИЧЕСКОЙ И ЭКСПЕРИМЕНТАЛЬНОЙ ФИЗИКИ
\end{center}

\vspace{5mm}
\begin{flushright}
На правах рукописи

\end{flushright}

\vspace{15mm}
\begin{center}
{ЗУБКОВ МИХАИЛ АЛЕКСАНДРОВИЧ}
\end{center}

\begin{center}
{\bf ПРИМЕНЕНИЕ НЕПЕРТУРБАТИВНЫХ МЕТОДОВ К ИССЛЕДОВАНИЮ ТЕОРЕТИКО-ПОЛЕВЫХ
МОДЕЛЕЙ СИЛЬНЫХ, ЭЛЕКТРОСЛАБЫХ И ГРАВИТАЦИОННЫХ ВЗАИМОДЕЙСТВИЙ}

\vspace{5mm}
\parbox{10cm}

(Специальность 01.04.02 - теоретическая физика)

\vspace{25mm} ДИССЕРТАЦИЯ\\ на соискание уч\"еной степени

доктора физико-математических наук

\end{center}

\vspace{55mm}
\begin{center}
{Москва -- 2011 г.}
\end{center}
\end{rm}
\newpage
\def\t#1#2{\leftskip=0pt\noindent #1\dotfill\hskip-10pt\hbox to 0mm{\hbox to \rightskip{\hfill #2}\hss}\par}
\def\tt#1#2{\leftskip=15mm\noindent\kern-\leftskip\hbox to\leftskip{Часть #1. \hfill} \ignorespaces  {\bf #2} \par}
\def\tth#1#2#3{\leftskip=7mm\noindent\kern-7mm\hbox to7mm{#1.\hfill}\ignorespaces#2\dotfill\hbox to 0mm{\hbox to \rightskip{\hfill #3}\hss}\par}
\def\ttth#1#2#3{\leftskip=22mm\noindent\kern-7mm\hbox to19mm{#1.\hfill}\ignorespaces#2\dotfill\hskip-30pt\hbox to 0mm{\hbox to \rightskip{\hfill #3}\hss}\par}
\newcommand{\cZ}{{\cal{Z}}}
\newcommand{\cD}{{\cal{D}}}
\newcommand{\cC}{{\cal{C}}}
\newcommand{\Z}{{Z \!\!\! Z}}
\newcommand{\beq}{\begin{equation}}
\newcommand{\eeq}{\end{equation}}
\newcommand{\beqn}{\begin{eqnarray}}
\newcommand{\eeqn}{\end{eqnarray}}
\newcommand{\CK}[1]{\mbox{\scriptsize c}_{\mbox{$\scriptstyle #1$}}}
\newcommand{\nsum}[2]{\sum_{ #1(\CK{#2}) \in \Z }}
\newcommand{\ndsum}[2]{\sum_{\stackrel{\scriptstyle #1(\CK{#2}) \in \Z}
{\delta #1=0}}}
\newcommand{\nddsum}[2]{\sum_{\stackrel{\scriptstyle \dual #1(\dual\CK{#2})
\in \Z} {\delta \dual #1=0}}}
\newcommand{\dd}{\mbox{d}}
\newcommand{\dual}{\mbox{}^{\ast}}
\newcommand{\bbbz}{{\mathchoice {\hbox{$\sf\textstyle Z\kern-0.4em Z$}}
{\hbox{$\sf\textstyle Z\kern-0.4em Z$}} {\hbox{$\sf\scriptstyle Z\kern-0.3em
Z$}} {\hbox{$\sf\scriptscriptstyle Z\kern-0.2em Z$}}}}
\newcommand{\LL}{{I\!\! L}}
\newcommand{\intpi}{\int\limits_{-\pi}^{+\pi} {\cD}}
\newcommand{\intinf}{\int\limits_{-\infty}^{+\infty} {\cD}}
\newcommand{\expb}[1]{\exp\left\{ #1 \right\} }
\newcommand{\ie}{\hbox{\it i.e.}{}}
\newcommand{\etc}{\hbox{\it etc.}{}}
\newcommand{\eg}{\hbox{\it e.g.}{}}
\newcommand{\cf}{\hbox{\it cf.}{}}
\newcommand{\etal}{\hbox{\it et al.}{}}
\newcommand{\re}[1]{(\ref{#1})}
\newcommand{\half}{\frac 12}
\newcommand{\const}{\mbox{const.}\cdot}
\newcommand\Appendix[1]{\par
\setcounter{section}{0}
 \setcounter{equation}{0}
 \renewcommand{\thesection}{Appendix \Alph{section}}
\section{#1}
 \def\theequation{\Alph{section}.\arabic{equation}}}
\newcommand\appendixn[1]{\par
 \setcounter{equation}{0}
 \renewcommand{\thesection}{Appendix \Alph{section}}
\section{#1}
 \def\theequation{\Alph{section}.\arabic{equation}}}
\newcommand{\Tr}{{\mathrm{Tr}}}
\newcommand{\mod}{{\mathrm{mod}}}
\newcommand{\dD}{{\cal D}}
\newcommand{\eq}[1]{(\ref{#1})}

\def\bbbone{{\mathchoice {\rm 1\mskip-4mu l} {\rm 1\mskip-4mu l}
{\rm 1\mskip-4.5mu l} {\rm 1\mskip-5mu l}}}
\def\bbbc{{\mathchoice {\setbox0=\hbox{$\displaystyle\rm C$}\hbox{\hbox
to0pt{\kern0.4\wd0\vrule height0.9\ht0\hss}\box0}}
{\setbox0=\hbox{$\textstyle\rm C$}\hbox{\hbox to0pt{\kern0.4\wd0\vrule
height0.9\ht0\hss}\box0}} {\setbox0=\hbox{$\scriptstyle\rm C$}\hbox{\hbox
to0pt{\kern0.4\wd0\vrule height0.9\ht0\hss}\box0}}
{\setbox0=\hbox{$\scriptscriptstyle\rm C$}\hbox{\hbox to0pt{\kern0.4\wd0\vrule
height0.9\ht0\hss}\box0}}}}
\def\bbbe{{\mathchoice {\setbox0=\hbox{\smalletextfont e}\hbox{\raise
0.1\ht0\hbox to0pt{\kern0.4\wd0\vrule width0.3pt height0.7\ht0\hss}\box0}}
{\setbox0=\hbox{\smalletextfont e}\hbox{\raise 0.1\ht0\hbox
to0pt{\kern0.4\wd0\vrule width0.3pt height0.7\ht0\hss}\box0}}
{\setbox0=\hbox{\smallescriptfont e}\hbox{\raise 0.1\ht0\hbox
to0pt{\kern0.5\wd0\vrule width0.2pt height0.7\ht0\hss}\box0}}
{\setbox0=\hbox{\smallescriptscriptfont e}\hbox{\raise 0.1\ht0\hbox
to0pt{\kern0.4\wd0\vrule width0.2pt height0.7\ht0\hss}\box0}}}}

\begin{rm}

{ \centerline{\bf О Г Л А В Л Е Н И Е}
\bigskip
\rightskip=7mm


\t{Введение}{\pageref{ch0}}

\tt{1}{ Топологические дефекты в моделях сильного взаимодействия}

\tth{1}{Центральная доминантность в глюодинамике. Центральные
монополи.}{\pageref{ch1}}

\tth{2}{Простая центральная проекция глюодинамики}{\pageref{ch2}}

\tth{3}{Абелево представление для неабелевой петли Вильсона}{\pageref{ch3}}

\tt{2}{ Монополи Намбу и физика Электрослабых взаимодействий}

\tth{4}{Решеточная регуляризация калибровочного и Хиггсовского секторов
Стандартной Модели}{\pageref{ch4}}

\tth{5}{Монополи Намбу при конечной температуре}{\pageref{ch5}}

\tth{6}{Флуктуационная область в модели Вайнберга - Салама}{\pageref{ch6}}

\tt{3}{ Решеточные формулировки квантовой гравитации}

\tth{7}{Двумерная гравитация. Предел слабой связи и
дискретизация}{\pageref{ch7}}

\tth{8}{Четырехмерная квантовая гравитация. Дискретизация Телепараллелизма и
Пуанкаре - гравитации}{\pageref{ch8}}

\tth{9}{Многомерные динамические триангуляции}{\pageref{ch9}}

\tt{4}{ Возможные пути построения моделей новой физики на масштабе ТэВ}

\tth{10}{Малое объединение на масштабе ТэВ и $Z_6$ симметрия Стандартной
Модели}{\pageref{ch10}}

\tth{11}{Продолжение $Z_6$ симметрии Стандартной Модели на теории
техницвета}{\pageref{ch11}}

\tth{12}{Калибровочная теория группы Лоренца как возможный источник
динамического нарушения Электрослабой симметрии}{\pageref{ch12}}

\t{Заключение}{\pageref{ch13}} }

\newpage

{\Large \bf Введение}

 \label{ch0}

\vspace{1cm}

\frenchspacing                

Современная теоретическая физика фундаментальных взаимодействий, на наш взгляд,
имеет четыре основных направления:

1. Теория сильных взаимодействий, основанная на квантовой Хромодинамике;

2. Теория Электрослабых взаимодействий;

3. Квантовая гравитация;

4. Теории новой физики, появление которой ожидается на масштабе ТэВ, доступном
Большому Адронному Коллайдеру.

Первое из указанных направлений занимается изучением физики сильных
взаимодействий. В настоящее время считается, что КХД прекрасно описывает данный
класс явлений и не нуждается в доопределении. Однако, с вычислительной точки
зрения КХД - исключительно сложная наука. Ведущую роль в ее динамике играют
непертурбативные явления. Одним из методов их изучения являются решеточные
симуляции. Среди наиболее важных задач упомянем объяснение механизмов
невылетания и спонтанного нарушения киральной симметрии.

Второе направление связано с изучением явлений, в которых играют роль
Электрослабые взаимодействия. Здесь теория возмущений в Модели Вайнберга -
Салама превосходно описывает реальность при энергиях, не превышающих
существенно 100 ГэВ. Однако, проблема иерархий указывает на то, что при
энергиях выше 1 ТэВ может появиться новая теория, причем модель Вайнберга -
Салама должна быть ее низкоэнергитическим приближением. Традиционно считается,
что непертурбативный анализ при изучении Электрослабых взаимодействий не
требуется. Однако, известно, что при температурах, приближающихся к температуре
Электрослабого фазового перехода (кроссовера), теория возмущений работать
перестает. Это вызывает необходимость использования непертурбативных методов
при конечной температуре. При нулевых температурах необходимость использования
непертурбативных методов связана с существованием объектов, не описывающихся
теорией возмущений. Это, например,  - так называемые Z струны и монополи Намбу,
представляющие классические неустойчивые решения уравнений движения. Масса этих
объектов оценивается в районе нескольких ТэВ. Поэтому при характерных энергиях
процессов, много меньших ТэВ, их появление не может внести значительного вклада
в наблюдаемые величины. Ситуация существенно изменяется при приближении
характерной энергии процесса к 1 ТэВ. Здесь указанные объекты начинают играть
существенную роль. Поэтому  и оказывается необходимым применение
непертурбативных методов.

Третье направление связано с построением теории квантовой гравитации.
Построение такой теории представляется важным в связи с тем, что история
Вселенной как целого является экспериментом, доносящим до нас информацию о
ранних этапах ее развития, когда характерные энергии соответствовали шкале, на
которой квантовая гравитация может появиться. Существует огромное количество
различных моделей, претендующих на описание квантовой гравитации. Среди них
упомянем теорию струн, некоммутативные теории, матричные модели, петлевую
квантовую гравитацию. К сожалению, на сегодняшний день невозможно сделать выбор
какой - либо одной из этих моделей. Поэтому автор настоящей диссертации при
изучении данной темы ограничился минимальным выбором - квантовой теорией
Римановой геометрии.

Четвертое направление связано с тем, что, как было указано выше, при энергиях
порядка 1 ТэВ ожидается появление новой физики. Среди моделей этой новой физики
упомянем суперсимметричные модели, модели Малого Хиггса, Модели Малого
Объединения, Модели Техницвета.

Текст настоящей диссертации делится на четыре части, соответствующие каждому из
указанных выше направлений. Каждая из частей состоит из трех глав, в которых
автор описывает результаты своих исследований задач, относящихся к данному
направлению. При этом в конце каждой главы указывается список печатных работ, в
которых были опубликованы эти результаты. В конце каждой из частей размещается
библиография, соответствующая данной части. Исследования, относящиеся к разным
частям диссертации практически независимы друг от друга. В то же время главы
каждой из частей логически связаны. Поэтому в начале каждой части мы помещаем
аннотацию представленной в ней работы.  В конце текста диссертации в разделе
Заключение  указывается полный список работ, в которых опубликованы результаты
настоящей диссертации.

Поскольку далее в тексте будут широко использоваться обозначения и терминология
решеточных калибровочных теорий, мы их здесь кратко напомним.

Как правило, рассматриваются гиперкубические решетки, состоящие из $D$ - мерных
гиперкубов, склеенных вместе. Обычно также предполагается реализация
периодических граничных условий, что сводится к тому, что система склеенных
гиперкубов реализует топологию $D$ - мерного тора. Вершины гиперкубов
называются точками или узлами решетки. Ребра, соединяющие соседние точки
решетки, именуются линками решетки. Квадраты, составленные из четырех узлов и
четырех линков, называются плакетами.

Дуальная решетка определяется следующим образом. Ее узлы находятся в центрах
гиперкубов исходной решетки. Линками соединяются узлы, лежащие в центрах
соседних гиперкубов исходной решетки и т.д.

Калибровочные поля, как правило, на решетке определяются переменными,
прикрепленными к линкам. Произведение их вдоль линков одного плакета дает
плакетную переменную, реализующую на решетке напряженность калибровочного поля.
Скалярное поле обычно определено в узлах решетки.

Удобным формализмом для работы с Абелевыми калибровочными полями является
формализм дифференциальных форм на решетке. Дифференциальная форма ранга $k$ на
решетке - это функция $\phi_{k}$ определенная на $k$-мерном кубе $c_k$ решетки,
\eg\ скалярное (калибровочное) поле - это 0--форма (1--форма). Внешняя
производная {\it d} определяется следующим образом:

\beq (\dd \phi ) (c_{k+1}) =\sum_{\CK{k} \in \partial\CK{k+1}} \phi(c_{k}).
\label{def-dd} \eeq Здесь $\partial c_{k}$ - ориентированная граница
$k$-мерного куба $c_{k}$. Таким образом, оператор {\it d} увеличивает ранг
формы на единицу; $\dd \varphi$ - это линковая переменная, сконстрованная через
переменную $\varphi$, определенную на узлах, а $\dd A$ - это плакетная
переменная, сконструируемая из линковой переменной $A$.

Скалярное произведение определяется следующим образом: если $\varphi$ и $\psi$
- $k$-формы, то $(\varphi,\psi)=\sum_{c_k}\varphi(c_k)\psi(c_k)$, где
$\sum_{c_k}$ - сумма по всем кубам $c_k$. Каждой $k$--форме на $D$--мерной
решетке соответствует $(D-k)$--форма $\dual\Phi(\dual c_k)$ на дуальной
решетке, $\dual c_k$ - это $(D-k)$--мерный куб на дуальной решетке.
Кодифференциал $\delta=\dual \dd \dual$ удовлетворяет правилу интегрирования по
частям: $(\varphi,\delta\psi)=(\dd\varphi,\psi)$. Заметим, что $\delta
\Phi(c_k)$ - это $(k-1)$--форма и $\delta \Phi(c_0) = 0$.

Норма определяется как  $\|a\|^2=(a,a)$; тогда, например,  $\|\dd\varphi+2\pi
l\|^2$ предполагает суммирование по всем линкам. $\nsum{l}{1}$ означает сумму
по всем конфигурациям целых чисел $l$, прикрепленных к линкам $c_1$.

Действие для калибровочных полей инвариантно относительно калибровочных
преобразований $A' = A + \dd \alpha$, $\varphi' = \varphi + \alpha$ благодаря
свойству $\dd^2 = \delta^2 = 0$. Решеточный лапласиан определяется как $\Delta
= \dd\delta + \delta\dd$.

\end{rm}
\begin{rm}

\part{Топологические дефекты в моделях сильного взаимодействия}

Одним из наиболее важных явлений физики сильных взаимодействий является
невылетание цвета. Материал настояшей части диссертации относится к описанию
конфайнмента с точки зрения различных Абелевых проекций теории. Мы изучаем КХД
без динамических фермионов. В первой главе рассматриваются свойства центральных
вихрей в Максимальной Центральной Проекции глюодинамики. Демонстрируется их
связь с явлением невылетания, изучаются перколяционные свойства. Вводится новое
понятие - центральный монополь. Показано, что в теории при конечной температуре
Центральные монополи и центральные вихри сконденсированы в фазе конфайнмента.
Во второй главе вводится понятие простой центральной проекции, являющейся
альтернативным методом выделения центральных вихрей из полевых конфигураций
неабелевых калибровочных теорий. Рассматриваются фрактальные свойства вихрей и
центральных монополей в этой проекции. Демонстрируется связь этих объектов с
конфайнментом. В третьей главе решается техническая задача об Абелевом
представлении для неабелевой петли Вильсона, которое может быть полезно при
исследовании различных абелевых проекций. Получено Абелево представление для
неабелевой петли Вильсона на решетке, являющееся аналогом непрерывной формулы
Дьяконова - Петрова. Разрешается вопрос о справедливости непрерывного
представления, связанный с определением меры по калибровочным преобразованиям.

\chapter{Центральная доминантность в глюодинамике. Центральные монополи.}
\label{ch1}

\section{Центральная проекция $SU(2)$ глюодинамики и Центральные Монополи}

Исследование конфайнмента в $SU(N)$ калибровочных теориях во многих работах
основывается на частичной фиксации калибровки до некоторой абелевой группы.
Примером подобной фиксации калибровки, предложенной в \cite{CenterGaugeFirst},
является так называемая Центральная проекция, когда оставшейся группой является
$Z_N$. В центральной калибровке $SU(N)$ калибровочная теория редуцируется к
 $Z_N$ калибровочной теории, которая содержит вихревые струны в качестве топологических дефектов.
Решеточные вычисления \cite{NumerousCenter} в Максимальной Центральной
калибровке показывают, что динамика этих дефектов играет важную роль в
невылетании цвета. Автором настоящей диссертации также предложено понятие
Центрального монополя, связанного с вихревой струной. Ниже приводится его
определение и результаты численных исследований, демонстрирующих его связь с
конфайнментом. При этом для простоты мы ограничиваемся изучением глюодинамики
без динамических фермионов.

Прежде всего, рассмотрим случай $SU(2)$ глюодинамики, которая является
упрощенной моделью реалистической $SU(3)$ глюодинамики, сохраняя при этом ее
основные свойства.  Максимальная Центральная проекция делает линковые матрицы
$U$ максимально близкими к центральным элементам группы $SU(2)$. Эта калибровка
определяется следующим образом \cite{CenterGaugeFirst}: вначале фиксируется
Максимальная Аделева калибровка посредством максимизации функционала (см.
\cite{MaA})  $\sum\nolimits_{l} \Tr (U_l \sigma^3 U^+_l \sigma^3)$ относительно
калибровочных преобразований $U^{(\Omega)}_{x,\mu} = \Omega^+_x U_{x,\mu}
\Omega_{x+\hat\mu}$ (сумма берется по линкам $l$ решетки, $\sigma^a$ - матрицы
Паули). Далее, максимизируется функционал $\sum\nolimits_{l} \cos^2
\arg({(U_l)}^{11} )$ относительно оставшихся  $U(1)$ калибровочных
преобразований, таким образом линковая матрица становится близкой к центральным
элементам $\pm \bbbone$.

Центральные вихри определяются следующим образом \cite{CenterGaugeFirst}. После
фиксации Максимальной Центральной проекции мы определяем $Z_2$ плакетную
переменную $\sigma_P$: \beqn \sigma_P \equiv {(\dd n)}_P = n_1 + n_2 - n_3 -
n_4\,, \label{sigma} \eeqn где линки $1,\dots,4$ формируют границу плакета $P$
и $n_l = {\mathrm{sign}} ({\mathrm{Tr}} U_l)$. Мировая поверхность струны
центрального вихря определяется на дуальной решетке как совокупность $\dual
\sigma$ плакетов дуальных ненулевым плакетам  $\sigma_P$, мировая поверхность
$\dual \sigma$ замкнута на дуальной решетке ($\delta \dual \sigma = 0$).

Взаимодействие центрального вихря с петлей Вильсона является топологическим.
Чтобы увидеть это представим $SU(2)$ калибровочное поле $U_l$ в Максимальной
Центральной калибровке как произведение $\Z_2$ переменной  $\exp\{i \pi n_l\}$,
$n_l=0,1$, и $SU(2) \slash \Z_2$ переменной $V_l$, $\Tr V_l \ge 0$: $U_l =
\exp\{i \pi n_l \} \cdot V_l$. Принимая во внимание \eq{sigma} мы представляем
петлю Вильсона для контура  $\cC$ как: \beqn W_\cC = \Tr \prod\limits_{l \in C}
U_l = \exp\Bigl\{i \pi \LL (\cC,\sigma) \Bigr\} \Tr \prod\limits_{l \in C}
V_l\,, \label{WC1} \eeqn где $\LL(\cC,\sigma)$ - число зацеплений кварковой
траектории $\cC$ за мировую поверхность струны $\dual \sigma$~\cite{forms}:
\beqn \LL (\cC,\sigma) = (\sigma,m[\cC]) = (\sigma,\Delta^{-1} \dd \cC)\,,
\label{LL} \eeqn $m[\cC]$ - это поверхность, натянутая на контур  $\cC$:
$\delta m[\cC] = \cC$. Последнее выражение для $\LL$ есть четырехмерный аналог
формулы Гаусса, описывающей зацепление замкнутой поверхности $\dual \sigma$ и
замкнутой петли $\cC$. Известно ~\cite{CenterGaugeFirst,LinkingConf}, что
топологическое взаимодействие $\exp\{i \pi \LL\}$ дает очень хорошее
приближение к выражению для натяжения струны.

Благодаря $\Z_2$ периодичности теория содержит монополеподобные возбуждения,
которые названы нами "Центральные монополи". Их мировые траектории определяются
следующим образом: \beqn j = \frac{1}{2} \dd \Bigl[(\dd n) \, \mod \, 2 \Bigr]
\equiv \frac{1}{2} \dd \Bigl[\sigma \, \mod \, 2 \Bigr]\,. \eeqn  $\Z_2$ заряд
Центральных монополей сохраняется, монопольные траектории замкнуты:~$\delta
\dual j = 0$.
\begin{figure}[!htb]
\begin{minipage}{13.5cm}
\begin{center}
\begin{tabular}{cc}
\epsfig{file=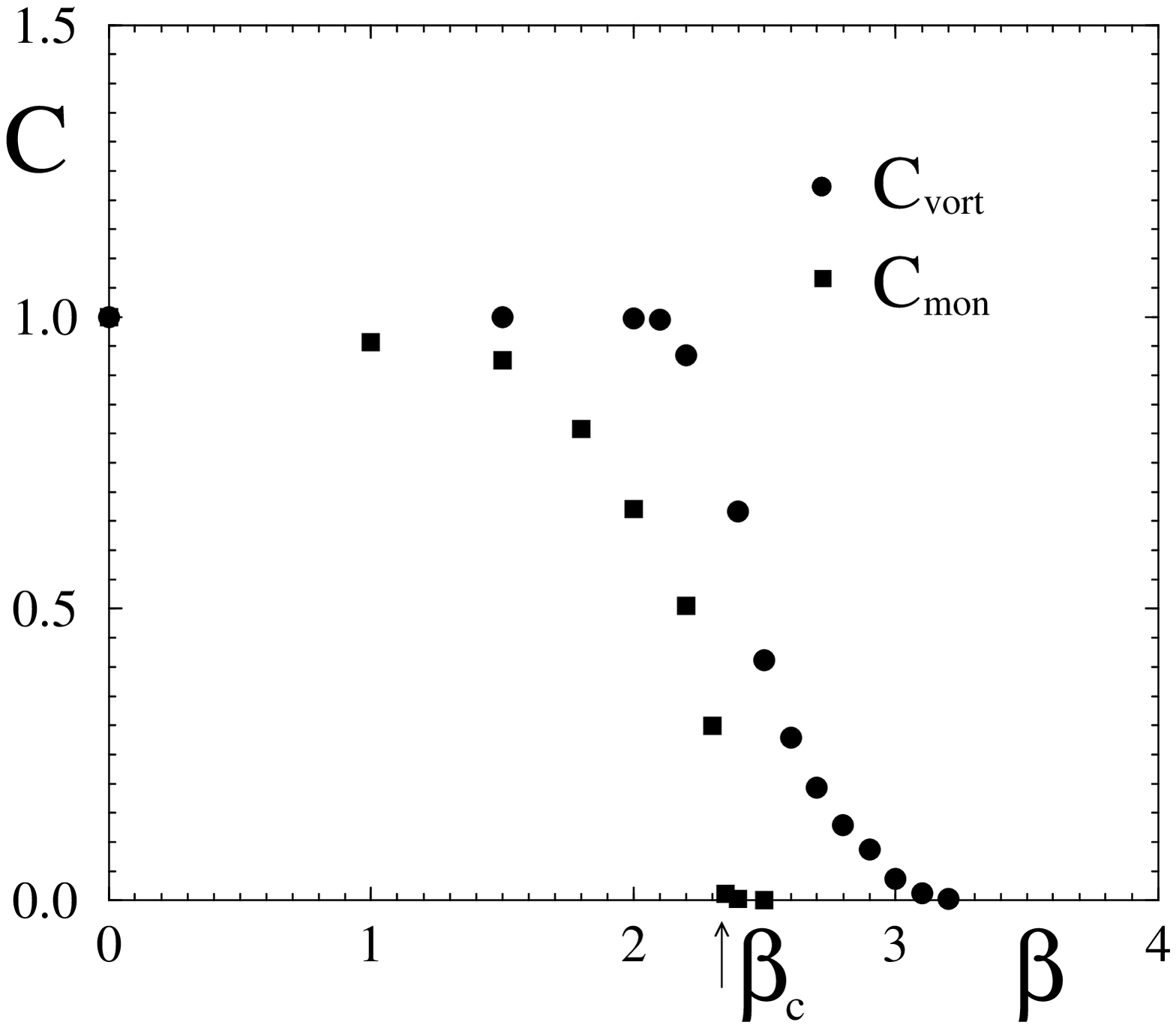,width=6.2cm,height=6.5cm,angle=0}&
\epsfig{file=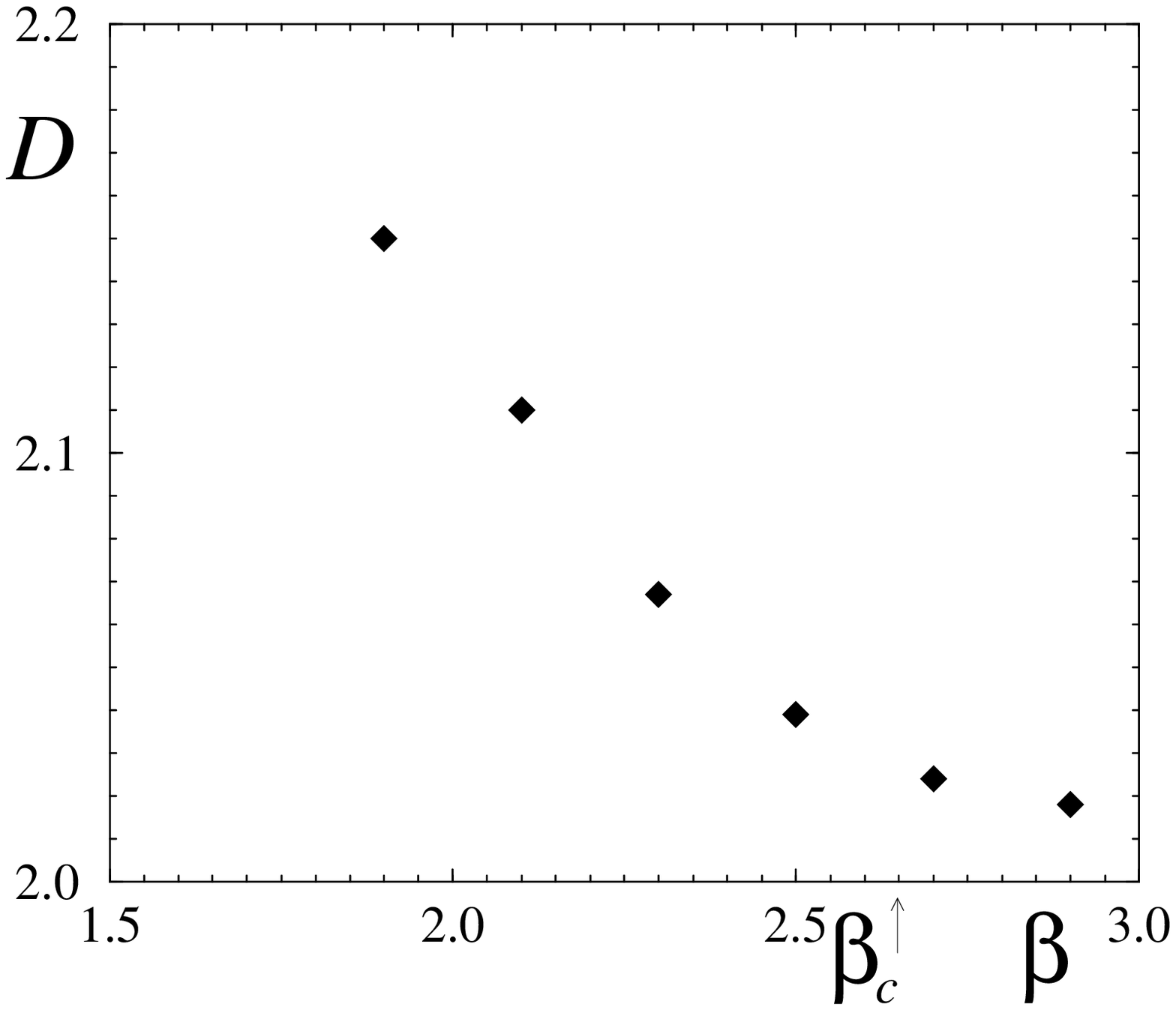,width=6.2cm,height=6.5cm,angle=0}\\
(a) & (b)\\
\end{tabular}
 \caption[]{\small (a) Перколяция центральных вихрей
$C_{\mathrm{vort}}$ и Центральных монополей $C_{\mathrm{mon}}$ {\it от} $\beta$
на решетке $16^3 \times 4$; (b) Фрактальная размерность $D$ центральных вихрей
на решетке $12^3 \times 8$.\label{f.1}}
\end{center}
\end{minipage}
\end{figure}
\vspace{-5mm}

Мы исследовали модель при конечной температуре, что соответствует асимметричной
решетке, у которой  размер по одному из измерений (мнимое время) меньше, чем по
трем оставшимся. Важное динамическое свойство центральных монополей это
перколяция $C_{\mathrm{mon}}$, которая определяется как вероятность двум разным
точкам решетки быть связанными монопольной траекторией ~\cite{Percolation}. Мы
видим, что $C_{\mathrm{mon}}$ зануляется в фазе деконфайнмента и отлична от
нуля в фазе конфайнмента ($C_{\mathrm{mon}}$ показана квадратами на Рис
\ref{f.1}(a)). Мы заключаем, что фаза конфайнмента сопровождается конденсацией
Центральных монополей. Монополи являются дуальными абелевыми степенями свободы
и их конденсация означает, что конфайнмент соответствует фазе дуального
сверхпроводника в Максимальной Центральной проекции.

Перколяционная вероятность для центральных вихрей $C_{\mathrm{vort}}$
определяется аналогично перколяции монополей. Следует отметить, что в данной
главе нами используется определение $C_{\mathrm{vort}}$ отличное от
общепринятого (и использующегося в других главах) \footnote{Определим
вероятность того, что две точки $x$, $y$ связаны вихревым кластером как
$\rho(x,y)$. Определение, использующееся в настоящей главе: $C_{\mathrm{vort}}
= {\rm lim}_{|x-y|\rightarrow \infty}\rho(x,y)$. Общепринятое определение:
$\tilde{C}_{\mathrm{vort}} = \frac{1}{N_T^2}\sum_{x_0,y_0} {\rm
lim}_{|\bar{x}-\bar{y}|\rightarrow \infty}\rho([\bar{x},x_0],[\bar{y}, y_0])$.
Здесь $4$ - вектор $x = [\bar{x},x_0]$ имеет "пространственную" компоненту
$\bar x$ и "временную" компоненту $x_0$. $N_T$ - размер решетки по "временному"
направлению. }.

Величина $C_{\mathrm{vort}}$ {\it от} $\beta$ на решетке $16^3\times 4$
представлена на Рис. \ref{f.1}(a) кругами. Явно видно, что в фазе конфайнмента
вихри перколируют.

На Рис. \ref{f.1}(b) мы представляем фрактальную размерность вихревой мировой
поверхности $D = 1+2A/L$ на решетке $12^3 \times 8$. Здесь $A$ - число плакетов
и  $L$ число линков на поверхности струны. Фрактальная размерность $D$ высока в
фазе конфайнмента. В фазе деконфайнмента она близка к  $2$ поскольку мы имеем
разреженный ансамбль струн.

\section{{Центральная Доминантность и Центральные монополи в $SU(3)$ глюодинамике } }

 Как было сказано выше, центральная доминантность, открытая в работах Д. Гринсайта
 наблюдается в так называемой Максимальной Центральной проекции. Струноподобные объекты взаимодействуют с кварками
 посредством сил Ааронова - Бома. Известно, что эти силы в значительной степени
  ответственны за конфайнмент  \cite{Tomboulis_}.
Конфайнмент традиционно описывается механизмом дуального сверхпроводника. Этот
механизм явно реализуется в  $SU(2)$ глюодинамике в Максимальной Абелевой
проекции \cite{Wiese_}, в которой появляются монополи, сконденсированные в фазе
конфайнмента. Благодаря этой конденсации силовые линии концентрируются в
струне, соединяющей кварк и антикварк. Эта струна имеет ненулевое натяжение,
что приводит к невылетанию цвета. Многие полагают, что та же картина должна
наблюдаться и в $SU(3)$ теории. Однако, в  $SU(3)$ глюодинамике после фиксации
Максимальной Абелевой проекции появляется не один, а два монополя, что
существенно усложняет общую картину  \cite{Suzuki}.

  В соответствии с гипотезой Центральной Доминантности центральные вихри ответственны за
  конфайнмент. Таким образом, возникает вопрос, какова связь между
центральными вихрями и картиной дуального сверхпроводника. Для того, чтобы
ответить на этот вопрос мы конструируем монополеподобные объекты из центральных
вихрей и называем их Центральными Монополями. Оказывается, что их конденсат
является параметром порядка для перехода конфайнмент - деконфайнмент, что дает
надежду полагать, что именно эти объекты могут играть роль скалярных частиц в
механизме дуального сверхпроводника.

Мы рассматриваем $SU(3)$ глюодинамику с действием Вильсона  $S(U) = \beta
\sum_{\mathrm{plaq}} (1-1/3 \mathrm{Re} \, \mathrm{Tr} U_{\mathrm{plaq}})$.
Здесь сумма по плакетам решетки. Если данные плакет состоит из линков
$[xy]$,$[yz]$,$[zw]$,$[wx]$ то $U_{\mathrm{plaq}} = U_{[xy]}
 U_{[yz]}  U_{[zw]}  U_{[wx]} $.

Максимальная Центральная проекция делает линковую матрицу   $U$ возможно более
близкой к элементам центра $Z_3$ группы $SU(3)$: $ Z_3 = \{{\rm
diag}(\mathrm{e}^{(2\pi i /3) N}, \mathrm{e}^{(2\pi i /3) N}, \mathrm{e}^{(2\pi
i /3) N}\}$, где $N \in \{1, 0, -1\}$.  Мы используем так называемую непрямую
(indirect)  версию Максимальной Центральной проекции, которая работает
следующим образом.

Прежде всего, максимизируем функционал
\begin{equation}
Q_1 = \sum_{\mathrm{links}} (|U_{11}| + |U_{22}| + |U_{33}|)
\end{equation}
по отношению к калибровочным преобразованиям  $U_{xy} \rightarrow g^{-1}_x
U_{xy} g_y$, таким образом фиксируя Максимальную Абелеву калибровку. Как
следствие, линковые матрицы становятся почти диагональны. Далее, для того чтобы
сделать эти матрицы максимально близкими к элементам центра $SU(3)$, делаем
фазы этих диагональных элементов матриц максимально близкими друг к другу. Это
достигается минимизацией функционала
\begin{eqnarray}
 Q_2 & = & \sum_{\mathrm{links}}[(1-\cos({\rm Arg}(U_{11})-{\rm Arg}(U_{22})))
                    +(1-\cos({\rm Arg}(U_{11})-{\rm Arg}(U_{33}))) \nonumber \\
 &&                    +(1-\cos({\rm Arg}(U_{22})-{\rm Arg}(U_{33})))].
\end{eqnarray}
по отношению к калибровочным преобразованиям.

Центральные вихри определяются следующим образом. Определяем целочисленную
линковую переменную  $N$:

\begin{eqnarray}
N_{xy}=0 &{\rm :}& ({\rm Arg}(U_{11})+{\rm Arg}(U_{22})+{\rm Arg}(U_{33}))/3
 \in \; ]-\pi/3, \pi/3], \nonumber \\
N_{xy}=1 &{\rm :}& ({\rm Arg}(U_{11})+{\rm Arg}(U_{22})+{\rm Arg}(U_{33}))/3
 \in \; ]\pi/3, \pi], \nonumber \\
N_{xy}=-1 &{\rm :}& ({\rm Arg}(U_{11})+{\rm Arg}(U_{22})+{\rm Arg}(U_{33}))/3
 \in \; ]-\pi, -\pi/3].
\end{eqnarray}

 Таким образом, $N=0$ если $U$ близко к $1$,  $N=1$ если $U$ близко к
 \noindent $\mathrm{e}^{2\pi i/3}$, и   $N=-1$ если $U$ близко к
$\mathrm{e}^{-2\pi i/3}$.

Далее, определяем плакетную переменную:

\begin{equation}
 \sigma_{xywz} =  N_{xy}+N_{yw}-N_{zw}-N_{xz}
\end{equation}

Мы вводим дуальную решетку, и определяем переменную $\sigma^*$ дуальную к
 $ \sigma$: если плакет $^*\Omega$ дуален плакету $\Omega$, то
$\sigma^*_{^*\Omega} = \sigma_{ \Omega}$.  Можно легко проверить, что  $\sigma$
представляет замкнутую поверхность. Эта поверхность и есть мировая поверхность
центрального вихря.

Мы выражаем $SU(3)$ калибровочное поле  $U$ как произведение $\exp((2\pi i / 3)
N)$ и $V$, где  $V$ - это $SU(3)/Z_3$ переменная $({\rm Arg}(V_{11})+{\rm
Arg}(V_{22})+{\rm Arg}(V_{33}))/3 \in \; ]-\pi/3, \pi/3]$. Тогда $U=\exp((2\pi
i/ 3) N) V$

Далее, петля Вильсона вдоль контура $C$ записывается как:
\begin{equation}
 W_C = \Pi_C U =\exp((2\pi i / 3) L(C,\sigma)) \Pi_C V
\end{equation}

Здесь $(2\pi i/3) L(C,\sigma)$ представляет взаимодействие Ааронова - Бома.
Величина $L(C,\sigma)$ - это число зацепления контура $C$ и поверхности
$\sigma^*$.

Содержание гипотезы центральной доминантности заключается в том, что в
Максимальной Центральной калибровке взаимодействие Ааронова - Бома само по себе
вызывает конфайнмент и определяет натяжение струны.

Центральный монополь - это $Z_3$ аналог монополя в  $U(1)$ теории. В
электродинамике уравнения Максвелла $dF=0$ запрещают существование магнитных
зарядов. Но в компактной теории значения  F, отличающиеся друг от друга на
$2\pi$, эквивалентны. Таким образом, правильное натяжение струны это $F \, {\rm
mod} \, 2\pi$ и $^*d (F \, {\rm mod} \, 2\pi) = 2\pi j_m$, где $j_m$ -
монопольный ток.

Взаимодействие Ааронова - Бома между кварками и центральными вихрями зависит
только от  $[\sigma] \, {\rm mod} \, 3$. Здесь $\sigma$ - это $Z_3$ аналог
$U(1)$ напряженности поля. Переменная $[\sigma] \, {\rm mod} \, 3$ представляет
поверхность с границей. Эта граница - замкнутая линия.  Мы предполагаем что эта
линия представляет Центральный Монополь:
\begin{equation}
 3j_m = \; ^*d ([\sigma] \, {\rm mod} \, 3) =
 \delta  ([\sigma^*] \, {\rm mod} \, 3).
\end{equation}
(Здесь используются обозначения дифференциальных форм на решетке. См.,
например,  \cite{forms}.)

Мы предлагаем гипотезу, что Центральный Монополь может играть роль скаляра в
механизме дуального сверхпроводника. Это частично подтверждается результатами
численных рассчетов модели при конечной температуре, представленных ниже.

Используется решетка размера $16^3 \times 4$. Фазовый переход конфайнмент -
деконфайнмент для этой решетки имеет место для  $\beta = 5.69$ (см.
\cite{engels}). Наши численные результаты следующие:

\begin{enumerate}
\item
\begin{figure}[!htb]
\begin{center}
\begin{tabular}{cc}
\epsfig{file=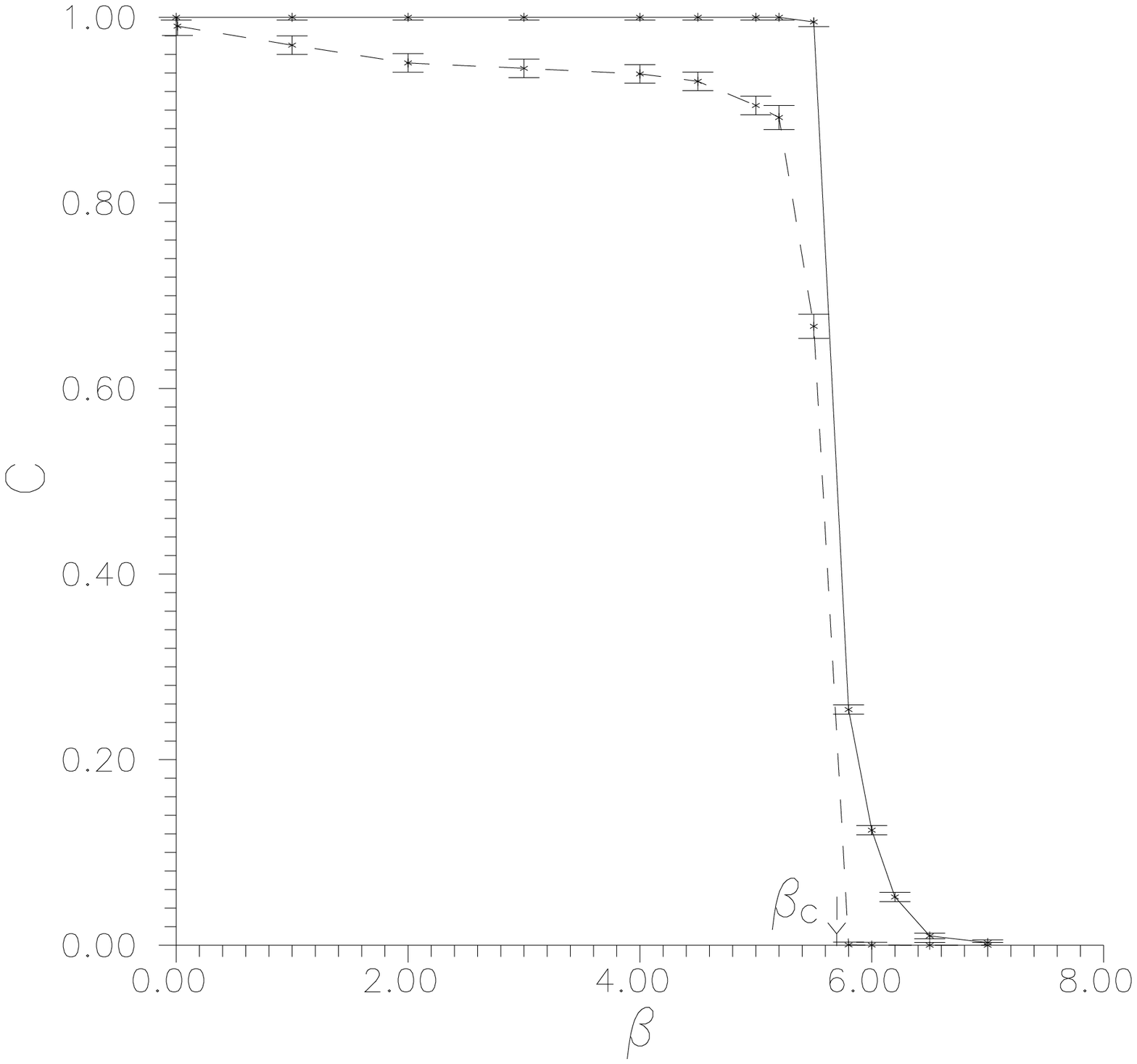,width=15.0cm,height=15.0cm,angle=0}\\
\end{tabular}
\end{center}
\caption[]{ Перколяция центральных вихрей (непрерывная линия) и Центральных
монополей (прерывистая линия).  \label{fig1}}
\end{figure}
Центральные вихри сконденсированы в фазе конфайнмента, и не сконденсированы при
высокой температуре. Это следует из рассмотрения вероятности того, что две
точки на решетке связаны струнной мировой поверхностью. Эта вероятность $
\rho(x,y)_{\mathrm{vort}} \rightarrow C_{\mathrm{vort}}(\beta)$ при
$|x-y|\rightarrow \infty$.  (Непрерывная линия на Рис. ~\ref{fig1}.)

\item
\begin{figure}[!htb]
\begin{center}
\begin{tabular}{cc}
\epsfig{file=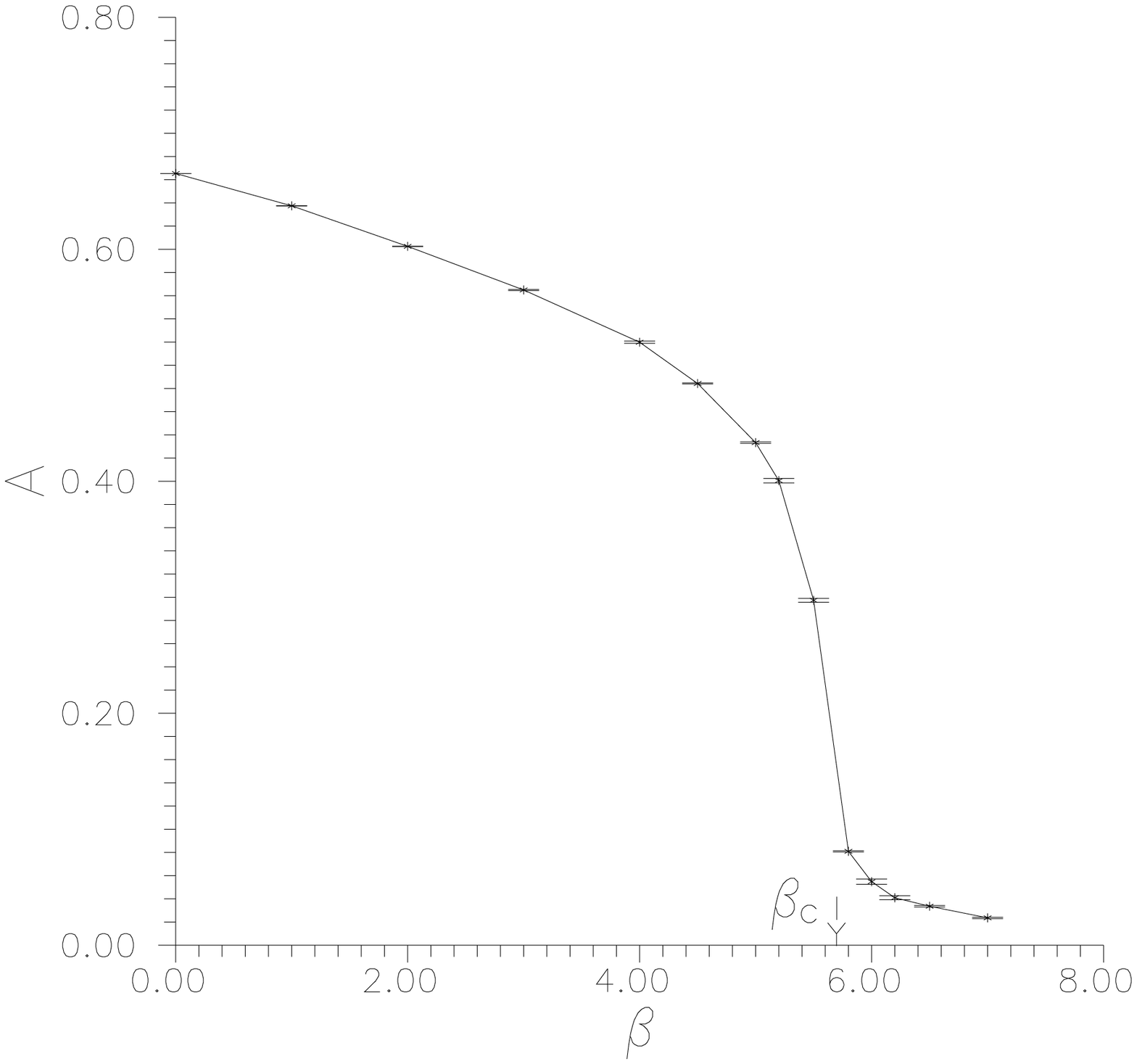,width=15.0cm,height=15.0cm,angle=0}\\
\end{tabular}
\end{center}
\caption[]{Плотность центральных вихрей.\label{fig2}}
\end{figure}

Плотность центральных вихрей представлена на рис. ~\ref{fig2}.

\item

\begin{figure}[!htb]
\begin{center}
\begin{tabular}{cc}
\epsfig{file=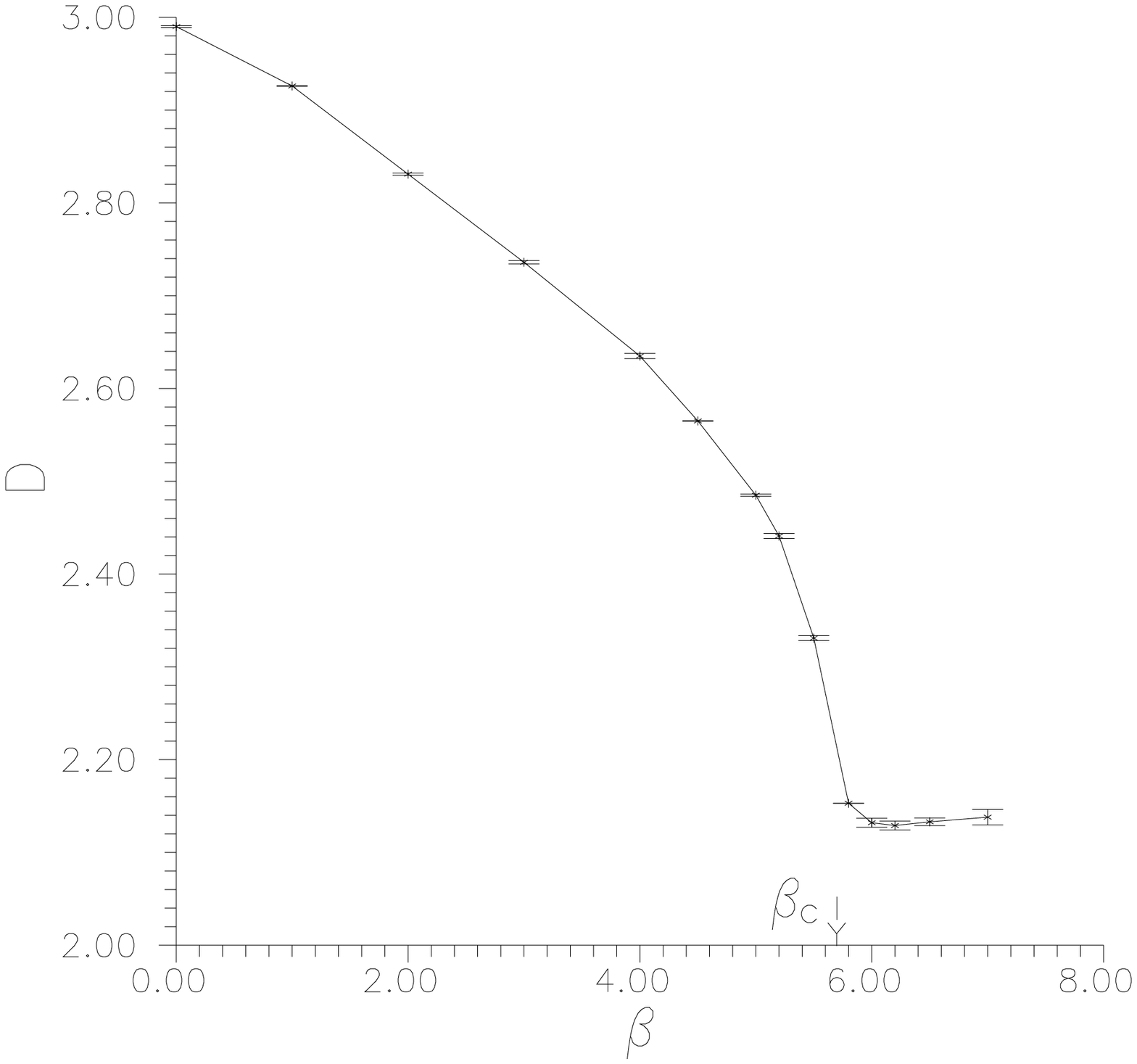,width=15.0cm,height=15.0cm,angle=0}\\
\end{tabular}
\end{center}
\caption[]{ Фрактальная размерность центральных вихрей.\label{fig3}}
\end{figure}

Мы определяем фрактальную размерность Центральных вихрей как
\begin{equation}
 D = 1+2A/L \label{D}
\end{equation}
где $A$ - это число плакетов, а $L$ - число линков струны. В таком виде это
определение существенно отличается от определения Хаусдорфовой размерности,
которое используется нами в следующей главе в п. 2.3. для изучения свойств
вихрей и монополей в так называемой Простой Центральной Проекции. Выражение
(\ref{D}) можно рассматривать как интерполирующую формулу. Легко проверить, что
(\ref{D}) дает правильное значение для фрактальной размерности D -  мерных
решеток, если считать, что $A$ - это полное число плакетов решетки, а $L$ -
полное число линков. Результаты численных исследований $D$ представлены на рис.
~\ref{fig3}.

\item Центральные монополи сконденсированы в фазе конфайнмента, и не
сконденсированы в фазе деконфайнмента. Это следует из рассмотрения вероятности
того, что две точки связаны монопольной линией. Для этой величины $
\rho(x,y)_{mon} \rightarrow C_{mon}(\beta)$ для $|x-y|\rightarrow \infty$. Мы
видим, что $C_{mon}$ равно $0$ в фазе деконфайнмента, и отлично от $0$ в фазе
конфайнмента. Конденсат как функция $\beta$ представлен на рис. ~\ref{fig1}
прерывистой линией.

\item В дополнение добавим что в соответствии с перколяционными свойствами
центральные вихри и центральные монополи распределены однородно. Это следует из
того, что вероятность двум точкам быть связанным мировой поверхностью (мировой
линией) этих объектов не зависит существенно от расстояния между точками.

\end{enumerate}

\section{Публикации}

Результаты настоящей главы опубликованы в работах:

"Aharonov-Bohm effect, center monopoles and center vortices in SU(2) lattice
gluodynamics", M.N. Chernodub, M.I. Polikarpov, A.I. Veselov, M.A. Zubkov,
Nucl.Phys.Proc.Suppl.73:575-577,1999, [hep-lat/9809158]

"Central dominance and the confinement mechanism in gluodynamics", B.L.G.
Bakker, A.I. Veselov, M.A. Zubkov, Phys.Lett.B471:214-219,1999,
[hep-lat/9902010]

"Central dominance and the confinement mechanism", B.L.G. Bakker, A.I. Veselov,
M.A. Zubkov, Nucl.Phys.Proc.Suppl.83:565-567,2000.

{
\chapter{Простая центральная проекция глюодинамики}
\label{ch2}
\vspace{1cm}

\section{Определение Простой Центральной Проекции}

Как было отмечено выше, Абелева проекция глюодинамики является одним из самых
популярных методов исследования механизма невылетания цвета. После фиксации
какой-либо Абелевой калибровки теория становится Абелевой, и возникает
возможность рассматривать механизм невылетания в упрощенном виде.

Абелевы проекции отличаются друг от друга выбором Абелевой подгруппы и методами
проектирования. Близость данной Абелевой проекции к решению проблемы
конфайнмента измеряется следующим образом. Предположим, что линковая групповая
переменная
 $g_{\rm link}\in G$ спроектирована на элемент Абелевой подгруппы $e_{\rm link}\in E \subset G$.
 Мы рассматриваем
\begin{equation}
 Z_C = {\rm Tr} \prod_{\rm link\in C} e_{\rm link}
 \label{eq.010}
\end{equation}
вместо полной петли Вильсона и извлекаем потенциал из $Z_C$, ( {\em
спроецированный} потенциал).  Если этот потенциал близок к исходному
удерживающему потенциалу на достаточно больших расстояниях, можно говорить о
том, что проекция пригодна для исследования механизма невылетания.

Для калибровочной группы $SU(2)$ ее Картанова подгруппа $U(1)$ и центральная
подгруппа $Z_2$ были рассмотрены подобным образом. Наиболее популярные проекции
это Максимальная Абелева и Максимальная Центральная проекция, рассмотренные в
предыдущей главе. Эти проекции достигаются минимизацией по отношению к
калибровочным преобразованиям расстояния между заданной конфигурацией линковых
полей и Картановой (Центральной) подгруппой группы  $SU(2)$.  В обоих случаях
потенциалы близки к полному удерживающему $SU(2)$ потенциалу, но, к сожалению,
не совпадают с ним в точности. Гринсайт и соавторы \cite{Gr2000,DFGGO1998}
указывают, что только центральный заряд $q= \pm 1$ подвержен невылетанию в
неабелевых калибровочных теориях.

Из работы Борнякова и соавторов \cite{BKPV2000} известно, что процедура
фиксации калибровки страдает от проблемы калибровочных копий (Грибовская
проблема).  В практических симуляциях были использованы прямая и непрямая
Центральные калибровки, и Лапласова Центральная калибровка \cite{FE1999}.
Только первые две страдают от проблемы Грибовских копий, в то время, как
последняя свободна от этой трудности. В соответствии с ~\cite{Gr2000, FGO1999a}
эта проблема исчезает для больших решеток и тогда, когда увеличивается
количество рассматриваемых Грибовских копий. Разумеется, это означает
необходимость увеличения вычислительных мощностей.

Автором настоящей диссертации также предложена новая Центральная проекция, не
связанная с процедурой частичной фиксации калибровки. Таким образом, все
объекты, существующие в этой проекции калибровочно инвариантны. Данная проекция
названа Простой Центральной проекцией (ПЦП). Ниже показывается, что
спроецированный таким образом потенциал близок к полному $SU(2)$ потенциалу на
больших расстояниях  (с точностью до члена, соответствующего перенормировке
массы).

В рамках ПЦП мы строим центральные вихри и центральные монополи, также
известные как нексусы ({\em nexuses}, см.  \cite{Vo1999}). Свойства монополей
ПЦП найденные в ходе численных исследований говорят о том, что эти объекты
могут быть кандидатами на роль куперовских пар в механизме дуального
сверхпроводника.

Мы рассматриваем $SU(2)$ глюодинамику с Вильсоновским действием
\begin{equation}
 S(U) = \beta
\sum_{\mathrm{plaq}} (1-1/2 \, \mathrm{Tr} U_{\mathrm{plaq}}).
 \label{eq.011}
\end{equation}
Здесь сумма по плакетам решетки. Плакетная переменная  $U_{\rm plaq}$
определена стандартным образом.

Прежде всего мы рассматриваем плакетную переменную
\begin{eqnarray}
 z_{\rm plaq} = 1  & {\rm если} &  ({\rm Tr} \, U_{\rm plaq} < 0), \nonumber\\
 z_{\rm plaq} = 0  & {\rm если} &  ({\rm Tr} \, U_{\rm plaq} > 0).
 \label{eq.020}
\end{eqnarray}
Мы можем представить $z$ как сумму замкнутой формы $d N$ для  $N\in \{0,1 \}$ и
формы $2m +q$. Здесь $N = N_{\rm link}$, $q\in \{0,1\}$, и $m\in {\sf Z} \!\!
{\sf Z}$.

\begin{equation}
 z = d N + 2m +q .
 \label{eq.030}
\end{equation}
Физические переменные, зависящие от $z$ могут быть выражены через
\begin{equation}
 {\rm sign} \, {\rm Tr} \, U_{\rm plaq} = \cos (\pi (dN +q)) .
 \label{eq.040}
\end{equation}
Далее мы будем говорить, что $N_{\rm link}$ - это и есть спроецированная
линковая переменная. Есть множество способов организовать проекцию.
Максимальная Центральная проекция использует калибровочную свободу для того,
чтобы сделать все линковые матрицы максимально близкими к   $e^{i\pi N}$. Таким
образом 1-форма $N$ фиксируется для каждой калибровочной конфигурации.

Существует несколько способов фиксации Максимальной Центральной проекции.
Прямая проекция  ({\em direct}) - это минимизация функционала
\begin{equation}
 R = \sum_x \sum_\mu {\rm Tr} [ U_\mu(x) ] {\rm Tr}  [U^\dag _\mu (x) ].
 \label{eq.041}
\end{equation}
В непрямом виде ({\em indirect}) проекция минимизирует функционал
\begin{equation}
 R' =
 \sum_x \sum_\mu {\rm Tr} [U_\mu (x) \sigma_3 U^\dagger _\mu (x) \sigma_3 ],
 \label{eq.042}
\end{equation}
и извлекает из $U_\mu(x)$ диагональную часть $A_\mu = \exp[i \theta_\mu(x)
\sigma_3]$, фиксируя таким образом калибровку. Оставшаяся Абелева симметррия
используется для того, чтобы сделать $A_\mu $ максимально близкой к элементу
$Z_2$ максимизируя
\begin{equation}
 R'' = \sum_x \sum_\mu \cos^2 \theta_\mu(x).
 \label{eq.043}
\end{equation}
Ясно, что обе процедуры сложны и требуют значительное количество временных
ресурсов поскольку число переменных, включенных в процедуры в случае группы
$SU(2)$ - три. Для любого метода 1-форма $N$ фиксируется для каждой
калибровочной конфигурации.  Ниже предлагается более простая процедура. А
именно, представим поверхность $\Sigma$, формируемую плакетами дуальными к
"негативным" плакетам (для которых  $z_{\rm plaq} = 1$). Эта поверхность имеет
границу. Мы увеличиваем поверхность, добавляя новую поверхность  $\Sigma_{\rm add}$ так, что:\\
1. результирующая поверхность $\Sigma^1=\Sigma+\Sigma_{\rm add}$ замкнута;\\
2. когда мы устраняем из поверхности $\Sigma_{\rm add}$ плакеты, несущие четные
числа $z_{\rm plaq} = \dots, -4,\,-2,\,2,\,4, \dots$, площадь оставшейся
поверхности минимальна для заданной границы.

Таким образом,  $\Sigma^1$ может быть представлена как замкнутая форма  $dN$ на
исходной решетке для целочисленной переменной  $N \in \{0,1\}$. И $N$ -
требуемая переменная ПЦП.

Численно данная процедура выглядит следующим образом. Для данной переменной $z$
мы должны выбрать такую $Z_2$ переменную $N$, что  $[dN]\,{\rm mod}\, 2$ близко
насколько возможно к $z$. Это означает, что мы минимизируем функционал

\begin{equation}
Q = \sum_{plaq} |(z - dN)\, {\rm mod}\, 2|
\end{equation}
по отношению к $N$.

Эта процедура работает следующим образом. Мы рассматриваем данный линк  $L$ и
сумму по плакетам

\begin{equation}
Q_{link} = \sum_{L \in {\rm plaq}} |(z-dN) \, {\rm mod} \,  2|
\end{equation}

Минимизируем эту сумму по отношению к одной линковой переменной  $N$. Все линки
обрабатываются подобным образом, и процедура повторяется пока не обнаруживается
 минимум.

 Мы называем описанную процедуру Простой Центральной Проекцией.
 Наш метод также находит локальные минимумы и Грибовские копии. Поскольку процедура много
 проще чем привычные Центральные проекции, включить в рассмотрение несколько Грибовских копий оказывается
 несложно.

 Наивно, напряженность поля на поверхностях $^*\Sigma^1$
стремится к бесконечности в непрерывном пределе. Это, однако, вовсе не
означает, что данные конфигурации исчезают в непрерывном пределе. Их
существование определяется балансом действия и энтропии. В работах Захарова,
Поликарпова, Губарева и соавторов показано \cite{PvortSing,PvortSing2}, что
действие также расходится в непрерывном пределе и на центральных вихрях
Максимальной Центральной Проекции. Тем не менее, показано, что эти объекты
выживают в непрерывном пределе. Переход к непрерывному пределу применительно к
вихрям  ПЦП рассматривается нами ниже в п. 2.3.  Указанные ниже результаты
позволяют надеяться, что эти поверхности выживают в непрерывном пределе так же,
как и  P - вихри и могут играть ключевую роль в описании конфайнмента.

В наших симуляциях мы используем решетки размера $24^4$. Для вычислений при
конечной температуре мы используем решетки $24^3 \times 4$. Для сравнения мы
также рассматриваем меньшие решетки размеров  $12^3 \times 4$ and $16^3 \times
4$. Некоторые результаты были проверены на больших решетках $32^3 \times 4$.
Как уже отмечалось, мы рассматриваем также проблему калибровочных копий.
Проверено, что для решеток линейного размера $L \leq 24$,  15 копий оказывается
достаточно.

Мы рассматриваем следующее определение спроецированной петли Вильсона
\begin{equation}
 W^{\rm SCP}_C = \frac{1}{8}\,  Z_C \, (3\pi/4)^{{\cal P}(C)/4},
 \label{eq.060}
\end{equation}
где ${\cal P}(C)$ - периметр петли $C$.

Интересно отметить, что это выражение близко к первому члену в разложении по
характерам из  \cite{FGO1999a, Og1999}. Наше выражение получается заменой
фактора  $1/4$ на $1/8$ и показателя степени ${\cal P}(C)$ на ${\cal P}/4$.
Однако, авторы ~\cite{FGO1999a, Og1999} получают свое выражение без
использования какой - либо проекции, при этом используя локальные операторы и
характеры калибровочной группы, и получают доминантность фундаментального
представления. В нашем случае мы производим проекцию, что приводит к появлению
нелокальных операторов. В результате мы не можем использовать тот же вывод, что
и в ~\cite{FGO1999a, Og1999}. Поскольку мы не можем дать последовательный вывод
(\ref{eq.060}), следует рассматривать это выражение как эмпирическое. Следует
отметить, что данное выражение было проверено для петель Вильсона размера до $6
\times 6$. Было бы полезно проверить, дает ли (\ref{eq.060}) хорошее
приближение для петель Вильсона большего размера.

\section{Численные результаты}
\begin{figure}[h]
\begin{center}
 \epsfig{figure=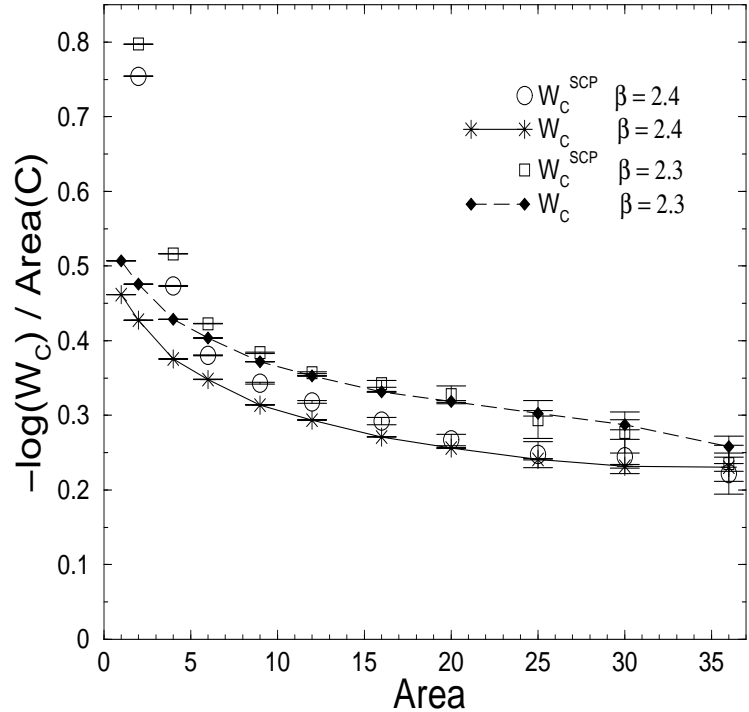,height=100mm,width=100mm}
 \caption{\label{2fig.01} $-\log (W_C)/{\rm Area}(C)$ для петель Вильсона
 $W_C$, и спроецированных петель Вильсона $W^{\rm SCP}_C$.
 $\beta$ = 2.3 и 2.4}
\end{center}
\end{figure}
\subsection{Спроецированная и обычная петли Вильсона}

Из рис. ~\ref{2fig.01} ясно, что $ W_C $ и $ W^{\rm SCP}_C$ совпадают друг с
другом для петель достаточно большого размера (мы представляем здесь $-\log W_C
/{{\rm Area}(C)}$ и $-\log W^{\rm SCP}_C/{{\rm Area}(C)}$ как функции площади
петли для значений $\beta$ = 2.3 и 2.4).
$W^{\rm SCP}_C$ отличается от  $Z_C$ фактором пропорциональным периметру. Таким
образом, мы видим, что спроецированный потенциал (извлеченный из $Z_C$)
отличается от полного потенциала (извлеченного из $W_C$) тoлько фактором
перенормировки массы. Другие члены (включая линейный удерживающий потенциал)
совпадают. Это и демонстрирует Центральную Доминантность.

\subsection{Центральные вихри}
\begin{figure}[h]
\begin{center}
 \epsfig{figure=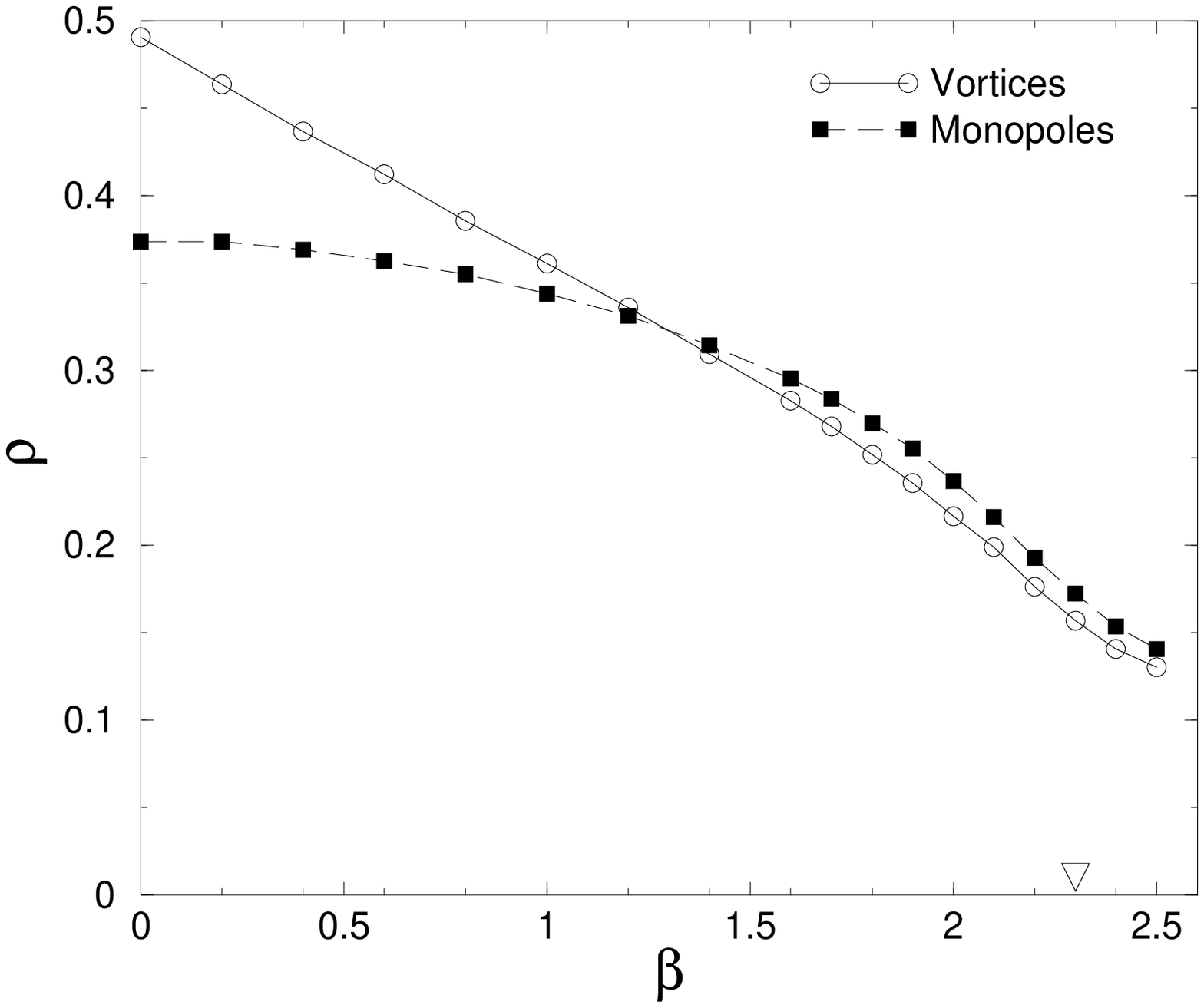,height=100mm,width=100mm}
 \caption{\label{2fig.02} Плотность центральных вихрей и монополей.
 Треугольник указывает положение фазового перехода.}
\end{center}
\end{figure}
Мы строим замкнутые двумерные центральные вихри следующим образом
\begin{equation}
 \sigma = \, ^*d N .
\end{equation}
Прежде всего, мы можем выразить $Z_C$ как
\begin{equation}
Z_C = \exp(i\pi L\!\!L(\sigma,C))
\end{equation}
где $L\!\!L$ - это число зацепления \cite{CPV1997}.

Таким образом, в соответствии с приведенным выше результатом взаимодействие
Ааронова - Бома между центральными вихрями и заряженными частицами приводит к
конфайнменту фундаментального заряда  \cite{CPV1997}.

Мы также исследуем свойства центральных вихрей в модели при конечной
температуре (несимметричная решетка $24^3 \times 4$).

Плотность вихрей $\rho$ показана на ~\ref{2fig.02}.
\begin{figure}[h]
\begin{center}
 \epsfig{figure=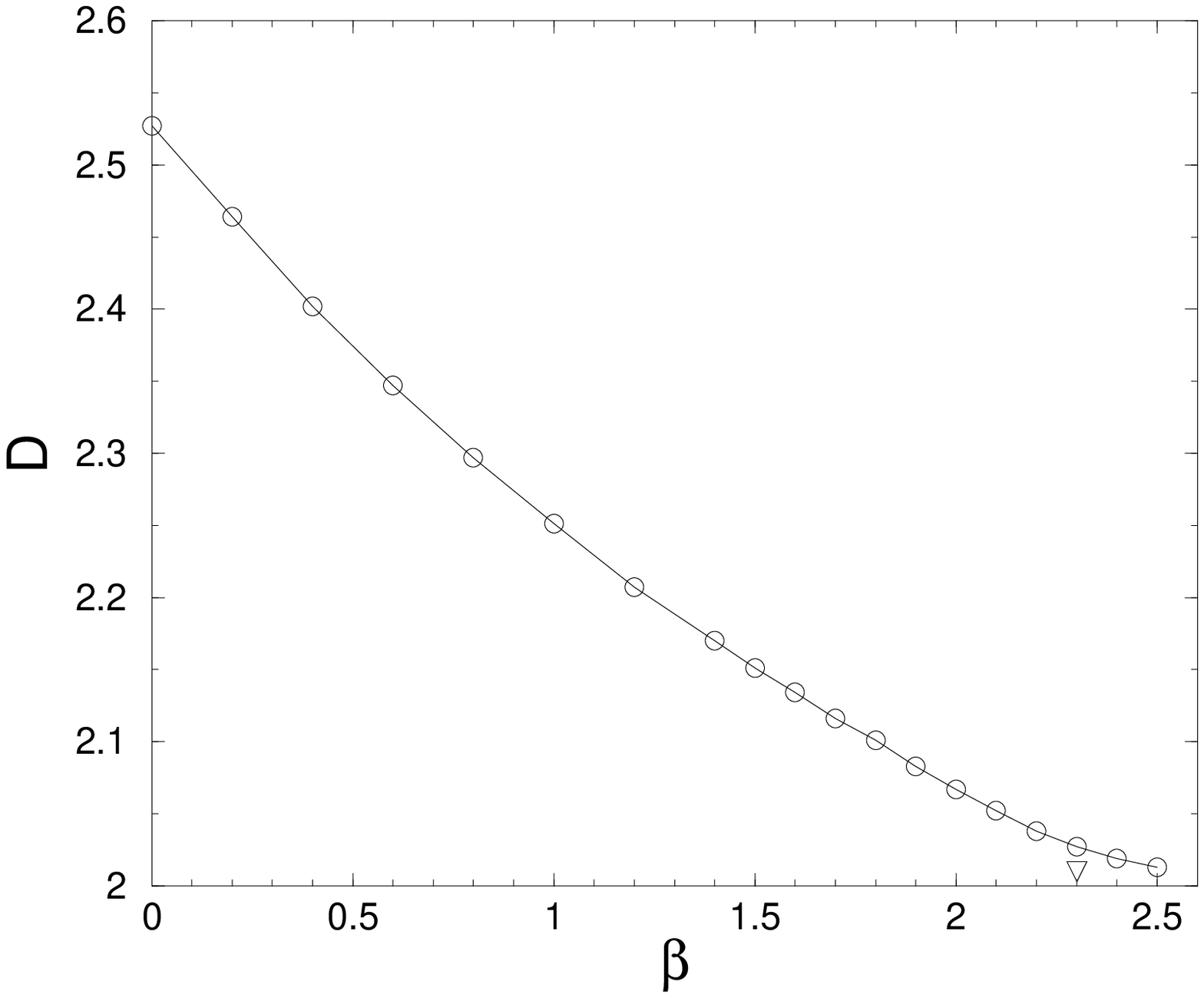,height=100mm,width=100mm}
 \caption{\label{2fig.03} Фрактальная размерность  $D$ центральных вихрей.
 Треугольник указывает положение фазового перехода.}
\end{center}
\end{figure}
Фрактальная размерность, определенная  как $D = 1 + 2 A/L$, где $A$ - это число
плакетов, а $L$ - число линков \cite{BVZ1999} на поверхности вихря, показана на
рис.~\ref{2fig.03}. Линк считается принадлежащим вихрю если хотя бы одна из
сторон куба, дуального к нему содержит плакет с зарядом  1. Следует отметить,
что данное определение отличается от определания Хаусдорфовой размерности,
используемого ниже в п. 2.3.

\subsection{Центральные монополи (nexuses)}
\begin{figure}[h]
\begin{center}
 \epsfig{figure=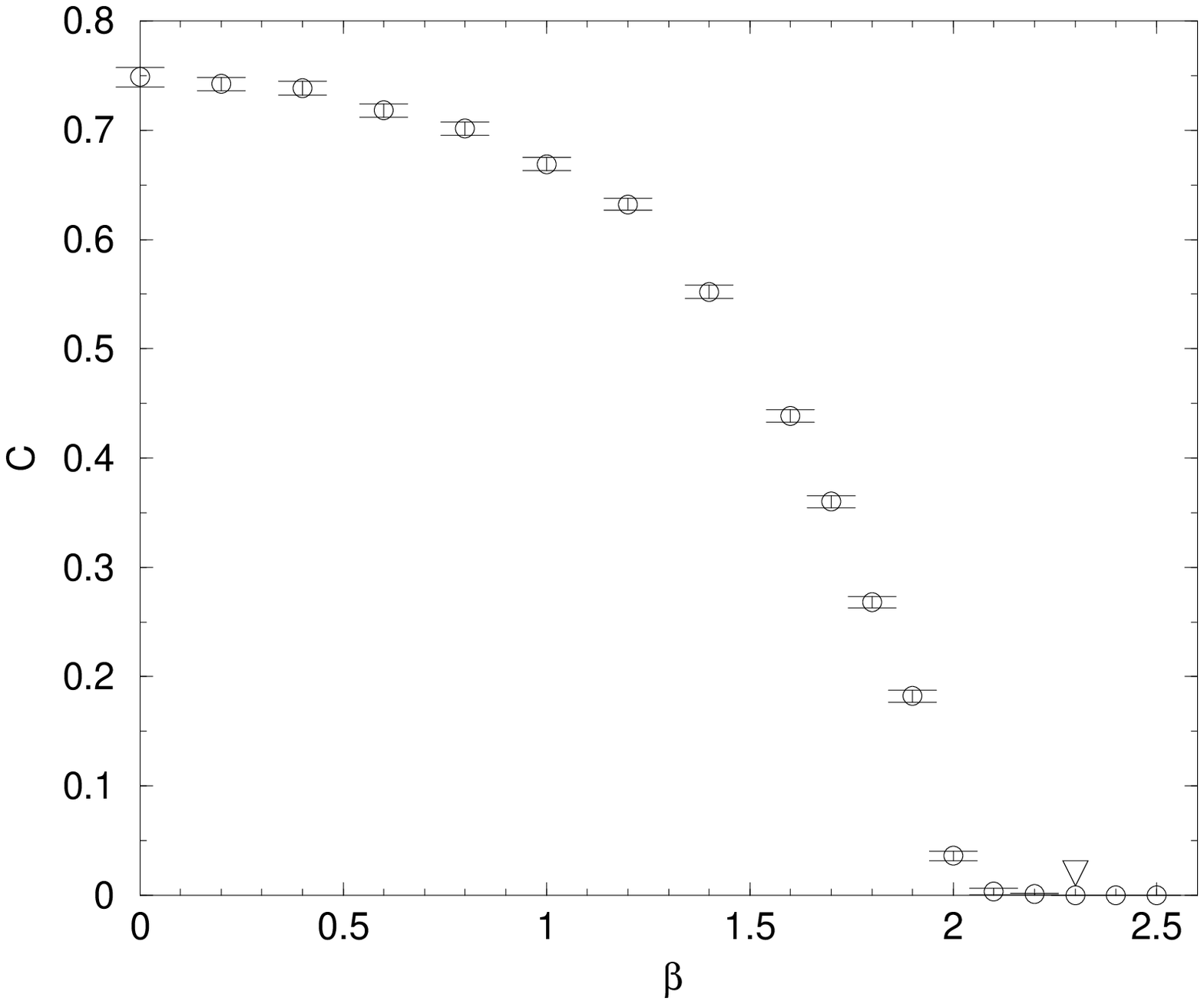,height=100mm,width=100mm}
 \caption{\label{2fig.04} Перколяция монопольных линий для решеток
 $24^3 \times 4$.
 Треугольник указывает положение фазового перехода. }
\end{center}
\end{figure}
Мы строим центральные монополи (nexuses) следующим образом
\begin{equation}
 j = \frac{1}{2} \; ^*d [d N]\; {\rm mod}\, 2 .
 \label{eq.070}
\end{equation}
Плотность Центральноых Монополей как функция $\beta$ представлена на
рис.~\ref{2fig.02}. Следует заметить, что монопольные линии замкнуты. Это
следует из того, что  $\delta j =\frac{1}{2} {}^*d{}^*{}^*d [dN]\; {\rm mod}\;
2 =0 $.

Мы исследовали перколяционные свойства $j$. Зависимость перколяционной
вероятности от $\beta$ для несимметричной решетки показана на рис.
~\ref{2fig.04}. Перколяционная вероятность зависит от размера решетки. Это
очевидно для решеток малого размера. С увеличением размера решетки эта
зависимость исчезает. Монопольный конденсат, таким образом, играет роль
параметра порядка. Мы ожидаем, что центральные монополи могут играть роль
куперовских пар в механизме дуального сверхпроводника.

\begin{figure}
\begin{center}
 \epsfig{figure=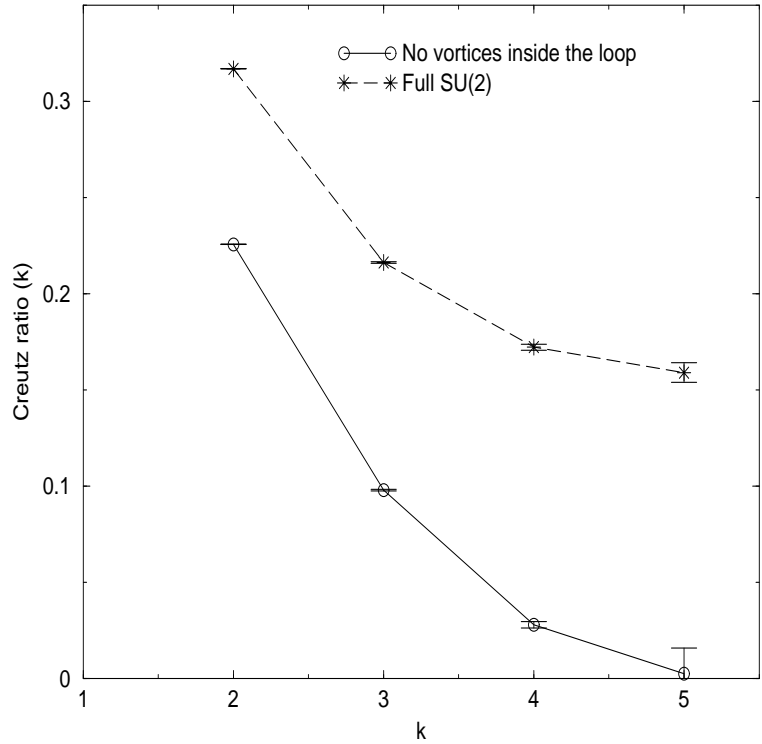,height=100mm,width=100mm,angle=-90}
 \caption{ Отношение Кройца для конфигураций, у которых внутри заданной
 петли
 Вильсона нет вихрей как функция размера петли (для вычисления отношения Кройца
 рассматриваются контуры $k\times k$, $(k+1)\times (k+1)$, $k\times (k+1)$). Отношение Кройца для полной SU(2) петли показано для сравнения.
  \label{fig.20}}
\end{center}
\end{figure}

\subsection{Отношение Кройца без простых центральных вихрей. }

Чтобы показать, что натяжение струны связано с центральными вихрями, мы
рассматриваем конфигурации, для которых нет вихрей внутри заданной петли
Вильсона. На рис. ~\ref{fig.20} зависимость отношения Кройца для таких
конфигураций от размера петли Вильсона представлено для  $\beta =2.3$ для
решетки $24^4$. Это отношение Кройца зануляется при больших размерах петли $k$.
Мы также нашли, что тот же результат имеет место для иных значений $\beta$. С
другой стороны, истинное натяжение струны не зануляется в непрерывном пределе.
Это означает, что данные вихри играют значительную роль в динамике.

\section{Исследование перехода к непрерывному пределу}

\subsection{Скейлинг и асимптотический скейлинг}

Непрерывный предел в решеточной теории достигается при приближении к точке
фазового перехода второго рода. Для рассматриваемой теории это точка  $\beta =
\infty$, что означает, что корреляционная длина  $r(\beta)$ стремится к
бесконечности когда $\beta \to \infty$. Физическая корреляционная длина
остается, разумеется, той же самой, но она становится бесконечной в решеточных
единицах. В этой ситуации любой физический объект конечной длины представляется
на решетке в виде бесконечного числа линков. Например, запишем физическую
корреляционную длину в решеточных единицах  $R_{\rm phys} = r(\beta) a(\beta)$,
где $a(\beta)$ - это длина линка. Для того, чтобы удержать физическую
корреляционную длину постоянной, длина ребра решетки должна вести себя как
$a(\beta) \sim 1/ r(\beta)$ и, следовательно, $a$ должно стремиться к  $0$
когда мы приближаемся к точке фазового перехода и наша решеточная модель
приближается к непрерывному пределу. Это означает, что размер решетки должен
вести себя как $L \sim 1/a$, если мы удерживаем физический размер решетки
независящим от $\beta$. Зависимость длины ребра решетки $a$ от $\beta$
называется скейлингом. Предположим, что некоторая физическая величина, которая
представляется некоторой решеточной величиной  $F_{\rm lat}$
 имеет размерность $D$ в единицах массы, тогда  $F_{\rm lat}
a^{-D} \rightarrow F_{\rm cont}$, где $ F_{\rm cont}$ - это эта же переменная в
непрерывном пределе. Таким образом, мы имеем для достаточно большого $\beta$:
\begin{equation}
F_{\rm lat} \sim a^D
\end{equation}
Хорошо известный пример подобного поведения - это поведение натяжения струны:
$\sigma_{\rm lat} a^{-2} \rightarrow \sigma_{\rm cont}$.

Ренормализационная группа в непрерывном пределе предсказывает (в двухпетлевом
приближении) следующую зависимость длины ребра решетки от  $\beta$
\cite{Creutz}
\begin{equation}
 \bar{a}(\beta) \sim \beta^{\frac{51}{121}}e^{-(3\pi^2/11)\beta}
 \label{a}
\end{equation}
Это поведение известно как асимптотический скейлинг.

Таким образом, мы бы хотели видеть, что для достаточно больших  $\beta$
скейлинг длины ребра решетки приближается к асимптотическому скейлингу. На
практике асимптотическая свобода в $SU(2)$ теории не достигается (во всяком
случае, для значений $\beta$ от $2.1$ до $2.7$, которые используются в
настоящей работе). Отклонение скейлинга от асимптотического скейлинга для этих
значений $\beta$ хорошо известно. Можно извлечь зависимость $a$ от  $\beta$ из
решеточного натяжения струны. В таблице ~\ref{tab.01} мы представляем данные из
~\cite{Scaling}.

\begin{table}
\caption{Поведение натяжения струны как функции $\beta$:
 $\sqrt{\sigma_{\rm lat}} \sim \sqrt{\sigma_{\rm cont}}a(\beta)$.}
\label{tab.01}
\begin{center}
\begin{tabular}{|c|c|c|c|}
\hline
  ~$N_\sigma$~ &~$N_\tau$~ &~$\beta$~ &~$\sqrt{\sigma_{\rm lat}}$~\\
\hline
 ~~8~&~10~&~2.20~&~0.4690(100)~\\
 ~10~&~10~&~2.30~&~0.3690(~30)~\\
 ~16~&~16~&~2.40~&~0.2660(~20)~\\
 ~32~&~32~&~2.50~&~0.1905(~~8)~\\
 ~20~&~20~&~2.60~&~0.1360(~40)~\\
 ~32~&~32~&~2.70~&~0.1015(~10)~\\
 ~48~&~56~&~2.85~&~0.0630(~30)~\\
\hline
\end{tabular}
\end{center}
\end{table}

Можно проверить, что $a(\beta)$, извлеченное из этих данных отклоняется от
$\bar{a}(\beta)$ для рассматриваемых значений $\beta$. Следует отметить, что
 $\sqrt{\sigma_{\rm cont}}a$ не зависит от размера решетки для существенно больших решеток.
 Данные в таблице ~\ref{tab.01} представлены для решеток размеров  $N_{\sigma}^3 N_{\tau}$.

\subsection{Фрактальные объекты в непрерывном пределе решеточной теории}

Рассмотрим определение фрактального объекта (см. также ~\cite{fractal1,
fractal}). Мы увидим, что из нашего анализа следует то, что объекты фрактальной
размерности $D>0$, выживают в непрерывном пределе.

Если одномерный объект выживает при переходе к непрерывному пределу, то он
должен иметь длину.  Мы можем ввести следующую характеристику этого объекта:
Средняя длина объекта, погруженного в 4 - объем единичного размера, которую мы
обозначаем  $\bar{l}$. Решеточная плотность этих объектов обозначена  $\rho$.
Тогда
\begin{equation}
 \rho = N/L^4,
\end{equation}
где $N$ - это общее число точек данного объекта внутри куба решеточного размера
$L$. 4 - куб единичного физического размера содержит  $L^4 \sim 1/a^4$ узлов
решетки. Длина линейного объекта состоящего из  $N$ точек это $N a$, так что
длина объекта, погруженного в четырехмерный куб решеточного размера  L это
$\rho L^4 a$. Таким образом, длина физического объекта скейлится как  $\bar{l}
\sim \rho a^{1-4}$. Для нас важно, что $\bar{l}$ это реальная физическая
характеристика непрерывного объекта и таким образом она не должна зависеть от
 $\beta$ при $\beta \rightarrow \infty$. Это означает, что решеточная плотность линейного
 объекта ведет себя как
\begin{equation}
 \rho \sim a^{4-1}  .
\end{equation}
Аналогичным образом для  2 - мерного объекта, выживающего в непрерывном пределе
\begin{equation}
 \rho \sim a^{4-2},
\end{equation}
где  $\rho$ - это снова решеточная плотность данных объектов.

Для любого целого $D$ мы имеем для $D$ - мерного объекта:
\begin{equation}
 \rho \sim a^{4-D}  .
\label{D}
\end{equation}
Таким образом, решеточный объект с плотностью, удовлетворяющей (\ref{D}) при
$\beta \rightarrow \infty$ может считаться выживающим в непрерывном пределе и
имеющим размерность $D$.

Если же наш объект удовлетворяет (\ref{D}) с нецелым $D>0$, мы можем говорить о
фрактальном объекте размерности $D$. Эта точка зрения становится очевидной
после демонстрации того, что указанное выше определение фрактальной размерности
находится в соответствии с определением Хаусдорфовой размерности множества
погруженного в четырехмерное пространство.

Хаусдорфова размерность объекта в 4 - х мерном пространстве определяется
следующим образом \cite{fractal1}: Рассмотрим 4-х мерный куб фиксированного
физического размера. Разделим его на $L^4$ подкубов. Число подкубов покрывающих
объект обозначим $N$. Если $N \sim L^D$, и $D>0$ при  $L \rightarrow \infty$,
мы говорим, что наш объект имеет Хаусдорфову размерность  $D$.

Как мы можем представить себе разделение куба некоторого физического размера на
разные количества подкубов, используя решеточную теорию? Ответ следующий.
Разделение на бесконечное количество подкубов представляется самим непрерывным
пределом (решеточная теория при $\beta = \infty$). Решеточная теория для
конечного $\beta$ не эквивалентна непрерывной теории. Но она приближается к ней
когда  $\beta$ становится большим. Вместо разделения куба на $L^4$ подкубов в
непрерывной теории мы можем использовать решеточную теорию, определенную на
решетке размера $L$. Мы уже видели, что размеры решеток, которые представляют
один и тот же физический объем, скейлятся как   $L \sim 1/a(\beta)$, где
$a(\beta)$ представляет собой длину ребра решетки. Таким образом, фрактальная
размерность некоторого объекта (с точностью до разницы между чистой непрерывной
теорией и ее решеточной версией при больших  $\beta$) может быть извлечена по
формуле
\begin{equation}
N \sim L^{D}, \label{LD}
\end{equation}
где $N$ - это число кубов, которые покрывают наш объект внутри решетки размера
$L$. Разница между двумя теориями изчезает при  $L \rightarrow \infty$ (что
предполагает что $\beta \rightarrow \infty$). Таким образом, ~(\ref{LD})
остается верным для $\beta \rightarrow \infty$, когда $L$ и $N$ рассматриваются
как функции от $\beta$ в то время, как $D$ остается независимым от $\beta$, и
мы можем рассматривать $D$ как фрактальную размерность нашего объекта,
существующего в непрерывной теории.

 Теперь покажем, что объект на решетке, чья плотность скейлится при $\beta \to \infty$ в соответствии с ~(\ref{D})
 с $D>0$ может рассматриваться как объект в непрерывной теории Хаусдорфовой размерности  $D$. Число $N$
 подкубов покрывающих элементы нашего объекта включено в определение решеточной плотности:  $\rho =
N/L^4$.  Таким образом, число элементарных кубов покрывающих наш объект внутри
4 - х мерного куба некоторого фиксированного физического размера может быть
представлено как
\begin{equation}
 N \sim \rho L^4,
\end{equation}
где $L \sim 1/a(\beta)$. Мы имеем из ~(\ref{D}):
\begin{equation} \rho = N/L^4 \sim a^{4-D}.
\end{equation}
Таким образом,
\begin{equation}
 N \sim L^{D}.
\end{equation}
Здесь $D$ не зависит от $\beta$. В соответствии с вышесказанным  мы
рассматриваем $D$ как Хаусдорфову размерность.

Итак, мы заключаем, что если решеточная плотность некоторого объекта
удовлетворяет (\ref{D}) для  $\beta \rightarrow \infty$, мы можем говорить, что
этот объект выживает в непрерывном пределе и имеет фрактальную размерность $D$.
Причем определение этой размерности соответствует классическому определению
Хаусдорфовой размерности.

\subsection{Центральные вихри и монополи в непрерывном пределе}
\begin{figure}
\begin{center}
 \epsfig{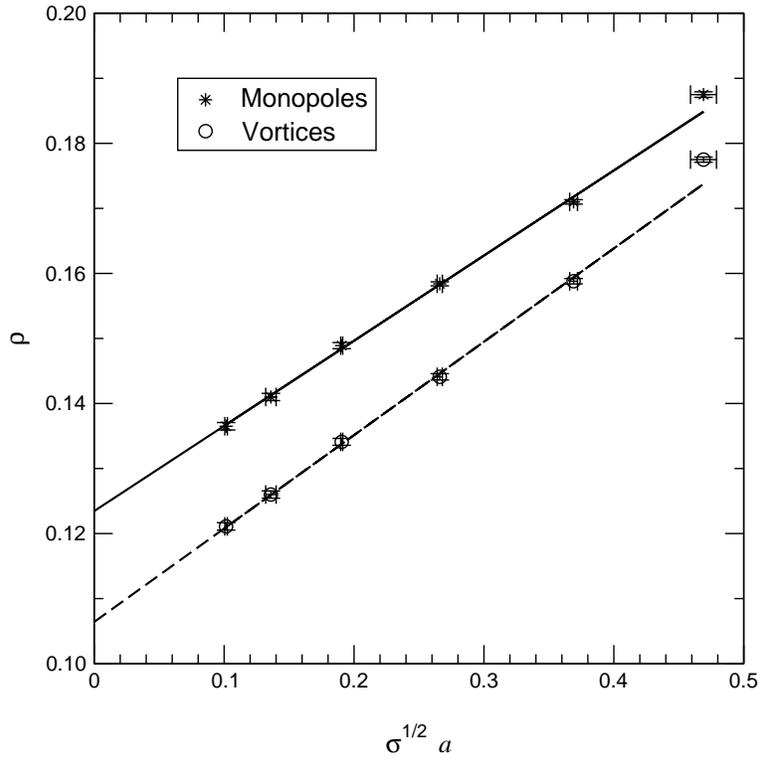}

\vspace{1ex}

 \caption{ Зависимость  $\rho$ от $a(\beta)$ для простых центральных вихрей и центральных монополей простой
 Центральной проекции. Решетка имеет размер $24^4$. Линейный фит представлен сплошной и прерывистой линиями.
 \label{fig.10}}
\end{center}
\end{figure}
Ниже мы представляем наши численные результаты. Симуляции проводились для
решеток размера $16^4$ и $24^4$. Мы не видим различия в результатах для этих
размеров решеток, что дает нам право полагать, что размер решетки не влияет на
рассматриваемые величины для решеток размера $16^4$ и больше.

Решеточная плотность вихрей и Центральных монополей в Простой Центральной
Проекции представлены на рис. \ref{fig.10}. Мы представляем $\rho$ как функцию
от $a(\beta)$. Значения $\sqrt{\sigma_{\rm cont}} a(\beta)$ для данных конечных
значений $\beta$ представлены в таблице \ref{tab.01}. Из ~\ref{fig.10} мы
находим, что зависимость линейна
\begin{equation}
\rho = \rho_c + \alpha\, a(\beta) .
\end{equation}
Здесь $\rho_c$ - это плотность при $a(\beta) = 0$, полученная экстраполяцией
данных из рис. ~\ref{fig.10}. Для центральных монополей мы находим: $\rho_c =
0.123\pm 0.001$ и для вихрей $\rho_c =0.106\pm 0.001 $. Ясно, что  $\rho
(\beta)$ не стремится к  $0$ при $\beta \to \infty$. Таким образом, мы имеем и
для вихрей и для монополей
\begin{equation}
 \rho \sim a^{4-D}
\end{equation}
с $D = 4$. Таким образом, мы находим для наших объектов специфическое поведение
плотности. Она не зануляется при $\beta \to \infty$. Это означает в
соответствии со сказанным выше, что эти объекты выживают в непрерывном пределе
с фрактальной размерностью $4$. Сам факт того, что фрактальная размерность
монопольных линий и мировых поверхностей вихрей оказалась максимально возможной
(равной 4) не должен удивлять, поскольку в соответствии с нашими результатами
эти объекты образуют перколирующие кластеры. Тем не менее, следует обратить
внимание на то, что интерполирующая формула для фрактальной размерности
(\ref{D}) дает значение существенно меньшее, чем Хаусдорфова размерность.
Качественно это соответствует тому, что (\ref{D}) определяет фрактальную
размерность вихрей как она видится на расстояниях порядка длины ребра решетки,
в то время, как Хаусдорфова размерность соответствует макроскопическому
масштабу всего решеточного объема. Подобная ситуация, однако, так же не
является новой и хорошо известна в теории динамических триангуляций, где разные
определения фрактальной размерности также дают разные результаты \cite{Triang}.
В частности, Хаусдорфова размерность двумерных динамических триангуляций
оказывается близка к 4 в то время, как так называемая спектральная размерность
близка к 2. Отличие между двумя определениями фрактальной размерности для
динамических триангуляций обусловлено возникновением так называемых "дочерних
вселенных" (baby universes). Спектральная размерность будучи параметром
локальным определяет размерность двумерной поверхности на масштабах существенно
меньше среднего размера "дочерних вселенных". Именно она и оказывается близка к
2. Однако, если смотреть на кластер, образуемый поверхностью "со стороны", то
есть с расстояний, превышающих средний размер "дочерних вселенных", то мы
увидим сильно перколирующий объект с фрактальной размерностью близкой к 4. Как
мы полагаем, совершенно такая же картина имеет место и в случае с нашими
вихревыми поверхностями.

Следует отметить, что среднее действие вблизи монопольных линий больше среднего
действия по всей решетке. Превышение составляет между  5\% и 8\% в интервале
$2.1 < \beta < 2.6$. Это означает, что данные объекты несут энергию. Монополи
формируют один большой кластер и несколько маленьких. Такая ситуация аналогична
ситуации в Максимальной Абелевой калибровке \cite{BCGPSVZ}. Следуя этой ссылке
мы называем большие монопольные петли инфракрасными, а маленькие -
ультрафиолетовыми. Последние - нефизические. Мы находим, что их количество
составляет от $1$ до $3$\% от всех монополей.

\section{Публикации}

Результаты настоящей главы опубликованы в работах:

"The Simple center projection of SU(2) gauge theory", B.L.G Bakker, A.I.
Veselov, M.A. Zubkov. Phys.Lett.B497:159-164,2001, [hep-lat/0007022]

 "The simple center projection of SU(2) gauge theory", B.L.G. Bakker, A.I.
Veselov, M.A. Zubkov. Nucl.Phys.Proc.Suppl.94:478-481,2001.

 "Evidence for the reality of singular configurations in SU(2) gauge theory",
B.L.G. Bakker, A.I. Veselov, M.A. Zubkov, Phys.Lett.B544:374-379,2002,
[hep-lat/0205027]

\chapter{Абелево представление для неабелевой петли Вильсона}
\label{ch3}

\section{Постановка проблемы}

В Абелевых проекциях неабелевых калибровочных теорий Неабелевы переменные
проектируются на абелевы, что значительно упрощает исследование \cite{tH}.

Одна из наиболее важных величин, с которыми приходится иметь дело
исследователям неабелевых калибровочных теорий - это петля Вильсона
\begin{equation}
W_q[{\cal C}] = \chi_q ({\rm P} {\rm exp} \,(i \int_{\cal C} A_{\mu} dx_{\mu}))
\end{equation}
Здесь $A$ - калибровочное поле  (принадлежащее алгебре Ли соответствующей
калибровочной группы), $\cal C$ - замкнутый контур, $\rm P$ означает
упорядочение вдоль контура, а $\chi_q$ - характер в представлении $q$
калибровочной группы. Для применений Абелевых проекций важно иметь
представление неабелевой петли Вильсона в Абелевом виде. Представление такого
вида было, действительно, предложено Дьяконовым и Петровым в  \cite{DP}. Для
случая неприводимого представления $q$ группы $SU(N)$ оно имеет вид:
\begin{equation}
W_q[{\cal C}] = \int D \mu_{{\cal C},q}(g) {\rm exp} \,(i \int_{\cal C} {\rm
Tr} \, A_{\mu}^g\, {\cal H}^q dx_{\mu}),\label{Wdp}
\end{equation}
где $g$ - это $SU(N)$ калибровочное преобразование, и $A_{\mu}^g = g A_{\mu}
g^+ - i g
\partial_{\mu} g^+$. Здесь ${\cal H}^q = \sum_{i = 1, ..., N-1} m_i H_i$, где
каждое $H_i$ обозначает базисный элемент Картановой подалгебры  $su(N)$, а
$(m_i)$ - это старший вес представления. Матрицы $H_i$ нормированы таким
образом, что ${\rm  Tr}\, H_i H_j = \delta_{ij}$. Калибровочное преобразование
определено на контуре $\cal C$. Мера $\mu_{{\cal C},q}(g)$ на пространстве
калибровочных преобразований строится используя инвариантную меру на группе
$SU(N)$. Ниже данная конструкция обсуждается детально.

И в монопольной и в $P$ - вихревой картинах удерживающие силы индуцируются
магнитным потоком заключенным внутри контура который соответствует мировой
линии заряженной частицы \cite{P}. В Абелевой теории теорема Стокса
непосредственно используется для вычисления магнитного потока. Для вычисления
магнитного потока, заключенного внутри контура соответствующего неабелевой
петле Вильсона Дьяконов и Петров предложили использовать представление
(\ref{Wdp}) в следующем виде:
\begin{equation}
W_q[{\cal C}] = \int D \mu_{{\cal C},q}(g) {\rm exp} \,(i \int_{{\cal M}(\cal
C)} {\rm  Tr} \, \partial_{[\mu}A_{\nu]}^g\, {\cal H}^q d x_{\mu}\wedge d
x_{\nu}),\label{Ws}
\end{equation}
где $g$ определено на поверхности ${\cal M}(\cal C)$, которая натягивается на
контур  $\cal C$. Интеграл от два - формы по этой поверхности обозначен
$\int_{{\cal M}(\cal C)} ... d x_{\mu} \wedge d x_{\nu}$. Выражение (\ref{Ws})
известно как неабелева теорема Стокса.

Вывод представлений (\ref{Wdp}) и (\ref{Ws}) был первоначально предложен в
терминах нерегуляризованной непрерывной теории. Некоторые проблемы с
регуляризацией позволили впоследствии авторам публикации \cite{Iv} оспорить
справедливость указанных представлений. Однако, последовательный вывод
решеточного варианта представления (\ref{Wdp}) данный автором настоящей
диссертации разрешил указанную проблему и подтвердил справедливость непрерывных
выражений (\ref{Wdp}) и (\ref{Ws}), верных с точностью до фактора
соответствующего перенормировке массы заряженной частицы. Именно появление
указанного бесконечного фактора было ошибочно воспринято авторами \cite{Iv} как
указание на ошибочность представления (\ref{Wdp}).

Ниже приводится вывод решеточного варианта формулы (\ref{Wdp}) и обсуждение его
непрерывного предела. Следует отметить, что в \cite{DP1,KT}) была дана
приближенная формулировка решеточного варианта представления (\ref{Wdp}).
Однако, для установления справедливости непрерывного выражения (\ref{Wdp})
необходимо именно точное решеточное представление, которое и выводится ниже
вместе с точным решеточным вариантом неабелевой теоремы Стокса.

\section{Неприводимые представления группы $SU(N)$}

Ниже кратко формулируются основные факты  теории представлений, используемые
далее.

1. Пространство $\cal V$ любого неприводимого представления группы $SU(N)$
состоит из тензоров $\Psi_{i_1 i_2 ... i_r}$. Симметрия $\Psi$ определяется
набором целых чисел $q_i \, (i = 1, ..., N-1)$ ($\sum_i q_i = r, \, q_i\geq 0
$), который называется сигнатурой представления. Элементы $\cal V$ - это
тензора
\begin{equation}
\Psi_{i^1_1 ...i^1_{q_1} ...i^{c_1}_1 ... i^{c_1}_{q_{c_1}}}
\end{equation}
Для любого $k \in \{1, ..., q_1\}$  число индексов $i^k_l$ обозначено как
$c_k$. Упорядочим эти индексы следующим образом:
\begin{equation}
\begin{array}{ccccc}
i^1_1 & ... & i^1_i & ... & i^1_{q_1} \\
i^2_1 & ... & i^2_{q_2} \\
... \\
i^{c_1}_1 & ...
\end{array} \label{Jung}
\end{equation}
Тогда $q_i$ - это длина $i$-й строки в то время как высота $l$-го столбца
$c_l$.

2. Рассмотрим произвольный тензор $\Psi_{i^1_1 ...i^1_{q_1} ...i^{c_1}_1 ...
i^{c_1}_{q_{c_1}}}$. Тензор требуемой симметрии получается посредством
следующего преобразования:
\begin{eqnarray}
\Psi_{i^1_1 ...i^1_{q_1} ...i^{c_1}_1 ... i^{c_1}_{q_{c_1}}} \rightarrow {\cal
S}^q \Psi_{i^1_1 ...i^1_{q_1} ...i^{c_1}_1 ... i^{c_1}_{q_{c_1}}} \nonumber\\
= \sum_{S} (-)^{P(S)} \Psi_{S[i^1_1 ...i^1_{q_1} ...i^{c_1}_1 ...
i^{c_1}_{q_{c_1}}]},\label{o}
\end{eqnarray}
где сумма берется по перестановкам $S$ индексов. Каждая перестановка состоит из
упорядоченного произведения ${\cal Q} \times {\cal P}$ некоторой перестановки
$\cal P$ внутри строк $(i^k_1 ...i^k_{q_k})$ таблицы (\ref{Jung}) и некоторой
перестановки $\cal Q$ внутри столюцов  $(i^1_l ...i^{c_l}_{l})$. $P(S) = 0(1)$
если $\cal Q$ четная (нечетная).

3. Пусть $U_{ij}$ - это $N\times N$ матрица, которая представляет некоторый
элемент $SU(N)$. Тогда действие этого элемента на $\cal V$ дается:
\begin{eqnarray}
\Psi_{i_1 i_2 ... i_r}  \rightarrow {\cal D}_q (U)^{j_1 j_2 ... j_r}_{i_1 i_2
... i_r} \Psi_{j_1 j_2 ... i_r} \nonumber\\
= U_{i_1 j_1} U_{i_2 j_2}... U_{i_r j_r} \Psi_{j_1 j_2 ... j_r}\label{q}
\end{eqnarray}
Мы также записываем:
\begin{equation}
|\Psi\rangle \rightarrow {\cal D}_q (U) |\Psi\rangle
\end{equation}

4. Представим инфинитезимальный элемент  $SU(N)$ как $ U = {\rm exp} (i A s)$.
Здесь $s$ - это малый параметр, а $A$ - матрица, которая представляет
определенный элемент алгебры Ли  $su(N)$. Тогда (\ref{q}) определяет
представление $su(N)$. Мы обозначаем соответствующую матрицу ${\cal D}_q(A)$:
\begin{equation}
{\cal D}_q (U) = {\rm exp} (i {\cal D}_q(A) s)
\end{equation}

5. Используя Картанов базис, $A \in su(N)$ может быть представлена как
\begin{eqnarray}
A = \sum_{i = 1,...,N-1} H^i {\rm Tr}\, A H^i + \nonumber\\ \sum_{i<j; i,j =
1,...,N-1}(E^{ij} ({\rm Tr} A E^{ij})^* + (E^{ij})^+ {\rm Tr} A
E^{ij})\label{Car}
\end{eqnarray}
Картановы элементы это
\begin{equation}
 H^i = \frac{1}{\sqrt{i(i+1)}} {\rm diag}\, (1, ..., 1, -i, 0,
 ...)
\end{equation}
Сдвиговый оператор определяется как:
\begin{equation}
 (E^{ij})_{ab} = \delta_{ai}\delta_{bj}
\end{equation}

6. Нормированный вектор $|\Lambda^a\rangle$ ($\langle
\Lambda^a|\Lambda^a\rangle = 1$), являющийся собственным для любого оператора
${\cal D}_q (H^i)$ называется весовым вектором. Здесь  $\Lambda^a \, (a = 1,
..., N-1)$ - соответствующие собственные значения:
\begin{equation}
{\cal D}_q (H^i) |\Lambda^a\rangle = \Lambda^i | \Lambda^a\rangle
\end{equation}

Сдвиговый оператор действует на весовой вектор следующим образом:
\begin{eqnarray}
{\cal D}_q (E^{ij})|\Lambda^a\rangle &\sim& |\Lambda^a + (\alpha^{ij})^a\rangle
\nonumber\\ {\cal D}_q (E^{ij})^+|\Lambda^a\rangle &\sim& |\Lambda^a
-(\alpha^{ij})^a\rangle
\end{eqnarray}
Здесь $|a\rangle \sim |b\rangle$ означает, что $|a\rangle = C |b\rangle$ для
некоторой постоянной $C$. Числа $\alpha^{ij}$ называются корнями $su(N)$. Они
входят в следующие коммутационные соотношения:
\begin{eqnarray}
&& [H^a, E^{ij}] =  (\alpha^{ij})^a E^{ij} \nonumber\\
&& [H^a,(E^{ij})^+]  =  -(\alpha^{ij})^a (E^{ij})^+
\end{eqnarray}

7. Весовой вектор, аннигилируемый каждым из операторов  $E^{ij}$ называется
старшим вектором. Соответствующий набор чисел $\Lambda^a$ называется старшим
весом и обозначается $m^a$.

Мы обозначаем старший вектор $|0\rangle$. С точностью до нормирующего фактора
он соответствует тензору
\begin{equation}
\Psi^0_{i^1_1 ...i^1_{q_1} ...i^{c_1}_1 ... i^{c_1}_{q_{c_1}}}
 \sim  {\cal S}^q \delta^1_{i^1_1} ... \delta^1_{i^1_{q_1}}
  ...\delta^{c_1}_{i^{c_1}_1} ... \delta^{c_1}_{i^{c_1}_{q_{c_1}}}
\end{equation}

8. Старший вес представления $q$ равен:
\begin{equation}
m_{i} = \frac{1}{\sqrt{i(i+1)}}(q_1 + q_2 + ... + q_{i-1} - i q_{i})
\end{equation}

9. Система когерентных состояний соответствующая представлению $q$ это
следующее подмножество $\cal V$:
\begin{equation}
\{| g \rangle \, :\, | g \rangle = e^{i \phi(g)} {\cal D}_q(g) | 0 \rangle \,
(g \in SU(N)) \},
\end{equation}
где $\phi(g)$ - это действительно - значная функция на $SU(N)$.

10. Система когерентных состояний полна:
\begin{equation}
\int d g |g \rangle \langle g | = \frac{1}{D(q)}
\end{equation}
Здесь $D(q)$ - это размерность представления. Мера на $SU(N)$ нормирована таким
образом, что  $\int d g = 1$.

\section{Абелево представление неабелевой петли Вильсона на решетке}

Рассмотрим решеточную  $SU(N)$ калибровочную модель со скалярным полем  $\Phi$,
принадлежащим к представлению $q$ группы $SU(N)$. В дополнение скалярное поле
рассматривается в присутствии внешнего $U(1)$ калибровочного поля $\theta\in
]-\pi, \pi]$ (определенного на линках решетки).
 Статистическая сумма имеет вид:
\begin{equation}
Z[\theta] = \langle \int D\Phi exp(\sum_{xy} \Phi^+_x {\cal D}_q(U_{xy})
e^{i\theta} \Phi_y - V(\Phi))\rangle   \label{F}
\end{equation}
Здесь $U \in SU(N)$, а среднее вычисляется в некоторой $SU(N)$ калибровочной
теории. Мы полагаем потенциал $V$ бесконечно глубоким и имеющим минимум при
$\Phi = |0\rangle$ c $V(|0\rangle  ) = 0$, где $|0\rangle$ - старший вектор
представления. Мы также требуем, чтобы
 $V$ был инвариантным относительно преобразования
  $\Phi \rightarrow e^{i\alpha} {\cal D}_q(g) \Phi$,
где  $\alpha$ - произвольное действительнозначное поле.

Идея заключается в том, чтобы представить статсумму двумя способами как сумму
по мировым линиям скалярной частицы. Первый способ будет производящим
функционалом для неабелевой петли Вильсона  (в представлении $q$). Второй
способ дает производящий функционал для Абелева представления петли Вильсона.
Наконец, благодаря произвольности меры интегрирования по $SU(N)$ полям мы
получим эквивалентность обоих представлений для петли Вильсона.

\subsection{Производящий функционал $Z[\theta]$ для петель Вильсона}

Мы можем редуцировать интегрирование по  $\Phi$ к интегрированию по $g \in
SU(N)$ и по действительному полю $\alpha$. Оба поля определены в узлах решетки.
Тогда мы можем переписать (\ref{F})
 как сумму по мировым линиям скаляра следующим образом:
\begin{eqnarray}
&& Z[\theta]  = \langle \int D\alpha Dg exp( \sum_{xy} \langle 0|{\cal
D}_q(g)^+_x {\cal D}_q(U_{xy})\nonumber\\&&
e^{i\theta_{xy} + i\alpha_y - i\alpha_x}{\cal D}_q(g)_y|0\rangle   )\rangle  \nonumber\\
& = &  \langle \int Dg \sum_{\cal C} (2\pi)^{n(j({\cal C}))} e^{i(j({\cal
C}),\theta)} \nonumber\\&&{\rm  Tr}\Pi_{\{xy\}\in {\cal C}} \frac{1}{|j({\cal
C})_{xy}|!} \langle g_x|{\cal D}_q(U_{xy})|g_y\rangle ^{|j({\cal C})_{xy}|}
\rangle \label{Z1}
\end{eqnarray}

Сумма здесь по замкнутым путям $\cal C$. Замкнутый решеточный путь определяется
как отображение ${\cal C}: Z_N = \{0, 1, ..., N-1\} \rightarrow X$ (где $X$ -
это множество узлов решетки, а $N$ - целое число) такое, что для каждого $k \in
Z_N$ точки ${\cal C}(k)$ и ${\cal C}({[k+1]{\rm mod}\,N})$ связаны линком.
Произведение по всем таким линкам обозначается как $\Pi_{\{xy\}\in {\cal C}}$,
где каждый линк считается один раз и предполагается, что  $x = {\cal C}(k)$ и $
y = {\cal C}([k+1]{\rm mod}\, N)$ для некоторого $k$. $j({\cal C})$ -
целочисленное линковое поле которое соответствует пути ${\cal C}$. $|j_{xy}|$
отсчитывает сколько раз данный линк $\{xy\}$ встречается когда мы движемся
вдоль пути. Знак $j_{xy}$ отражает ориентацию пути $\cal C$. Таким образом,
млучай $|j|
> 1$ соответствует наложению разных частей $\cal C$ друг на друга. Следует отметить, что
 $j$ не содержаит полной информации об ориентации $\cal C$.
Поэтому разные пути могут соответствовать той же $1$ - форме $j({\cal C})$. Это
может произойти если ${\cal C}$ имеет самопересечения. Однако, в (\ref{Z1})
стоит сумма по  $\cal C$ которые имеют {\it разные} $j({\cal C})$. Это
означает, что  каждый член в сумме зависит только от  $j({\cal C})$ (а не от
самого $\cal C$). $n(j)$ - это число точек, которые входят в путь
представляемый $j$. Также мы используем следующее обозначение:
\begin{equation}
(j,\theta) = \sum_{xy} j_{xy} \theta_{xy},
\end{equation}
где сумма - по линкам решетки.

Мы обозначаем $\langle i|{\cal D}_q(g)|0\rangle   = {\cal D}_q(g)_{i0};\langle
0|{\cal D}_q(g)^+|i\rangle   = {\cal D}_q(g)^+_{0i}$, где $|i\rangle  $ -
базисный вектор пространства $\cal V$ представления $q$. Интегрирование по $Dg$
может быть произведено с использованием следующих выражений:
\begin{eqnarray}
&& \int dg {\cal D}_q(g)^+_{i0}{\cal D}_q(g)_{0j} = \frac{\bf 1}{M_1(q)}\delta_{ij}\nonumber\\
&& \int dg {\cal D}_q(g)^+_{i_10}{\cal D}_q(g)_{0j_1}{\cal D}_q(g)^+_{i_20}{\cal D}_q(g)_{0j_2} = \nonumber\\
&&\frac{\bf 1}{M_2(q)}(\delta_{i_1j_1}\delta_{i_2j_2}+\delta_{i_1j_2}\delta_{i_2j_1})\nonumber\\
&& ... \label{Int}
\end{eqnarray}
Здесь мера $dg$ нормирована так, что $\int dg = 1$. Первое выражение отражает
полноту системы когерентных состояний и предполагает, что  $M_1(q) = D(q)$, где
$D(q)$ - размерность представления. Другие уравнения следуют из калибровочной
инвариантности. Соответствующие коэффициенты $M_i(q)$ зависят от представления
и могут быть вычислены в явном виде используя выражения для базисных векторов
 $\cal V$ и технику разложения сильной связи (см., например, \cite{Creutz}).

Мы получаем:
\begin{equation}
 Z[\theta]  =  \langle \sum_{{\cal C}}C(j({\cal C}))
e^{i(j({\cal C}),\theta)} W_q[{\cal C}] \rangle \label{W1}
\end{equation}
где сумма - по  {\it всем разным} $\cal C$. $W_q[{\cal C}]$ - это петля
Вильсона (в предствалении $q$) соответствующая пути ${\cal C}: Z_N \rightarrow
X$:
\begin{eqnarray}
W_q[{\cal C}]& = &{\rm Tr} {\cal D}_q(U_{{\cal C}(0) {\cal C}(1)}){\cal
D}_q(U_{{\cal C}(1) {\cal C}(2)})...\nonumber\\&&...{\cal D}_q(U_{{\cal C}(N-1)
{\cal C}(0)})
\end{eqnarray}
Мы также обозначаем
\begin{equation}
 C(j) =
\frac{(2\pi)^{n(j({\cal C}))}}{(M_1(q))^{n_1(j)}(M_2(q))^{n_2(j)}...}
\label{C1}
\end{equation}
Здесь $n_1$ - число точек пути ${\cal C}$, в которых нет самопересечений, $n_2$
- число путей, в которых есть одно самопересечение  (т.е. есть два разных $k_a
\in Z_N, (a = 1,2)$, таких, что ${\cal C}(k_1) = {\cal C}(k_2)$), $n_3$ - число
точек, в которых ${\cal C}$ пересекает сам себя трижды и т.д.  ($n = \sum_i
n_i$).

Как мы уже отмечали, $j$ не содержит полную информацию об ориентации  $\cal C$.
Поэтому мы извлекаем из (\ref{W1}) сумму по петлям Вильсона, которые имеют одно
и то же $j({\cal C})$. Выражение для вакуумного среднего  $W_q^{l} =
\sum_{j[{\cal C}] = l}
 W_q[{\cal C}]  $ может быть получено, используя преобразование Фурье:
\begin{eqnarray}
\langle W_q^l\rangle   = \sum_{j[{\cal C}] = l} \langle W_q[{\cal C}]\rangle
\nonumber\\ = \frac{1}{C[{l}]}\int D \theta exp(- i(l, \theta)) Z[\theta]
\end{eqnarray}

\subsection{$Z[\theta]$ как производящий функционал для Абелевого представления
петли Вильсона}

Теперь зафиксируем калибровку, в которой $\Phi_x = e^{i\alpha_x}|0\rangle $
($\alpha$ - действительное поле). Мы имеем:

\begin{eqnarray}
&&Z[\theta] = \langle \int D\alpha \,{\rm exp}( \sum_{xy} \langle 0|{\cal
D}_q(U_{xy})|0\rangle   e^{i(\theta_{xy} + \alpha_x - \alpha_y)}) \nonumber\\
&& = \langle \sum_{\cal C} \bar{C}(j({\cal C}))
 \,{\rm  exp}(i (j({\cal C}),\theta)) \nonumber\\&&
 \Pi_{\{xy\}\in{\cal C}}\frac{1}{|j({\cal C})_{xy}|!}\langle 0|{\cal D}_q(U_{xy})
 |0\rangle ^{|j({\cal C})_{xy}|}  \rangle \label{W2}
\end{eqnarray}

Здесь сумма по $\cal C$, которые могут иметь разные  $j({\cal C})$, и
\begin{equation}
\bar{C}(j) = (2\pi)^{n(j)}
\end{equation}

\subsection{Эквивалентность петли Вильсона и ее Абелева представления}
 Вспомним, что (\ref{W1}) и (\ref{W2}) - это разные представления одной и той же величины.
 Тогда мы получим:
\begin{eqnarray}
&&\langle W_q^l\rangle   = \frac{1}{C(l)}\int D \theta \,{\rm exp}\,(-i (l,
\theta)) Z[\theta]\nonumber\\
&& =  \sum_{j[{\cal C}] = l}\frac{\bar{C}(l)}{C(l)}\langle \Pi_{\{xy\}\in {\cal
C}}\frac{1}{|j({\cal C})_{xy}|!}\langle 0|{\cal D}_q(U_{xy})|0\rangle
^{|j({\cal C})_{xy}|}  \rangle \nonumber\\
&&= \sum_{j[{\cal C}] = l}\tilde{C}(l)\langle \int Dg \Pi_{\{xy\}\in {\cal C}}
\langle 0|{\cal D}_q(U^g_{xy})|0\rangle ^{|j({\cal C})_{xy}|} \rangle ,
\label{Rp}
\end{eqnarray}
где мы использовали калибровочную инвариантность операции $\langle ... \rangle
$ и ввели новый комбинаторный фактор
\begin{equation}
\tilde{C}(l) = (M_1(q)^{n_1(l)}M_2(q)^{n_2(l)}...)\Pi_{\rm
links}\frac{1}{|l_{\rm link}|!},
\end{equation}
где сумма по всем линкам решетки. Калибровочное преобразование линковой
переменной $U_{xy}$ обозначена $U_{xy}^g = g U_{xy} g^+$. Мера $D g$ определена
так, что $\int D g = 1$.

Благодаря произвольности усреднения $\langle ... \rangle$ по калибровочному
полю $U$ мы имеем:
\begin{equation}
W_q^l = \sum_{j[{\cal C}] = l}\tilde{C}[l] \int D g \Pi_{\{xy\}\in {\cal
C}}\langle 0|{\cal D}_q(U^g_{xy})|0\rangle ^{|l_{xy}|} \label{fin1}
\end{equation}

Таким образом, сумма $SU(N)$ петель Вильсона, соответствующих одной и той же
$1$ - форме $l$ может быть представлена через Абелево представление
$\Pi_{\{xy\}\in {\cal C}}\langle 0|{\cal D}_q(U^g_{xy})|0\rangle ^{|l_{xy}|}$.
Для петель Вильсона без самопересечений имеем:
\begin{equation}
W^{l[{\cal C}]} = W[\cal C]
\end{equation}
Тогда $\tilde{C}[l] = D(q)^{n(l)} = D(q)^{|{\cal C}|}$, где  $|{\cal C}|$ -
длина контура. В этом случае (\ref{fin1}) имеет вид:
\begin{equation}
W_q[{\cal C}] = D(q)^{|{\cal C}|} \int D g \Pi_{{xy}\in {\cal C}}\langle
0|{\cal D}_q(U^g_{xy})|0\rangle ,\label{fin2}
\end{equation}

Выражение (\ref{fin2}) и есть наш основной результат. Далее мы получим явное
выражение для матричных элементов, входящих в  (\ref{fin2}) и обсудим свойства
симметрии (\ref{fin2}).

\section{Свойства полученного представления}

Ниже выводится выражение для $\langle 0|{\cal D}_q(U^g_{xy})|0\rangle$
входящего в (\ref{fin2}). Также мы вводим новые обозначения для того, чтобы
продемонстрировать Абелеву природу (\ref{fin2}).

Используем явное выражение для старшего вектора и для матрицы ${\cal D}_q(U)$
приведенные выше. Мы можем представить  $\langle 0|{\cal D}_q(U)|0\rangle$ как
\begin{equation}
\langle 0|{\cal D}_q(U)|0\rangle = {\cal U}(U),
\end{equation}
где
\begin{eqnarray}
{\cal U}(U) = \frac{1}{K}\sum_{S,\tilde{S}} (-)^{P(S)+P(\tilde{S})}
 U_{\tilde{S}^1_1 S^1_1} ... U_{\tilde{S}^1_{q_1} S^1_{q_1}}\nonumber\\
 ... U_{\tilde{S}^{c_1}_1 S^{c_1}_1}...U_{\tilde{S}^{c_1}_{q_{c_1}}
 S^{c_1}_{q_{c_1}}}\label{Up}
\end{eqnarray}
а нормировочный фактор $K$ выбран так, что ${\cal U}(1) = 1$. Здесь сумма
берется по $S^j_{i}$ и $\bar{S}^j_{i}$ входящим в следующие таблицы:
\begin{equation}
\begin{array}{cc}
\begin{array}{ccccc}
S^1_1 & ... & S^1_i & ... & S^1_{q_1} \\
S^2_1 & ... & S^2_{q_2} \\
... \\
S^{c_1}_1 & ...
\end{array}\,\,\, & \,\,\,
\begin{array}{ccccc}
\bar{S}^1_1 & ... & \bar{S}^1_i & ... & \bar{S}^1_{q_1} \\
\bar{S}^2_1 & ... & \bar{S}^2_{q_2} \\
... \\
\bar{S}^{c_1}_1 & ...
\end{array}
\end{array}
\end{equation}
(Строки содержат $q_1 , q_2, ..., q_{c_1}$ элементов в то время, как столбцы
содержат $c_1 , c_2, ..., c_{q_1}$ элементов.) Каждая из двух таблиц получена
посредством перестановки внутри строк и перестановки внутри столбцов из
следующей таблицы:
\begin{equation}
\begin{array}{ccccc}
1 & ... & 1 & ... & 1 \\
2 & ... & 2 \\
... \\
c_1 & ...
\end{array}
\end{equation}
$P(S) = 0(1)$ если соответствующая перестановка внутри  столбцов четная
(нечетная).

Мы можем переписать (\ref{Up}) в компактном виде
\begin{equation}
{\cal U}(U) = \omega_{c_1}(U) \omega_{c_2}(U) ... \omega_{c_{q_1}}(U),\label{U}
\end{equation}
где
\begin{equation}
\omega_M (U) = {\rm det} \left(
\begin{array}{ccc}
 U_{1 1} & ...& U_{1 M}\\
  ... & ... & ... \\
 U_{M 1} & ...& U_{M M}
\end{array}\right)\label{U}
\end{equation}

Для того, чтобы продемонстрировать $U(1)$ природу выведенного выражения для
петли Вильсона (\ref{fin2}) мы переписываем его следующим образом:
\begin{equation}
W_q[{\cal C}] =  \int D g {\rm exp} \,(i({\cal A}^g,j({\cal C})) + {\rm log}\,
(D(q)) |{\cal C}|), \label{fin3}
\end{equation}
где
\begin{equation}
{\cal A}^g_{xy} = - i {\rm log}\, {\cal U}(U^g_{xy})
\end{equation}
это комплексно - значное линковое поле. Ниже мы покажем, что его действительная
часть может рассматриваться как Абелево калибровочное поле. (Следует отметить,
что мнимая часть зануляется в непрерывном пределе. Это будет показано ниже
когда мы будем рассматривать непрерывный предел полученных выражений.)

Экспонента в (\ref{fin3}) инвариантна относительно $U(N-c_1)$ - преобразования
\begin{equation}
U_{xy} \rightarrow h^+(\Omega_x)U_{xy}h(\Omega_y)\,
\end{equation}
где $\Omega_x \in U(N - c_1)$ определено в точках пути $\cal C$,
\begin{equation}
h(\Omega) = \left( \begin{array}{ccccc}
e^{i\alpha} &  0           & ... & ...            & ... \\
0           &  e^{i\alpha} & ... & ...            & ... \\
...         &  ...         & ... & ...            & ... \\
...         &  ...         & 0   & e^{i\alpha} & 0   \\
...         &  ...         & ... & 0              & \Omega
\end{array} \right),
\end{equation}
и
\begin{equation}
{\rm  det}\Omega = e^{-i c_1 \alpha}
\end{equation}

Это преобразование действует на ${\cal A}^g$ следующим образом:
\begin{equation}
{\cal A}_{xy}^g \rightarrow {\cal A}_{xy}^{hg} = {\cal A}_{xy}^g + r (\alpha_x
- \alpha_y),
\end{equation}
где $r = \sum_i q^i$ - это ранг представления. Это демонстрирует, что
отмеченная $U(N-c_1)$ симметрия индуцирует $U(1)$ калибровочную симметрию, а
действительная  часть ${\cal A}^g$ играет роль соответствующего калибровочного
поля.

По аналогии с непрерывной теорией мы получаем решеточный вариант неабелевой
теоремы Стокса:
\begin{equation}
W_q[{\cal C}] =  \int D g {\rm exp} \,(i(d {\cal A}^g,m({\cal C})) + {\rm
log}\, (D(q)) |{\cal C}|), \label{fin4}
\end{equation}
где $m({\cal C})$ - это целочисленная плакетная переменная которая представляет
поверхность, натянутую на контур $\cal C$. $d {\cal A}^g$ - это плакетная
переменная, которая определена по аналогии с Абелевой напряженностью поля:
\begin{equation}
[d {\cal A}]_{xyzw} = {\cal A}_{xy} + {\cal A}_{yz} - {\cal A}_{wz} - {\cal
A}_{xw},
\end{equation}
где мы предполагаем, что направления $(xy)$ и $(yz)$ положительны. Сумма по
плакетам определена как
\begin{equation}
(m,d{\cal A}) = \sum_{xyzw} m_{xyzw} [d{\cal A}]_{xyzw}
\end{equation}

Отметим два частных случая, когда выражение (\ref{U}) имеет наиболее простую
форму.

Для симметричного ($q_1 = r, q_2 = ... = q_{N-1} = 0$) представления мы имеем:
${\cal U} = U_{11}^r$. Тогда выражение для петли Вильсона принимает вид:
\begin{equation}
W_r[{\cal C}] =  \int D g {\rm exp} \,(r({\rm log}U^g_{11} ,j({\cal C})) + {\rm
log}\, (D(q)) |{\cal C}|)
\end{equation}
(Присоединенное представление - это пример такого представления,
соответствующий $r = 2$. Фундаментальное представление соответствует $r = 1$.)

Другой простой случай соответствует абсолютно антисимметричному представлению
($q_1 = q_2 = ... = q_{r} = 1, q_{r+1} = ... = q_{N-1} = 0$). В этом случае мы
имеем:
\begin{equation}
{\cal U} = {\rm det} \left( \begin{array}{ccc}
 U_{1 1} & ...& U_{1 r}\\
  ... & ... & ... \\
 U_{r 1} & ...& U_{r r}
\end{array}\right)
\end{equation}

\section{Непрерывный предел}

Для того, чтобы рассмотреть непрерывный предел решеточной теории, следует
рассматривать окрестность фазового перехода второго рода, где решеточные поля
переходят в гладкие поля непрерывной теории.

Рассмотрим выражение (\ref{fin3}) в решеточной $SU(N)$ модели имеющей фазовый
переход второго рода. Изучим следующий коррелятор:
\begin{equation}
<O(U)W_q[{\cal C}]>
\end{equation}
Здесь $O(U)$ - это калибровочно - инвариантный функционал. Мы предполагаем, что
 $\cal C$ не имеет самопересечений. Далее:

\begin{eqnarray}
\langle O(U)W_q[{\cal C}]\rangle = \langle O(U) {\rm exp} \,(i({\cal A},j({\cal
C}))\nonumber\\ + {\rm log}\, (D(q)) |{\cal C}|)\rangle ,\label{Av}
\end{eqnarray}
Здесь интегрирование по калибровочным преобразованиям включено в среднее
$\langle ...\rangle$. Предположим, что сингулярные калибровочные поля не
возникают в непрерывном пределе данной модели. Тогда мы можем использовать
соответствие:
\begin{equation}
U_{xy} \rightarrow (1 + i A_{\mu} \Delta x), \label{UA}
\end{equation}
где $A$ - это непрерывное калибровочное поле, а $\Delta x$ - бесконечно малое
расстояние. Согласно (\ref{UA}) $\langle 0|{\cal D}_q(U)|0\rangle$
соответствует
\begin{equation}
\langle 0|{\cal D}_q(U_{xy})|0\rangle   \rightarrow 1 + i \langle 0| {\cal
D}_q(A_{\mu})|0\rangle \Delta x
\end{equation}
Теперь ясно, что
\begin{equation}
{\cal A}_{xy}\rightarrow \langle 0| {\cal D}_q(A_{\mu})|0\rangle \Delta x
\end{equation}
а мнимая часть ${\cal A}$ зануляется в непрерывном пределе.

Вспомним, что в окрестности фазового перехода второго рода  ${\cal C} = {\cal
C}_{\rm phys} \frac{1}{a}$, где длина ребра решетки $a$ стремится к нулю, а
физическая длина контура равна  ${\cal C}_{\rm phys} = \int_{\cal C}
\sqrt{dx_{\mu}dx_{\mu}}$. Рассмотрим такое $O$, которое стремится к гладкому
функционалу от $A$ в непрерывном пределе. Тогда мы можем переписать непрерывный
предел (\ref{Av}) следующим образом:
\begin{eqnarray}
\langle O(U)W_q[{\cal C}]\rangle = \langle O(A) {\rm exp} \,(i \int \langle 0|
{\cal D}_q(A_{\mu}) | 0 \rangle \, d x_{\mu} \, \nonumber\\+ \, m_q \int_{\cal
C} \sqrt{dx_{\mu}dx_{\mu}}) \rangle
\end{eqnarray}
где постоянная  $m_q$ расходится как
\begin{equation}
m_q  \sim {\bf M} {\rm log} D(q),\label{Mq}
\end{equation}
Здесь ${\bf M}$ - это ультрафиолетовое обрезание (которое в случае решеточной
регуляризации равно обратной длине ребра решетки).

В калибровочной теории любая петля Вильсона соответствует заряженной частице.
Расходимость $m_q$ в (\ref{Mu}) должна быть включена в перенормировочную схему
для массы этой частицы.

Восстановим интегрирование по калибровочным преобразованиям. Полученная
непрерывная теория калибровочно инвариантна. Тогда мы можем переписать
последнее выражение как:
\begin{eqnarray}
\langle O(U)W_q[{\cal C}]\rangle &=&  \langle O(A) \int D g {\rm exp} \,( i
\int \langle 0| {\cal D}_q(A_{\mu}) | 0 \rangle \, d x_{\mu} \, \nonumber\\&& +
\, m_q \int_{\cal C} \sqrt{dx_{\mu}dx_{\mu}}) \rangle,
\end{eqnarray}
где $D g $  - это мера на пространстве всех $\it гладких$ калибровочных
преобразований, определенная так, что $\int D g = 1$.

Благодаря произвольности среднего $ \langle ... \rangle$ мы имеем:
\begin{equation}
W_q[{\cal C}] = \int D \mu_{\cal C}(g) {\rm exp} \,(i \int \langle 0| {\cal
D}_q(A_{\mu}) | 0 \rangle \, d x_{\mu} ),\label{final}
\end{equation}
где
\begin{equation}
D \mu_{{\cal C},q}(g) = {\cal F}({\cal C}) Dg = {\rm exp} ( m_q \int_{\cal C}
\sqrt{dx_{\mu}dx_{\mu}}) Dg \label{Mu}
\end{equation}
Эта мера отличается от привычной, нормируемой так, что  $\int Dg = 1$.
Формально ${\cal F} \rightarrow \infty$ в непрерывном пределе. Поэтому если мы
подставим $Dg$ в (\ref{Wdp}) вместо $D \mu_{{\cal C},q}(g)$, то интеграл
занулится в непрерывном пределе. Именно это и было обнаружено авторами
\cite{Iv} для некоторых частных случаев $A$. Наш анализ показывает, что
включение в рассмотрение расходящейся величины $m_q$ восстанавливает
справедливость абелева представления для неабелевой петли Вильсона,
предложенного Дьяконовым и Петровым.

Действительно, используя картаново представление (\ref{Car}) для $A^g$, мы
получаем
\begin{eqnarray}
\langle 0|{\cal D}_q(A^g)|0\rangle   =  \langle 0| \sum_{i = 1,...,N-1} H^i
{\rm Tr}\, A^g H^i |0\rangle \nonumber\\ =   {\rm Tr}\, A^g {\cal H}^q
\end{eqnarray}

Таким образом, представление (\ref{final}) имеет вид (\ref{Wdp}) изначально
предложенный Дьяконовым и Петровым \cite{DP}, но с мерой интегрирования по $g$
определенной в (\ref{Mu}).

Используя выражение для Картановых элементов $H^i$ и для старшего веса  $m^i$
легко вычислить ${\cal H}^q $ (что не было сделано в \cite{DP}). Его ненулевые
элементы это:
\begin{equation}
 {\cal H}^{q}_{ii} = q_i - \frac{1}{N} \sum_{k = 1, ..., N} q_k \, (i = 1, ..., N),\label{H}
\end{equation}
где мы полагаем $q_N = 0$.

Тогда мы можем переписать (\ref{final}) в наиболее простом виде:
\begin{equation}
W_q[{\cal C}] = \int D \mu_{\cal C}(g) {\rm exp} \,(i \int \sum_{i = 1, ...,
N-1} ([A^g_{\mu}]_{ii} q_i) d x_{\mu} )\label{final1}
\end{equation}

Следует заметить, что это выражение может быть получено и непосредственно из
(\ref{fin3}) (с $\cal U$ данным в (\ref{U})).

\section{Выводы}

Таким образом, нами выведено Абелево представление (\ref{fin3}) для $SU(N)$
петли Вильсона на решетке. Это представление содержит комплексно - значное
линковое поле, являющееся функцией матричных элементов $SU(N)$ калибровочного
поля. (\ref{fin3}) обладает $U(1)$ калибровочной симметрией. Действительная
часть отмеченного линкового поля может рассматриваться как соответствующее
калибровочное поле. Мнимая часть зануляется в непрерывном пределе.

Неабелева теорема Стокса на решетке следует естественным образом из
(\ref{fin3}). Она может быть использована для вычисления магнитного потока
заключенного внутри контура соответствующего петле Вильсона. Непрерывный предел
выведенного выражения совпадает с исходным выражением, полученным Дьяконовым и
Петровым, но с мерой интегрирования, отличающейся от общепринятой, и
определенной в (\ref{Mu}). Именно непонимание того, что мера интегрирования
имеет такой вид и привела авторов \cite{Iv}  к неправильному выводу об
ошибочности представления (\ref{Wdp}).

\section{Публикации}

Результаты настоящей главы опубликованы в работе:

 "Abelian representation of nonAbelian Wilson loop and nonAbelian Stokes
theorem on the lattice", M.A. Zubkov, Phys.Rev.D68:054503,2003,
[hep-lat/0212001]

\part{Монополи Намбу и физика Электрослабых взаимодействий}

Как указывалось во введении, применение непертурбативных методов в
Электрослабой теории становится необходимым при конечной температуре при
приближении к температуре Электрослабого перехода. Также оказывается
необходимым применение непертурбативных методов и при нулевой температуре,
когда характерная энергия процесса приближается к 1 ТэВ. Это связано с тем, что
при таких энергиях необходимо учитывать появление монополей Намбу и Z - струн,
чья масса находится на шкале ТэВ. В четвертой главе нами рассматривается
решеточная реализация модели Вайнберга - Салама при нефизически больших
значениях постоянной тонкой структуры. Изучается фазовая структура модели и
свойства различных топологических дефектов, к  которым относятся и монополи
Намбу. В пятой главе изучаются свойства монополей Намбу в теории при конечной
температуре. Показывается, что они сконденсированы при температурах больших
температуры Электрослабого перехода. В шестой главе рассматривается модель
Вайнберга - Салама при нулевой температуре и при реалистических значениях
констант связи. Показано, что в окрестности фазового перехода в решеточной
модели, где предполагается производить переход к непрерывной физике,
расположена флуктуационная область, где флуктуации скалярного поля становятся
большими. Получены указания на то, что в этой области невозможно применять
обычную теорию возмущений, что подтверждается исследованием свойств монополей
Намбу, которые являются зародышами нефизической фазы внутри физической.
Показано, что в флуктуационной области расстояние между монополями становится
близко к их размерам.

\chapter{Решеточная регуляризация калибровочного и Хиггсовского секторов Стандартной Модели}
\label{ch4}


\section{Решеточная формулировка Стандартной Модели без динамических фермионов}

Ниже мы не интересуемся фермионным сектором модели, поскольку нами
рассматривается Стандартная Модель в пренебрежении вкладом динамических
фермионов. Для решения задачи об исследовании свойств монополей Намбу при
конечной температуре мы выбираем решеточную регуляризацию $SU(3)\times
SU(2)\times U(1)$ калибровочной модели со скалярным полем таким образом, что в
решеточной формулировке явно присутствует дополнительная $Z_6$ симметрия,
обсуждаемая в главе 10. В то же время для исследования фазовой диаграммы модели
Вайнберга - Салама при нулевой температуре при реалистических константах связи
нами использована общеупотребительная регуляризация, в которой отсутствует
дополнительная дискретная симметрия. Обе решеточные формулировки имеют один и
тот же наивный непрерывный предел. Наши исследования также показывают, что для
реалистических значений констант связи решеточные результаты, полученные с
помощью указанных регуляризаций, совпадают. В то же время, решеточная
регуляризация, уважающая $Z_6$ симметрию имеет ряд технических особенностей,
отличающих ее от общепринятой в нефизической области констант связи.

В настоящей главе мы описываем решеточную регуляризацию калибровочного и
Хиггсовского секторов стандартной модели, определяем измеряемые величины и
представляем результаты численного исследования решеточных моделей при нулевой
температуре в области нефизически больших значений электрослабой константы
связи. Подобное исследование необходимо должно предварять исследование модели
при конечной температуре (представленное в главе 5) и в области реалистических
значений констант связи (представленное в главе 6), поскольку будучи технически
более простым позволяет установить основные качественные свойства модели.

Решеточная модель содержит следующие переменные:

1. Решеточные калибровочные поля (определенные на линках решетки):
\begin{eqnarray}
 \Gamma \in SU(3), \quad U \in SU(2), \quad e^{i\theta} \in U(1).
\end{eqnarray}

2. Скалярный дублет $\Phi^{\alpha}, \;\alpha = 1,2$ (определенный на узлах
решетки).
Действие имеет вид:
\begin{equation}
 S = S_g + S_H,
\end{equation}
где мы обозначаем посредством $S_g$ калибровочную часть действия, а действие
для скалярного поля обозначено $S_H$.

Мы выбираем $S_H$ в виде
\begin{equation}
 S_H = \sum_{xy} |U_{xy}e^{-i\theta_{xy}}\Phi_y - \Phi_x|^2
     + \sum_x V(|\Phi_x|),
\end{equation}
где $V(r)$ - потенциал, имеющий минимум при ненулевом значении  $r =
\sqrt{\gamma/2}$.

Аналог непрерывного преобразования (\ref{symlat}) - решеточное преобразование:
\begin{eqnarray}
 U & \rightarrow & U e^{-i\pi N}, \nonumber\\
 \theta & \rightarrow & \theta +  \pi N, \nonumber\\
 \Gamma & \rightarrow & \Gamma e^{(2\pi i/3)N},
\label{sym}
\end{eqnarray}
где $N$ - произвольная целочисленная линковая переменная. Она представляет
трехмерную гиперповерхность на дуальной решетке.

Для численных исследований модели с $Z_6$ симметрией мы выбираем следующее
решеточное действие:
\begin{eqnarray}
 S_g & = & \beta \sum_{\rm plaquettes}
 (2(1-\mbox{${\small \frac{1}{2}}$} {\rm Tr}\, U_p \cos \theta_p)+
 \nonumber \\
 && +(1-\cos 2\theta_p) +\nonumber \\
 && +6(1-\mbox{${\small \frac{1}{6}}$} {\rm Re Tr}
 \,\Gamma_p {\rm Tr}\, U_p {\rm exp} (i\theta_p/3))+
 \nonumber\\
 && +3(1-\mbox{${\small \frac{1}{3}}$} {\rm Re Tr}
 \, \Gamma_p {\rm exp} (-2i\theta_p/3)) +\nonumber \\
 && +3(1-\mbox{${\small \frac{1}{3}}$} {\rm Re Tr}
 \, \Gamma_p {\rm exp} (4i\theta_p/3))),\label{Act_}
\end{eqnarray}
где сумма - по элементарным плакетам решетки. Каждый член в ~(\ref{Act_})
соответствует оператору параллельного переноса вдоль границы плакета в
представлениях соответствующих фермионам Стандартной Модели (см. Главу 10).
Очевидным образом выражение (\ref{Act_}) инвариантно относительно решеточной
реализации $Z_6$ - симметрии.

Наивно ~(\ref{Act_}) имеет тот же непрерывный предел, что и следующее действие
(при соответствующем выборе констант $\beta^0_i$):
\begin{eqnarray}
 S^0_g & = & \sum_{\rm plaquettes}
 \{\beta^0_1(1-\mbox{${\small \frac{1}{2}}$} {\rm Tr}\, U_p )
 \nonumber \\
 && + \beta^0_2 (1-\cos \theta_p) \nonumber \\
 && + \beta^0_3 (1-\mbox{${\small \frac{1}{3}}$} {\rm Re Tr}
 \,\Gamma_p )\},\label{Act0}
\end{eqnarray}
Однако, (\ref{Act_}) сохраняет симметрию (\ref{sym}) в то время, как
(\ref{Act0}) - нет.

Формально действие (\ref{Act_}) получается из решеточного действия объединенной
$SU(5)$ модели при пренебрежении X и Y - бозонами (вместе с A, Z и W
составляющими набор возбуждений $SU(5)$ калибровочного поля). Этим и обусловлен
выбор коэффициентов в (\ref{Act_}). А именно, (\ref{Act_}) возникает, если мы
выберем решеточное действие для объединенной $SU(5)$ модели в виде
\begin{equation} S = \beta
\sum_{\rm plaq}(1 - {\rm Re}\,\chi_{F} ({\cal V}_{\rm plaq})) + \beta \sum_{\rm
plaq}(1 - {\rm Re}\,\chi_{A} ({\cal V}_{\rm plaq})),\label{act0}
\end{equation} где ${\cal V}\in SU(5)$, a $\chi_{F}$ - характер фундаментального представления,   $\chi_{A}$
 - характер антисимметричного
представления (фермионы объединенной модели расположены в этих двух
представлениях).

Ясно, что значение угла Вайнберга, полученное из действия (\ref{Act_})
оказывается равным углу Вайнберга в $SU(5)$ модели объединения ${\rm cos}^2 \,
\theta_W = \frac{5}{8}$. Кроме того, из (\ref{Act_}) следует, что Электрослабая
и цветная константы связи близки друг к другу. Для того, чтобы прояснить, как
это получается, рассмотрим наивный непрерывный предел (\ref{Act_}). Мы полагаем
\begin{eqnarray}
 \Gamma_{x,\mu} = e^{iC_{\mu}(x)a}, \quad U_{x,\mu} = e^{iA_{\mu}(x)a}, \quad e^{i\theta_{x,\mu}} = e^{iB_{\mu}(x)a}
\end{eqnarray}
Здесь $a$ - длина ребра решетки. Следует отметить, что при таком определении
поле $B_{\mu}=\frac{\tilde{B_{\mu}}}{2}$, где $\tilde{B_{\mu}}$ - общепринятое
в литературе обозначение для $U(1)$ - поля. В непрерывном пределе (\ref{Act_})
должно перейти в
\begin{eqnarray}
 S_g & = & \int d^4x
 \{\frac{1}{2g_2^2}  {\rm Tr}\, [ 2 \times  \sum_{i>j}G^2_{ij}]
 \nonumber \\
 && + \frac{1}{4g_1^2}  [ 2 \times \sum_{i>j}\tilde{F}^2_{ij}]
  + \frac{1}{2g_3^2} {\rm Tr} [ 2 \times \sum_{i>j}R^2_{ij}]\},\label{Act0c}
\end{eqnarray}
Здесь $\tilde{F}_{ij} = \partial_{i}\tilde{B}_j - \partial_{j}\tilde{B}_i = 2
(\partial_{i}{B}_j - \partial_{j}{B}_i) = 2 F_{ij}$, ${G}_{ij} =
\partial_{i}{A}_j -
\partial_{j}{A}_i - i[A_i,A_j]$, и ${R}_{ij} = \partial_{i}{C}_j -
\partial_{j}{C}_i - i[C_i,C_j]$.
Мы также имеем соответствие между плакетными переменными и напряженностями
поля:
\begin{eqnarray}
 {\rm Tr} \Gamma_{x,\mu,\nu} &=& {\rm Tr}[1-\frac{1}{2}R^2_{\mu \nu}a^4],
\nonumber\\  \quad {\rm Tr} U_{x,\mu\nu} &=& {\rm Tr}[1-\frac{1}{2}G^2_{\mu
\nu}a^4],
 \nonumber\\  \quad {\rm cos} \, N {\theta_{x,\mu\nu}} &=& [1-\frac{N^2}{2}{F}^2_{\mu \nu}a^4]
\end{eqnarray}

Теперь для установления соответствия между константами непрерывной теории
$g_{1,2,3}$ и $\beta$ следует подставить выражения для напряженностей поля в
(\ref{Act_}) и сравнить с (\ref{Act0c}). Имеем:
\begin{eqnarray}
 \frac{1}{g^2_1} = \frac{5}{3} \times 2 \beta , \quad \frac{1}{g^2_2} = \frac{1}{g^2_3}= 2\beta
\end{eqnarray}

Таким образом,
\begin{eqnarray}
 {\rm tg} \theta_W &=& \frac{g_1}{g_2} = \sqrt{\frac{3}{5}},\nonumber\\
  \quad \alpha_s &=&
 \frac{g_3^2}{4\pi}= \frac{1}{8\pi \beta},\nonumber\\ \quad \alpha &=&
 \frac{e^2}{4\pi}= \frac{[\frac{1}{g^2_1}+\frac{1}{g^1_1}]^{-1}}{4\pi}= \frac{3}{64\pi \beta}
\end{eqnarray}

Исследуемая нами область фазовой диаграммы соответствует значениям $\beta \sim
0.7$. Таким образом, мы изучаем модель, в которой затравочные значения констант
связи: ${\rm sin}^2 \theta_W \sim 0.38$; $\alpha_s \sim \frac{1}{20}$; $\alpha
\sim \frac{1}{50}$. При этом затравочное значение массы Хиггса бесконечно. Из -
за вычислительных сложностей мы не исследовали значение перенормированной массы
Хиггса в нашей модели. Однако, для ее оценки можно использовать результаты
предыдущих исследований $SU(2)$ модели Хиггса. А именно, значение $\beta \sim
0.7$ в нашей модели соответствуют значениям $\beta \sim 5.6$ в $SU(2)$ модели
Хиггса. Известно \cite{Montvay}, что при $\lambda = \infty, \beta \sim 2.3$ в
этой модели $M_H \sim 200$ ГэВ в физически интересной области вблизи фазового
перехода между фазой Хиггса и симметричной фазой. В то же время, при $\lambda =
\infty, \beta \sim 8$ в той же области $M_H \le 800$ ГэВ \cite{13,14}. Таким
образом, область значений констант связи, изучаемая в нашей решеточной модели,
должна соответствовать значениям $M_H$ между $200$ и $800$ ГэВ.

Указанные значения следует сравнить с экспериментальными ${\rm sin}^2
\theta_W(100 {\rm Gev}) \sim 0.23$; $\alpha_s(100 {\rm Gev})  \sim
\frac{1}{10}$; $\alpha(100 {\rm Gev})  \sim \frac{1}{128}$. Таким образом,
исследуя данную модель, мы находимся в некотором отдалении от реалистической
области. Тем не менее, для качественного исследования свойств решеточной модели
изучение этой области представляется приемлемым.

Основная особенность действия (\ref{Act_}) заключается в том, что в нем смешаны
цветные и Электрослабые решеточные поля, так что не представляется возможным
симулировать Электрослабую модель в такой регуляризации, если не разыгрываются
одновременно и $SU(3)$ поля \footnote{\rm Исторически основным мотивом к
исследованию решеточной модели с дополнительной $Z_6$ симметрией было
предположение о том, что эта модель может привести к результатам, отличным от
результатов полученных с применением общепринятой регуляризации Стандартной
Модели, не уважающей $Z_6$ симметрию. Это предположение было основано на
аналогии с теорией динамических фермионов. Известно, что модель, реализующая
решеточный вариант киральной симметрии оказывается существенно ближе к описанию
непрерывной физики, чем та, которая не реализует эту симметрию. Таким образом,
представляло интерес сравнение двух решеточных моделей: с $Z_6$ симметрией и
без нее.  В результате наших исследований мы выяснили, что обе решеточные
модели приводят к одним и тем же результатам в физически интересной области
констант связи. Таким образом, сравнение обеих моделей представляет лишь
академический интерес. Однако, как это часто бывает, исследование монополей
Намбу, которое изначально представлялось дополнительной линией исследования,
привело к получению важных физических результатов, изложенных в главах 5 и 6.
Таким образом, одновременная симуляция цветных и Электрослабых полей оказалась
артефактом исходной мотивации. Впрочем, фазовые диаграммы двух моделей
оказываются различными в области больших значений электрослабой константы
связи, причем поведение модели, имеющей дополнительную симметрию обнаруживает
весьма интересные свойства. Эти свойства, как нам кажется,  могут оказаться
важными для развития
 техники решеточной теории как таковой.}.
По этой причине нами также рассматривается модель с дополнительной $Z_2$
симметрией, в которую не включаются $SU(3)$ поля.

\section{Модель без включения $SU(3)$ полей}
\subsection{Определение модели}
\begin{figure}
\begin{center}
 \epsfig{figure=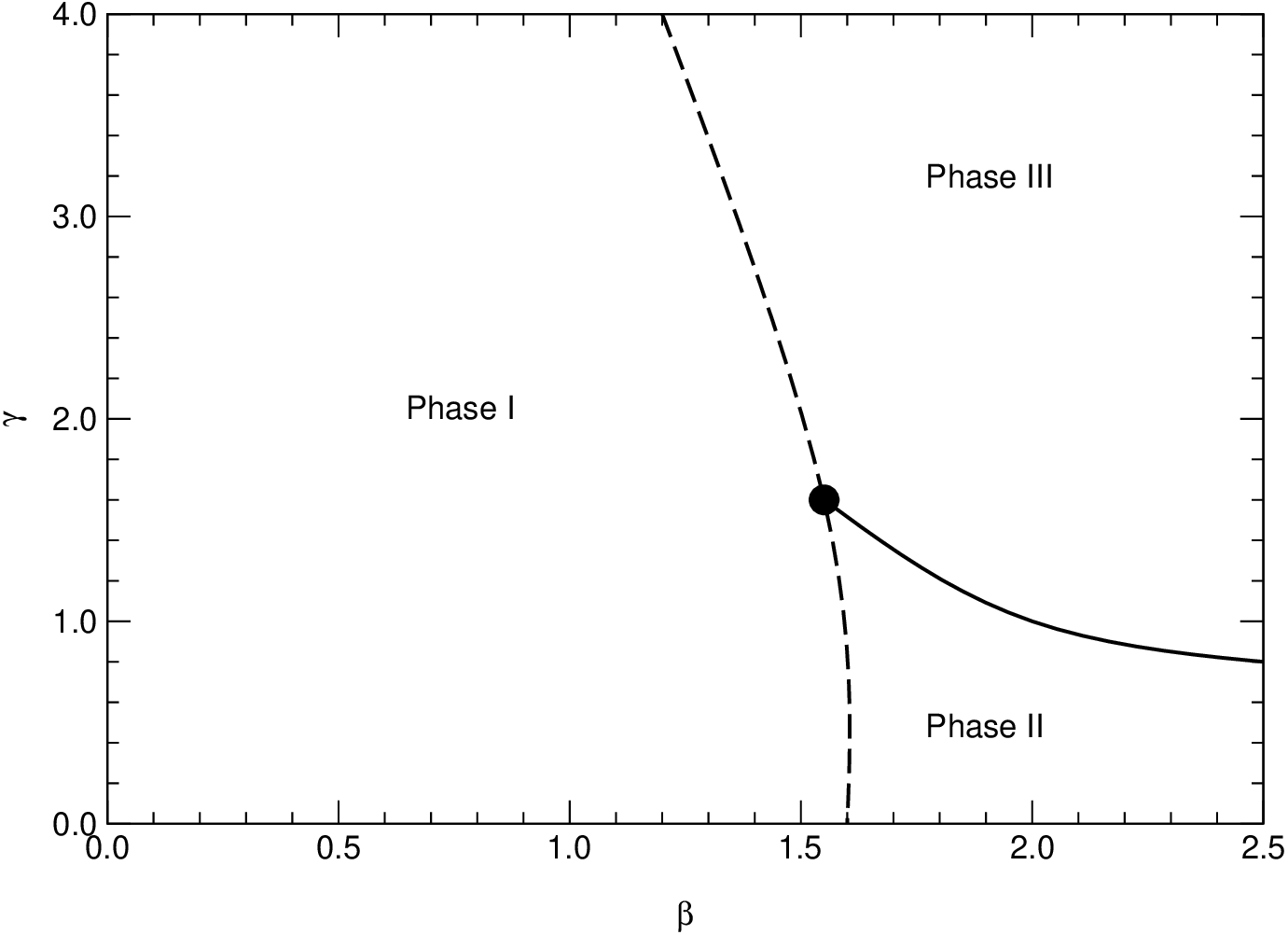,height=60mm,width=80mm,angle=0}
 \caption{\label{4fig.1} Фазовая диаграмма в плоскости
 $(\beta, \gamma)$.}
\end{center}
\end{figure}

  Потенциал для скалярного поля мы рассматриваем в Лондоновском
пределе, то есть в пределе бесконечной затравочной массы Хиггса. Мы выбираем
действие в виде
\begin{eqnarray}
 S & = &  \beta \!\! \sum_{\rm plaquettes} \!\!
 ((1- \mbox{${\small \frac{1}{2}}$} \,{\rm Tr}\, U_p \cos \theta_p)
 + \mbox{${\small \frac{1}{2}}$} (1-\cos 2\theta_p)) +\nonumber\\
 && +\sum_{xy} |U_{xy}e^{i\theta_{xy}}\Phi_y - \Phi_x|^2 +
 V(|\Phi|).\label{S_num}
\end{eqnarray}
$\Phi$ - поле Хиггса, а $V$ - бесконечно глубокий потенциал, приводящий к
вакуумному среднему $\langle |\Phi| \rangle = \sqrt{\gamma/2}$. Наивный
непрерывный предел ~(\ref{S_num}) дает значение угла Вайнберга $\theta_W =
\pi/6$, также имеем $\alpha = \frac{1}{4\pi \beta}$. Изучаемая нами область
констант связи соответствует $\beta \sim 2.0$. Таким образом $\alpha \sim
\frac{1}{25}$. Значение перенормированной массы Хиггса в нашей модели мы
подробно не исследовали. Однако, в качестве оценки для нее можно использовать
результат \cite{Montvay}, где изучалась $SU(2)$ модель Хиггса. Для $\beta \sim
2.3$ в области вблизи фазового перехода между фазой Хиггса и симметричной
фазой, представляющей наибольший интерес, значение массы Хиггса, полученное в
\cite{Montvay} составляет $M_H \sim 2.5 M_W \sim 200$ ГэВ. Реалистическая
область $\alpha \sim \frac{1}{100}$ при значениях $M_H \sim 800, 300, 100$ ГэВ
и $M_H \sim 150$ ГэВ изучается нами в главе 6. В унитарной калибровке получаем:
\begin{eqnarray}
 S & = & \beta \!\! \sum_{\rm plaquettes}\!\!
 ((1-\mbox{${\small \frac{1}{2}}$} \, {\rm Tr}\, U_p \cos \theta_p)
 + \mbox{${\small \frac{1}{2}}$} (1-\cos 2\theta_p)+\nonumber\\
 && + \gamma \sum_{xy}(1 - Re(U^{11}_{xy} e^{i\theta_{xy}}))).
\end{eqnarray}
(Здесь мы в качестве унитарной калибровки использовали $\Phi_1=const,
\Phi_2=0$, а не общепринятый выбор $\Phi_2=const, \Phi_1=0$. )

  Мы также
исследуем обычную $SU(2) \times U(1)$ модель с действием
\begin{eqnarray}
 S_g & = & \beta \!\! \sum_{\rm plaquettes}\!\!
 ((1-\mbox{${\small \frac{1}{2}}$} \, {\rm Tr}\, U_p )
 + 3 (1-\cos \theta_p))+\nonumber\\
 && + \gamma \sum_{xy}(1 - Re(U^{11}_{xy} e^{i\theta_{xy}})).\label{Susual}
\end{eqnarray}
которое также приводит к $\theta_W = \pi/6$.

Следующие переменные рассматриваются как рождающие квант $U(1)$ поля, $Z$ -
бозон, и $W$ - бозон:
\begin{eqnarray}
 A_{xy} & = & A^{\mu}_{x} \; = \,[-{\rm Arg} U_{xy}^{11} + \theta_{xy}]
 \,{\rm mod} \,2\pi, \nonumber\\
 Z_{xy} & = & Z^{\mu}_{x} \; = -\,{\rm sin}\,[{\rm Arg} U_{xy}^{11} + \theta_{xy}]
 \,{\rm mod} \,2\pi, \nonumber\\
 W_{xy} & = & W^{\mu}_{x} \,= \,U_{xy}^{12} e^{-i\theta_{xy}}.
\end{eqnarray}
Здесь, $\mu$ представляет направление $(xy)$. После фиксации унитарной
калибровки остается $U(1)$ симметрия:
\begin{eqnarray}
 U_{xy} & \rightarrow & g^\dag_x U_{xy} g_y, \nonumber\\
 \theta_{xy} & \rightarrow & \theta_{xy} +  \alpha_y/2 - \alpha_x/2,
\end{eqnarray}
где $g_x = {\rm diag} (e^{i\alpha_x/2},e^{-i\alpha_x/2})$. Поля $A$, $Z$, и $W$
преобразуются следующим образом:
\begin{eqnarray}
 A_{xy} & \rightarrow & A_{xy} - \alpha_y + \alpha_x, \nonumber\\
 Z_{xy} & \rightarrow & Z_{xy}, \nonumber\\
 W_{xy} & \rightarrow & W_{xy}e^{-i\alpha_x}.
\label{T}
\end{eqnarray}

Следует отметить, что поле  $A$ не может рассматриваться как обычное
электромагнитное поле, поскольку набор переменных  $A$, $Z$, и $W$ не
диагонализует кинетическую часть действия в его наивном непрерывном пределе. В
нашей модели электромагнитное поле $A_{\rm EM}$ должно быть определено как
\begin{equation}
 A_{\rm EM}  =  A - Z^{\prime} + 2 \,{\rm sin}^2\, \theta_W Z^{\prime},
\label{A_em}
\end{equation}
где  $Z^{\prime} = -[ {\rm Arg} U_{xy}^{11} + \theta_{xy}]{\rm mod} 2\pi$.

Как и любая другая компактная калибровочная теория, наша модель содержит
монополи.

Мы исследуем два вида монополей. $U(1)$ монополи, извлеченные из $2\theta$
определяются как
\begin{equation}
 j_{2\theta} = \frac{1}{2\pi} {}^*d([d 2\theta]{\rm mod}2\pi).
\end{equation}
Кроме того, мы строим монополи из поля $A$. В следующей главе будет показано,
что они могут рассматриваться как квантовые монополи Намбу.
\begin{equation}
 j_{A} = \frac{1}{2\pi} {}^*d([d A]{\rm mod}2\pi) .
\end{equation}
\begin{figure}
\begin{center}
 \epsfig{figure=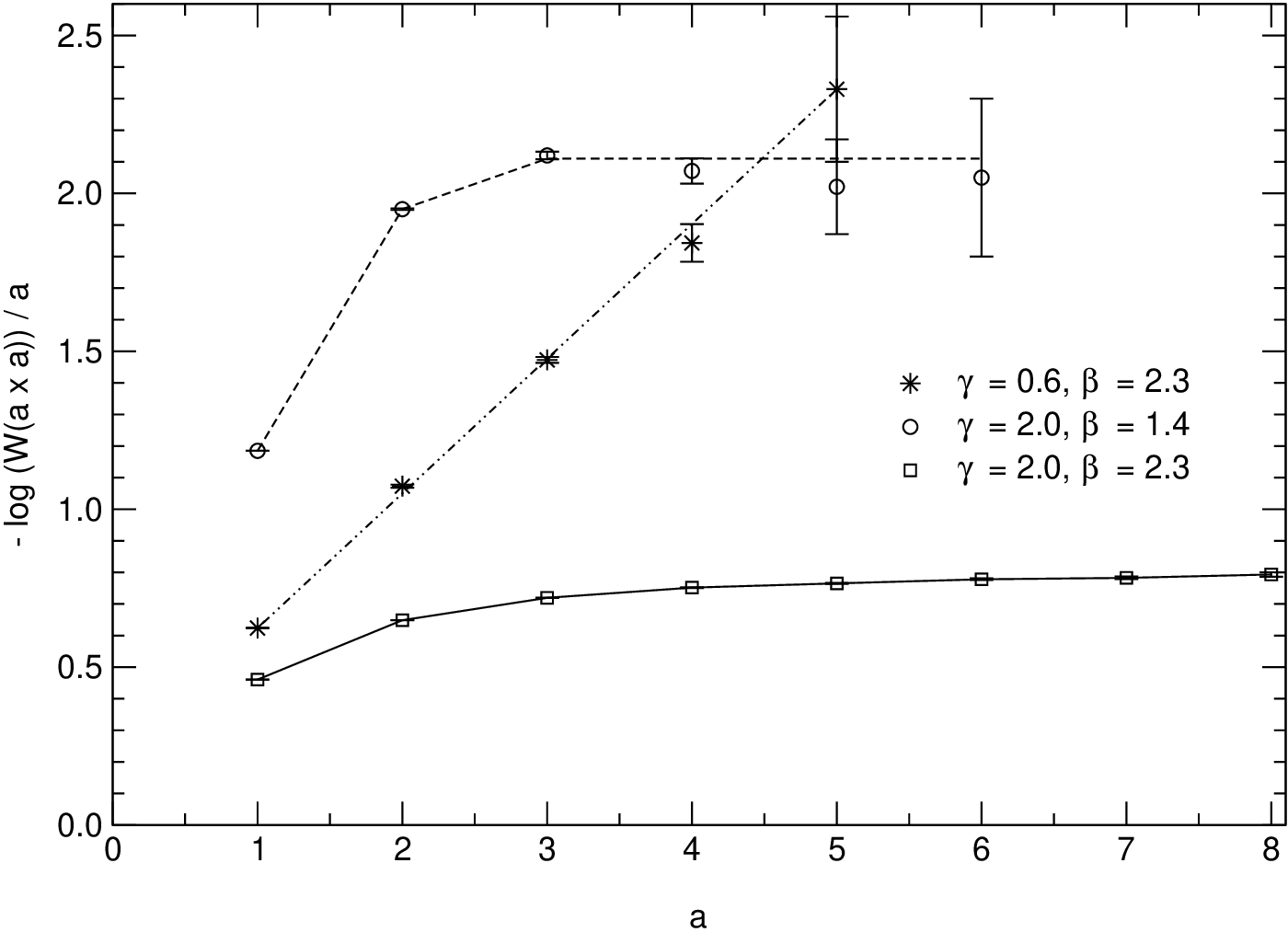,height=60mm,width=80mm,angle=0}
 \caption{\label{4fig.5} ${\cal V}_L(a)$ в трех точках, принадлежащих трем разным фазам модели.}
\end{center}
\end{figure}

\begin{figure}
\begin{center}
 \epsfig{figure=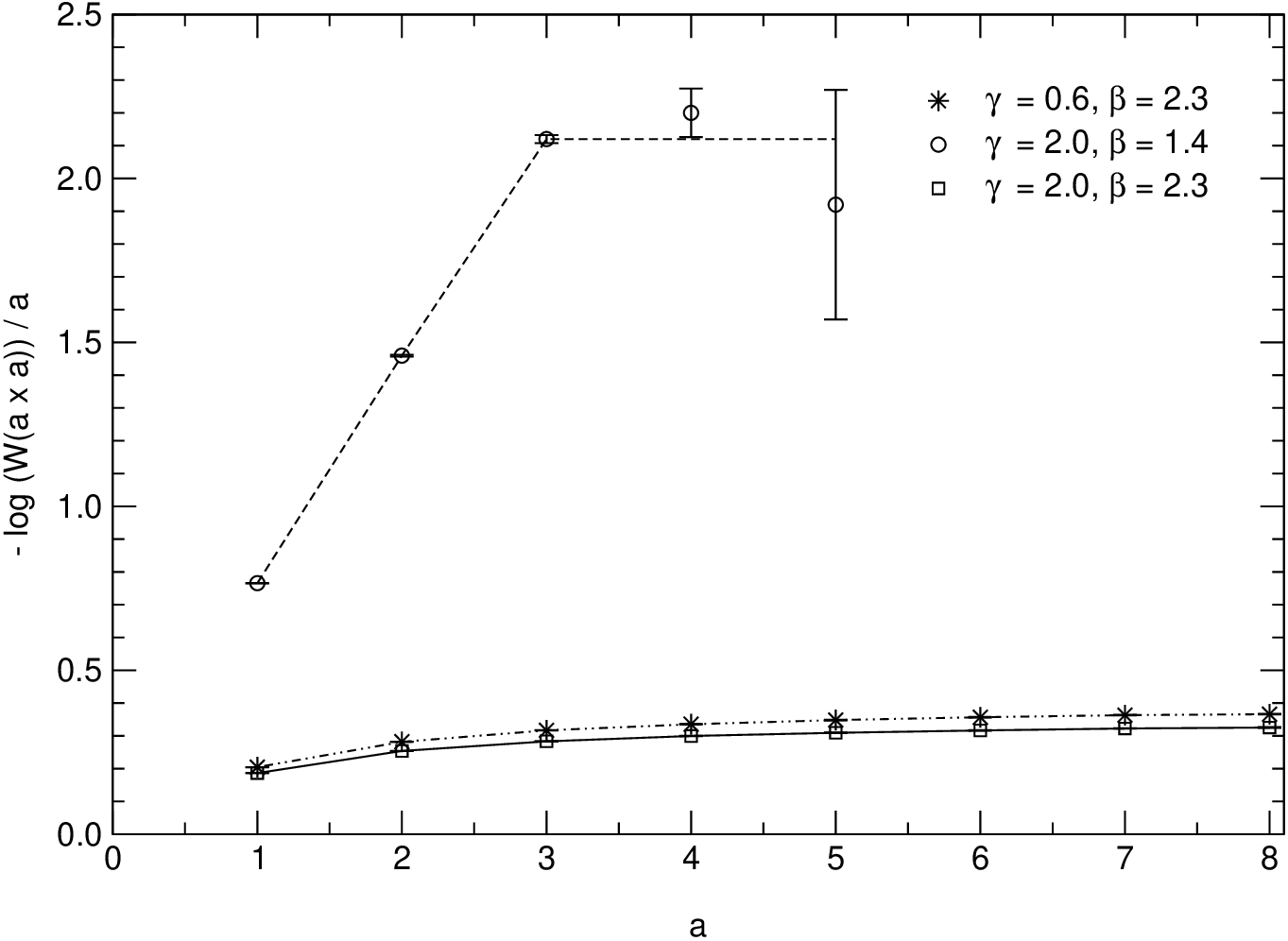,height=60mm,width=80mm,angle=0}
 \caption{\label{4fig.6}  ${\cal V}_R(a)$ в трех точках, принадлежащих трем разным фазам модели.}
\end{center}
\end{figure}

Плотность монополей определяется как:
\begin{equation}
 \rho = \left\langle \frac{\sum_{\rm links}|j_{\rm link}|}{4L^4} \right\rangle,
\end{equation}
где $L$ - размер решетки.
Для того, чтобы выявить динамику внешних заряженных частиц, мы рассматриваем
Петли Вильсона в представлениях левополяризованных и правополяризованных
лептонов:
\begin{eqnarray}
 {\cal W}^{\rm L}(l) & = & \langle {\rm Re} {\rm Tr} \,\Pi_{(xy) \in l}
 U_{xy} e^{-i\theta_{xy}}\rangle, \nonumber\\
 {\cal W}^{\rm R}(l) & = & \langle {\rm Re} \Pi_{(xy) \in l}
 \, e^{-2i\theta_{xy}}\rangle .
\end{eqnarray}
Здесь $l$ обозначает замкнутый контур на решетке. Мы рассматриваем следующие
величины, построенные из прямоугольных петель Вильсона размера $a\times a$:
\begin{equation}
{\cal V}_{R,L}(a) = - \log {\cal W}^{R,L}(a\times a)/a.
\end{equation}
Линейное поведение ${\cal V}(a)$ означало бы существование струны с ненулевым
натяжением между соответствующими зарядами.

\subsection{Численные результаты}

\begin{figure}
\begin{center}
 \epsfig{figure=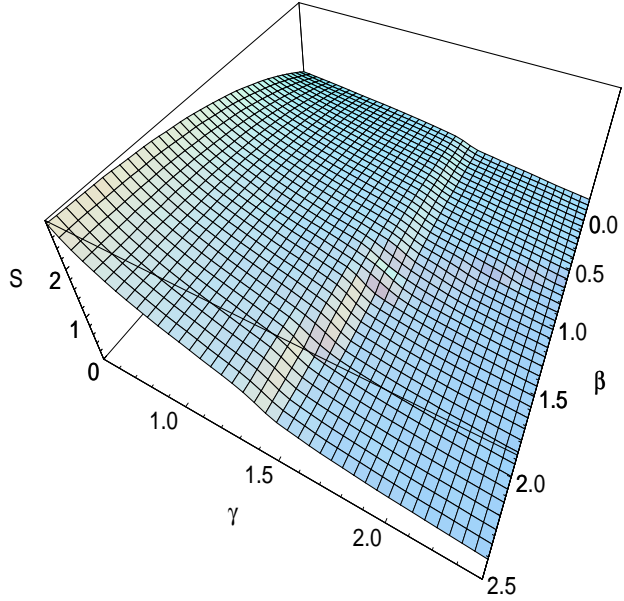,height=80mm,width=80mm,angle=0}
 \caption{\label{4fig.2} Действие
 $\bar{S} =  \langle S \rangle /(6\beta L^4)$ }
\end{center}
\end{figure}

В наших исследованиях мы используем решетки $L^4$ для $L = 6$, $L = 12$, и $L =
16$ с симметричными граничными условиями. Ниже представлены результаты
исследования модели с действием (\ref{S_num}).

Качественно поведение $Z_2$ симметричной модели описывается фазовой диаграммой
представленной на рис. ~\ref{4fig.1}. Модель содержит три фазы. Первая (I) -
фаза "конфайнмента", в которой динамика внешних заряженных частиц подобна КХД с
динамическим фермионами. Слово конфайнмент мы здесь заключаем в кавычки,
поскольку струна между зарядами рвется на некотором расстоянии и конфайнмента
как такового не возникает. Во второй фазе (II) только поведение
левополяризованных частиц обнаруживает наличие струны с ненулевым натяжением
между зарядами. Последняя фаза (III) - это фаза Хиггса, в которой струна не
возникает вовсе.

\begin{figure}
\begin{center}
 \epsfig{figure=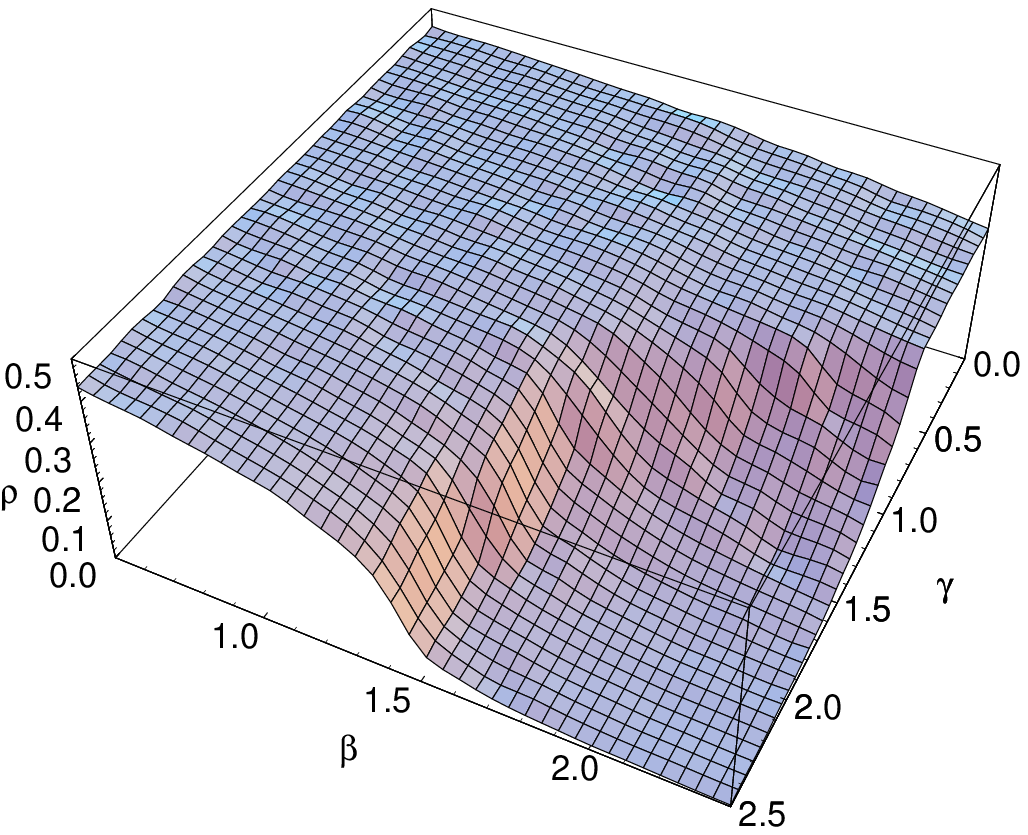,height=80mm,width=80mm,angle=0}
 \caption{\label{4fig.3} Плотность электромагнитных монополей. Уменьшается в
 фазе Хиггса.}
\end{center}
\end{figure}

\begin{figure}
\begin{center}
 \epsfig{figure=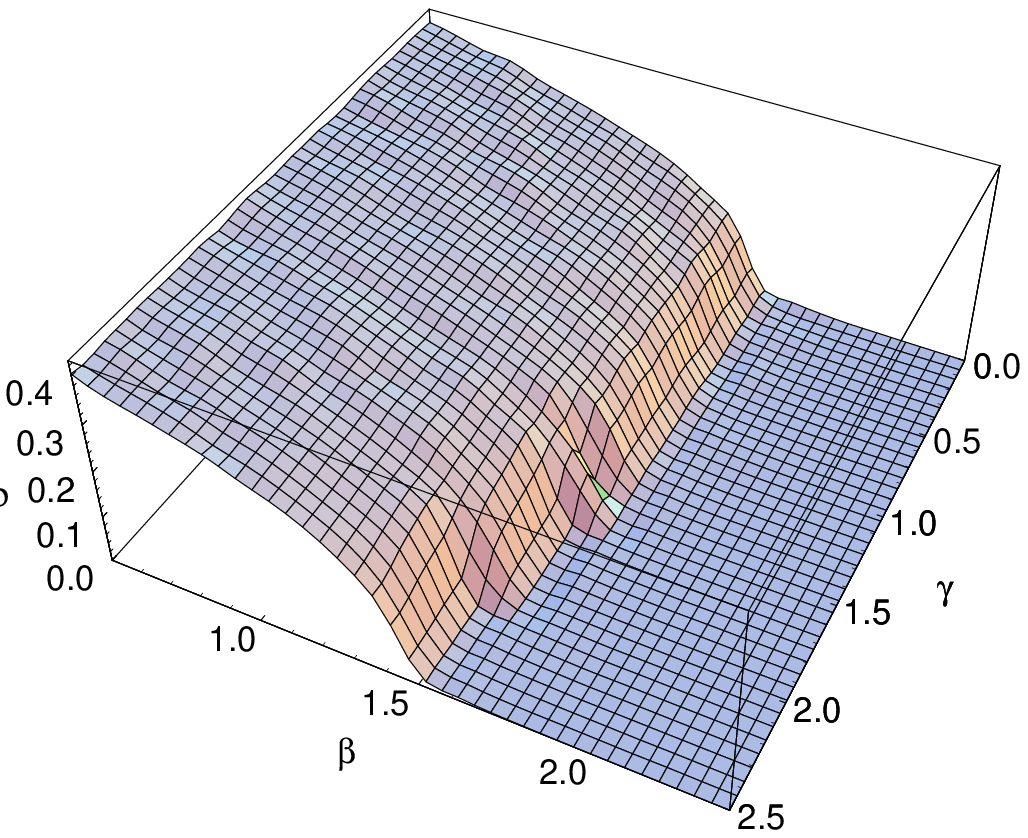,height=80mm,width=80mm,angle=0}
 \caption{\label{4fig.4} Плотность $U(1)$ - монополей. Она уменьшается, когда поведение правых
 внешних частиц не проявляет удерживающих сил.}
\end{center}
\end{figure}

 Указанные свойства различных фаз проиллюстрированы рис.
~\ref{4fig.5} и рис. \ref{4fig.6}, где представлены ${\cal V}_L(a)$ и ${\cal
V}_R(a)$ для трех типичных точек фазовой диаграммы, принадлежащих указанным
трем фазам. Ясно, что в фазе Хиггса форма ${\cal V}(a)$ исключает возникновение
удерживающего потенциала.  То же поведение обнаружено в фазе II для ${\cal
V}_R(a)$. С другой стороны, в фазе II форма ${\cal V}_L(a)$ сигнализирует
наличие линейного потенциала для достаточно малых расстояний (до пяти
решеточных единиц). Однако, как и для КХД с динамическими фермионами, или
$SU(2)$ модели Хиггса \cite{Montvay,EW_T}, эти результаты не означают наличие
конфайнмента. Струна между зарядами может быть разорвана виртуальными
заряженными скалярными частицами. Таким образом, ${\cal V}(a)$ может оставаться
линейным только на малых расстояниях, в то время, как начиная с некоторых
расстояний он перестает расти, указывая на разрыв струны. К сожалению, точность
наших измерений не позволяет нам наблюдать это явление в деталях. Однако, оно
иллюстрируется формой  ${\cal V}_L(a)$ и ${\cal V}_R(a)$ в фазе I, показанных
на рис.~\ref{4fig.5} и рис.~\ref{4fig.6}.

Фазовая структура модели проявляется также в поведении среднего действия
$\bar{S} = \langle S \rangle /(6\beta L^4)$, представленного на рис.
~\ref{4fig.2}. Оно оказывается неоднородным в некоторой окрестности линии
фазового перехода.

Связь между свойствами монополей и фазовой структурой иллюстрируется рис.
~\ref{4fig.3} и \ref{4fig.4}, которые показывают плотность монополей как
функцию констант связи. Плотность монополей Намбу уменьшается при движении от
фазовой линии и падает до нуля в фазе Хиггса, в то время, как плотность
$U(1)$ монополей (гиперзарядовых монополей) падает до нуля в фазах  II и III.
Мы видим, что поведение $U(1)$ монополей связано с динамикой правых частиц, а
поведение монополей Намбу связано с динамикой левых лептонов.

Следует отметить, что $SU(2)$ модель Хиггса имеет сходную фазовую структуру за
исключением отсутствия в ней линии фазового перехода, разделяющего фазы
 I и II. Известно, что в этой модели две фазы в действительности не являются разными.
 Линия фазового перехода заканчивается в некоторой точке и далее продолжается как кроссовер.
 Можно было бы ожидать, что линия фазового  перехода в нашей модели также является линией кроссовера.
 Однако, мы не исследовали этот вопрос детально.

В нашей модели линии фазового перехода соединяются в тройной точке, формируя
общую линию. Очевидно, это есть следствие дополнительной $Z_2$ симметрии,
связывающей $SU(2)$ и $U(1)$ возбуждения. Такая картина, разумеется, не
возникает в модели с действием (\ref{Susual}) (см. также  \cite{SU2U1}).

Следует также отметить, что фазовая диаграмма нашей модели может содержать
нефизическую область, соответствующую нефизической области чистой  $SU(2)$
модели (которая наблюдается при $\beta < \beta_c$, где $\beta_c$ - точка
кроссовера). Наше исследование показывает, что если эта область значений
констант связи и существует, то она должна быть далеко от фазы Хиггса,
представляющей для нас основной интерес.  В действительности, эта область может
появиться при  $\beta < 2.25$ и $\gamma < 0.5$.

\begin{figure}
\begin{center}
 \epsfig{figure=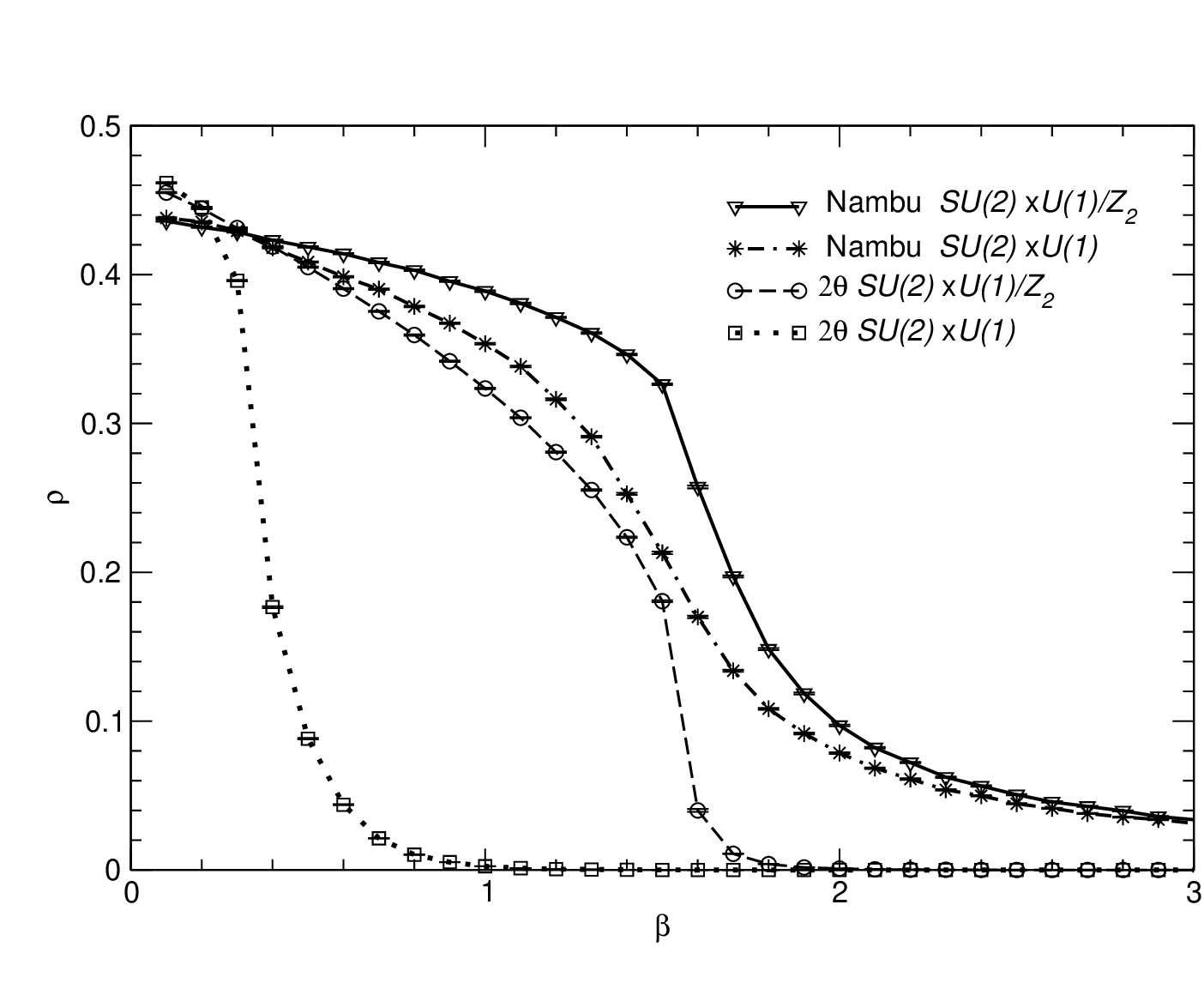,height=60mm,width=80mm,angle=0}
\caption{\label{4fig.10_} Плотности гиперзарядовых монополей и Намбу монополей
(извлеченных из $Z^{\prime}$) как функция $\beta$ для $\gamma = 1.5$ для обеих
моделей.}
\end{center}
\end{figure}

Различие между моделями с действием (\ref{Susual}) и (\ref{S_num})
иллюстрируется поведением плотности монополей, вычисленной в обеих моделях и
представленной на рис. \ref{4fig.10_}.

\section{Модель с включением $SU(3)$ переменных}

\begin{figure}
\begin{center}
\epsfig{figure=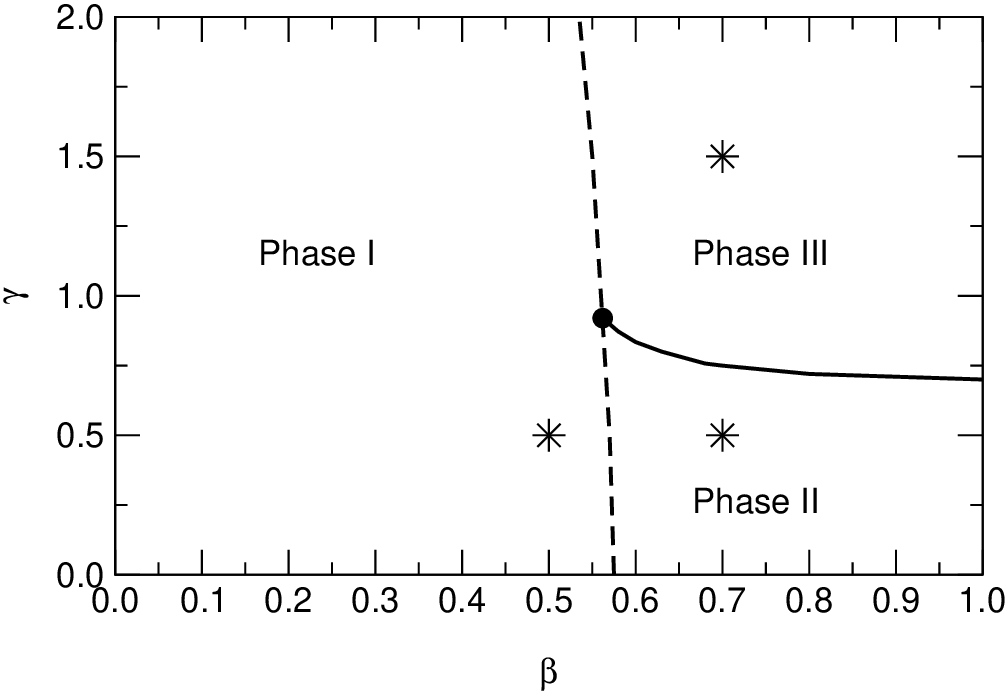,height=100mm,angle=0} \caption{\label{fig.1__}
Фазовая диаграмма модели в плоскости
 $(\beta, \gamma)$.}
\end{center}
\end{figure}

\subsection{Исследуемая модель и вычисляемые величины} \label{sect.3}

Мы рассматриваем модель с действием для калибровочного поля (\ref{Act_}).
Потенциал для скалярного поля рассматривается в его простейшем виде в
Лондоновском пределе. После фиксации унитарной калибровки имеем:
\begin{eqnarray}
 S_H & = & \gamma \sum_{xy}[1 - Re(U^{11}_{xy} e^{i\theta_{xy}})].
\end{eqnarray}
Так же как и для модели без цветных полей мы строим величины, соответствующие
полям $A$, полю  $Z$ - бозона и полю $W$  - бозона:
\begin{eqnarray}
 A_{xy} & = & A^{\mu}_{x} \; = \,[-{\rm Arg} U_{xy}^{11} + \theta_{xy}]
 \,{\rm mod} \,2\pi, \nonumber\\
 Z^{\prime}_{xy} & = & Z^{\mu}_{x} \; = -[{\rm Arg} U_{xy}^{11} + \theta_{xy}]
 \,{\rm mod} \,2\pi, \nonumber\\
 W_{xy} & = & W^{\mu}_{x} \,= \,U_{xy}^{12} e^{-i\theta_{xy}}.\label{AZW}
\end{eqnarray}
Здесь $\mu$ представляет направление $(xy)$.

Для того, чтобы извлекать необходимую информацию из $SU(3)$ полей мы
используем так называемую непрямую Максимальную Центральную проекцию (см. главу
1).

Максимальная Центральная проекция делает линковые матрицы  $\Gamma$ возможно
более близкими к элементам центра $Z_3$ группы $SU(3)$: $ Z_3 = \{{\rm
diag}(\mathrm{e}^{(2\pi i /3) N}, \mathrm{e}^{(2\pi i /3) N}, \mathrm{e}^{(2\pi
i /3) N}\}$, где $N \in \{1, 0, -1\}$.  Эта процедура работает следующим
образом.

Прежде всего, максимизируем функционал
\begin{equation}
Q_1 = \sum_{\mathrm{links}} (|\Gamma_{11}| + |\Gamma_{22}| + |\Gamma_{33}|)
\end{equation}
по отношению к калибровочным преобразованиям $\Gamma_{xy} \rightarrow g^\dag_x
\Gamma_{xy} g_y$, фиксируя Максимальную Абелеву проекцию.

Далее, делаем полученные линковые матрицы возможно более близкими к центру
 $SU(3)$, делая фазы диагональных элементов максимально близкими друг к другу.
Это достигается минимизацией функционала
\begin{eqnarray}
 Q_2 & = &
 \sum_{\mathrm{links}}\{[1-\cos({\rm Arg}(\Gamma_{11})-{\rm Arg}(\Gamma_{22}))]
     + [1-\cos({\rm Arg}(\Gamma_{11})-{\rm Arg}(\Gamma_{33}))]
\nonumber \\
 &&  +\; [1-\cos({\rm Arg}(\Gamma_{22})-{\rm Arg}(\Gamma_{33}))]\}.
\end{eqnarray}
по отношению к калибровочным преобразованиям. Это калибровочное условие
инвариантно относительно центральной подгруппы  $Z_3$ группы $SU(3)$.
\begin{figure}
 \epsfig{figure=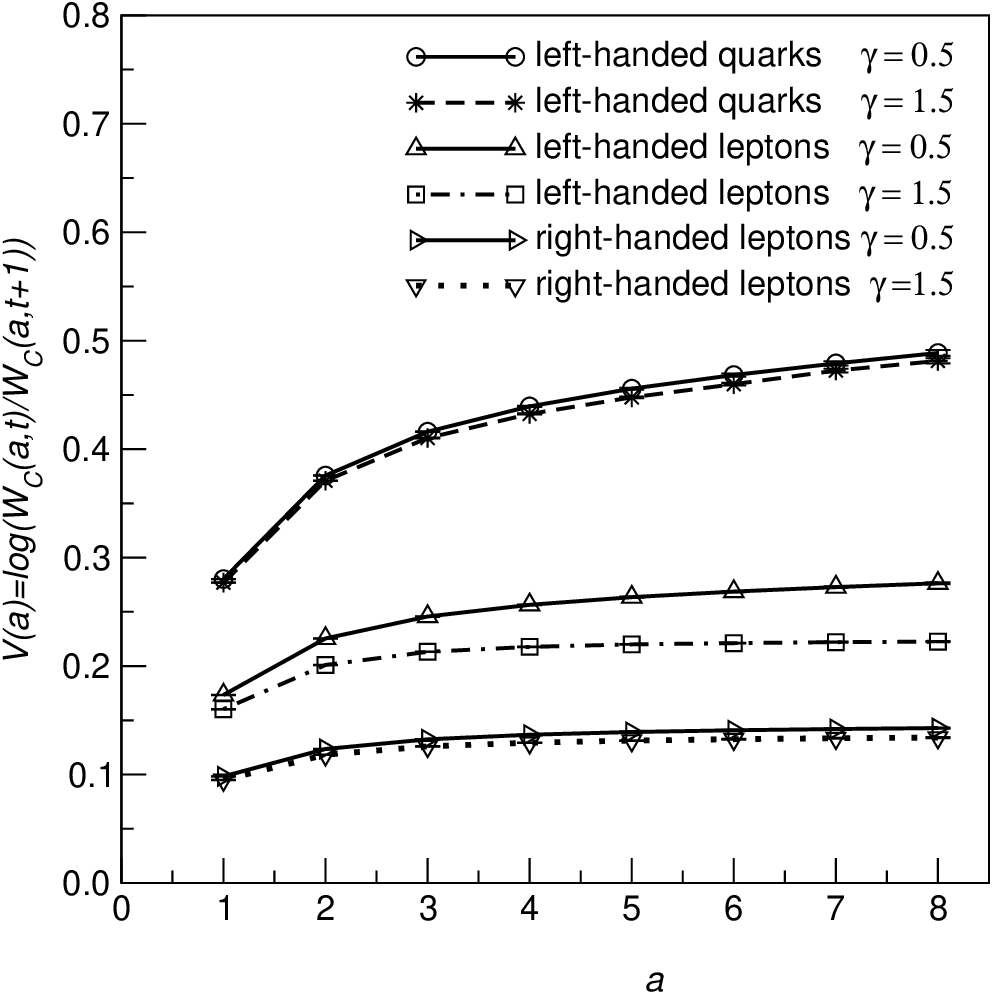,height=100mm,angle=0}
 \caption{\label{fig.2__} ${\cal V}_L(a)$ вычисленный при $\beta = 0.7$. Здесь
 потенциалы извлекаются из
  $ {\cal W}^{\rm L}_{ {\rm quarks}}$(левые кварки), $ {\cal W}^{\rm L}_{ {\rm lept}}$ (левые лептоны),
и $ {\cal W}^{\rm R}_{ {\rm lept}}$ (правые лептоны).}
\end{figure}

В нашей модели $SU(3)$ поля связаны с $U(1)$ и $SU(2)$ полями посредством
центра калибровочной группы. Поэтому вместо центральных вихрей и центральных
монополей мы определяем различные типы монополей. Определение этих полей
включает следующее целочисленное поле  $N$ (определенное после фиксации
Максимальной Центральной калибровки):
\begin{eqnarray}
 N_{xy}=0 &{\rm :}& ({\rm Arg}(\Gamma_{11})+{\rm Arg}(\Gamma_{22})+{\rm Arg}
 (\Gamma_{33}))/3 \in \; ]-\pi/3, \pi/3],
\nonumber \\
 N_{xy}=1 &{\rm :}& ({\rm Arg}(\Gamma_{11})+{\rm Arg}(\Gamma_{22})+{\rm Arg}
 (\Gamma_{33}))/3 \in \; ]\pi/3, \pi],
\nonumber \\
 N_{xy}=-1 &{\rm  :}& ({\rm Arg}(\Gamma_{11})+{\rm Arg}(\Gamma_{22})+{\rm Arg}
 (\Gamma_{33}))/3 \in \; ]-\pi, -\pi/3].
\end{eqnarray}
Другими словами, $N=0$ если $\Gamma$ близка к $1$,  $N=1$ если $\Gamma$ близка
к \noindent $\mathrm{e}^{2\pi i/3}$ и  $N=-1$ если $\Gamma$ близка к
$\mathrm{e}^{-2\pi i/3}$.

Далее, мы определяем следующие линковые поля
\begin{eqnarray}
 C^1_{xy} & = & \,\left[\frac{2\pi}{3}N_{xy} +
 {\rm Arg} U_{xy}^{11} + \frac{1}{3}\theta_{xy}\right] \,{\rm mod} \,2\pi,
\nonumber\\
 C^2_{xy} & = &  \,\left[\frac{2\pi}{3}N_{xy} - \frac{2}{3}\theta_{xy}\right]
 \,{\rm mod} \,2\pi,
\nonumber\\
 C^2_{xy} & = & \,\left[\frac{2\pi}{3}N_{xy} + \frac{4}{3}\theta_{xy}\right]
 \,{\rm mod} \,2\pi.
\end{eqnarray}
Эти поля соответствуют последним трем членам в ~(\ref{Act_}). Их построение
следует из представления $\Gamma$ как произведения $\exp((2\pi i / 3) N)$ и
$V$, где $V$ - это $SU(3)/Z_3$ переменная $({\rm Arg}(V_{11})+{\rm
Arg}(V_{22})+{\rm Arg}(V_{33}))/3 \in \; ]-\pi/3, \pi/3]$. Таким образом,
$\Gamma=\exp((2\pi i/ 3) N) V$. Мы ожидаем, что (\ref{Act_}) подавляет $V_{\rm
plaq}$ и $C^i_{\rm plaq}, i = 1,2,3$ в то время, как поля $N$,
$\frac{\theta}{3}$, и $U_{11}$ (будучи рассматриваемые независимо друг от
друга), как ожидается, неупорядочены. Это ожидание подтверждается результатами
вычислений.

Мы исследуем пять типов монополей. Монополи, несущие информацию о цветных
полях, извлекаются из $C^i$:
\begin{equation}
 j_{C^i} = \frac{1}{2\pi} {}^*d([d C^i]{\rm mod}2\pi) .
\end{equation}

Чистые $U(1)$ монополи, соответствующие второму члену в (\ref{Act_}),
извлекаются из $2\theta$:
\begin{equation}
 j_{2\theta} = \frac{1}{2\pi} {}^*d([d 2\theta]{\rm mod}2\pi).
\end{equation}
Монополи Намбу:
\begin{equation}
 j_{A} = \frac{1}{2\pi} {}^*d([d A]{\rm mod}2\pi) .
\end{equation}
(В следующей главе эти объекты рассматриваются более подробно. Причем
устанавливается их соответствие с классическими монополями Намбу.)

Плотность монополей определяется как:
\begin{equation}
 \rho = \left\langle \frac{\sum_{\rm links}|j_{\rm link}|}{4L^4} \right\rangle,
\end{equation}
где $L$ - размер решетки.
Чтобы понять динамику внешних заряженных частиц, мы рассматриваем петли
Вильсона, определенные в фермионных представлениях кварков и лептонов:
\begin{eqnarray}
 {\cal W}^{\rm L}_{\rm lept}(l) & = &
 \langle {\rm Re} {\rm Tr} \,\Pi_{(xy) \in l} U_{xy} e^{-i\theta_{xy}}\rangle,
\nonumber\\
 {\cal W}^{\rm R}_{\rm lept}(l) & = &
 \langle {\rm Re} \Pi_{(xy) \in l} \, e^{-2i\theta_{xy}}\rangle ,
\nonumber\\
 {\cal W}^{\rm L}_{ {\rm quarks}}(l) & = & \langle {\rm Re} \Pi_{(xy) \in l}
 \, \Gamma_{xy} \, U_{xy}\, e^{\frac{i}{3}\theta_{xy}}\rangle ,
\nonumber\\
 {\cal W}^{\rm R}_{{\rm down} \, {\rm quarks}}(l) & = &
 \langle {\rm Re} \Pi_{(xy) \in l} \, \Gamma_{xy} \,
 e^{-\frac{2i}{3}\theta_{xy}}\rangle ,
\nonumber\\
 {\cal W}^{\rm R}_{{\rm up} \, {\rm quarks}}(l) & = &
 \langle {\rm Re} \Pi_{(xy) \in l} \, \Gamma_{xy} \,
 e^{\frac{4i}{3}\theta_{xy}}\rangle .
\label{WL_}
\end{eqnarray}
Здесь $l$ обозначает замкнутый контур на решетке. Мы рассматриваем следующую
величину, извлекаемую из прямоугольных петель Вильсона размера $a\times t$:
\begin{equation}
 {\cal V}(a) = \lim_{t \rightarrow \infty} \, {\rm log}\,
 \frac{  {\cal W}(a\times t)}{{\cal W}(a\times (t+1))}.
\end{equation}
Линейное поведение ${\cal V}(a)$ указывало бы на существование струны с
ненулевым натяжением между зарядами.

\subsection{Численные результаты} \label{sect.6}
\begin{figure}
 \epsfig{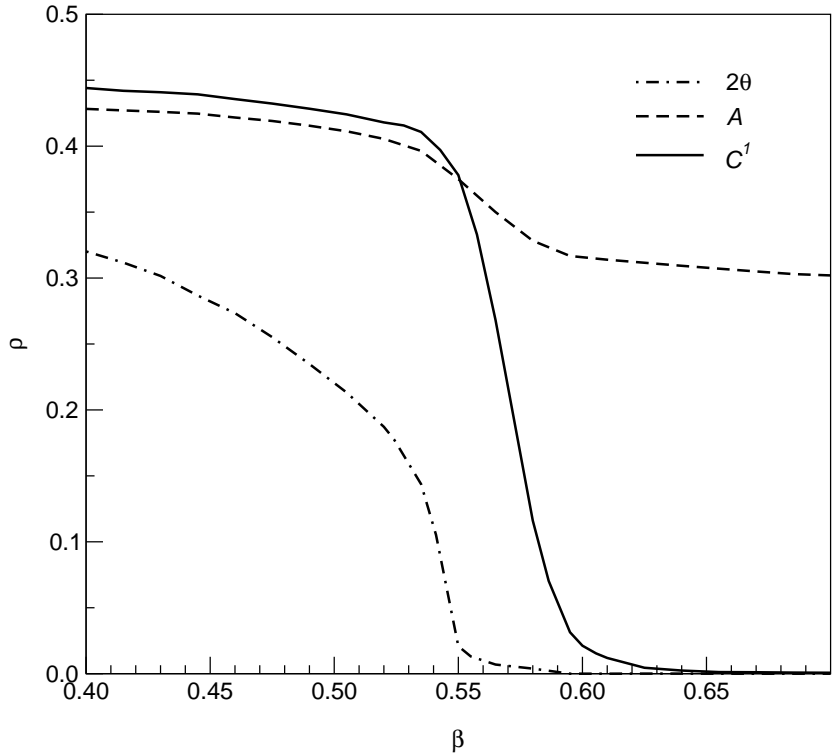}
 \caption{\label{fig.3__} Плотность монополей (построенных из линковых полей
 $A, 2\theta$, и $C^1$)
относительно $\beta$ при $\gamma = 0.5$.}
\end{figure}
В наших исследованиях мы используем решетки $L^4$ при $L = 6$, $L = 12$, и $L =
16$ с симметричными граничными условиями.

Фазовая диаграмма представлена на рис. ~\ref{fig.1__}. Модель содержит три
фазы. Первая (I) - это фаза, в которой динамика внешних лептонов определяется
"удерживающими" силами, т.е. аналогично поведению зарядов в КХД с динамическими
фермионами.  Во второй фазе (II) поведение левых лептонов определяется
"удерживающими" силами, в то время, как поведение правых лептонов - нет.
Последняя фаза  (III) - фаза Хиггса, в которой нет "удерживающих" сил. Во всех
трех фазах имеет место конфайнмент кварков.

Это иллюстрируется рис. ~\ref{fig.2__}, где показан ${\cal V}(a)$ извлеченный
из петель Вильсона рис.~(\ref{WL_}) для двух типических точек, принадлежащих
фазе
 II ($\gamma = 0.5$) и III
($\gamma = 1.5$) модели (поведение всех потенциалов в фазе I - "удерживающее").
Мы представляем здесь потенциал только для одной цветной петли Вильсона, а
именно, для $ {\cal W}^{\rm L}_{ {\rm quarks}}$, поскольку натяжение струны,
извлеченное из других двух потенциалов совпадает с натяжением струны,
извлеченного из потенциала, представленного на рис. Разумеется, так и должно
быть - натяжение струны для всех типов кварков должно быть одинаковым.

Применяя линейный фит к лептонному потенциалу при $a \ge 5$ мы находим, что
только в случае левых лептонов значение натяжения струны много больше, чем
статистическая ошибка в фазе  II. Для левых лептонов в фазе Хиггса и правых
лептонов в обеих фазах, натяжение струны - ноль в пределах статистических
ошибок. Как и в КХД с динамическими фермионами или в $SU(2)$ модели Хиггса
\cite{Montvay,EW_T}, наши результаты не означают наличия конфайнмента левых
лептонов в фазе II. Струна между ними должна разорваться из - за наличия
виртуальных заряженых скалярных частиц.  ${\cal V}(a)$ может быть линейным
только на достаточно малых расстояниях. К сожалению, вычислительные возможности
не позволяют нам наблюдать разрыв струны.

Связь между свойствами монополей и фазовой структурой модели иллюстрируется
рис. ~\ref{fig.3__}, который показывает плотность монополей как функцию $\beta$
для $\gamma = 0.5$. Снова мы представляем здесь только один тип монополей,
который имеет цветное происхождение. А именно, рассматривается  $j_{C^1}$.
(Поведение остальных аналогично.) Видно, что плотность $2\theta$ - монополей
так же, как и $C^1$ - монополей быстро падает в фазе II, в то время, как
плотность монополей Намбу - нет.

Согласно нашим вычислениям плотность монополей Намбу падает в фазе III. Цветные
монополи и $2\theta$ - монополи так же быстро зануляются в фазе III. Таким
образом, цветные монополи чувствуют фазовый переход, который согласно нашей
интуиции соответствует $U(1)$ переменным. Это происходит из-за $Z_6$ симметрии,
связывающей  $U(1)$ переменные с центром группы $SU(3)$.

В нашей модели оба фазовых перехода встречаются в тройной точке, формируя общую
линию. Это, очевидно, также есть следствие $Z_6$ симметрии. Разумеется, ничего
подобного не наблюдается в обычной решеточной  $SU(3) \otimes SU(2) \otimes
U(1)$ модели с Хиггсом: $SU(2) \otimes U(1)$ часть исследована, например, в
\cite{SU2U1}. Что касается $SU(3)$ теории, она не имеет фазового перехода
вообще для конечных $\beta$ при нулевой температуре.

\section{Публикации}

Результаты настоящей главы опубликованы в работах:

"A hidden symmetry in the Standard Model",  By B.L.G. Bakker, A.I. Veselov,
M.A. Zubkov. Phys.Lett.B583:379-382,2004, [hep-lat/0301011]

"An Additional symmetry in the Weinberg -  Salam model", B.L.G. Bakker, A.I.
Veselov, M.A. Zubkov, Yad.Fiz.68:1045-1053,2005, Phys.Atom.Nucl.68:
1007-1015,2005, [hep-lat/0402004]

"Standard model with the additional Z(6) symmetry on the lattice", B.L.G.
Bakker, A.I. Veselov, M.A. Zubkov, Phys.Lett.B620:156-163,2005,
[hep-lat/0502006]

\chapter{Монополи Намбу при конечной температуре}
\label{ch5}

 В настоящей главе мы излагаем результаты нашего исследования
поведения Монополей Намбу при конечной температуре. Показано, что Электрослабый
фазовый переход (кроссовер) сопровождается конденсацией Монополей Намбу. На
возможность возникновения этого явления было указано М. Чернодубом в
\cite{Chernodub_Nambu}. Однако, впервые доказательство существования этого
интересного явления было дано в работах автора настоящей диссертации и его
соавторов, результаты которых излагаются ниже.

В предыдущей главе нами было дано описание решеточной дискретизации
Калибровочного и Хиггсовского секторов Стандартной Модели. Также было дано
общее описание фазовой структуры решеточной модели при нулевой температуре (в
формулировке, учитывающей $Z_6$ симметрию Стандартной Модели), основанное на
результатах численных исследований модели при $\alpha \sim 1/50$, $\alpha_s
\sim 1/20$. В физической области $\alpha \sim 1/100$ электрослабой константы
связи решеточная модель с дополнительной дискретной симметрией приводит к тем
же результатам, что и решеточная модель без этой симметрии. В настоящей главе
для качественного исследования свойств монополей Намбу при конечных
температурах мы используем решеточную дискретизацию с $Z_6$ симметрией.
Симуляция этой решеточной модели необходимо включает  разыгрывание как
Электрослабых, так и цветных полей. Сама решеточная модель имеет ряд
особенностей, заслуживающих изучение самих по себе. При этом нами не ставится
задача получения точных количественных характеристик изучаемых явлений. Нашей
целью является выявление общих закономерностей. Скорость выполнения численных
симуляций существенно зависит от значений электрослабой константы связи. При
этом в реалистической области малых значений постоянной тонкой структуры
$\alpha \sim 1/128$ симуляции существенно усложнены. По этой причине мы
исследуем область значений констант связи, соответствующую значениям $\alpha
\sim 1/50$, $\alpha_s \sim 1/20$. Значение угла Вайнберга в нашей модели
соответствует ${\rm sin}^2 \theta_W \sim 0.38$, а значение массы Хиггса $200$
ГэВ  $ < M_H < 800$ ГэВ (см. предыдущую главу). Для изучения свойств монополей
Намбу при конечной температуре использование этой области представляется нам
достаточным. В следующей главе будут представлены результаты симуляции модели
при нулевой температуре и при значениях $\alpha \sim 1/150$.

Мы находим ряд особенностей фазовой диаграммы рассматриваемой модели. В
частности, фазовый переход конфайнмент - деконфайнмент для полей $SU(2)$ и
$SU(3)$ совпадает. (Строго говоря, в фазе конфайнмента нет конфайнманта
лептонов, но обнаруживаются удерживающие силы между ними. При этом струна
рвется на некоторых расстояниях благодаря наличию виртуальных скалярных
заряженных частиц.) Линия фазового перехода между фазой Хиггса и симметричной
фазой деконфайнмента встречается с линией фазового перехода конфайнмент -
деконфайнмент, формируя тройную точку. Переход между  фазой Хиггса и
симметричной фазой соответствует конечнотемпературному Электрослабому фазовому
переходу. Мы видим, что монополи Намбу сконденсированы при $T>T_c$ в то время,
как при $T<T_c$ их конденсат зануляется.


При доступных для современных исследований значениях энергиях теория возмущений
при нулевой температуре работает прекрасно. Однако, в \cite{M_W_T} было
показано, что при конечной температуре теория возмушений перестает работать
вблизи Электрослабого фазового перехода уже для масс Хиггса выше  $60$ ГэВ.
Поэтому существующее в настоящее время ограничение на массу Хиггса  $M_H
> 114$ ГэВ требует использования непертурбативной техники исследования высокотемпературной
физики. Именно поэтому наше использование решеточной техники и оправдано при
изучении Электрослабого перехода.

\section{Монополи Намбу}

\begin{figure}
\begin{center}
\epsfig{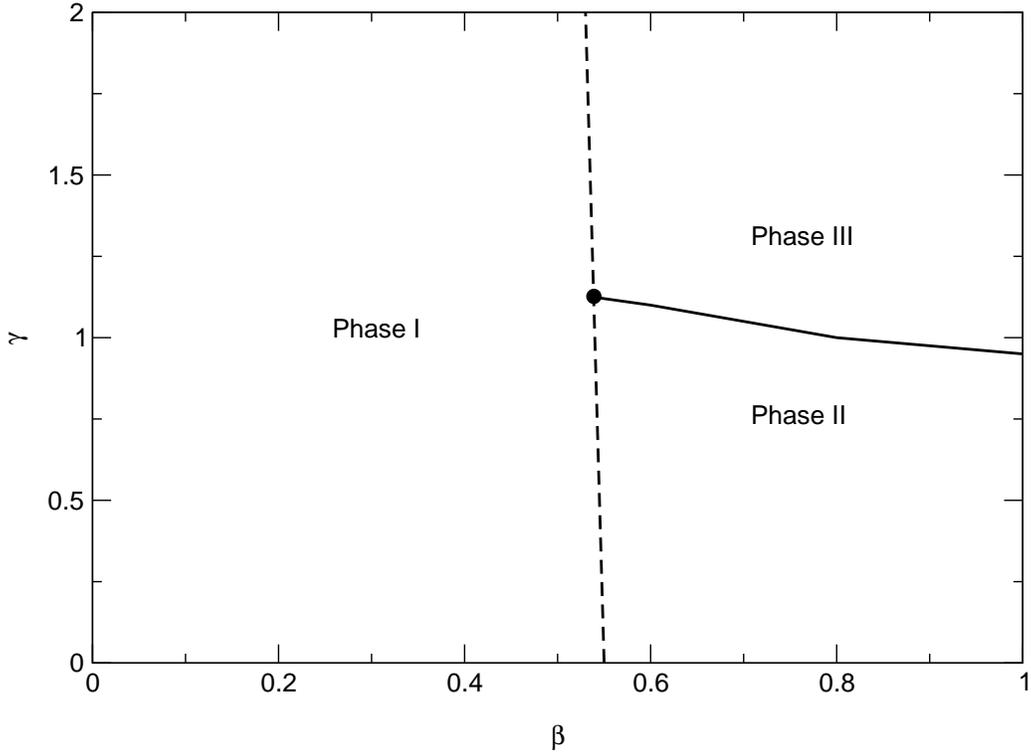}

\caption{\label{f5.1}Фазовая диаграмма модели в плоскости $(\beta, \gamma)$-
для размера решетки по мнимому времени $N_T = 2$.}

\end{center}
\end{figure}

Ниже дается объяснение того, что такое монополь Намбу с классической и с
квантовой точек зрения. Прежде всего, определим непрерывные Электрослабые поля
как они появляются в модели Вайнберга - Салама. Пусть  $\Phi$ - скалярное поле
Электрослабой теории. Мы определяем поле  $Z$ - бозона $Z^{\mu}$ и
электромагнитное поле $A_{\rm EM}^{\mu}$  следующим образом:
\begin{eqnarray}
 Z^{\mu} = - \frac{1}{\sqrt{\Phi^+ \Phi}} \Phi^+ A^{\mu} \Phi -  B^{\mu},
\nonumber\\
 A_{\rm EM}^{\mu} =  2 B^{\mu}  + 2 \,{\rm sin}^2\, \theta_W
 Z^{\mu},\label{FSM}
\end{eqnarray}

где $A^{\mu}$ и $B^{\mu}$ - это соответственно $SU(2)$ и $U(1)$ калибровочные
поля Стндартной Модели.

Далее, фиксируем унитарную калибровку $\Phi_2=const.$, $\Phi_1 = 0$. Тогда
\begin{eqnarray}
 Z^{\mu} =  \frac{g_z}{2}[\frac{\tilde{A_3}^{\mu}}{g_2}{\rm cos}\theta_W -  \frac{\tilde{B}^{\mu}}{g_1}{\rm sin}\theta_W] = \frac{1}{2}\tilde{Z}^{\mu},
\nonumber\\
 A_{\rm EM}^{\mu} =  e[\frac{\tilde{A_3}^{\mu}}{g_2}{\rm sin}\theta_W +  \frac{\tilde{B}^{\mu}}{g_1}{\rm cos}\theta_W] = \tilde{A}^{\mu},
\end{eqnarray}
где $\frac{\tilde{A}_3}{g_2} = \frac{1}{g_2}{\rm Tr}\, A \sigma^3$,
$\frac{\tilde{B}}{g_1} = 2 B/g_1$,  $\frac{\tilde{Z}}{g_z}$,
$\frac{\tilde{A}}{e}$ - привычные поля Стандартной Модели, а $g_z =
\sqrt{g_1^2+g_2^2}$.

В области пространства - времени, в которой $\Phi = 0$ обычные поля Стандартной
Модели (\ref{FSM}) не определены. Такие области формируют двумерные
поверхности, являющиеся мировыми поверхностями так называемых Z - струн, если
при обходе вдоль малого контура вокруг струны поле $Z$ набирает ненулевую фазу,
кратную $2\pi$. Монополи Намбу определяются как концы  $Z$-струны \cite{Nambu}.
С классической точки зрения $Z$-струна есть нестабильное решение уравнений
движения, характеризующаяся магнитным потоком, извлекаемым из поля $Z$-бозона.
А именно, для малых контуров $\cal C$ вокруг $Z$ - струны должно быть
\begin{equation}
 \int_{\cal C} Z^{\mu} dx^{\mu} \sim 2\pi;\,
 \int_{\cal C} A_{\rm EM}^{\mu} dx^{\mu} \sim 0;\,
 \int_{\cal C} B^{\mu} dx^{\mu} \sim 2\pi {\rm sin}^2\, \theta_W .
\end{equation}
Струна оканчивается в месте нахождения монополя Намбу. Поток поля гиперзаряда
$B$ сохраняется в этой точке \footnote{На классическом уровне монополеподобные
объекты с ненулевым потоком поля $B$ имеют бесконечную собственную энергию. На
квантовом уровне такие объекты присутствуют в решеточной теории при конечных
значениях длины ребра решетки. Однако, в физически интересной области плотность
этих объектов также зануляется (см. Главу 6).}. Поэтому монополь Намбу несет
электромагнитный поток $4\pi {\rm sin}^2\, \theta_W$. Явный вид классического
решения уравнений движения модели Вайнберга - Салама, соответствующего $Z$ -
струне, так же как и подробное обсуждение вопросов стабильности струн
интересующийся читатель может найти в \cite{Nambu}.

Размер монополя Намбу оценивался в \cite{Nambu} и оказался близок к обратной
массе $Z$-бозона, в то же время его масса должна быть в районе нескольких ТэВ.
В соответствии с \cite{Nambu} монополи Намбу появляются только в форме
связанного состояния монополь - антимонопольной пары.

В решеточной теории классические решения, соответствующие $Z$-струне должны
формироваться вокруг $2$-мерного топологического дефекта, который представлен
целочисленным полем, определенным на дуальной решетке
\begin{equation}
 \Sigma = \frac{1}{2\pi}^*([d Z^{\prime}]_{{\rm mod} 2\pi} - d Z^{\prime})
\end{equation}
Тогда, $\Sigma$ может рассматриваться как мировая поверхность {\it квантовой}
$Z$-струны \cite{Chernodub_Nambu,Chernodub}.

Далее, мировые линии монополей - это границы мировой поверхности  $Z$- струны:
\begin{equation}
 j_Z = \delta \Sigma
\label{jN}
\end{equation}
Решеточная модель становится $U(1)$ калибровочной моделью после фиксации
унитарной калибровки. Соответствующее компактное $U(1)$ поле представлено в
~(\ref{AZW}).  Поэтому можно извлекать монопольные траектории непосредственно
из $A$:
\begin{equation}
 j_{A} = \frac{1}{2\pi} {}^*d([d A]{\rm mod}2\pi)
\label{Am}
\end{equation}
И $j_Z$, и $j_A$ представляют объекты, несущие магнитный заряд. Поэтому важно
установить соотношение между ними. Имеем
\begin{equation}
 A  =  [ Z^{\prime} + 2 \theta]{\rm mod}2\pi .
\end{equation}
или непрерывных обозначениях
\begin{equation}
 A^{\mu}  =  Z^{\mu} + 2 B^{\mu},
\end{equation}
где $B$ - гиперзарядовое поле. Магнитный поток, соответствующий $B$ зануляется.
Поэтому полный $Z$ поток, исходящий из центра монополя равен полному потоку $A$
поля. (Как $A$, так и $Z$ не определены внутри монополя.) Таким образом, в
непрерывном пределе положение монополя Намбу должно совпадать с положением
монополя, извлеченного из поля  $A$. В результате мы должны рассматривать
~(\ref{Am}) как иное определение квантового монополя Намбу. Интересно, что
определение  (\ref{Am}) не связано непосредственно ни с какой наблюдаемой
струной, поскольку соответствующая струна Дирака, соединяющая решеточные
монополи, извлеченные из $A$ - невидима.

\section{Численные результаты}

\begin{figure}
\begin{center}
\epsfig{figure=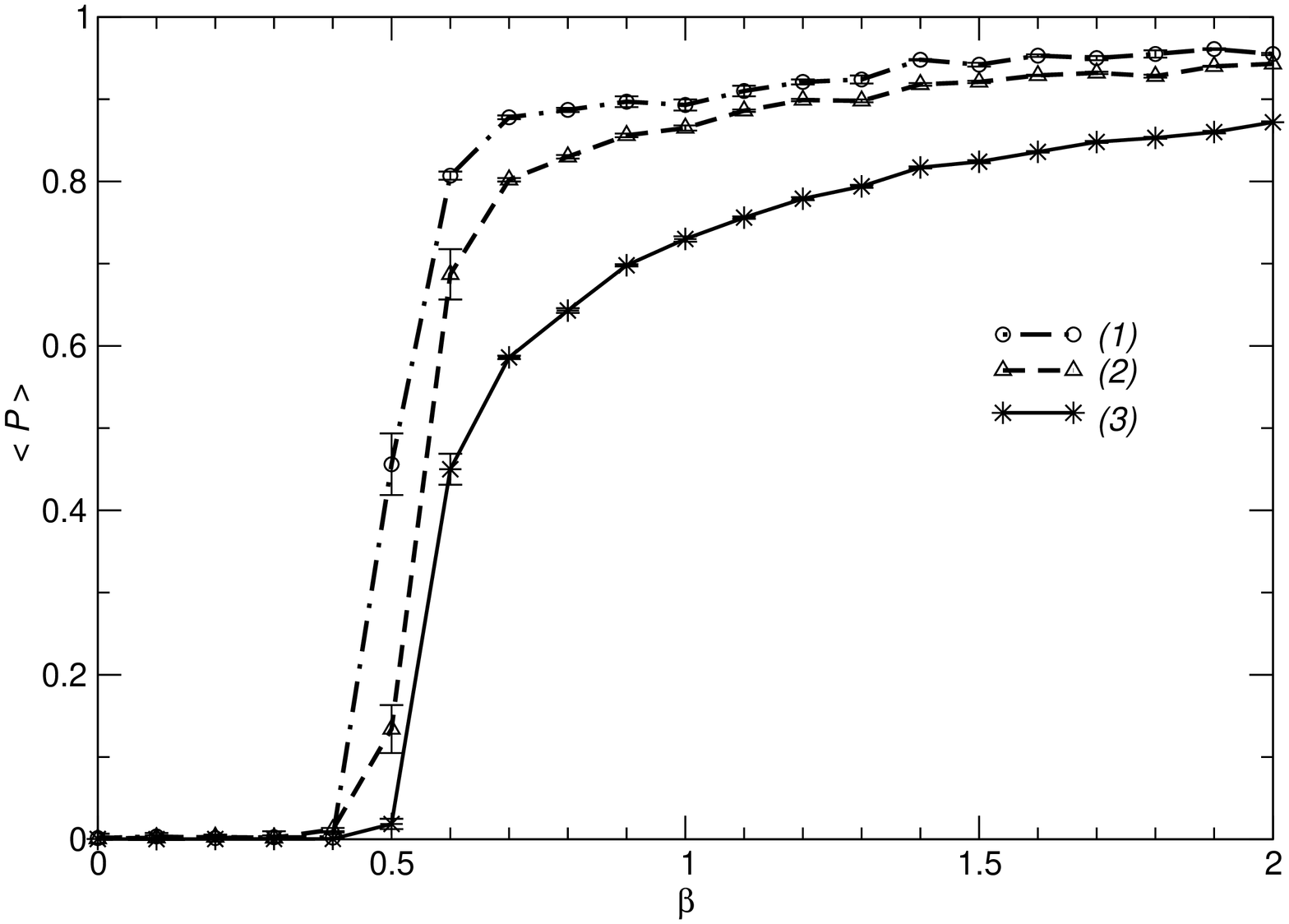,height=100mm,angle=-90}

\vspace{4ex}

\caption{\label{f5.2} Поляковские линии как функция $\beta$ для фиксированного
значения $\gamma = 0.5$. (1) соответствует ${\cal P}^{\rm R}_{\rm lept}$; (2)
соответствует ${\cal P}^{\rm L}_{\rm lept}$; (3) соответствует ${\cal P}^{\rm
L}_{\rm quarks}$.}

\end{center}
\end{figure}

Фазовая диаграмма нашей модели (Рис. \ref{f5.1}) существенно отличается от
фазовой диаграммы обычной решеточной регуляризации. А именно, наша модель
содержит три фазы. Первая (I) - фаза конфайнмента, где $SU(3)$ поля создают
удерживающие силы для кварков, а также наблюдаются удерживающие силы между
лептонами. Впрочем, струна между лептонами рвется на некоторых расстояниях. В
следующей фазе (II) удерживающих сил нет, но монополи Намбу сконденсированы.
Последняя часть диаграммы (III)  соответствует низкотемпературной физике, где
поле Хиггса сконденсировано. В этой части фазовой диаграммы монополи Намбу не
сконденсированы и их плотность быстро падает до нуля. Далее мы будем называть
части диаграммы II и III фазами, хотя переход между ними вполне может оказаться
кроссовером.

Увеличение температуры соответствует переходу из фазя III в фазу II. Физическая
температура выражается как $T = \frac{1}{a N_T}$, где $N_T$ - это размер
решетки по направлению мнимого времени, $a$ - длина ребра решетки, зависящая от
значений констант связи. Мы обозначаем значение температуры в точке перехода
как $T_c$. Здесь  конденсат монополей Намбу играет роль параметра порядка. Мы
не наблюдаем зануление конденсата монополей Намбу в фазе II с увеличением
размера решетки. Таким образом, мы полагаем, что монополи сконденсированы в
непрерывной теории при  $T>T_c$. Это наблюдение находится в соответствии с
предположением, сделанным в рамках $SU(2)$ модели Хиггса в
\cite{Chernodub_Nambu} \footnote{В \cite{Chernodub_Nambu} было показано, что в
определенном пределе значений констант связи  $SU(2)$ модель Хиггса становится
эквивалентной модели Джорджи - Глэшоу. Тогда монополи т'Хофта - Полякова могут
быть отождествлены с монополями Намбу. И, как следствие, конденсация монополей
т'Хофта - Полякова в симметричной фазе модели Джорджи - Глэшоу означает, что,
как минимум, в этом конкретном пределе значений констант связи монополи Намбу
сконденсированы также и в  $SU(2)$ модели Хиггса.}. В \cite{EW_T} было
показано, что электрослабый фазовый переход является в действительности
кроссовером при допустимых значениях массы Хиггса. Это привело к заключению о
том, что барионная асимметрия не могла возникнуть во время электрослабого
фазового перехода, как было предложено в \cite{Rubakov}. Наше исследование
показывает, однако, что структура вакуума ниже и выше перехода существенно
различна. Электрослабый переход является перколяционным и конденсат монополей
Намбу является параметром порядка. Ситуация здесь может быть аналогична
ситуации в $3D$ компактной  $U(1)$ решеточной модели Хиггса \cite{Janke}, где
фазовая линия состоит из линии перехода первого рода при малых константах связи
поля Хиггса и заканчивается в критической точке. Далее фазовая линия
продолжается как линия Кертежа, на которой термодинамические величины
неразрывны. Следует отметить, что в $3D$ $SU(2)$ модели Хиггса \cite{Chernodub}
было найдено, что Z-вихри перколируют при $T>T_c$ в то время, как при $T<T_c$
перколяция отсутствует. По этой причине в ~\cite{Chernodub} этот переход был
назван перколяционным. Это также находится в соответствии с нашими
наблюдениями, поскольку в $4D$ Z-вихри оканчиваются на монополях Намбу.

\begin{figure}
\begin{center}
\epsfig{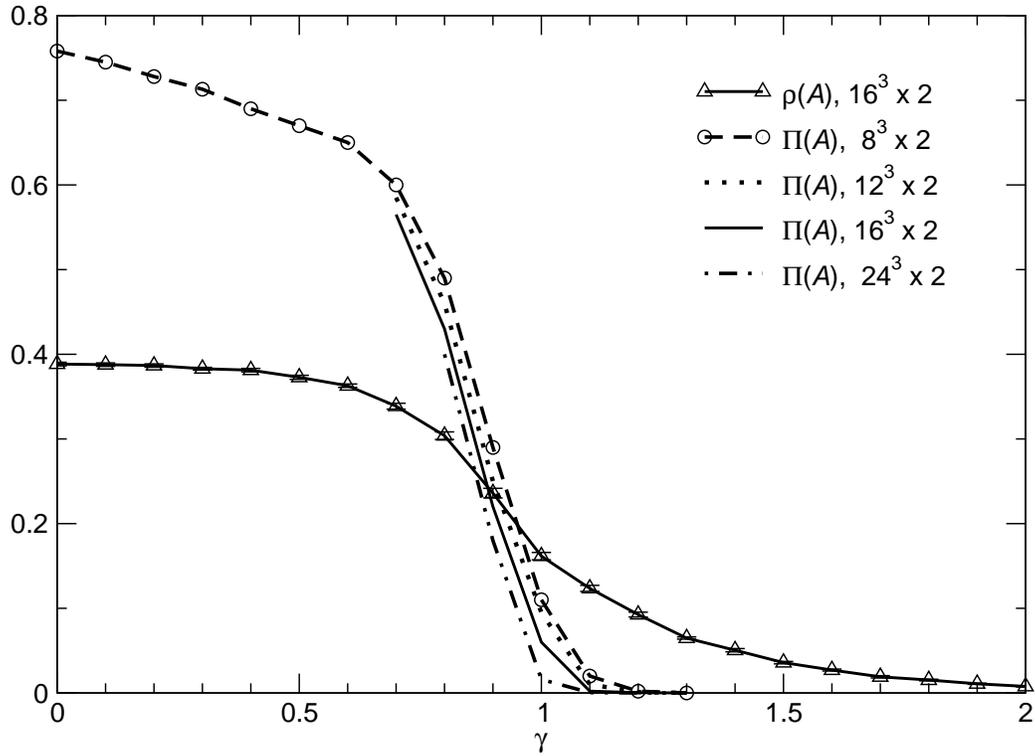}

\vspace{3ex} \caption{\label{f5.3} Плотность монополей Намбу $\rho(A)$ и
перколяционная вероятность $\Pi(A)$ как функция $\gamma$ для $\beta=0.6$.}

\end{center}
\end{figure}

Ниже рассматривается модель с включением $SU(3)$ полей, описанная в предыдущей
главе. (Там же были представлены результаты исследования этой модели при
нулевой температуре.) Мы используем несимметричные решетки с размером вдоль
направления мнимого времени $2$ и $4$ и пространственным размером от $8^3$ до
$24^3$. Калибровочная часть действия имеет вид
\begin{eqnarray}
 S_g & = & \beta \sum_{\rm plaquettes}  \{2(1-\mbox{${\small
 \frac{1}{2}}$} {\rm Tr}\, U_p \cos \theta_p) +
\nonumber \\
 && +\;(1-\cos 2\theta_p)
\nonumber \\
 && +\;6[1-\mbox{${\small \frac{1}{6}}$} {\rm Re Tr}  \,
 \Gamma_p {\rm Tr}\, U_p {\rm exp} (i\theta_p/3)]
\nonumber\\
 && +\;3[1-\mbox{${\small \frac{1}{3}}$} {\rm Re Tr}  \,
 \Gamma_p {\rm exp} (-2i\theta_p/3)]
\nonumber \\
 && +\;3[1-\mbox{${\small \frac{1}{3}}$} {\rm Re Tr}  \,
 \Gamma_p {\rm exp} (4i\theta_p/3)]\},
\label{Act_5}
\end{eqnarray}
где сумма -  по плакетам, а
\begin{eqnarray}
 \Gamma \in SU(3), \quad U \in SU(2), \quad e^{i\theta} \in U(1).
\end{eqnarray}
Каждый член в ~(\ref{Act_5}), соответствует оператору параллельного переноса
вдоль границы $\partial p$ плакета $p$.

Действие для скалярного поля рассматривается в его простейшем виде в
Лондоновском пределе. После фиксации унитарной калибровки мы получаем:
\begin{eqnarray}
 S_H & = & \gamma \sum_{xy}[1 - Re(U^{11}_{xy} e^{i\theta_{xy}})].
\end{eqnarray}
Здесь $\gamma = v^2$, где $v/\sqrt{2}$ - это затравочное вакуумное среднее
скалярного поля. Мы пренедрегаем вкладом динамических фермионов. Поэтому полное
действие нашей модели имеет вид $S = S_g + S_H$.

Соотношения между затравочными константами связи $\alpha_s$ (константа сильного
взаимодействия кварков), $\alpha$ (постоянная тонкой структуры), а также
$\theta_W$ (Угол Вайнберга) в нашей модели равны соответствующим выражениям в
$SU(5)$ модели объединения.
 Мы в нашем численном исследовании
изучаем область фазовой диаграммы, в которой  значения $\alpha$ в районе $1/50$
и $\alpha_s \sim 1/20$. Это связано с техническими проблемами, возникающими при
симуляции модели с большим значением $\beta$, соответствующем малому значению
$\alpha$. Кроме того, если константы связи, соответствующие разным подргуппам
калибровочной группы на порядок отличаются друг от друга (как это имеет место в
реальной Стандартной Модели, где $\alpha_s \sim 10 \alpha$), то возникают
дополнительные трудности в работе численного алгоритма. Эта проблема не
возникает в рассматриваемой нами области констант связи, гда $\alpha_s \sim
\alpha$. Не смотря на то, что мы, таким образом, изучаем область фазовой
диаграммы, находящуюся на некотором отдалении от физической (например,
используемые нами значения $\alpha$ в $2.5$ раза меньше ее физического
значения), как ожидается, качественно поведение монополей Намбу при
реалистических значениях констант связи не должно отличаться от обнаруженного
нами.

В оправдание нашего выбора значений констант связи следует сказать, что
большинство численных симуляций Электрослабой модели, выполненных к настоящему
моменту, использовали нереалистические значения констант связи. Вообще говоря,
последовательное изучение решеточной модели должно включать изучение ее в
области констант связи, где алгоритм работает быстро и позволяет изучать все
области фазовой диаграммы. При этом качественно свойства модели внутри одной
фазы совпадают для малых и больших значениях констант связи. Такое исследование
должно необходимо предварять исследование при реалистических значениях. Именно
этот предварительный этап исследования мы и выполнили при изучении свойств
монополей Намбу при конечных температурах. Нами также изучались свойства
монополей Намбу при более реалистических значениях постоянной тонкой структуры,
но при нулевой температуре. Эти результаты будут обсуждаться в следующей главе.

Все перенормированные константы связи должны вычисляться с использованием
статических кварковых и лептонных потенциалов. Тогда линии постоянной физики
(ЛПФ) в пространстве затравочных параметров (включая размер решетки по мнимому
времени) определяются как линии, на которых перенормированные константы связи
(при нулевой температуре) в физических единицах постоянны \cite{Montvay,EW_T}.
Мы не исследуем последовательно приближение решеточной модели  к непрерывной
физике \footnote{Следует отметить, что такое исследование может потребовать
включения в рассмотрение при исследовании линий постоянной физики различных
затравочных констант связи для $SU(2)$, $SU(3)$, $U(1)$ полей так же, как и
затравочной массы Хиггсовского бозона.}. Рассмотрение ЛПФ также необходимо для
определения соответствия между фазовой диаграммой в плоскости  $\beta$ -
$\gamma$ и привычной фазовой диаграммой в плоскости  $T - M_H$. Однако,
возникновение температуры на фазовой диаграмме представленной на рис.
\ref{f5.1} может быть понято на основании древесного приближения для длины
ребра решетки и массы Z - бозона. А именно, в фазе III имеем $M_Z = g_z v/2$ в
решеточных единицах. Здесь $g_z = g_2/{\rm cos} \, \theta_W$, $8 \beta =
\frac{4}{g_2^2}$, и ${\rm cos}^2 \, \theta_W = \frac{5}{8}$. Тогда $M_Z \sim
\sqrt{\frac{\gamma}{5\beta}}$. Далее, длина ребра решетки в физических единицах
ракна  $a = M_Z /M^{\rm phys}_Z$, где $M^{\rm phys}_Z$ - около $90$ ГэВ. Тогда,
скажем, при $\beta = 0.6$, $\gamma = 1.5$ древесное приближение для длины ребра
решетки дает $a \sim (130\, {\rm GeV})^{-1}$, а при $\beta = 0.8$, $\gamma =
1.0$ имеем $a \sim (180\, {\rm GeV})^{-1}$. Наконец, температура оценивается
как $T = \frac{1}{N_Ta} \sim \sqrt{\frac{5\beta}{\gamma N_T^2}}\, 90\, {\rm
GeV}$. Хотя последнее выражение должно быть модифицировано с использованием
решеточной ренормализационной группы, оно показывает, что температура
увеличивается с увеличением  $\beta$ и уменьшением $\gamma$ и $N_T$. В
древесном приближении ЛПФ соответствует постоянному $\beta$, фиксирующему
значение $\alpha$, $\alpha_s$ и $\theta_W$. Поэтому при заданном $N_T$
увеличение температуры соответствует уменьшению $\gamma$, то есть переходу на
фазовой диаграмме (в плоскости $\beta - \gamma$ из фазы III в фазу II). Таким
образом, вертикальная линия на Рис. \ref{f5.1}, соответствующая постоянному
$\beta$, в древесном приближении соответствует $M_H = \infty$, $\alpha =
\frac{3}{64\pi \beta} = const$, $\alpha_s = \frac{1}{8\pi \beta} = const$,
${\rm sin}^2 \theta_W = 3/8$. При этом движение сверху вниз по этой линии
соответствует уменьшению длины ребра решетки и увеличению температуры.
Квантовые поправки несколько изменяют эту картину. Все перечисленные физические
величины меняются при движении вдоль указанной линии. Однако, как показывают
наши исследования, изменения $\alpha_s$, $\alpha$, и ${\rm sin}^2 \theta_W $
малы. Что касается перенормированной массы Хиггса, она уже не бесконечна. Как
будет показано в следующей главе в области значений $\alpha$ вокруг ее
физического значения $1/128$ вблизи линии фазового перехода между фазами II и
III значение массы Хиггса составляет $\sim 9 M_Z$.

В нашей модели затравочный заряд $e^2 = g^2 {\rm sin}^2 \, \theta_W =
\frac{3}{16 \beta}$; его экспериментальное значение $\alpha =
\frac{e^2(M_Z)}{4\pi}\sim \frac{1}{128}$. Поэтому для получения реалистического
значения постоянной тонкой структуры  $\beta \sim \frac{6}{\pi} \sim 1.9$. Наша
оценка для критического $\gamma$ при $N_T = 2$, $\beta = 1.9$ есть $\gamma_c
\sim 0.9$. Поэтому в Лондоновском пределе для Хиггсовского потенциала c помощью
древесного приближения можно оценить критическую температуру при $\beta = 1.9$
как $T_c \sim 150$ ГэВ. Разумеется, это весьма грубая оценка для критической
температуры \footnote{Это значение следует сравнить со значением, вычисленным в
$SU(2)$ модели Хиггса в \cite{Rum}. Там при $\alpha \sim 1/100$, $M_H = 120$
ГэВ $T_c$ было обнаружено в районе $210$ ГэВ, а для $M_H = 180$ ГэВ $T_c \sim
250$ GeV}. Как уже было сказано выше, вычислительные мощности не позволили нам
выполнить иследование модели в окрестности $\beta \sim 1.9$. Вместо этого мы
исследовали область $\beta \in (0.6, 0.8)$ в фазе III. Это соответствует
значениям $\alpha$ в районе $1/50$ и $\alpha_s \sim 1/20$. В этой области
грубая древесная оценка дает значение критической температуры в районе $90$
ГэВ.

Мы изучаем различные Поляковские линии, соответствующие бесконечно тяжелым
кваркам и лептонам:
\begin{eqnarray}
 {\cal P}^{\rm L}_{\rm lept} & = &
 \langle {\rm Re} {\rm Tr} \,\Pi_{(xy) \in l} U_{xy}  e^{-i\theta_{xy}}\rangle, \nonumber\\
 {\cal P}^{\rm R}_{\rm lept} & = & \langle {\rm Re} \Pi_{(xy) \in l} \,
 e^{-2i\theta_{xy}}\rangle ,
\nonumber\\
 {\cal P}^{\rm L}_{ {\rm quarks}} & = & \langle {\rm Re} \Pi_{(xy) \in l}  \,
 \Gamma_{xy} \, U_{xy}\, e^{\frac{i}{3}\theta_{xy}}\rangle ,
\nonumber\\
 {\cal P}^{\rm R}_{{\rm down} \, {\rm quarks}} & =  &
 \langle {\rm Re} \Pi_{(xy) \in l} \, \Gamma_{xy} \,
 e^{-\frac{2i}{3}\theta_{xy}}\rangle ,
\nonumber\\
 {\cal P}^{\rm R}_{{\rm up} \, {\rm quarks}} & = &
 \langle {\rm Re} \Pi_{(xy) \in l} \, \Gamma_{xy} \,
 e^{\frac{4i}{3}\theta_{xy}}\rangle .
\label{WL}
\end{eqnarray}
Здесь $l$ обозначает линию на решетке, направленную вдоль оси мнимого времени,
замкнутую через граничные условия.

Мы выяснили, что левее вертикальной линии фазового перехода все Поляковские
линии зануляются в то время, как справа от этой линии все они отличны от нуля
(рис.~\ref{f5.2}).

Нами исследуется несколько видов монополей, построенных из цветных полей (см.
предыдущую главу). Также исследуются монополи, извлекаемые из поля $2\theta$
(Монополи гиперзаряда), и монополи Намбу. Мы находим, что плотность цветных
монополей и монополей гиперзаряда резко падает справа от вертикальной линии
фазового перехода на рис. \ref{f5.1}. Это (вместе с поведением Поляковских
линий) говорит о том, что данная линия является линией перехода конфайнмент -
деконфайнмент, общей для $SU(2), SU(3)$ и $U(1)$ полей. Поведение полного
действия и монопольных плотностей обнаруживают здесь гистерезис, указывающий,
что это фазовый переход первого рода.

Плотность монополей Намбу отлична от нуля в фазе II. В фазе III их плотность
быстро падает. Мы видим, что перколяционная вероятность монополей Намбу
является параметром порядка для перехода между фазами II и III (см. рис.
\ref{f5.3}). Положение этого перехода также совпадает с максимумом
восприимчивости $\chi = \langle H^2 \rangle - \langle H\rangle^2$, извлекаемой
из оператора рождения Хиггсовского бозона $H_x = \sum_{y} |W_{xy}|^2$ или $H_x
= \sum_{y} Z^2_{xy}$. В то же время мы не наблюдаем резкого изменения действия
вблизи перехода. Корреляционные длины для различных корреляторов полей не
стремятся к бесконечности в точке перехода. Все это указывает на то, что мы
имеем дело с кроссовером, или с переходом, названным в \cite{Chernodub}
перколяционным.

Исследование различных размеров решеток   (от $8^3\times 2$ до $24^3\times 4$)
показывает, что в фазе II ни плотность, ни перколяция монополей Намбу не
уменьшаются с увеличением размера решетки. При этом ряд точек на фазовой
диаграмме в этой фазе проверялся нами вплоть до значений $\beta \sim 2.0$,
соответствующих $\alpha \sim 1/130$, где также не обнаружили зануление
конденсата. Таким образом, мы обнаружили, что в рассматриваемой модели монополи
Намбу сконденсированы при $T>T_c$.

\section{Публикации}

Результаты настоящей главы опубликованы в работе:

"Z(6) symmetry, electroweak transition, and magnetic monopoles at high
temperature. By B.L.G. Bakker", A.I. Veselov, M.A. Zubkov,
Phys.Lett.B642:147-152,2006, [hep-lat/0606010]

\chapter{Флуктуационная область в модели Вайнберга - Салама}
\label{ch6}

\section{Введение}

В этой главе мы излагаем результаты наших численных исследований модели
Вайнберга - Салама при нулевой температуре и при реалистических значениях
констант связи. Нами исследуется подробно область фазовой диаграммы,
соответствующая затравочным значениям ($\alpha \sim 1/150, \theta_W = \pi/6$,
$\lambda = 0.0025$, что соответствует  $M_H \sim 150$ ГэВ)\footnote{Отличие
затравочного значения постоянной тонкой структуры от эксперимнтального значения
$\alpha(M_Z) \sim 1/128$ обусловлено тем, что на старте исследования при
вычислении перенормированной постоянной тонкой структуры на  решетках малого
размера нами не учитывались последовательно эффекты конечного объема. В
результате вычисленная перенормированная $\alpha_R$ содержала зависимость от
размера решетки, а ее значение для решеток используемого размера отличалось
существенно от затравочного. Это и обусловило выбор затравочного значения
$\alpha \sim 1/150$ с тем, чтобы значение $\alpha_R$ на решетках используемого
размера было близко к $1/128$.  Последовательный учет эффектов конечного объема
(см. далее текст данной главы) в  $\alpha_R$ показал, что эта величина может
быть вычислена с хорошей точностью даже на решетках очень небольшого размера, а
ее отклонение от затравочного значения весьма незначительно.}. Кроме того, ряд
результатов получен также для и для некоторых других затравочных значений
констант. Однако, прежде, чем изложить наши результаты, мы произведем небольшой
экскурс в историю и кратко расскажем о том, какие основные численные результаты
были получены в рамках предыдущих исследований решеточной модели Вайнберга -
Салама.

Первые численные исследования Модели Вайнберга - Салама проводились в 80 - х
годах 20 века и в основном ограничивались моделью без $U(1)$ составляющей, то
есть $SU(2)$ моделью Хиггса при нулевой температуре  (см., например,
 \cite{Jersak,Montvay,12,13,14} и ссылки внутри текстов этих работ).
В этих исследованиях была установлена фазовая структура решеточной модели. При
малых $\lambda$ был обнаружен фазовый переход первого рода между физической
Хиггсовской фазой (где сконденсировано скалярное поле) и нефизической
симметричной фазой, где конденсат отсутствует. При этом установлено так
называемое абсолютное ограничение на массу Хиггса (достигаемое при $\lambda =
\infty$): $M_H \le 10 M_W$ (см., например, \cite{12,13,14}). Исследование
модели с включением $U(1)$ полей в рамках $SU(2)\otimes U(1)$ модели Хиггса
производилось в \cite{SU2U1}, где была дана фазовая диаграмма модели. Однако, в
\cite{SU2U1} численное исследование было ограничено нефизически большими
значениями $\alpha$. До работ, изложенных в настоящей главе исследования модели
с включением $U(1)$ поля при реалистических значениях констант связи не
производилось.

Следующий этап численных исследований модели Вайнберга - Салама был мотивирован
возможной связью физики Электрослабого перехода при конечной температуре с
космологией, в частности, с проблемой барионной асимметрии (см., например,
\cite{1,2,3,4,5,6,7,8,9,10,11,EW_T} ). Одно из важных достижений этих
исследований  -  обнаружение точки на фазовой диаграмме (в плоскости $T -
M_H$), в которой фазовый переход первого рода переходит в кроссовер. Было
обнаружено, что при экспериментально допустимых значениях массы Хиггса
Электрослабый переход не является фазовым переходом первого или второго рода, а
относится к разряду кроссоверов.

В  $4D$ решеточных исследованиях $SU(2)$ модели Хиггса шкала фиксируется
значением массы $W$-бозона. А именно, масса в решеточных единицах  $M_W = a \,
\times \, 80 $ ГэВ, где $a$ это значение длины ребра решетки. Ультрафиолетовое
обрезание тогда равно $\Lambda = \frac{\pi}{a}$. Для справочных целей в таблице
\ref{6Table} мы приводим значения ультрафиолетового обрезания в некоторых
численных исследованиях $SU(2)$ модели Хиггса. Во всех цитируемых работах
$\alpha \sim 1/100$. Из указанной таблицы мы можем почерпнуть максимальное
достигнутое до настоящего времени значение ультрафиолетового обрезания в районе
$1.5$ ТэВ. (В \cite{1} длина ребра решетки различна для пространственного
направления и для направления вдоль мнимого времени.)

\begin{table}
\label{6Table} \caption{Значения ултрафиолетового обрезания в некоторых работах
по численному исследованию $SU(2)$ модели Хиггса.}
\begin{center}
\begin{tabular}{|c|c|c|}
\hline
{\bf ссылка}  & $\frac{\pi}{a}$ (ГэВ) & {\bf $M_H$} (ГэВ)\\
\hline
\cite{1}  & 440 (1800 для "временного" направления) & 80 \\
\hline \cite{2}  & 880 (для "временного" направления) & 80 \\
\hline \cite{3}  & 880 & 34 \\
\hline \cite{4}  & 350 & 16 \\
\hline \cite{5}  & 280 (1100 для "временного" направления) & 34 \\
\hline \cite{6}  & 880 & 48 \\
\hline \cite{7}  & 440 & 35 \\
\hline \cite{8}  & 880 & 20 , 50 \\
\hline \cite{9}  & 600 & 50 \\
\hline \cite{10}  & 820 & 57 - 85 \\
\hline \cite{11}  & 630 - 940 & 47 - 108 \\
\hline \cite{12}  & 1250 & 480 \\
\hline \cite{13}  & 1030 -  1470 & 280 - 720
( $\lambda =\infty$ и конечное $\lambda$) \\
\hline \cite{14}  & 780 -  1470 &  720
($\lambda =\infty$ и конечное  $\lambda$) \\
\hline
\end{tabular}
\end{center}
\end{table}

Отдельно заслуживает внимания вопрос о необходимости применять непертурбативные
и, в частности, решеточные методы к исследованию Электрослабой теории. К
сожалению в настоящее время достаточно широко распространено заблуждение, что в
Электрослабой теории все может быть вычислено с помощью обычной теории
возмущений. Однако, как уже упоминалось в предыдущей главе, известно, что в
теории при конечной температуре теория возмущений перестает работать при
температурах вблизи температуры Электрослабого перехода для физически
допустимых значений массы Хиггса \cite{M_W_T}. В теории при нулевой температуре
необходимость применения непертурбативных методов следует из того, что в теории
существуют нетривиальные классические решения (а именно, Z - вихри и монополи
Намбу). Разумеется, эти объекты не могут быть изучены на основе теории
возмущений вокруг тривиального вакуума. Масса монополя Намбу оценивалась в
\cite{Nambu} и оказалась в районе ТэВ. Поэтому при энергиях много меньших $1$
ТэВ теория возмущений должна быть применима, что и было продемонстрировано в
многочисленных исследованиях, приводящих к замечательному совпадению с
экспериментом. Однако, при повышении энергии роль монополей Намбу и Z вихрей
повышается, что и обусловливает применение непертурбативных методов. В работах
автора настоящей диссертации и его соавторов обнаружены указания на то, что
 в области энергий порядка $1$ ТэВ монополи Намбу доминируют в вакууме и,
 соответственно, теория возмущений вокруг тривиального вакуума не может быть
 применима.

Следует отметить, что на неприменимость  самой модели Вайнберга - Салама в этой
области энергий указывает так называемая проблема иерархий \cite{TEV}. А
именно, массовый параметр  $\mu^2$ скалярного поля получает квадратично
расходящуюся поправку в однопетлевом приближении теории возмущений. Поэтому
затравочный массовый параметр ($\mu^2= - \lambda_c v^2$, где $v$ - это
вакуумное среднее скалярного поля) должен быть задан бесконечным, чтобы
перенормированный параметр $\mu^2_R$ оставался отрицательным и конечным. Это
приводит к необходимости "тонкой подстройки" затравочного параметра и
Ультрафиолетового обрезания, которая считается во многих работах неестественной
\cite{TEV}. Поэтому многие полагают, что необходимо задать конечное значение
ультрафиолетового обрезания $\Lambda$. Из требования, что однопетлевая поправка
к  $\mu^2$ не должна превышать  $10 |\mu_R^2|$ выводим  $\Lambda \sim 1$ TeV.
Таким образом, из соображений "естественности" следует, что некая новая физика
должна появиться на масштабе $1$ ТэВ.

Следует отметить, что в самой теории возмущений есть также более строгое
ограничение на значение Ультрафиолетового обрезания. Оно появляется как
следствие проблемы тривиальности, связанной с полюсом Ландау в
перенормированной константе самодействия поля Хиггса  $\lambda$ и в
перенормированной постоянной тонкой структуры $\alpha$. Полюс Ландау в
постоянной тонкой структуры связан с фермионными петлями и не имеет
непосредственного отношения к исследуемому нами пределу модели, в котором мы
пренебрегаем динамическими фермионами. Благодаря полюсу Ландау
перенормированная  $\lambda$ зануляется и единственный способ удержать ее
равной ее измеряемому значению заключается в том, что необходимо наложить
ограничение сверху на Ультрафиолетовое обрезание $\Lambda$. В частности, при
значениях массы Хиггса более $800$ ГэВ $\Lambda < 1$ ТэВ, а при $M_H \sim 300$
ГэВ $\Lambda < 1000$ ТэВ. При $M_H \sim 200$ ГэВ значение $\Lambda$ практически
оказывается неограниченным.

Физическая шкала фиксируется нами, используя массу $Z$-бозона $M^{\rm phys}_Z
\sim 90$ ГэВ. Тогда длина ребра решетки равна $a \sim [90\,{\rm GeV}]^{-1}
M_Z$, где $M_Z$ - это масса $Z$ бозона в решеточных единицах. Внутри физической
фазы теории линии постоянной физики (ЛПФ) определяются как соответствующие
постоянным перенормированным физическим постоянным.  (постоянная тонкой
структуры $\alpha$, угол Вайнберга $\theta_W$, и отношение массы Хиггса к массе
 Z-бозона $\eta = M_H/M_Z$). Точки ЛПФ параметризуются длиной ребра решетки.
 При увеличении ультрафиолетового обрезания при движении вдоль ЛПФ, соответствующей реалистическим значениям
 $\alpha$, $\theta_W$, и $\eta$ происходит приближение к фазовому переходу между физической фазой и
 нефизической фазой.

\section{Описание решеточной модели и древесное приближение}

Ниже используются следующие решеточные переменные:

1. Калибровочное поле ${\cal U} = (U, \theta)$, где
\begin{eqnarray}
 \quad U = \left( \begin{array}{c c}
 U^{11} & U^{12}  \\
 -[U^{12}]^* & [U^{11}]^*
 \end{array}\right)
 \in SU(2), \quad e^{i\theta} \in U(1),
\end{eqnarray}
определенное на линках решетки.

2. Скалярный дублет
\begin{equation}
 \Phi_{\alpha}, \;\alpha = 1,2.
\end{equation}

Действие мы рассматриваем в следующем виде
\begin{eqnarray}
 S & = & \beta \!\! \sum_{\rm plaquettes}\!\!
 ((1-\mbox{${\small \frac{1}{2}}$} \, {\rm Tr}\, U_p )
 + \frac{1}{{\rm tg}^2 \theta_W} (1-\cos \theta_p))+\nonumber\\
 && - \gamma \sum_{xy} Re(\Phi^+U_{xy} e^{i\theta_{xy}}\Phi) + \sum_x (|\Phi_x|^2 +
 \lambda(|\Phi_x|^2-1)^2), \label{S}
\end{eqnarray}
где плакетные переменные определяются как $U_p = U_{xy} U_{yz} U_{wz}^*
U_{xw}^*$, и $\theta_p = \theta_{xy} + \theta_{yz} - \theta_{wz} - \theta_{xw}$
для плакетах, составленных из точек $x,y,z,w$. Здесь $\lambda$ - постоянная
скалярного поля, и $\gamma = 2\kappa$, где $\kappa$ соответствует константе,
используемой в предыдущих исследованиях $SU(2)$ модели Хиггса. $\theta_W$ -
угол Вайнберга. Затравочная постоянная тонкой структуры $\alpha$ выражается
через $\beta$ и $\theta_W$ как
\begin{equation}
\alpha = \frac{{\rm tg}^2 \theta_W}{\pi \beta(1+{\rm tg}^2
\theta_W)}.\label{alpha}
\end{equation}

 Перенормированный угол Вайнберга должен выражаться через отношение решеточных масс:
   ${\rm cos} \, \theta_W = M_W/M_Z$. Перенормированная постоянная тонкой структуры
   должна быть извлечена из потенциала бесконечно тяжелых внешних заряженных частиц.

Решеточная модель с действием (\ref{S}) исследовалась во многих работах. При
этом большинство из них имело дело с  $SU(2)$ моделью Хиггса, то есть при
$\theta_W= \pi/2$. Система с произвольным $\theta_W$ численно исследовалась при
нефизически большом значении $\alpha$ в \cite{SU2U1}.

При конечном $\lambda$ для того, чтобы исследовать ЛПФ необходимо менять
$\lambda$ одновременно с $\gamma$ для того, чтобы удержать отношение масс
$M_H/M_W$ постоянным. В древесном приближении ЛПФ соответствует постоянным
затравочных $\beta$ и $\theta_W$.

Древесное приближение в решеточной теории получается следующим образом. Мы
переопределяем скалярное поле как  $\tilde{\Phi} = \sqrt{\frac{\gamma}{2}}
\Phi$ и получаем:

\begin{eqnarray}
 S & = & \beta \!\! \sum_{\rm plaquettes}\!\!
 ((1-\mbox{${\small \frac{1}{2}}$} \, {\rm Tr}\, U_p )
 + \frac{1}{{\rm tg}^2 \theta_W} (1-\cos \theta_p))+\nonumber\\
 && + \sum_{xy} |\tilde{\Phi}_x - U_{xy} e^{i\theta_{xy}}\tilde{\Phi}_y|^2 + \sum_x (\mu^2 |\tilde{\Phi}_x|^2 +
 \tilde{\lambda} |\tilde{\Phi}_x|^4) + \omega , \label{S2}
\end{eqnarray}
где $\mu^2 = - 2(4+(2\lambda-1)/\gamma)$, $\tilde{\lambda} =
4\frac{\lambda}{\gamma^2}$, и $\omega = \lambda V$. Здесь  $V = L^3\times N_T$
- объем решетки, $L$ - пространственный размер, $N_T$ - размер по направлению
мнимого времени.

Для отрицательных  $\mu^2$ мы фиксируем унитарную калибровку
$\tilde{\Phi}_2=0$, ${\rm Im}\, \tilde{\Phi}_1 = 0$, и вводим вакуумное
значение $\tilde{\Phi}$: $\tilde{v} = \sqrt{\frac{\gamma}{2}}v =
\frac{|\mu|}{\sqrt{2\tilde{\lambda}}}$. При этом вакуумное значение поля $\Phi$
обозначаем $v$.  Мы также вводим скалярное поле $\sigma$ вместо $\tilde{\Phi}$:
$\tilde{\Phi}_1 = \tilde{v} + \sigma$. Обозначим $V_{xy} =
(U^{11}_{xy}e^{i\theta_{xy}} - 1)$, и получаем:
\begin{eqnarray}
 S & = & \beta \!\! \sum_{\rm plaquettes}\!\!
 ((1-\mbox{${\small \frac{1}{2}}$} \, {\rm Tr}\, U_p )
 + \frac{1}{{\rm tg}^2 \theta_W} (1-\cos \theta_p))+\nonumber\\
 && + \sum_{xy} ((\sigma_x - \sigma_y)^2 + |V_{xy}|^2 \tilde{v}^2)  + \sum_x 2|\mu|^2 \sigma_x^2 \nonumber\\
 && + \sum_{xy} ((\sigma^2_y+2\tilde{v} \sigma_y)|V_{xy}|^2 - 2(\sigma_x - \sigma_y){\rm Re} V_{xy} (\sigma_y +\tilde{v}) ) + \nonumber\\
 && + \sum_x  \tilde{\lambda} \sigma_x^2 (\sigma_x^2 + 4 \tilde{v} \sigma_x) +  \tilde{\omega} , \label{S2}
\end{eqnarray}
где $\tilde{\omega} = \omega - \tilde{\lambda} \tilde{v}^4 V$.

Теперь мы легко получаем древесные оценки:
\begin{eqnarray}
v &=& \sqrt{2\frac{\gamma - \gamma_c}{\lambda}}\nonumber\\
M_H &=& v\sqrt{\frac{8\lambda}{\gamma}}\nonumber\\
M_Z &=& v\sqrt{\frac{\gamma}{\beta \, {\rm cos}^2 \theta_W}}\nonumber\\
\gamma_c &=& \frac{1 - 2\lambda}{4}\nonumber\\
M_W &=& {\rm cos}\theta_W M_Z\nonumber\\
M_H/M_W &=& \sqrt{8\lambda \beta/\gamma^2};\nonumber\\
\Lambda &=& \sqrt{\frac{\pi^2\lambda\beta}{2\gamma(\gamma-\gamma_c)}} \, [80\,
{\rm GeV}];\label{tree}
\end{eqnarray}
Постоянная тонкой структуры дается формулой (\ref{alpha}) и должна зависеть от
 $\lambda$ и $\gamma$. Из (\ref{tree}) получаем, что ЛПФ на древесном уровне соответствует фиксированным $\beta =
\frac{{\rm tg}^2 \theta_W}{\pi \alpha(1+{\rm tg}^2 \theta_W)} \sim 10 $ и $\eta
= M_H/M_W$, и дается уравнением $\lambda(\gamma) = \frac{\eta^2}{8\beta}
\gamma^2$.

Важный частный случай - это $\lambda = \infty$, когда древесные оценки дают
\begin{eqnarray}
M_H &=& \infty; \nonumber\\
M_W &=& \sqrt{\frac{\gamma}{\beta}}; \nonumber\\
M_Z &=& \sqrt{\frac{\gamma}{\beta}}{\rm cos}^{-1}\theta_W; \nonumber\\
\Lambda &=& \sqrt{\frac{\pi^2\beta}{\gamma}} \, [80\, {\rm GeV}];\label{treei}
\end{eqnarray}

В $SU(2)$ модели Хиггса для $\lambda << 0.1$ древесные оценки для  $M_H/M_W$
дают значения, которые отличаются от перенормированного отношения примерно на
 20\%\cite{11}.
Древесная оценка для ультрафиолетового обрезания около  $1$ ТэВ при $\lambda =
\infty,\gamma = 1, \beta = 15$ близка к нашему численному результату. В $SU(2)$
модели Хиггса при $\lambda = \infty$ критическое $\gamma_c = 0.63$ для $\beta =
8$ \cite{14}. В этой точке древесная оценка дает  $\Lambda = 900$ ГэВ в то
время, как прямые измерения дают $\Lambda \in [850; 1500]$ ГэВ для  $\gamma \in
[0.64; 0.95]$ \cite{14}.

При конечном  $\Lambda$ имеем оценку для критического $\gamma$: $\gamma_c =
(1-2\lambda)/4$. При малом $\lambda$ эта формула дает значения, близкие к
результатам численных расчетов \cite{12,13,14}. В частности, $\gamma_c
\rightarrow 0.25$ ($\kappa_c \rightarrow 0.125$) при $\lambda << 1$. Однако,
эта формула явно не работает пр $\lambda
> 1/2$. Из \cite{Montvay,12,13,14} мы знаем, что критическое значение в
$SU(2)$ модели Хиггса в  $2 - 4$ раза меньше для $\lambda =0$, чем для
$\lambda = \infty$.

При $\lambda < 1/2$ древесные оценки предсказывают, что ультрафиолетовое
обрезание возрастает вдоль ЛПФ с уменьшением $\gamma$, причем в точке перехода
она становится бесконечной. Тем самым древесное приближение предсказывает
существование фазового перехода второго рода. Однако, благодаря
непертурбативным эффектам эта картина может модифицироваться. А именно, фазовый
переход, может вырождаться в кросовер.

\section{Численное исследование модели при $\theta_W=\pi/6$, $M_H \sim 150$ ГэВ, $\alpha \sim 1/150$}
}

Ниже излагаются результаты численного исследования решеточной модели Вайнберга
- Салама без динамических фермионов для $\theta_W=\pi/6$, $M_H \sim 150$ ГэВ,
$\alpha \sim 1/150$. При этом мы показываем, что в окрестности перехода между
физической и нефизической фазами существует флуктуационная область, где
флуктуации скалярного поля значительно увеличиваются. Монополи Намбу могут
расматриваться как зародыши нефизической фазы внутри физической. Мы выяснили,
что в флуктуационной области расстояния между монополями Намбу становятся
сравнимы с их размерами. Это свидетельствует о том, что в этой области теория
возмущений вокруг тривиального вакуума может оказаться неприменима.
Рассматриваются решетки $8^3\times 16$, $12^3\times 16$, $16^4$, $16^3\times
32$. Максимальное значение ультрафиолетового обрезания в физической фазе, вне
Флуктуационной области, которое мы можем получить на этих решетках не может
превысить значения $\Lambda \sim 1 $ ТэВ. В то же самое время теория возмущений
предсказывает, что благодаря полюсу Ландау в константе самодействия скалярного
поля максимально допустимое значение ультрафиолетового обрезания существенно
выше.

Хорошо известно то, что в ряде феноменологических моделей, описывающих
конденсированные среды, существует окрестность конечнотемпературного фазового
перехода, называемая флуктуационной областью, в которой флуктуации параметра
порядка становятся большими. Вклад этих флуктуаций в термодинамические величины
становится больше, чем вклад древесного приближения. Таким образом, теория
возмущений в этой области неприменима. Одним из примеров таких моделей является
модель Гинзбурга - Ландау, описывающая сверхпроводимость.

Наше основное предположение заключается в том, что решеточная модель Вайнберга
- Салама сходна с упомянутыми выше феноменологическим моделями в том смысле,
что в ней также существует окрестность перехода между Хиггсовской фазой и
нефизической симметричной фазой, которую можно назвать флуктуационной областью.
В соответствии с численными результатами решеточная модель приближается в
непрерывной при движении на ее фазовой диаграмме вдоль ЛПФ при уменьшении длины
ребра решетки. В соответствии с численными результатами это движение вдоль ЛПФ
соответствует приближению к точке перехода между физической (где сконденсирован
Хиггс) и нефизической фазами. Таким образом, именно при приближении к
флуктуационной области и возникает непрерывная физика. В этой области, как мы
полагаем, теория возмущений не может быть применима. Таким образом, вопрос о
переходе к непрерывной теории необходимо включает в себя непертурбативное
исследование.

Как уже говорилось выше, согласно общепринятой точке зрения, верхний предел
ультрафиолетового обрезания $\Lambda$ в Электрослабой теории зависит от массы
Хиггса. Он увеличивается с уменьшением массы Хиггса. При массе Хиггса около $1$
ТэВ $\Lambda$ становится порядка $M_H$. В то же время для $M_H \sim 200$ Gev
значение $\Lambda$ может быть почти бесконечным\footnote{Здесь не
рассматривается ограничение на массу Хиггса снизу, которое связано с проблемой
стабильности вакуума и с фермионными петлями. }. Этот вывод был сделан на
основе теории возмущений.

Мы вычисляем эффективный потенциал $V(|\Phi|)$ для поля Хиггса $\Phi$ тремя
различными способами. В физической фазе эффективный потенциал всех трех видов
имеет минимум при некотором ненулевом значении $\phi_m$ поля $|\Phi|$. Это
показывает наличие спонтанного нарушения симметрии, как оно и должно быть.
Однако, существует окрестность фазового перехода, в которой флуктуации поля
Хиггса становятся  значительными, и указанные три потенциала дают разные
значения для вакуумного значения $\phi_m$. Мы ожидаем, что в этой области
теория возмущений вокруг тривиального вакуума $\Phi = (\phi_m,0)^T$  может быть
неприменима. Мы называем эту область флуктуационной.

Наше предположение также подтверждается изучением квантовых монополей Намбу,
которые по сути являются зародышами нефизической фазы внутри физической (внутри
них $|\Phi|=0$). Мы видим, что их плотность растет при приближении к фазовому
переходу (кроссоверу). Внутри флуктуационной области (ФО) эти объекты
расположены столь плотно, что невозможно вообще говорить о них, как об
отдельных объектах. А именно, среднее расстояние между ними становится порядка
их размера. Разумеется, такие конфигурации не могут иметь ничего общего с
тривиальным вакуумом теории возмущений.

Наша оценка для максимального значения ультрафиолетового обрезания, которое
достигается на границе флуктуационной области, - около  $1$ ТэВ.

\begin{figure}
\begin{center}
 \epsfig{figure=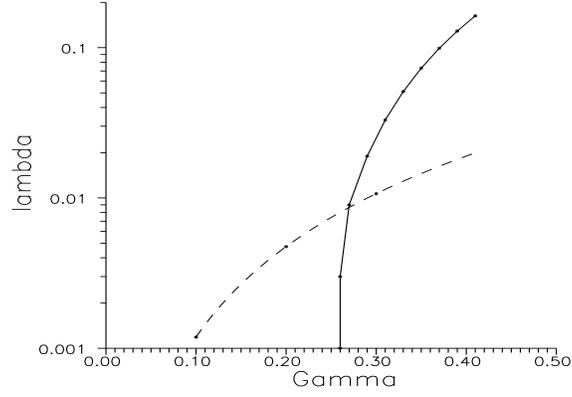,height=60mm,width=80mm,angle=0}
\caption{\label{62fig.2} Фазовая диаграмма модели в плоскости
 $(\gamma, \lambda)$ при $\beta = 12$. Пунктирная линия - древесная оценка для линии постоянной физики,
 соответствующей затравочной $M^0_H = 270$ ГэВ. Непрерывная линия - линия перехода между физической Хиггсовской
 фазой и нефизической симметричной фазой.   }
\end{center}
\end{figure}

\subsection{Фазовая диаграмма}

В нашем исследовании мы фиксируем затравочный угол Вайнберга равным  $30^o$.
Симуляции производились на решетках размера  $8^3\times 16$, $12^3\times 16$,
$16^4$. Также производилась проверка ряда результатов на решетке $16^3\times
32$.

 На трехмерной ($\beta, \gamma, \lambda$) фазовой диаграмме поверхности
 фазового перехода двумерны. ЛПФ в древесном приближении - это линии ($\frac{\lambda}{\gamma^2} = \frac{1}{8
\beta} \frac{M^2_H}{M^2_W} = {\rm const}$; $\beta = \frac{1}{4\pi \alpha}={\rm
const}$). Мы полагаем, что отклонение ЛПФ от древесного приближения вблизи
фазового перехода могут быть значительными. Однако, качественно их поведение
остается тем же. А именно, ультрафиолетовое обрезание растет с убыванием
$\gamma$, и его максимальное значение достигается в точке перехода. Плотность
монополей Намбу  в решеточных единицах также растет вдоль ЛПФ с ростом
ультрафиолетового обрезания.

При $\beta = 12$ фазовая диаграмма представлена на рис. \ref{62fig.2}.
Физическая фаза Хиггса расположена справа от линии перехода. Положение перехода
локализуется нескольким различными способами. При этом на решетках
рассмотренного размера эти точки не совпадают друг с другом в пределах
статистической погрешности, что указывает на то, что рассматриваемый переход
является кроссовером. Следующая переменная может рассматриваться как поле $Z$
бозона:
\begin{equation} Z_{xy} = Z^{\mu}_{x} \;
 = - {\rm sin} \,[{\rm Arg} (\Phi_x^+U_{xy} e^{i\theta_{xy}}\Phi_y) ]. \label{Z1_}
\end{equation}

Следует отметить, что перенормированная масса Хиггса не отклоняется существенно
от ее затравочного значения. Например, для $\lambda = 0.0025, \gamma = 0.261$
 имеем $M_H = 170 \pm 30$ ГэВ (затравочное значение $M^0_H \sim 150$ ГэВ).

\subsection{Ультрафиолетовый эффективный потенциал скалярного поля}

\begin{figure}
\begin{center}
 \epsfig{figure=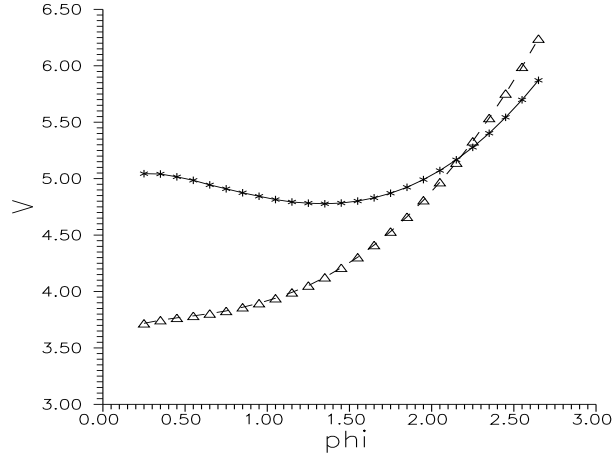,height=60mm,width=80mm,angle=0}
\caption{\label{62fig.0} Эффективный потенциал при  $\lambda =0.0025$ и $\beta
= 12$. Кресты и непрерывная линия  соответствуют $\gamma = 0.262$.
Треугольники соответствуют $\gamma_c = 0.26$. Статистические неопределенности -
порядка размера символов. }
\end{center}
\end{figure}

Эффективный потенциал для  значения скалярного поля в некоторой фиксированной
точке пространства - времени $|\Phi_x|$ мы вычисляем, используя гистограммный
метод. Вычисления проводились на решетке $8^3\times 16$. Существенного отличия
от полученной формы потенциала не обнаружено также и на решетках большего
размера. Вероятность $h(\phi)$ обнаружить значение $|\Phi_x|$ в интервале
$[\phi-0.05;\phi+0.05)$ вычислялась для $\phi = 0.05 + N*0.1$, $N = 0,1,2, ...$
Эта вероятность связана с эффективным потенциалом как $ h(\phi) = \phi^3
e^{-V(\phi)}$. Таким образом, мы извлекаем потенциал из $h(\phi)$ как
\begin{equation}
V^{u-v}(\phi) = - {\rm log}\, h(\phi) + 3 \, {\rm log} \, \phi \label{CEP}
\end{equation}
Следует отметить, что $h(0.05)$ вычисляется как вероятность обнаружить значение
 $|\Phi|$ в интервале $[0;0.1]$. Внутри этого интервала  ${\rm log}\,
\phi$ плохо определен. Поэтому мы исключаем точку $\phi = 0.05$ из наших
данных. Вместо этого мы вычисляем $V(0)$ используя экстраполяцию данных при
$0.15 \le \phi \le 2.0$. Экстраполяция производится с использованием
полиномиального фита со степенями $\phi$ вплоть до третьей (среднее отклонение
фита от данных - около $1$ процента). Далее, мы вводим величину $H = V(0) -
V(\phi_m)$, которую называем высотой барьера (здесь $\phi_m$ - это то значение,
при котором $V$ достигает минимума).

В нулевом порядке теории возмущений вычисляемый ультрафиолетовый эффективный
потенциал  имеет вид
\begin{equation}
V^{u-v}(\phi) =  \frac{\gamma}{2G_{m_H}(0)} (\phi - v )^2 \label{UVPOT}
\end{equation}
Здесь $N$ - объем решетки, и потенциал Юкавы при нулевом расстоянии обозначен $
G_{m_H}(0) = \frac{1}{N} \sum_p (4 sin^2p/2 + m_H^2)^{-1}$. Даже на бесконечной
решетке это значение остается конечным. Это означает, что ультрафиолетовые
флуктуации $\delta \phi = \sqrt{G_{m_H}(0)/\gamma}$ выраженные в решеточных
единицах остаются конечными. Когда физический объем решетки удерживается
постоянным $Na^4
>> M_H^{-4}$ в то время, как длина ребра решетки стремится к нулю, и
$m_H=M_H a \rightarrow 0$ (здесь  $M_H$ - масса Хиггса в ГэВ, $a$ - длина ребра
решетки), значение $G_{m_H}(0)$ остается конечным. Так мы приходим к обычному
предсказанию непрерывной теории $\delta \phi^{phys} =
\sqrt{G^{phys}_{M_H}(0)/\gamma}\sim \Lambda = \frac{\pi}{a}$, где $\phi^{phys}$
- скалярное поле, выраженное в физических единицах, а  $G^{phys}_{M_H}(x)$ -
пропагатор в физических единицах (здесь, впрочем, в решеточной регуляризации).
Таким образом, на уровне теории возмущений рассматриваемый ультрафиолетовый
потенциал позволяет вычислить вакуумное значение скалярного поля как положение
минимума этого потенциала. Следует отметить, однако, что это утверждение может
показаться несколько странным, поскольку конденсат есть по своей природе
величина инфракрасная. Тем не менее, наши результаты показывают, что в глубине
физической фазы, действительно вычисляемое таким образом вакуумное значение
скалярного поля отлично от нуля, в то время, как в глубине нефизической фазы
оно равно нулю.

 Значения $\phi_m$ и $H$ увеличиваются с ростом $\gamma$. При $\gamma \le
0.26$ минимум потенциала достигается при $\phi = 0$. Эта точка соответствует
также максимуму флуктуации поля $H_x = \sum_{y} Z^2_{xy}$,  рассматриваемой для
фиксированной точки пространства - времени. При $\gamma > 0.26$ минимум
потенциала наблюдается при $\phi_m \ne 0$. Эти данные указывают на точку
$\gamma_c=0.26$ как на положение перехода между двумя фазами.

Важно понять, какие значения высоты барьера являются большими, а какие - нет.
Наше предложение - сравнивать  $H = V(0) - V(\phi_m)$ с $H_{\rm fluct} =
V(\phi_m + \delta \phi) - V(\phi_m)$, где $\delta \phi$ - это флуктуации
$|\Phi|$.

Существует значение $\gamma$ (мы называем его
 $\gamma_{c2}$) такое, что при $\gamma_c < \gamma < \gamma_{c2}$ высота барьера
 $H$ - порядка $H_{\rm fluct}$ в то время, как для $\gamma_{c2} << \gamma$
высота барьера значительно превышает $H_{\rm fluct}$. Грубая оценка этого
псевдокритического значения:  $\gamma_{c2} \sim 0.262$.

Флуктуации поля $|\Phi|$ можно оценить как $\delta \phi \sim 0.6$ для всех
рассматриваемых значений $\gamma$ при $\lambda = 0.0025$, $\beta = 12$. Из
наших данных следует, что $\phi_m >> \delta \phi$ при $\gamma_{c2} << \gamma$ в
то время, как $\phi_m \sim \delta \phi$ при $\gamma_{c2} > \gamma$.

То, что флуктуации скалярного поля оказываются сравнимы с его вакуумным
значением в области $\gamma_c < \gamma < \gamma_{c2}$ указывает на то, что
 обычная теория возмущений вокруг
тривиального вакуума спонтанно нарушенной теории может быть неприменима в этой
области. Далее мы приведем также и иные указания на это.

\subsection{Инфракрасный эффективный потенциал}

Ультрафмолетовый потенциал, описанный выше, определяется как

\begin{equation}
exp(-V^{u-v}(\phi)) = <\delta(\phi - |\Phi_x|)>\label{uv}
\end{equation}

Также в унитарной калибровке $\Phi =
\left(\begin{array}{c}h\\0\end{array}\right), h \in R$ можно определить и
инфракрасный эффективный потенциал
\begin{equation}
exp(-V^{i-r}(\phi) )= <\delta(\phi - |\frac{1}{N} \sum_x h_x|)>, \label{ir}
\end{equation}
где $N$ - число точек решетки.

Теперь, однако,  присутствует остаточная $Z_2$ калибровочная инвариантность
$h_x \rightarrow (-1)^{n_x} h_x, \, Z \rightarrow [Z + \pi d n]{\rm mod}\, 2
\pi$, где $Z_{xy} = Z^{\mu}_{x} \;
 = - [{\rm Arg} (\Phi_x^+U_{xy} e^{i\theta_{xy}}\Phi_y) ]$ - поле Z - бозона.
Для того чтобы придать смысл потенциалу необходимо эту остаточную калибровочную
инвариантность зафиксировать.

Возможный способ сделать это - минимизировать
\begin{equation}
\sum_{links}(1-{\rm cos}\, Z) \rightarrow min \label{Z_}
\end{equation}
 по отношению к указанным  $Z_2$ калибровочным преобразованиям.
 Далее мы называем эту калибровку $Z$ - версией унитарной калибровки, а соответствующий потенциал (\ref{ir}) -
 UZ  потенциалом.
Этот потенциал имеет минимум при $\phi = 0$ для $\gamma \le 0.2575$ (см. Рис.
\ref{uz}).
 При $\gamma \ge 0.258$ данный потенциал имеет минимум для ненулевого значения  $\phi$.
Таким образом, данный потенциал указывает на
$\gamma_{c}^{\prime}=0.25775\pm0.00025$ как на положение перехода между двумя
фазами.

\begin{figure}
\begin{center}
 \epsfig{figure=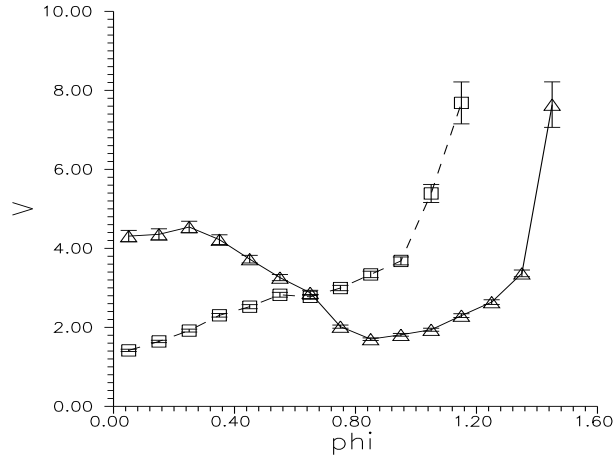,height=60mm,width=80mm,angle=0}
\caption{\label{uz} Инфракрасный потенциал (\ref{ir}) UZ при $\gamma = 0.2575$
(квадраты), и $\gamma = 0.258$ (треугольники);  $\lambda =0.0025$ , $\beta =
12$. (Решетка $8^3\times 16$.)}
\end{center}
\end{figure}

Иной способ определить унитарную калибровку с $h_x \in R$ - минимизировать
дивиргенцию  $Z$ по отношению к $Z_2$ преобразованиям:
\begin{equation}
\sum_x [\delta Z]^2 \rightarrow min \label{dz}
\end{equation}
Далее мы называем эту калибровку $DZ$ - версией Унитарной калибовки, а
соответствующий потенциал (\ref{ir}) -  UDZ потенциалом. Этот потенциал имеет
минимум при  $\phi = 0$ для $\gamma \le \gamma_c=0.26$. При $\gamma
> \gamma_{c}$  он имеет минимум при ненулевом значении  $\phi$ (см. Рис.\ref{c.4.fig.4}).
Таким образом, форма этого потенциала меняется в той же точке фазовой
диаграммы, что и форма ультрафиолетового потенциала.

\begin{figure}
\begin{center}
 \epsfig{figure=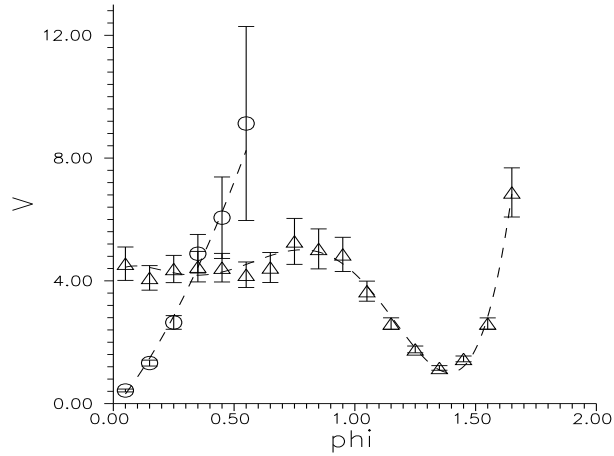,height=60mm,width=80mm,angle=0}
\caption{\label{c.4.fig.4} Инфракрасный потенциал (\ref{ir}) UDZ при $\gamma =
0.262$ (треугольники), и при $\gamma = 0.26$ (круги), $\lambda =0.0025$ ,
$\beta = 12$. (Решетка $8^3\times 16$.)}
\end{center}
\end{figure}

Следует отметить, что для рассматриваемых решеток размера от $8^3\times 16$ до
$16^3\times 32$ значения $\gamma_c, \gamma_c^{\prime}$ практически не меняются.
Древесное приближение для инфракрасного потенциала обоих рассматриваемых типов
дает:
\begin{equation}
V^{i-r}(\phi) =  N \lambda (\phi^2 - v^2 )^2 \label{IRPOT}
\end{equation}
Таким образом, на основании теории возмущений они оба должны менять свою форму
одновременно, в точке фазового перехода между двумя фазами. Тот факт, что этого
не происходит, свидетельствует о том, в некоторой окрестности фазового перехода
теория возмущений вокруг тривиального вакуума, судя по всему, не может быть
применима. Сам фазовый переход  может является кроссовером. Впрочем, мы не
исключаем возможности того, что на решетках бесконечного размера фазовый
переход второго рода может иметь место в точке $\gamma_c^{\prime}$.

\subsection{Среднее действие и флуктуации}

На Рис. \ref{LAQ} представлены данные линковой части действия
$\frac{1}{4N}\sum_{xy}H_x^+U_{xy} e^{i\theta_{xy}}H_y$ (часть действия наиболее
чувствительная к изучаемому фазовому переходу). Квадраты соответствуют
холодному старту ($450 000$ шагов Алгоритма Метрополиса). Кресты соответствуют
горячему старту ($350 000$ шагов). Видно, что обе линии соединились. Наши
данные демонстрируют отсутствие сигнала двух состояний, что говорит о том, что
мы не имеем дело с фазовым переходом первого рода. Данные среднего действия
указывают, что в точке, близкой к $\gamma_c^{\prime}$ имеет место разрыв
производной термодинамических потенциалов, что говорит о том, что в этой точке
на решетках бесконечного размера может иметь место фазовый переход второго
рода.

\begin{figure}
\begin{center}
 \epsfig{figure=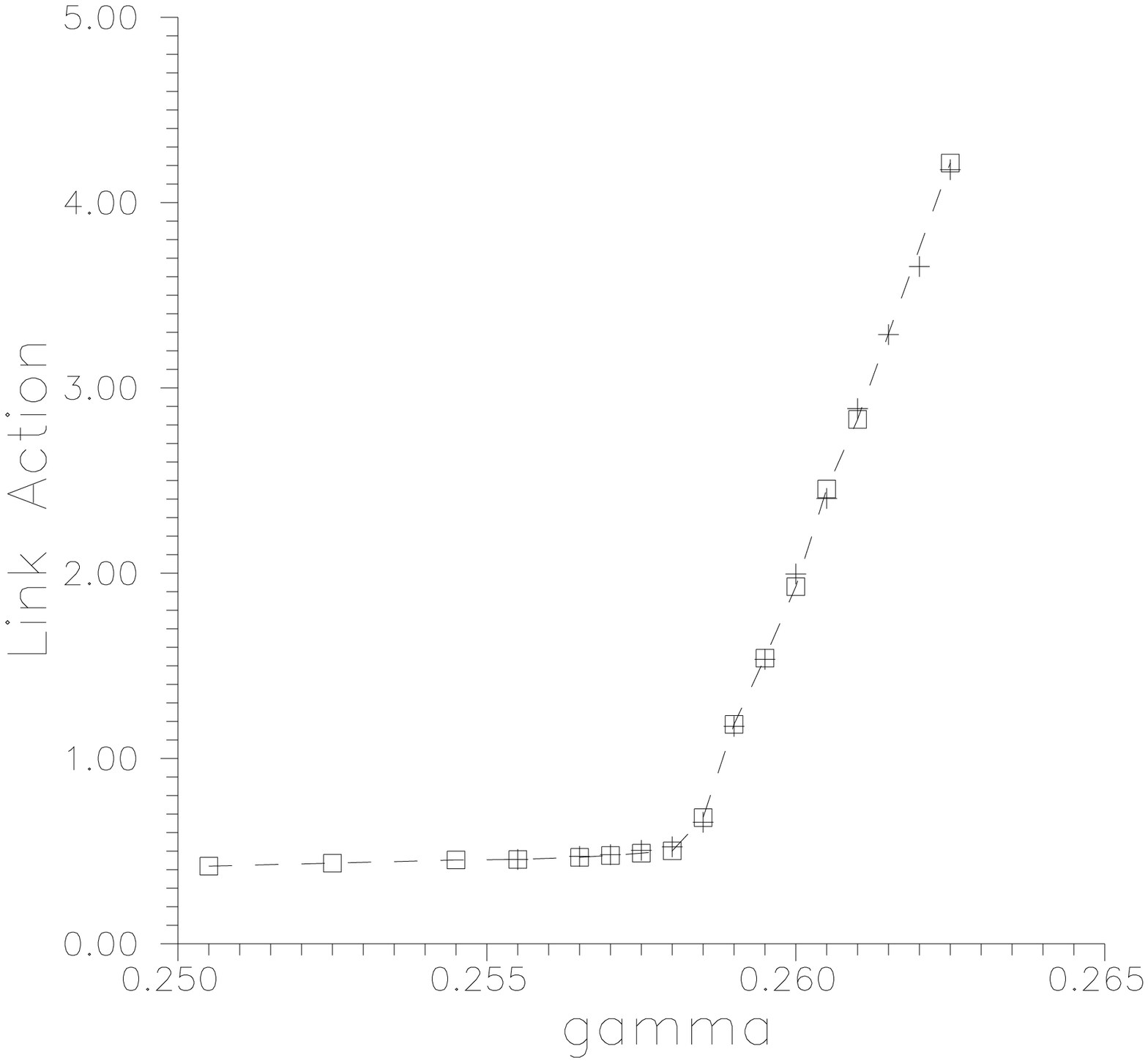,height=60mm,width=80mm,angle=0}
\caption{\label{LAQ} Линковая часть действия как функция  $\gamma$ при $\lambda
=0.0025$ , $\beta = 12$. Холодный старт соответствует квадратам. Горячий старт
соответствует крестам. Статистические ошибки - примерно размера используемых
символов.}
\end{center}
\end{figure}

Флуктуации поля $|\Phi_x|$ в фиксированной точке пространства - времени
представлены на Рис. \ref{dphi25}. Ясно, что максимум флуктуаций достигается
при значении $\gamma$, существенно большем, чем $\gamma_c, \gamma_c^{\prime}$.
В то же время низшие порядки теории возмущений предсказывают, что указанные
значения должны совпадать, и максимум флуктуаций достигается в точке фазового
перехода. Это является еще одним указанием на то, что в некоторой окрестности
перехода теория возмущений неприменима.

\begin{figure}
\begin{center}
 \epsfig{figure=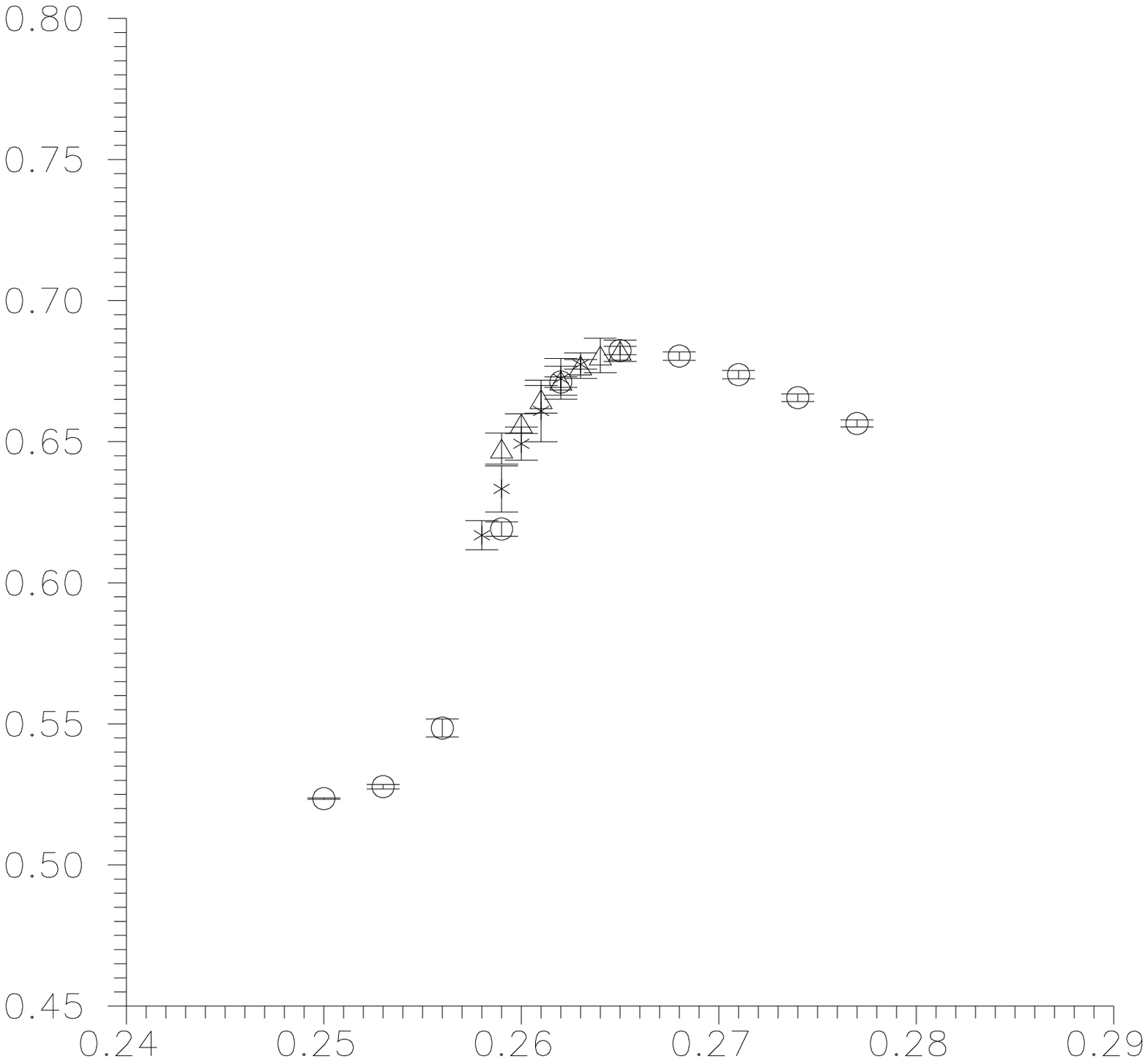,height=60mm,width=80mm,angle=0}
\end{center}
\begin{picture}(0,0)(0,0)
\end{picture}
\begin{center}
\caption{\label{dphi25} Флуктуации $\delta \phi$ как функция $\gamma$ при
$\lambda =0.0025$,  $\beta = 12$. Круги соответствуют решетке  $8^3\times16$.
Кресты соответствуют решетке $12^3\times 16$. Треугольники соответствуют
решетке $16^4$.}
\end{center}
\end{figure}

\subsection{Массы и длина ребра решетки}

Для оценки масс $Z$-бозона и бозона Хиггса мы используем корреляторы
\begin{equation}
\frac{1}{N^6} \sum_{\bar{x},\bar{y}} \langle \sum_{\mu} Z^{\mu}_{x} Z^{\mu}_{y}
\rangle   \sim
  e^{-M_{Z}|x_0-y_0|}+ e^{-M_{Z}(L - |x_0-y_0|)}
\label{corZ}
\end{equation}
и
\begin{equation}
  \frac{1}{N^6}\sum_{\bar{x},\bar{y}}(\langle H_{x} H_{y}\rangle - \langle H\rangle^2)
   \sim
  e^{-M_{H}|x_0-y_0|}+ e^{-M_{H}(L - |x_0-y_0|)},
\label{cor2}
\end{equation}
 Здесь сумма $\sum_{\bar{x},\bar{y}}$ - по трехмерным компонентам 4 - векторов
 $x$ и $y$, а $x_0, y_0$ обозначают их "временные" компоненты.  $N$ размер решетки в
 "пространственном направлении". Причем все величины рассматриваются в версии
 Унитарной калибровки, в которой действительное скалярное поле неотрицательно.

В решеточных вычислениях мы используем два вида операторов, рождающих бозон
Хиггса:  $ H_x = |\Phi|$ и $H_x = \sum_{y} Z^2_{xy}$. В обоих случаях  $H_x$
определен в точке $x$, сумма $\sum_y$ - по  соседним точкам $y$.

После фиксации унитарной калибровки ($\Phi_2 = 0$; $\Phi_1 \in {\cal R}$;
$\Phi_1 \ge 0$), решеточная Электрослабая модель становится решеточной  $U(1)$
калибровочной теорией с $U(1)$ калибровочным полем
\begin{equation}
 A_{xy}  =  A^{\mu}_{x} \;
 = \,[Z^{\prime}  + 2\theta_{xy}]  \,{\rm mod}
 \,2\pi, \label{A}
\end{equation}
где новое решеточное поле $Z$ - бозона  (отличающееся от (\ref{Z1_}))
определяется как
\begin{equation}
Z^{\prime} =  -{\rm Arg} (\Phi_x^+U_{xy} e^{i\theta_{xy}}\Phi_y) .\label{Z2}
\end{equation}

Обычное электромагнитное поле связано с $A$ как $ A_{\rm EM} = A - Z^{\prime} +
2 \,{\rm sin}^2\, \theta_W Z^{\prime}$.

Физическая шкала дается в нашей решеточной теории значением массы $Z$-бозона
$M^{phys}_Z \sim 91$ ГэВ. Тогда длина ребра решетки равна $a \sim [91 {\rm
GeV}]^{-1} M_Z$, где $M_Z$ - масса $Z$ бозона в решеточных единицах.
Аналогичные вычисления для $\lambda = \infty$ были нами описаны выше. Было
показано, что масса $W$ - бозона в отличие от массы $M_Z$ зависит от размера
решетки благодаря наличию фотонного облака. Поэтому нами используется масса $Z$
- бозона для фиксации шкалы. Ниже мы не исследуем массу $W$ - бозона.

Наши данные  показывают, что
 $\Lambda= \frac{\pi}{a} = (\pi \times 91~{\rm GeV})/M_Z$ медленно растет с уменьшением
 $\gamma$ при любом фиксированном $\lambda$. Мы исследовали более подробно окрестность фазового перехода
 для $\lambda = 0.0025$ и $\beta =
12$. Было обнаружено, что при $\gamma > \gamma^{\prime}_c$ значение $\Lambda$
на решетках вплоть до размера $16^4$ не превышает величины в несколько ТэВ. В
то же время при $\gamma \le \gamma^{\prime}_c$ решеточная масса $Z$ - бозона не
может быть вычислена из - за больших статистических ошибок в корреляторе. На
рис. \ref{mz25} зависимость $M_Z$ в решеточных единицах от $\gamma$
представлена для  $\lambda =0.0025$ и $\beta = 12$.

В Хиггсовском канале ситуация сложнее с вычислительной точки зрения. Из-за
недостаточной статистики мы не можем оценить массу Хиггса для всех
рассматриваемых значений $\gamma$. В настоящий момент получено, что для решетки
$8^3\times16$ при $\lambda = 0.0025, \gamma = 0.261$
 имеем $M_H = 170 \pm 30$ ГэВ (затравочное значение $M^0_H \sim 150$ ГэВ).

В принципе, в этом канале могут возникнуть связанные состояния двух
калибровочных бозонов. Следуя предыдущим работам по исследованию $SU(2)$ модели
Хиггса (см., например, \cite{Montvay}), мы интерпретируем массу в данном канале
при малых "временных" расстояниях как массу Хиггса и ожидаем появления массы
связанного состояния двух калибровочных бозонов на больших расстояниях.

В указанной точке ($\gamma = 0.261$, $\lambda =0.0025$, $\beta = 12$) мы
собрали достаточно статистики, чтобы вычислить коррелятор   вплоть до
"временных" расстояний $|x_0-y_0| = 4$. Следует отметить, что для вычисления
массы $Z$ - бозона мы фитируем коррелятор
 (\ref{corZ}) для $8 \ge |x_0-y_0| \ge 1$.

\begin{figure}
\begin{center}
 \epsfig{figure=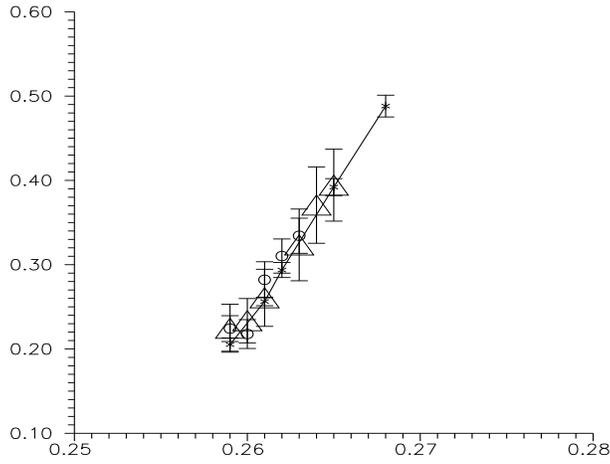,height=60mm,width=80mm,angle=0}
\caption{\label{mz25} Масса Z - бозона в решеточных единицах при $\lambda
=0.0025$, $\beta = 12$. Круги соответствуют решетке $12^3\times 16$.
Треугольники - $16^4$. Кресты - решетке $8^3\times 16$.}
\end{center}
\end{figure}

\subsection{Перенормированная постоянная тонкой структуры}

Для вычислния перенормированной постоянной тонкой структуры $\alpha_R =
e^2/4\pi$ (где $e$ - электрический заряд) мы используем потенциал для
бесконечно тяжелого внешнего фермиона. Рассматривается коррелятор двух
Поляковских линий, соответствующих внешним правым лептонам:
\begin{equation}
 {\cal C}(|\bar{x}-\bar{y}|)  =
 \langle {\rm Re} \,\Pi_{t} e^{2i\theta_{(\bar{x},t)(\bar{x},t+1)}}\,\Pi_{t} e^{-2i\theta_{(\bar{y},t)(\bar{y},t+1)}}\rangle.
\end{equation}
Потенциал извлекается из этого коррелятора следующим образом:
\begin{equation}
 {\cal V}(R) = -\frac{1}{L} { \rm log}\,  {\cal C}(R)
\end{equation}
Здесь $L$ - размер решетки в направлении мнимого "времени".

Благодаря обмену виртуальными фотонами на больших расстояниях можно ожидать
появления Кулоновского взаимодействия
\begin{eqnarray}
 {\cal V}(r) & = & -\alpha_R \, {\cal U}_0(r)+ const,\,
\nonumber\\
{\cal U}_0(r) & = & -\frac{ \pi}{N^3}\sum_{\bar{p}\ne 0} \frac{e^{i p_3
r}}{{\rm sin}^2 p_1/2 + {\rm sin}^2 p_2/2 + {\rm sin}^2
 p_3/2}
 \label{V1}
\end{eqnarray}
Здесь $N$ - размер решетки, $p_i = \frac{2\pi}{L} k_i, k_i = 0, ..., L-1$.

Однако, на решетках конечного размера появляются эффекты конечного объема,
упоминавшиеся в начале главы. Эти эффекты возникают благодаря обмену
виртуальным Z - бозоном, чья масса вблизи фазового перехода - имеет порядок
величины $\sim 0.1$. Поэтому мы используем следующий фит потенциала, который
учитывает эти эффекты:
\begin{eqnarray}
 {\cal V}(r) & = & -\alpha_R \, [{\cal U}_0(r)+\frac{1}{3}{\cal U}_{m_Z}(r)] +   const,\,
\nonumber\\
{\cal U}_m(r) & = & -\frac{ \pi}{N^3}\sum_{\bar{p}} \frac{e^{i p_3 r}}{{\rm
sin}^2 p_1/2 + {\rm sin}^2 p_2/2 + {\rm sin}^2
 p_3/2 +  {\rm sh}^2 m/2}
 \label{V2}
\end{eqnarray}

Мы подставляем в (\ref{V2}) вычисляемую нами массу  $Z$ - бозона. Результат
представлен на Рис. \ref{alpha}. Его следует сравнить с затравочным значением
постоянной тонкой структуры $\alpha^{(0)} \sim \frac{1}{151}$ и с однопетлевым
выражением (где мы полагаем затравочное значение $\alpha$ живущим на шкале
$\sim 1$ ТэВ, а перенормированное значение соответствующим электрослабой шкале
$M_Z$): $\alpha^{(1)}(M_Z/{1 \, {\rm TeV}}) \sim \frac{1}{149.7}$.

 Рис. \ref{alpha} демонстрирует, что перенормированная постоянная тонкой структуры
 вычисленная указанным способом близка к однопетлевому значению (когда
 обрезание $\Lambda$ в $\alpha^{(1)}(M_Z/\Lambda)$ - близко к $1$ ТэВ).
Это также подтверждает косвенно правильность вычисленных нами значений массы
$Z$ - бозона.

\begin{figure}
\begin{center}
 \epsfig{figure=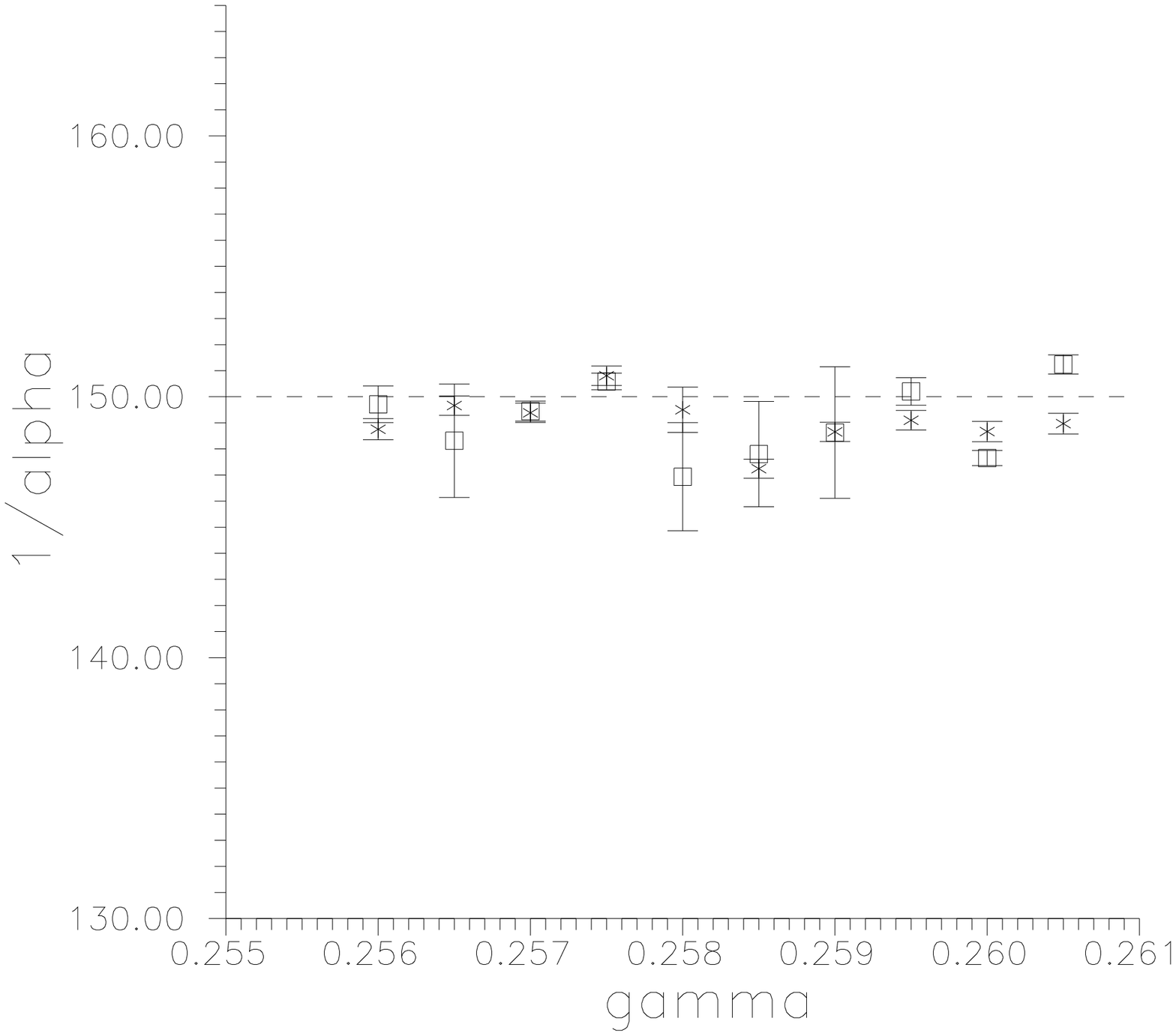,height=60mm,width=80mm,angle=0}
\caption{\label{alpha} Обратная перенормированная постоянная тонкой структуры
как функция $\gamma$ при $\lambda =0.0025$ , $\beta = 12$. Кресты соответствуют
решетке $8^3\times 16$, квадраты - решетке $16^3\times 32$. }
\end{center}
\end{figure}

\subsection{Плотность монополей Намбу }

На Рис.\ref{fig.rho} мы представляем плотность монополей Намбу как функцию от
$\gamma$ при $\lambda = 0.0025$, $\beta = 12$. Значение плотности монополей при
 $\gamma_c = 0.26$ - около $0.1$. В этой точке значение ультрафиолетового обрезания
$\Lambda \sim 1.2 \pm 0.2$ ТэВ.

В соответствии с классическими представлениями размер монополя Намбу - порядка
$M^{-1}_H$. Тогда для  $a^{-1} \sim 400$ ГэВ и  $M_H \sim 150$ ГэВ ожидаемый
размер монополя - около $2$-х решеточных единиц.

Монопольная плотность в районе $0.1$ означает, что среди $10$ точек существуют
 $4$ точки, занятые монополем. Среднее расстояние между монополями, таким образом, меньше, чем
 длина ребра решетки и невозможно вообще говорить о данных конфигурациях как о представляющих физический монополь
 Намбу.

При $\gamma = \gamma_{c2}\sim 0.262$ плотность монополей Намбу - около $0.03$.
Это значит, что среди $7$ точек найдется одна, занятая монополем. Среднее
расстояние между монополями, таким образом, - примерно две длины ребра решетки
или  $\sim \frac{1}{160\, {\rm Gev}}$. Таким образом, монопольная плотность в
физических единицах - около  $[{160\, {\rm Gev}}]^3$. Мы видим, что при этом
значении $\gamma$ среднее расстояние между монополями Намбу - порядка их
размера.

\begin{figure}
\begin{center}
 \epsfig{figure=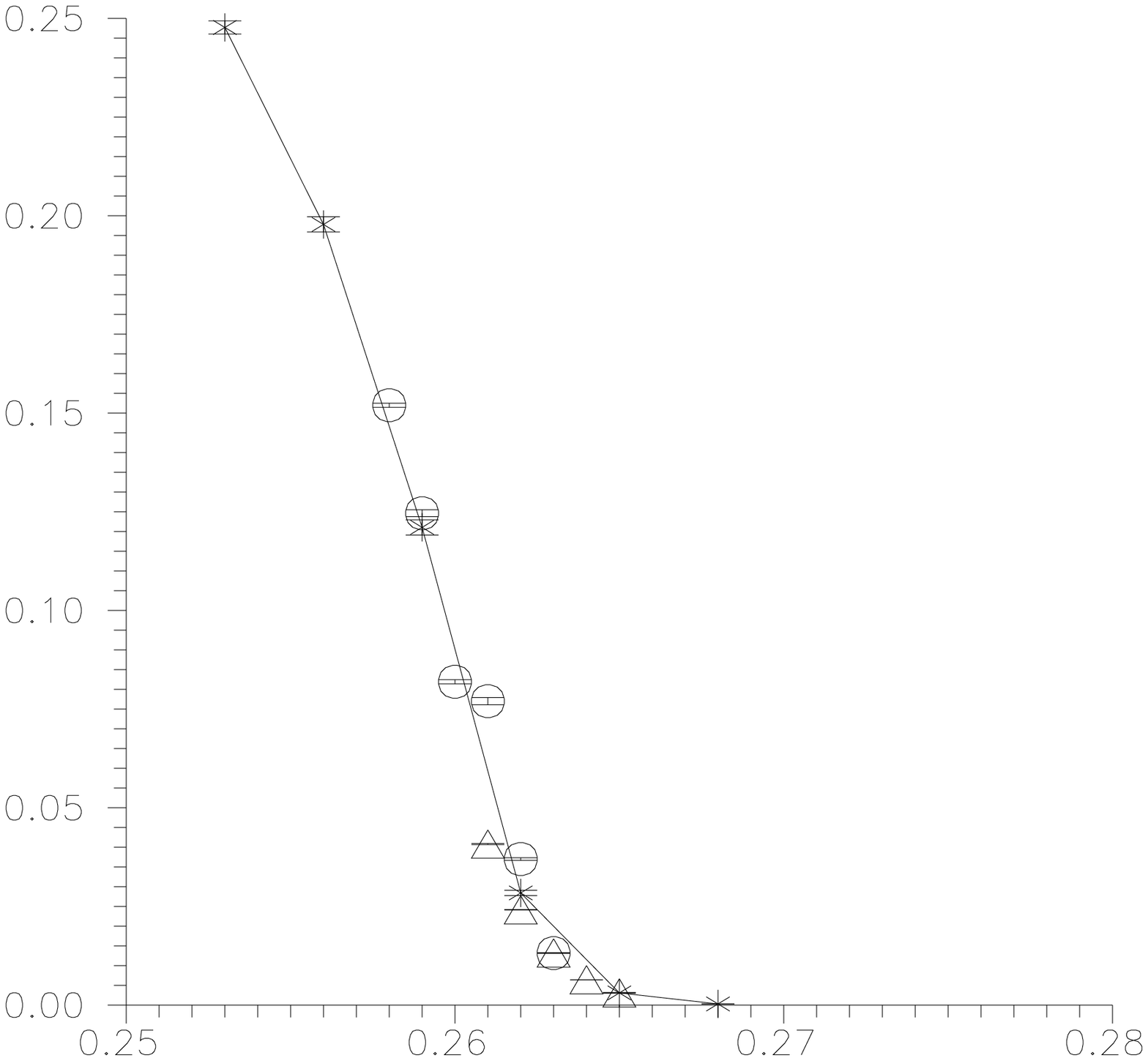,height=60mm,width=80mm,angle=0}
\caption{\label{fig.rho} Монопольная плотность как функция $\gamma$ при
$\lambda =0.0025$ , $\beta = 12$. Кресты соответствуют решетке $8^3\times 16$,
круги  - решетке $12^3\times 16$, треугольники - $16^4$. }
\end{center}
\end{figure}

Суммируем указанные наблюдения. Внутри ФО рассматриваемые конфигурации не
представляют отдельно расположенных монополей Намбу. Вместо этого, они должны
рассматриваться как представляющие собрание плотно расположенных монополей. С
другой стороны, при $\gamma
>> \gamma_{c2}$ рассматриваемые конфигурации представляют отдельно расположенные монополи, поскольку
их размеры в этой области существенно меньше средних расстояний между ними.
Другими словами, вне ФО вакуум представляет собой разреженный газ монополей
Намбу, а внутри ФО - жидкость, составленную из монополеподобных объектов.

Внутри $Z$ - струны, соединяющей монополи Намбу, а равно и внутри самих
монополей $|\Phi| = 0$. Это означает, что монополь Намбу вместе с $Z$ - струной
может рассматриваться как зародыш нефизической фазы внутри физической. Мы
видим, что плотность этих зародышей растет при приближении к точке перехода.
Внутри ФО две фазы перемешаны, что связано с большим значением монопольной
плотности.

Таким образом, мы приходим к выводу, что  динамика решеточной модели Вайнберга
- Салама внутри ФО не имеет ничего общего с наивными представлениями о вакууме
обычной теории возмущений, где флуктуации скалярного поля вокруг тривиального
вакуума (калибровочное поле равно нулю, скалярное поле равно $(\phi_m,0)^T$),
существенно меньше значения $ \phi_m $.

\section{Некоторые результаты для других значений затравочных констант}

Как уже говорилось, нами исследуется подробно область фазовой диаграммы,
соответствующая затравочным значениям ($\alpha \sim 1/150, \theta_W = \pi/6$,
$\lambda = 0.0025$, что соответствует  $M_H \sim 150$ ГэВ). Однако, ряд
результатов получен также для следующих затравочных значений констант: ($\alpha
\sim 1/190, \theta_W = \pi/6, \lambda = \infty$, где перенормированная $M_H
\sim 800$ ГэВ); ($\alpha \sim 1/150, \theta_W = \pi/6, \lambda = 0.009, 0.001$,
что соответствует $M_H \sim 300, 100$ ГэВ). Ниже мы кратко останавливаемся на
некоторых из этих результатов.

\subsection{Численное исследование модели при $\theta_W=\pi/6$, $\lambda = \infty$, $\beta \sim 15$}

 При
малых значениях $\beta$ решеточная модель с группой $SU(2)\otimes U(1)/Z_2$
была нами исследована в главе 4. Фазовая диаграмма  решеточной $SU(2)\otimes
U(1)/Z_2$-симметричной модели (то есть в модели c $Z_2$ симметрией)
представлена на рис. \ref{4fig.1}. Модель с калибровочной группой
 $SU(2)\otimes U(1)$ была исследована в \cite{SU2U1} (также для малых $\beta$).
 На фазовой диаграмме есть две линии. Одна из них представляет фазовый переход
  конфайнмент - деконфайнмент, соответствующий $U(1)$ полям.
  Другая (горизонтальная) линия представляет переход между
  фазой Хиггса и симметричной фазой. В $SU(2)\otimes U(1)/Z_2$ модели обе линии встречаются, формируя тройную
  точку. В $SU(2)\otimes U(1)$ модели две линии не встречаются. В то же время в
  области больших значений $\beta$, соответствующей значениям констант связи,
  исследуемым в настоящей главе, фазовые диаграммы обеих моделей совпадают.

В обеих моделях есть три фазы. Первая расположена на левой стороне фазовой
диаграммы. В этой фазе удерживающие силы наблюдаются между как левыми, так и
правыми внешними фермионами. Однако, благодаря наличию скалярного поля струна,
соединяющая заряженные частицы, рвется и удерживающие силы исчезают, начиная с
некоторых расстояний (см. Главу 4). В этой фазе монополи Намбу и гиперзарядовые
монополи сконденсированы. Вторая фаза расположена под горизонтальной линией
фазового перехода и справа от вертикальной линии. В этой фазе удерживающие силы
наблюдаются только между левыми фермионами. Гиперзарядовые монополи не
сконденсированы в этой фазе, и их плотность быстро убывает.

Реальная физика соответствует фазам обеих моделей, расположенным в верхнем
правом углу фазовой диаграммы. В этой фазе ни монополи Намбу, ни гиперзарядовые
монополи не сконденсированы. Удерживающие силы не наблюдаются. Как было
отмечено в главе 4, плотность гиперзарядовых монополей чувствительна к $U(1)$
переходу конфайнмент - деконфайнмент, в то время, как плотность монополей Намбу
чувствует переход между Хиггсовской фазой и симметричной фазой. Положение
горизонтальной линии перехода было локализовано в точках, где зануляется
перколяционная вероятность монополей Намбу.  Эта линия соответствует также
максимуму флуктуаций операторов (\ref{HW}) и (\ref{HZ}), рассматриваемых в
фиксированной точке пространства - времени.

Все вычисления проводились на решетках линейного размера от $8$ до $16$  при
значениях $\beta$ в окрестности $\beta = 15$, что соответствует затравочному
значению $\alpha \sim 1/190$. Некоторые точки были проверены на решетке $24^4$.
В основном мы не видим существенной разницы между вычисляемыми величинами для
этих размеров решетки.


\begin{figure}
\begin{center}
 \epsfig{figure=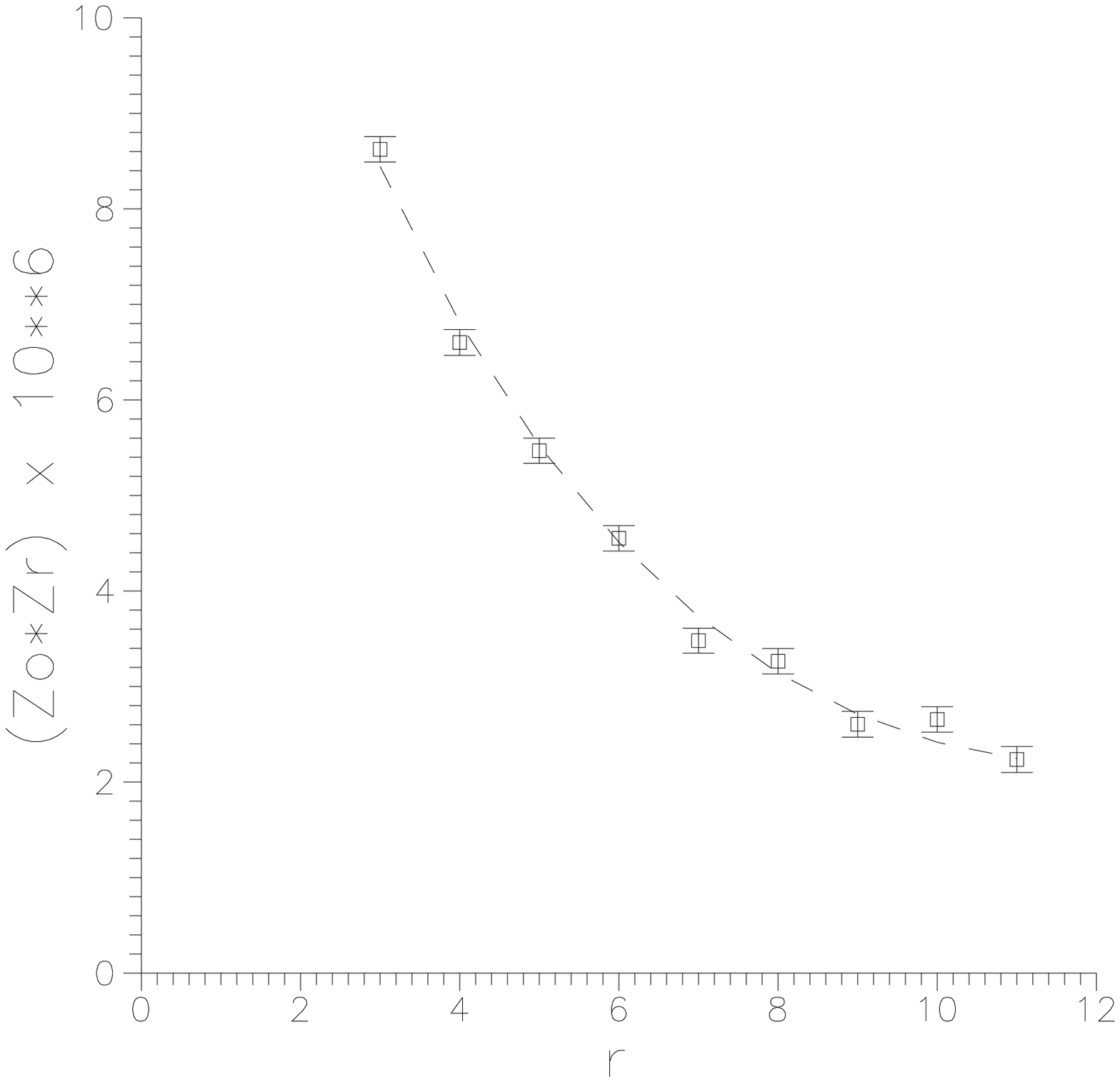,height=60mm,width=80mm,angle=0}
\caption{\label{6fig.2}  $\frac{1}{4 R^6} \sum_{\bar{x}, \bar{y}} \sum_{\mu}
<Z^{\mu}_{x} Z^{\mu}_{y} >$ как функция $r =|x_0-y_0|$. Здесь $R = 16$ - размер
решетки в "пространственном" направлении.}
\end{center}
\end{figure}

Физическая шкала фиксируется в нашей решеточной теории массой $Z$-бозона
$M^{\rm phys}_Z \sim 90$ GeV. Длина ребра решетки выражается как  $a \sim
[90\,{\rm GeV}]^{-1} M_Z$, где $M_Z$ - это масса $Z$ бозона в решеточных
единицах.

\begin{figure}
\begin{center}
 \epsfig{figure=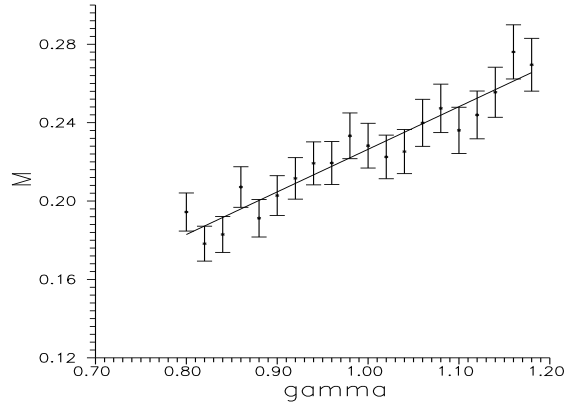,height=60mm,width=80mm,angle=0}
 \vspace{1.5ex}
\caption{\label{6fig.3}  $M_Z$ в решеточных единицах как функция $\gamma$ при
$\beta = 15$.}
\end{center}
\end{figure}


Помимо массы Z бозона мы вычисляем также и массу $W$ бозона, заряженного по
отношению к $U(1)$. Поэтому мы фиксируем решеточную калибровку Ландау для
вычисления пропагатора $W$ бозона. Эта калибровка фиксируется посредством
минимизации (по отношению к $U(1)$ калибровочным преобразованиям) следующего
функционала:
\begin{equation}
 F  =  \sum_{xy}(1 - \cos(A_{xy})).
\end{equation}
Тогда мы извлекаем массу $W$ бозона из коррелятора
\begin{equation}
\frac{1}{N^6} \sum_{\bar{x},\bar{y}} \langle \sum_{\mu} W^{\mu}_{x}
(W^{\mu}_{y})^{\dagger} \rangle   \sim
  e^{-M_{W}|x_0-y_0|}+ e^{-M_{W}(L - |x_0-y_0|)}
\label{corW}
\end{equation}


В области $\beta \in (10,20)$, $\gamma \in (1,2)$ мы не видим различия между
двумя решеточными версиями электрослабой теории.

Масса $Z$ - бозона меняется очень медленно с изменением $\beta$. Зависимость от
 $\gamma$ -  существеннее. $M_Z$ в решеточных единицах растет с ростом $\gamma$.

Для вычисления масс Z-бозона и $W$ - бозона мы используем решетки размера
$6^3\times 12$, $8^3\times 16$, $12^3\times 24$, и $16^3\times 24$. Зависимость
коррелятора Z-бозона  от $r = x_0-y_0$ представлена на рис. \ref{6fig.2} для
$\gamma = 1, \beta = 15$ на решетке $16^3\times 24$. Из этого рисунка мы
извлекаем массу $Z$-бозона равной $0.22\pm 0.02$. Следует отметить, что мы не
наблюдаем зависимости $M_Z$ от размера решетки.

При $\beta = 15$ мы локализуем положение перехода между симметричной и
Хиггсовской фазами вблизи $\gamma_c = 0.92 \pm 0.02$.


Мы находим, что масса $W$ - бозона содержит зависимость от размера решетки,
которую мы считаем артефактом. Мы полагаем, что эта зависимость возникает
благодаря фотонному облаку, окружающему $W$ - бозон. Поэтому мы используем
именно его для вычисления ультрафиолетового обрезания в нашей модели.
Зависимость массы $Z$-бозона от $\gamma$ при $\beta = 15$ на решетке $8^3\times
24$ представлена на рис. \ref{6fig.3}. Линейный фит -  $M_Z = 0.009 + 0.217
\gamma$.

Основываясь на этих данных мы заключаем, что масса Z - бозона в решеточных
единицах на решетках рассмотренного нами размера в физической фазе не может
превосходить $0.21\pm 0.01$ для $\beta = 15$ поскольку положение фазового
перехода мы локализуем при $\gamma = 0.92 \pm 0.02$.


Масса Хиггса в решеточных единицах измеряется с использованием коррелятора
\begin{equation}
  \sum_{\bar{x},\bar{y}}(\langle H_{x} H_{y}\rangle - \langle H \rangle^2)
  \sim
  e^{-M_{H}|x_0-y_0|}+ e^{-M_{H}(L - |x_0-y_0|)},
\label{corH}
\end{equation}
где $H$ - оператор рождения Хиггсовского бозона.

Мы используем три различных оператора:
\begin{equation}
H_x = \sum_{y} |W_{xy}|^2,\label{HW}
\end{equation}

\begin{equation}
H_x = \sum_{y} Z^2_{xy}\label{HZ}
\end{equation}
и
\begin{equation}
H_x = \sum_{y} Re(U^{11}_{xy} e^{i\theta_{xy}})
\end{equation}

Здесь $H_x$ определен в точке $x$, а сумма $\sum_y$ берется по соседним точкам
 $y$.



Из - за недостаточной статистики мы не можем извлечь из наших данных  $M_H$ с
хорошей точностью. Наша грубая оценка дает при  $\beta=15, \gamma = 1$ значение
$M_H/M_Z \sim 9\pm 2$. Эта оценка находится в соответствии исследованием
$SU(2)$ модели Хиггса \cite{12,13,14}, выполненным возле точки перехода между
фазами в Лондоновском пределе при реалистических значениях $\beta$. В
действительности, так же, как и в \cite{12} мы основываем нашу оценку на
рассмотрении коррелятора на "малых временных" расстояниях  ($ \le 3$).

В  \cite{14} было обнаружено, что на больших расстояниях второй массовый
параметр, близкий к $2 M_W$ вносит вклад в коррелятор. В нашем исследовании
точность измерения не позволяет нам извлекать информацию из $H$-$H$ коррелятора
при "временных" расстояниях $\ge 4$. Поэтому мы  не видим сигнала от связанного
состояния двух калибровочных бозонов. (В \cite{14} для оценки массы Хиггса в
этой ситуации значение массы  $\sim 2 M_W$ было интерпретировано как масса
связанного состояния двух калибровочных бозонов, и другая масса в данном канале
рассматривалась, как масса Хиггса.)


Как это объяснялось в главе 5, мировые линии квантовых монополей Намбу могут
извлекаться двумя способами:
\begin{equation}
 j_Z = \delta \Sigma = \frac{1}{2\pi} {}^*d([d Z^{\prime}]{\rm mod}2\pi)
\end{equation}
и
\begin{equation}
 j_A = \delta \Sigma = \frac{1}{2\pi} {}^*d([d A]{\rm mod}2\pi).
\end{equation}
Плотность монополей определяется как
\begin{equation}
 \rho = \left\langle \frac{\sum_{\rm links}|j_{\rm link}|}{4L^4}
 \right\rangle,
\label{rho}
\end{equation}
Где $L$ - размер решетки (в решеточных единицах).

Для изучения явления конденсации мы используем перколяционную вероятность
$\Pi(A)$.

\begin{figure}
\begin{center}
 \epsfig{figure=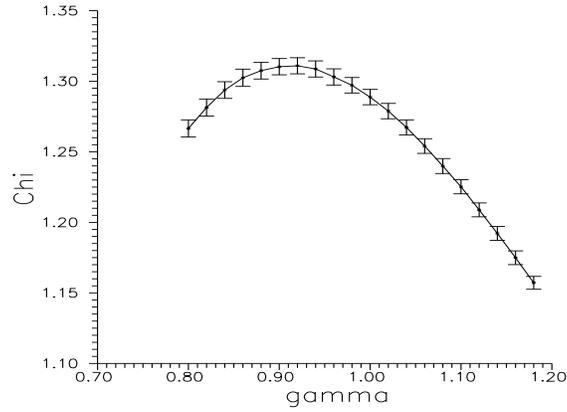,height=60mm,width=80mm,angle=0}
\vspace{1.5ex} \caption{\label{fig.61} $\chi = \langle H_Z^2 \rangle -
 \langle H_Z\rangle^2$ при $\beta = 15$ на решетке  $8^3\times 16$. }
\end{center}
\end{figure}

И $j_A$, и  $j_Z$ представляют один и тот же объект. Однако, этот объект может
иметь размер, превышающий  длину ребра решетки. Поэтому два различных
определения могут давать разные монопольные токи. Разница между двумя токами
есть $j_Z - j_A$. Поэтому плотность $j_A-j_Z$ измеряет степень того, насколько
 $j_A$ отличается от $j_Z$. Для того, чтобы изучить разницу между двумя определениями монополей Намбу
 мы используем величину  $\rho(j_A-j_Z)$,
построенную из $j_Z-j_A$ так же, как и (\ref{rho}).

 Плотности и перколяционные
вероятности для обоих определений совпадают в то время, как точное расположение
монополей, полученное двумя разными способами отличается примерно на  $30$\%,
то есть, мы находим, что $2\rho(j_A-j_Z)/(\rho(j_A)+\rho(j_z))\sim 0.3$. Это
означает, что два рассматриваемых способа действительно относятся к одному и
тому же объекту, но локализуют его в  $30$\% случаев по разному. То есть,
размер монополя Намбу в нашем исследовании превышает длину ребра решетки.

Перколяционная вероятность является параметром порядка для перехода между
симметричной и Хиггсовской фазами.

Мы также исследовали воприимчивость  $\chi = \langle H^2 \rangle - \langle
H\rangle^2$ извлеченную из $H_Z = \sum_{y} Z^2_{xy}$ и из $H_W = \sum_{y}
|W_{xy}|^2$.  Зависимость $\chi$ от $\gamma$ вдоль вертикальной линии $\beta =
15$ представлена на рис.  \ref{fig.61}. Положение перехода совпадает с
максимумом обеих восприимчивостей.

Наше исследование показывает, что плотность монополей Намбу растет с
уменьшением  $\gamma$ (для $\gamma
> 1$). Следует отметить, что монопольная плотность, как и масса $Z$ - бозона не зависит существенно от размера
 решетки.  Хорошим фитом при $\beta = 15$ является
\begin{equation}
 \rho \sim e^{2.08 - 4.6 \gamma} .\label{rho1}
\end{equation}
Монопольная плотность, выраженная в физических единицах определяется как
$\rho_c = \frac{\rho}{a^{3}}$, где $a$ - длина ребра решетки. Таким образом,
плотность монополей Намбу в физических единицах увеличивается с уменьшением
 $a$.


Итак, прежде всего, мы выяснили, что два решеточных определения (с группами
$SU(2)\otimes U(1)/Z_2$ и $SU(2)\otimes U(1)$) ведут к одним и тем же
результатам при физических значениях постоянной тонкой структуры.

Наше исследование показывает, что значения длины ребра решетки меньше, чем
$(430\pm 40\, {\rm ГэВ})^{-1}$ не могут быть достигнуты в принципе для решеток
рассматриваемого размера если потенциал скалярного поля выбран в Лондоновском
пределе, что соответствует перенормированной массе Хиггса $M_H \sim 800$ ГэВ.
Это говорит о том, что на данных решетках при данной массе Хиггса
ультрафиолетовое обрезание не может быть больше, чем ${\Lambda}_c =
\frac{\pi}{a_c} \sim 1.4$ ТэВ.

Также мы выяснили, что перколяция монополей Намбу является параметром порядка
для рассматриваемого перехода, что указывает на то, что мы имеем дело с так
называемым "перколяционным переходом", или переходом  Кертежа \cite{Kertesz}.

\subsection{Численное исследование модели при $\theta_W=\pi/6$, $\lambda = 0.001, 0.009$, $\beta \sim 12$}

Для данных значений констант связи инфракрасный эффективный потенциал нами не
вычислялся. Поэтому мы не располагаем данными о положении точки
$\gamma_c^{\prime}$ для $\lambda = 0.001, 0.009$. Положение точки $\gamma_c$
нами определяется по ультрафиолетовому потенциалу также, как это сделано для
$\lambda = 0.0025$. Основываясь на данных, полученных для $\lambda = 0.0025$ мы
полагаем, что $\gamma_c$ не является точкой истинного фазового перехода, а
является лишь одной из выделенных точек кроссовера. Также, мы не исключаем
возможности того, что в точках, где зануляется масса Z - бозона может иметь
место фазовый переход второго рода. Однако, эти точки не могут совпадать с
$\gamma_c$ поскольку в $\gamma_c$ вычисленная нами масса Z - бозона отлична от
нуля.

 Было обнаружено, что при $\lambda = 0.009$ в точке $\gamma_c = 0.274 \pm
 0.001$ и $\lambda = 0.001$ в точке $\gamma_c = 0.256 \pm
 0.001$
значение $\Lambda$ равно $1.4 \pm 0.2$ ТэВ. Сравнение результатов для решеток
разного размера не показывает изменения этого значения. На рис. \ref{62fig.3}
зависимость $M_Z$ в решеточных единицах от $\gamma$ представлена для  $\lambda
=0.009$ и $\beta = 12$. На рис. \ref{62fig.3_} зависимость $M_Z$ в решеточных
единицах от $\gamma$ представлена для  $\lambda =0.009$ и $\beta = 12$.

В Хиггсовском канале для $\lambda =0.009$ мы представляем данные для двух точек
на решетке $8^3\times16$: ($\gamma = 0.274$, $\lambda =0.009$, $\beta = 12$) и
($\gamma = 0.290$, $\lambda =0.009$, $\beta = 12$). Первая точка соответствует
положению перехода в то время, как вторая расположена глубоко в фазе Хиггса.
Затравочное значение массы Хиггса равно  $270$ ГэВ.

В точке ($\gamma = 0.274$, $\lambda =0.009$, $\beta = 12$) мы собрали
достаточно статистики, чтобы вычислить коррелятор  вплоть до "временных"
расстояний $|x_0-y_0| = 4$. Значение $\gamma = 0.274$ соответствует положению
перехода. Найденная нами масса в решеточных единицах равна $M^L_H = 0.75 \pm
0.1$ в то время, как затравочная масса $M_H$ равна $M^0_H \sim 270$ ГэВ. В то
же время $M_Z^L = 0.23 \pm 0.007$. Таким образом, мы оцениваем массу Хиггса как
$M_H = 300 \pm 40$ Gev.

В точке ($\gamma = 0.29$, $\lambda =0.009$, $\beta = 12$) мы вычислили
коррелятор с разумной точностью вплоть до $|x_0-y_0| = 3$. В этой точке
затравочная масса Хиггса $M_H$ равна $M^0_H \sim 260$ ГэВ в то время, как
перенормированная масса Хиггса в решеточных единицах  $M^L_H = 1.2 \pm 0.3$. В
то же время $M_Z^L = 0.41 \pm 0.01$. Таким образом, в этой точке мы оцениваем
$M_H = 265 \pm 70$ Gev.

 Для того, чтобы вычислить массу Хиггса
 при $\gamma = 0.274$ мы используем данные коррелятора  при
 $4 \ge |x_0-y_0| \ge 0$. Для вычисления массы Хиггса при
$\gamma = 0.29$ мы используем коррелятор для $3 \ge |x_0-y_0| \ge 0$.

В Хиггсовском канале для $\lambda =0.001$ мы представляем данные полученные на
решетке $8^3\times16$ для $\gamma = 0.257$. Имеем $M_H = 90 \pm 20$ ГэВ
(затравочное значение  $M^0_H \sim 100$ ГэВ). В этой точке мы собрали
достаточно статистики, чтобы вычислить коррелятор вплоть до "временного"
расстояния $|x_0-y_0| = 8$.

\begin{figure}
\begin{center}
 \epsfig{figure=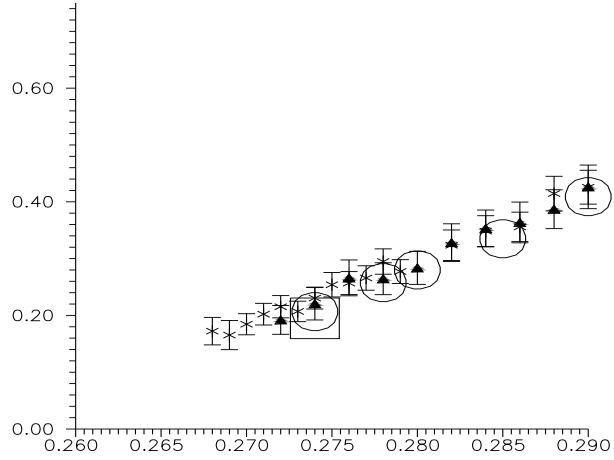,height=60mm,width=80mm,angle=0}
\caption{\label{62fig.3} Масса Z - бозона в решеточных единицах при $\lambda
=0.009$ and $\beta = 12$. Кресты соответствуют решетке $8^3\times 16$.
Треугольники -  $12^3\times 16$. Круги соответствуют решетке $16^4$, Квадрат
соответствует решетке $20^3\times 24$ (статистические ошибки - размера
используемых символов).}
\end{center}
\end{figure}

\begin{figure}
\begin{center}
 \epsfig{figure=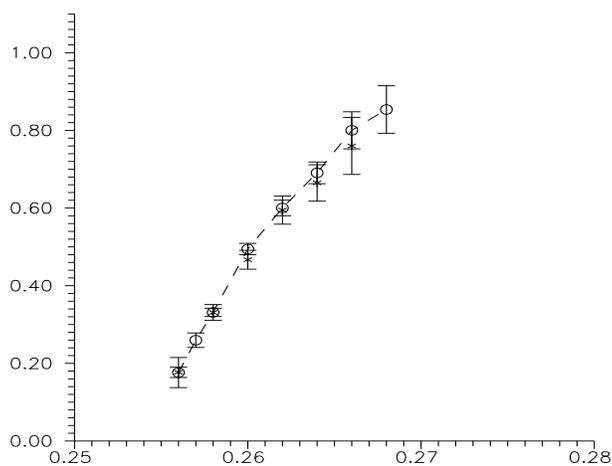,height=60mm,width=80mm,angle=0}
\caption{\label{62fig.3_} Масса Z - бозона в решеточных единицах при $\lambda
=0.001$ and $\beta = 12$. Кресты соответствуют решетке $8^3\times 16$. Круги  -
$12^3\times 16$. }
\end{center}
\end{figure}

Ультрафиолетовый эффективный потенциал для $\lambda = 0.001, 0.009$ ведет себя
вполне аналогично его поведению при $\lambda = 0.0025$.


\begin{figure}
\begin{center}
 \epsfig{figure=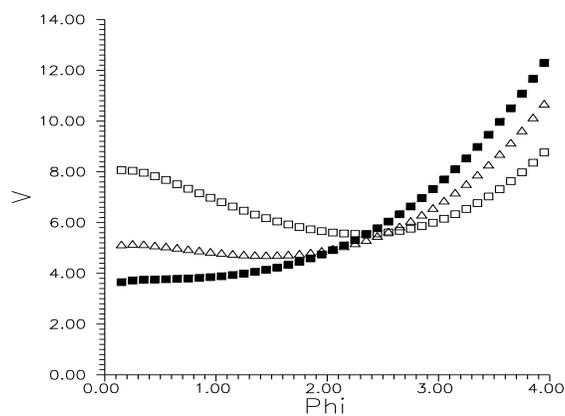,height=60mm,width=80mm,angle=0}
\caption{\label{62fig.1} Эффективный потенциал при  $\lambda =0.009$ и $\beta =
12$. Черные квадраты соответствуют $\gamma_c = 0.273$. Пустые квадраты
соответствуют $\gamma =0.29$. Треугольники соответствуют $\gamma = 0.279$.
Статистические неопределенности - порядка размера символов.  }
\end{center}
\end{figure}

Мы вводим величину $H = V(0) - V(\phi_m)$, которую называем высотой барьера
(здесь $\phi_m$ - это то значение, при котором $V$ достигает минимума).

\begin{table}
\caption{Значения $\phi_m$, $H$, $H_{\rm fluct}$, и плотности монополей Намбу
$\rho$ при некоторых значениях $\gamma$ для $\lambda = 0.009$, $\beta = 12$
(Решетка $8^3\times 16$.)}
\begin{center}
\begin{tabular}{|c|c|c|c|c|c|}
\hline
{\bf $\gamma$}  & {\bf $\phi_m$}  & {\bf $H$}& $H_{\rm fluct}$& $\rho$ \\
\hline $0.273$  & $0$ & $0$ & $0.1\pm 0.1$ & $0.098\pm 0.001$\\
\hline $0.274$  & $0$ & $0$ & $0.04 \pm 0.1$ & $0.081\pm 0.001$\\
\hline $0.275$  & $0.85\pm 0.1$ & $0.01 \pm 0.06$&  $0.15\pm 0.05$ & $0.067\pm 0.001$\\
\hline $0.276$  & $1.05\pm 0.1$ & $0.05\pm 0.06$& $0.16\pm 0.01$ & $0.054\pm 0.001$\\
\hline $0.277$  & $ 1.25 \pm 0.05$ &$ 0.19\pm 0.05$ &$0.25\pm 0.05$ & $0.044\pm 0.001$ \\
\hline $0.278$  & $ 1.35 \pm 0.1$  & $0.28 \pm 0.07$&$0.25\pm 0.06$& $0.035\pm 0.001$\\
\hline $0.279$  & $1.45\pm 0.05$ & $0.5 \pm 0.06$&  $0.25\pm 0.06$& $0.028\pm 0.001$\\
\hline $0.282 $ & $1.75\pm 0.05$ & $1.04\pm 0.07$&$0.31\pm 0.07$& $0.014\pm 0.001$\\
\hline $0.284 $ & $1.95\pm 0.05$ & $1.41\pm 0.08$&$0.38\pm 0.08$& $0.0082\pm 0.0005$\\
\hline $0.286 $ & $2.05\pm 0.05$ & $1.86\pm 0.08$&$0.35\pm 0.08$& $0.0049\pm 0.0002$\\
\hline $0.288 $ & $2.15\pm 0.05$ & $2.33\pm 0.08$&$0.32\pm 0.07$& $0.0029\pm 0.0002$\\
\hline $0.29 $ & $2.25\pm 0.05$ & $2.82\pm 0.08$&$0.44\pm 0.08$ & $0.0017\pm 0.0001$\\
\hline
\end{tabular}
\end{center}
\label{62Table}
\end{table}

В Таблице \ref{62Table} мы представляем значения $\phi_m$ и $H$ для  $\lambda =
0.009$, $\beta = 12$.  Видно, что значения $\phi_m$ и $H$ увеличиваются с
ростом $\gamma$. При $\gamma = 0.273$ минимум потенциала достигается при $\phi
= 0$. Эта точка соответствует максимуму восприимчивости построенной из поля
Хиггса. При $\gamma = 0.274$ мы также наблюдаем единственный минимум потенциала
при $\phi = 0$. При  $\gamma = 0.275$ минимум потенциала наблюдается при
$\phi_m = 0.85\pm 0.1$ с очень маленькой высотой барьера. Поэтому мы локализуем
положение перехода при $\gamma = 0.273\pm 0.002$.

Мы сравниваем  $H = V(0) - V(\phi_m)$ с $H_{\rm fluct} = V(\phi_m + \delta
\phi) - V(\phi_m)$, где $\delta \phi$ - это флуктуации $|\Phi|$. Из таблицы
\ref{62Table} видно, что существует значение $\gamma$ (мы называем его
 $\gamma_{c2}$) такое, что при $\gamma_c < \gamma < \gamma_{c2}$ высота барьера
 $H$ - порядка $H_{\rm fluct}$ в то время, как для $\gamma_{c2} << \gamma$
высота барьера значительно превышает $H_{\rm fluct}$. Грубая оценка этого
псевдокритического значения:  $\gamma_{c2} \sim 0.278$.

Флуктуации поля $|\Phi|$ можно оценить как $\delta \phi \sim 0.6$ для всех
рассматриваемых значений $\gamma$ при $\lambda = 0.009$, $\beta = 12$. Из наших
данных следует, что $\phi_m >> \delta \phi$ при $\gamma_{c2} << \gamma$ в то
время, как $\phi_m \sim \delta \phi$ при $\gamma_{c2} > \gamma$.

Таким же образом при $\lambda = 0.001$ мы получаем $\gamma_{c2} = 0.258\pm
0.001$. При $\gamma$ равном $\gamma_{c2}$ в рассмотренных случаях вычисленное
значение ультрафиолетового обрезания $1.0 \pm 0.1 $ ТэВ.


В Таблице \ref{62Table} мы представляем плотность монополей Намбу как функцию
от $\gamma$ при $\lambda = 0.009$, $\beta = 12$. Значение плотности монополей
при
 $\gamma_c = 0.273$ - около $0.1$. В этой точке значение ультрафиолетового обрезания
$\Lambda \sim 1.4 \pm 0.2$ ТэВ. На Рис.\ref{fig.rho.1} представлена монопольная
плотность для $\lambda = 0.001$.

\begin{figure}
\begin{center}
 \epsfig{figure=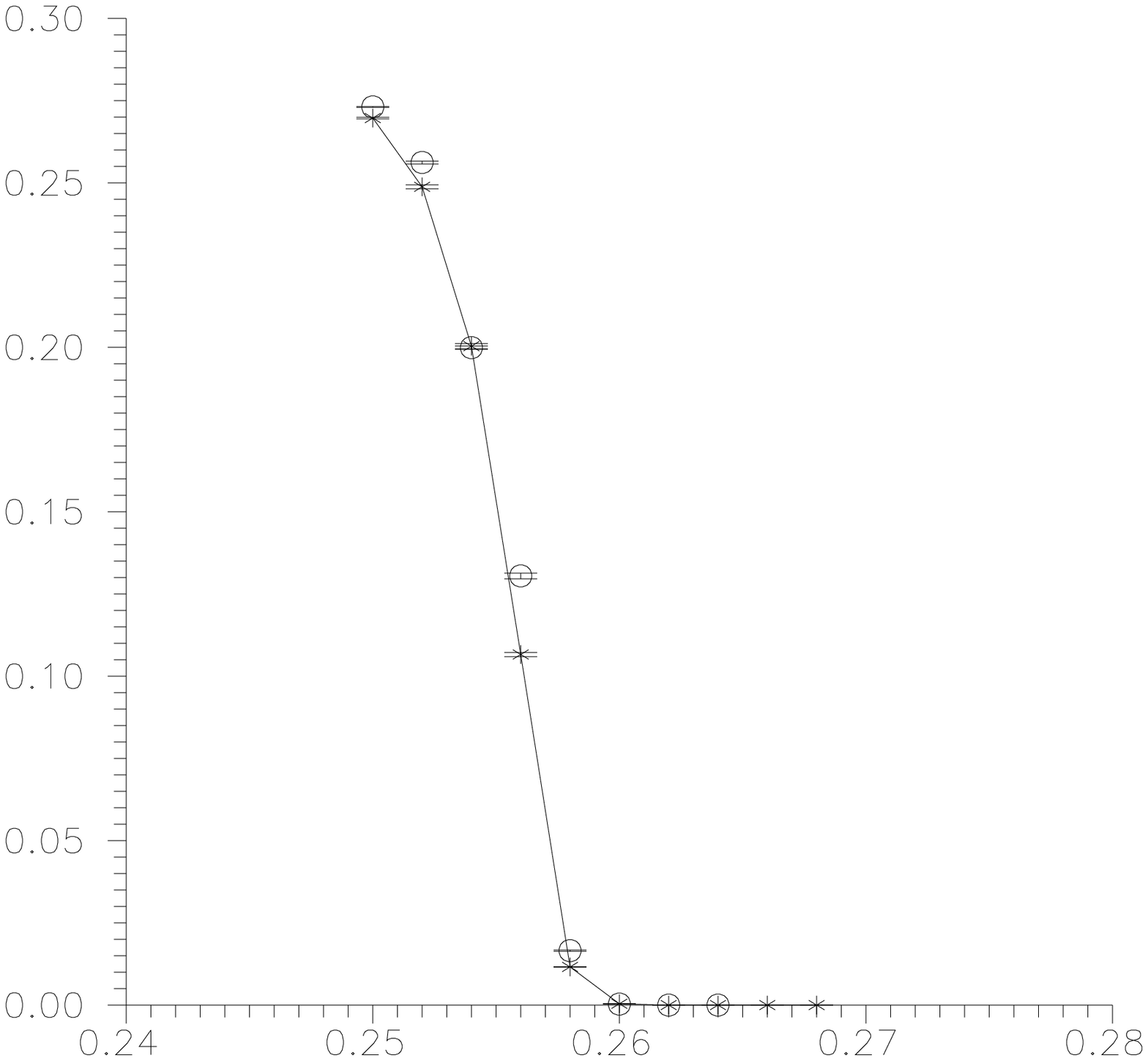,height=60mm,width=80mm,angle=0}
\caption{\label{fig.rho.1} Монопольная плотность как функция $\gamma$ при
$\lambda =0.0025$ , $\beta = 12$. Кресты соответствуют решетке $8^3\times 16$,
круги  - решетке $12^3\times 16$. }
\end{center}
\end{figure}

Снова для $\lambda = 0.008$ при $\gamma = \gamma_{c2}\sim 0.278$ и для $\lambda
= 0.001$ при $\gamma = \gamma_{c2}\sim 0.258$ мы видим, что среднее расстояние
между монополями Намбу - становится порядка их размера.

\section{Выводы}

В настоящей главе мы показали, что существует Флуктуационная Область (ФО) на
фазовой диаграмме решеточной модели Вайнберга - Салама. Эта область расположена
в окрестности фазового перехода (кроссовера) между физической Хиггсовской Фазой
и нефизической фазой. Как ультрафиолетовый, так и инфракрасный эффективные
потенциалы скалярного поля имеют минимум при одном и том же ненулевом значении
$\phi_m$ глубоко внутри физической фазы. Внутри ФО различные виды эффективного
потенциала дают разные значения $\phi_m$. Кроме того, они в разных точках
фазовой диаграммы качественно меняют свое поведение (приобретают такую форму,
что единственным их минимумом становится нулевое значение скалярного поля).

Скалярное поле равно нулю внутри классического монополя Намбу. Поэтому он
рассматривается нами как зародыш нефизической фазы внутри физической. Мы
исследовали свойства квантовых монополей Намбу. Внутри ФО они расположены столь
тесно, что расстояния между ними становятся порядка их размера. Это означает,
что внутри ФО флуктуации скалярного поля значительны и обе фазы перемешаны.

Что касается характера самого фазового перехода, мы пока не можем придти к
окончательному выводу о том, каковым он окажется на бесконечной решетке.
Наиболее вероятна, с нашей точки зрения, возможность того, что этот переход
является кроссовером для рассмотренных значений затравочных констант. Однако,
 мы не исключаем возможность существования фазового перехода второго рода при
$M_H\sim 300, 150, 100$ ГэВ. При этом мы полагаем, что наши выводы относительно
процессов внутри Флуктуационной Области, сделанные на основании исследования
модели на решетках конечного размера, останутся верны и для решетки
бесконечного размера.

Таким образом, вакуум модели Вайнберга - Салама внутри ФО существенно
отличается от тривиального вакуума, используемого в обычной теории возмущений.
Этот  результат мы рассматриваем как указание на то, что в ФО теория возмущений
может оказаться не применима.

Важно то, что переход к непрерывной физике производится при приближении к точке
перехода. Наши численные результаты говорят о том, что для $M_H\sim 300, 150,
100$ ГэВ при реалистическом значении постоянной тонкой структуры и угла
Вайнберга максимальное достижимое значение ультрафиолетового обрезания вне ФО
 - около  $1.0 \pm 0.1$ ТэВ. Наши оценки
были сделаны для решеток $8^3\times 16$, $12^3\times 16$, $16^4$ (результаты
при $M_H\sim 300$ ГэВ проверялись вплоть до размеров решетки $20^3\times 24$) и
не зависят от размера решетки. Мы также выяснили, что при правильном учете
эффектов конечного объема в перенормированной постоянной тонкой структуры
$\alpha$, ее значение оказывается близко к однопетлевому выражению, если в
последнем подставить в качестве обрезания масштаб $1$ ТэВ.

\section{Публикации}

Результаты настоящей главы опубликованы в работах:

"Nambu monopoles in lattice Electroweak theory",  B.L.G. Bakker, A.I. Veselov,
M.A. Zubkov. J.Phys.G36:075008,2009, [arXiv:0707.1017]

"Lattice study of monopoles in the Electroweak theory",  A.I. Veselov, B.L.G.
Bakker, M.A. Zubkov. PoS LAT2007:337,2007, [arXiv:0708.2864]

"Upper bound on the cutoff in lattice Electroweak theory", M.A. Zubkov, A.I.
Veselov. JHEP 0812:109,2008, [arXiv:0804.0140]

"Upper bound on the cutoff in the Standard Model", M.A. Zubkov, A.I. Veselov,
proceedings of 27th International Symposium on Lattice Field Theory (Lattice
2009), Beijing, China, 25-31 Jul 2009, [arXiv:0909.2840] [hep-lat]

"Monopoles in lattice Electroweak theory", B.L.G. Bakker, A.I. Veselov, M.A.
Zubkov, Proceedings of SPMTP08, [arXiv:0809.1757]

"The Fluctuational region on the phase diagram of lattice Weinberg-Salam
model",  M.A. Zubkov. Phys.Lett.B684:141-146,2010, [arXiv:0909.4106]

"The vicinity of the phase transition in the lattice Weinberg - Salam Model",
M.A.Zubkov, Proceedings of QUARKS-2010, [arXiv:1007.4885]

"How to approach continuum physics in lattice Weinberg - Salam model", M.A.
Zubkov, Phys.Rev.D82:093010,2010

"Effective constraint potential in lattice Weinberg - Salam model",
M.I.Polikarpov, M.A.Zubkov, Physics Letters B 700 (2011) pp. 336-342,
[arXiv:1104.1319]

\part{Решеточные формулировки квантовой гравитации}

В данной части нами рассматриваются подходы к построению квантовой теории
гравитации, основанные на динамической теории Римановой геометрии и ее
простейшего расширения - геометрии Римана - Картана. В седьмой главе мы
рассматриваем двумерную гравитацию. Прежде всего, решается простая задача о
поведении корреляторов в непрерывной двумерной гравитации с действием,
квадратичным по кривизне. Затем, рассматривается Редже - дискретизация
непрерывной модели и решается вопрос о мере интегрирования по полям. А именно,
показано, что при наложении на Редже - дискретизацию определенного условия мера
Лунда - Редже становится локальной. В восьмой главе мы рассматриваем
четырехмерную квантовую гравитацию. После рассмотрения Редже дискретизации и
демонстрации того, что присутствуют серьезные проблемы с мерой интегрирования
по полям в этой дискретизации, мы предлагаем две различных дискретизации, в
которых такие сложности отсутствуют. А именно, предложена калибровочно
инвариантная дискретизация телепараллелизма и калибровочно - инвариантная
дискретизация Пуанкаре - гравитации. Обе модели способны описывать квантовую
теорию Римановой геометрии. В девятой главе нами изучается дискретизация
Римановой геометрии, основанная на динамических триангуляциях. Эта
дискретизация отличается от рассмотренных в предыдущих двух главах тем, что в
ней особую роль играет энтропийный фактор. Действие непрерывной теории, по
сути, должно появиться динамически. Мы численно изучаем модель с затравочной
размерностью 10. Показано, что в этой модели в отличии от моделей с
затравочными размерностями  3, 4, и 5, энтропия принуждает модель пребывать  в
фазе ветвящихся полимеров (в низших размерностях она приводит модель в
хаотическую фазу).

\chapter{Двумерная гравитация. Предел слабой связи и дискретизация}
\label{ch7}
\section{Двумерная $R^2$ гравитация в пределе слабой связи }

\subsection{Определение модели и необходимость рассмотрения двумерного случая}

Ниже рассматривается двумерная $R^2$ квантовая гравитация в бесконечном
инвариантном объеме. Мы показываем, что в пределе слабой связи ее динамика
сводится к одномерной квантовой механике. Это позволяет получить простое
выражение для двухточечной функции Грина, которое может быть использовано для
тестирования различных решеточных дискретизаций.


Попытки квантовать гравитацию как модель Римановой геометрии с действием
Эйнштейна - Гильберта встречаются с трудностями, связанными с тем, что действие
неограничено снизу (после Виковского поворота). В результате, как ожидается, в
вакууме доминируют фрактальные структуры с большими значениями скалярной
кривизны. Эти ожидания действительно находят себе подтверждение в численных
исследованиях различных дискретизованных моделей  \cite{Loll}. Еще одна
трудность - это неперенормируемость  \cite{nrenorm}. Все это ведет к
необходимости включить в рассмотрение новые члены в действие, которые могут
сделать его ограниченным снизу и обеспечить перенормируемость теории. Это
действительно достигается если в действие включаются члены, квадратичные по
кривизне \cite{R2}. Кроме того, такая модель может быть построена таким
образом, что (см. следующую главу) в классическом пределе возникают уравнения
Эйнштейна в пустоте. Более того, теория становится асимптотически свободной. К
сожалению, непосредственное включение материи разрушает правильный классический
предел. Однако, в качестве альтернативы материя может рассматриваться как
сингулярности пространства - времени. В этом случае классические уравнения
Эйнштейна в присутствии материи возникают естественным путем
\cite{mass_singularities} (обсуждение этого также см. в следующей главе).
Проблема унитарности, возникающая в теории гравитации с вторыми производными от
метрики, остается в силе, однако, и она, в принципе может получить решение
непертурбативно \cite{unitarity}. Все указанное позволяет предположить, что
теория гравитации, основанная на Римановой геометрии и действии, квадратичном
по кривизне,  заслуживает исследования. Ряд шагов в этом направлении мы и
предпринимаем в настоящей  и  следующей главах. В главе 8, впрочем, в
рассмотрение включается простейшее расширение Римановой геометрии, основанное
на ненулевом кручении - геометрия Римана - Картана. (Это продиктовано
определенными трудностями в определении меры интегрирования по полям.) В главе
9 мы рассматриваем несколько иной путь квантования гравитации, основанный на
динамических триангуляциях.

Наиболее последовательным способом изучения непертурбативных свойств квантовой
теории является численное исследование, основанное на решеточной дискретизации.
Модель, описанная выше, будучи перенесена на решетку, теряет значительную долю
симметрии непрерывной теории. Эта симметрия, разумеется должна
восстанавливаться в непрерывном пределе. Поэтому необходимо сравнение
решеточных результатов с аналитическими. Четырехмерные модели весьма сложны и
их компьютерные симуляции требуют значительного времени и ресурсов. В связи с
этим особое значение приобретает рассмотрение упрощенных моделей, таких, как
двумерная  $R^2$ гравитация \footnote{См. также \cite{Odintsov}, где
рассматривалась $2D$ дилатонная гравитация (воспроизводящая $2D$ $R^2$
гравитацию).}.

\subsection{Динамические переменные и вычисления в пределе слабой связи}

При изучении $2D$ $R^2$ квантовой гравитации ыозникает вопрос: что следует
измерять? Часто используемой величиной является статсумма при фиксированном
инвариантном объеме  $V$ как функция этого объема.
\begin{eqnarray}
Z(V) & = & \int Dg {\rm exp} ( - \beta \int R^2 \sqrt{g} d^2 x)\delta(\int
\sqrt{g}
d^2 x - V)\nonumber\\
& = & \frac{1}{V}\int Dg {\rm exp} ( - \frac{\beta}{V} \int R^2 \sqrt{g} d^2
x)\delta(\int \sqrt{g} d^2 x - 1) , \label{Z}
\end{eqnarray}
где $\beta$ - обратная гравитационная постоянная.

В пределе $\beta/V >> 1$ эту величину можно оценить следующим образом. Прежде
всего, нетороидальная топология подавлена, поскольку только тор может нести
нулевую кривизну. Далее, при $\beta \rightarrow \infty$ система (\ref{Z})
становится эквивалентна $\cal N$ осцилляторам, где $\cal N$ - число степеней
свободы. Для дискретизованной модели это число -  $N \frac{D(D-1)}{2} = {\cal
N}$, где $N$ - число точек дискретизации. В результате $Z(V) \sim
\frac{1}{V}[\frac{V}{\beta}]^{N/2}$. С другое стороны, если использовать
размерную регуляризацию \cite{2DR2perturb} или технику конформной теории поля
\cite{R2CFT}, то получается иное выражение для статсуммы: $Z(V) \sim
\frac{1}{V}$. Это отличие как раз и отражает тот факт, что значительное
количество симметрии теряется при переходе к решеточной дискретизации. Таким
образом, величина (\ref{Z}) (также, как и любая другая величина, содержащая
зависимость от полного объема системы) неудобна для тестирования дискретизации
квантовой гравитации. В действительности, вычисленный в решеточной модели вклад
элементарных осцилляторов в эффективное действие должен быть соответствующим
образом вычтен из физических наблюдаемых при аккуратном рассмотрении
непрерывного предела (см., например, обсуждение ниже в п. 7.2.).

Более удобным является рассмотрение локальных переменных, которые не содержат в
себе отмеченных выше проблем свойственных глобальным переменным, зависящим от
объема. Локальные переменные зависят от метрики $g$. Ниже рассматривается
полный набор таких переменных, определенных в синхронной системе отсчета.
Динамика, записанная в терминах этих переменных оказывается неожиданно простой
в пределе слабой связи, то есть на расстояниях много меньших, чем
$\sqrt{\beta}$. Мы также предполагаем, что рассматриваемые расстояния много
меньше  $\sqrt{V}$, или, что то же - общий объем системы предполагается
бесконечным.

Зафиксируем точку $A$ и рассмотрим геодезические линии, заканчивающиеся в этой
точке. Вблизи к A эти линии выглядят как прямые. Зафиксируем одну из них
($l_0$) и параметризуем остальные $l_{\sigma}$ (инвариантным) углом $\sigma$
между $l_0$ и $l_s$. При $\beta \rightarrow \infty$ поверхность гладкая и
$\sigma \in [0, 2\pi[$.

Обозначим посредством $B_{\tau}(s) \in l_{\sigma}$ такую точку, что расстояние
между $A$ и $B_{\tau}(\sigma)$ равно $\tau = \rho(A,B_{\tau}(\sigma))$. Эти
точки формируют замкнутую кривую ${\cal C}_{\tau} = \{ B_{\tau}(\sigma) |
\sigma \in [0, 2\pi[ \}$. Длина части этой кривой, соединяющей
 $B_{\tau}(\sigma)$ и $B_{\tau}(0)$ обозначим $s(\sigma, \tau)$.

Функция $s(\sigma, \tau)$ полностью описывает локальные свойства метрики.
Параметризуем точки многообразия переменными $(\tau, \sigma)$. Тогда
$ g(\tau, \sigma) = \left(\begin{array}{cc} 1 & 0 \\
0 & e^{\phi(\tau, \sigma)}\end{array}\right),$
$s(\tau,\sigma) = \int_0^{\sigma} d \sigma_1 e^{\phi(\tau, \sigma_1)/2}$, и
\begin{equation}
R = - [\partial_{\tau}^2 \phi + \frac{1}{2} (\partial_{\tau}\phi)^2].\label{R7}
\end{equation}

Следует отметить, что 2D квантовая гравитация без члена с $R^2$ в действии в
синхронной системе отсчета была рассмотрена во множестве работ (см., например,
\cite{R2synchronic}).

Кривизна (\ref{R7}) зависит только от "временной" производной
$\partial_{\tau}$ (и не зависит от "пространственных" производных
$\partial_{\sigma}$). Это существенно упрощает динамику. Статсумма записывается
в виде:
\begin{equation}
Z = \int D\phi {\rm exp} ( - \beta \int [\partial_{\tau}^2 \phi + \frac{1}{2}
(\partial_{\tau}\phi)^2]^2 e^{\phi/2} d\sigma d\tau).\label{Z_1}
\end{equation}
(При $\beta \rightarrow \infty$ детерминант Фаддеева - Попова не существенен.)

Минимум экспоненциального фактора в (\ref{Z_1}) достигается, если
\begin{equation}
\partial_{\tau}^2 \phi + \frac{1}{2} (\partial_{\tau}\phi)^2 = 0 \label{R_1}
\end{equation}
Решая (\ref{R_1}) с соответствующими граничными условиями при $\tau = 0$
получаем:
$\phi_{cl} =  2\, {\rm log} \, \tau$.
Это предполагает, в частности, что длина ${\cal C}_{\tau}$ равна $2 \pi \tau$
как оно и должно быть для плоской геометрии.

Поэтому мы представляем
\begin{equation}
\phi =  2\, {\rm log} \, \tau + \omega (\tau, \sigma)
\end{equation}
и наша новая динамическая переменная - $\omega$, представляющая отклонение от
плоского случая. Ряд Фурье для $\omega$:
$\omega(\tau, \sigma) = \sum_n \omega_n(\tau) e^{i n \sigma}$, where
$\omega_{n} = \omega^+_{-n}$. Обозначим $\omega_n = \omega_n^{(1)} + i
\omega_n^{(2)}$ с действительной $\omega^{(k)}_n$. Очевидно, $\omega_0^{(2)} =
0$.
Гауссово приближение дает:
\begin{equation}
Z = \int \Pi_n D\omega_n {\rm exp} ( - 2\pi \beta \sum_{n\ge 0} \int
|\partial_{\tau}^2 \omega_n + \frac{2}{\tau} \partial_{\tau}\omega_n|^2 \tau
 d\tau).\label{Z2_}
\end{equation}

Ясно, что все гармоники распространяются независимо и наша задача - рассмотреть
распространение одной гармоники. С граничными условиями $\omega = 0$ при $\tau
= 0$ мы получаем следующее выражение для двухточечной функции Грина (вывод см.
ниже):
\begin{equation}
 <\omega^{(k)}_n(\tau) \omega^{(k^{'})}_{n^{'}}(\tau^{'})> = \frac{\delta_{n n^{'}}\delta_{k k^{'}}}{24 \pi \beta}\{ \tau^2(1 - \frac{\tau}{2\tau^{'}}) \theta(\tau^{'} - \tau) +
 (\tau^{'})^2(1 -  \frac{\tau^{'}}{2\tau}) \theta(\tau - \tau^{'})
 \}\label{FIN}
\end{equation}
(для $(k,n) \ne (2,0)$).

Этим и исчерпывается описание слабой связи в 2D $R^2$ гравитации в Гауссовом
приближении. Интересной особенностью (\ref{FIN}) является то, что корреляция
присутствует даже если одна из точек расположена  при  $\tau = \infty$.

Теперь мы можем вычислять, в частности, корреляцию длин  $|{\cal C}_{\tau}|$
кривых ${\cal C}_{\tau}$ при разных $\tau$:
\begin{equation}
 < [\frac{|{\cal C}_{\tau}|}{2\pi \tau} - 1] [\frac{|{\cal C}_{\tau^{'}}|}{2\pi \tau^{'}}-1] > = \frac{1}{96 \pi \beta}\{ \tau^2(1 - \frac{\tau}{2\tau^{'}}) \theta(\tau^{'} - \tau) +
 (\tau^{'})^2(1 -  \frac{\tau^{'}}{2\tau}) \theta(\tau - \tau^{'}) \}
\end{equation}
и флуктуацию \footnote{(\ref{C}) дает нам оценку применимости нашего
приближения: $\tau << \sqrt{192 \pi \beta} \sim 25 \sqrt{\beta}$ и $\tau <<
\sqrt{V/\pi}$.} этой длины как функцию  $\tau$:
\begin{equation}
 <[\frac{|{\cal C}_{\tau}|}{2\pi \tau}-1]^2> = \frac{1}{192 \pi \beta} \tau^2\label{C}
\end{equation}

Таким образом, мы определили динамические переменные, полностью описывающие
локальные свойства двумерной квантовой гравитации. В пределе слабой связи моды
фурье этих переменных распространяются независимо. Соответствующая
квантовомеханическая задача в Гауссовом приближении сводится к вычислению
двухточечной функции. Полученный нами результат для этой функции дается
формулой  (\ref{FIN}). Полученные выражения (\ref{FIN}) - (\ref{C})  могут
использоваться, в частности, для проверки различных решеточных дискретизаций
двумерной гравитации.

\subsection{Вывод выражения (\ref{FIN})}

 Выполним замену переменных  $t = \frac{1}{\tau}$. Результирующая статсумма
 для одной гармоники имеет вид
\begin{equation}
Z_0 = \int D\omega {\rm exp} ( - 2\pi \beta \int (\omega^{''})^2
 t^5 d t).\label{Z3}
\end{equation}

Производная $\partial_{\tau} \omega$ ограничена при $\tau = 0$ благодаря
гладкости многообразия. Тогда $\omega^{'} = - \frac{1}{t^2}\partial_{ \tau}
\omega$ должно стремиться к нулю при бесконечном t. Таким образом, система
(\ref{Z3}) дополняется граничными условиями  $\omega, \omega^{'} \rightarrow 0$
при $t \rightarrow \infty$.

Мы используем вспомогательную переменную $\eta = \omega^{'}$. Она
распространяется в соответствии с волновой функцией
\begin{equation}
\Psi(\eta, t) = \int D \bar{\eta} {\rm exp} [ - 2\pi \beta \int_0^{\infty}
(\bar{\eta}^{'})^2
 t^5 d t]\, \delta( \bar{\eta}(t) - \eta),\label{Psi}
\end{equation}
где $\bar{\eta} (\infty) = 0$.

Определим гамильтониан $H(t)$:
\begin{eqnarray}
 <\eta_1|e^{- H(t) \Delta t)}|\eta_2> & = & \int D \bar{\eta} {\rm exp} [ - 2\pi
\beta \int_t^{t + \Delta t} (\bar{\eta}^{'})^2
 t^5 d t]\nonumber\\ &&\delta(\bar{\eta}(t) - \eta_1)\, \delta(\bar{\eta}(t+\Delta t) - \eta_2).\label{H}
\end{eqnarray}

Стандартные методы дают
\begin{equation}
<\eta_1|e^{- H(t) \Delta t)}|\eta_2> = {\rm exp} ( - 2\pi \beta (\eta_2 -
\eta_1)^2 t^5 /\Delta t)
\end{equation}
и
\begin{equation}
<\eta_1|e^{- \int_{t_1}^{t_2} H(t) d t}|\eta_2> = {\rm exp} ( - 8\pi \beta
(\eta_2 - \eta_1)^2 /(\frac{1}{t_1^4} - \frac{1}{t_2^4}))
\end{equation}
Тогда
\begin{eqnarray}
\Psi(\eta, t)& = & \int d \eta_1 <\eta_1|e^{- \int_{0}^{t} H(t^{'}) d
t^{'}}|\eta> <\eta|e^{- \int_{t}^{\infty} H(t^{'}) d t^{'}}|0> \nonumber\\
& = &  <\eta|e^{- \int_{t}^{\infty} H(t^{'}) d t^{'}}|0>  =  {\rm exp} ( - 8\pi
t^4 \beta \eta^2 )
\end{eqnarray}

Получаем следующий результат для двухточечной функции:
\begin{eqnarray}
 <\eta(t) \eta(t^{'})> & = & \frac{1}{Z_0}\int d \eta d \eta^{'} \eta \eta^{'}
\nonumber \\ &&  {\rm exp} ( - 8\pi (t^{'})^4 \beta (\eta^{'})^2 - 8\pi \beta
(\eta - \eta^{'})^2 /(\frac{1}{t^4} - \frac{1}{(t^{'})^4}))
\end{eqnarray}
при $t^{'} > t$. Тогда
\begin{equation}
<\eta(t) \eta(t^{'})> =  \frac{1}{16\pi \beta (t^{'})^4} \theta(t^{'} - t) +
\frac{1}{16\pi \beta t^4} \theta(t - t^{'})
\end{equation}
Пусть $G(t, t^{'}) = <\omega(t) \omega(t^{'})>$. Тогда
\begin{equation}
\partial_{t}\partial_{t^{'}} G(t, t^{'}) = \frac{1}{16\pi \beta (t^{'})^4}
\theta(t^{'} - t) + \frac{1}{16\pi \beta ^4 t} \theta(t - t^{'})
\end{equation}
и
\begin{equation}
 G(t, t^{'}) = \int_t^{\infty} d \bar{t} \int_{t^{'}}^{\infty} d \bar{t^{'}}
\frac{1}{16\pi \beta (\bar{t}^{'})^4} \theta(\bar{t}^{'} - \bar{t}) +
\frac{1}{16\pi \beta ^4 \bar{t}} \theta(\bar{t} - \bar{t}^{'})
\end{equation}

Окончательное решение
\begin{equation}
 G(t, t^{'}) = \frac{1}{48 \pi \beta}\{\frac{2 t -  t^{'}}{ t ^3} \theta(t - t^{'}) +
\frac{2 t^{'} -  t}{ (t^{'}) ^3} \theta(t^{'}-t) \}
\end{equation}

Возвращаясь к переменной $\tau$, получаем:
\begin{equation}
 <\omega(\tau) \omega(\tau^{'})> = \frac{1}{24 \pi \beta}\{ \tau^2(1 - \frac{\tau}{2\tau^{'}}) \theta(\tau^{'} - \tau) +
 (\tau^{'})^2(1 -  \frac{\tau^{'}}{2\tau}) \theta(\tau - \tau^{'})
 \} \label{F}
\end{equation}

Это выражение можно проверить следующим образом. Для любой функции  $f(\tau)$
зануляющейся при $\tau = 0, \infty$ вместе с производными (вплоть до третьего
порядка) имеем:
\begin{eqnarray}
&& \int 4\pi \beta \, f(\tau) \,(\partial_{\tau}^2 - 2 \partial_{\tau}
\frac{1}{\tau})\tau (\partial_{\tau}^2 + \frac{2}{\tau}) <\omega(\tau)
\omega(\tau^{'})>  d \tau =
\nonumber\\
&& = \int \frac{1}{6}\{ 24\, \delta(\tau - \tau^{'}) \nonumber\\
&& + [-20 \tau +25 \frac{\tau^2}{\tau^{'}} - \frac{\tau^3}{(\tau^{'})^2}
 - 4 \frac{(\tau^{'})^2}{\tau}] \delta^{'}(\tau - \tau^{'})\nonumber\\
\nonumber\\
&& + [-10 \tau^2 + 7 \frac{\tau^3}{\tau^{'}} + 2 (\tau^{'})^2 +
\frac{\tau^3}{\tau^{'}}]\delta^{''}(\tau - \tau^{'})\nonumber\\
\nonumber\\
&& + [- \tau^3 +  \frac{\tau^4}{2\tau^{'}} + \tau (\tau^{'})^2 -
\frac{\tau^3}{2}]\delta^{'''}(\tau - \tau^{'})\} f(\tau) d \tau
\end{eqnarray}

Используя интегрирование по частям, это выражение переписываем в виде
\begin{eqnarray}
&& \int 4\pi \beta (\partial_{\tau}^2 - 2 \partial_{\tau} \frac{1}{\tau})\tau
(\partial_{\tau}^2 + \frac{2}{\tau}) <\omega(\tau) \omega(\tau^{'})> f(\tau) d
\tau =
\nonumber\\
&& = \int  \, \delta(\tau - \tau^{'}) f(\tau) d \tau
\end{eqnarray}

Это и является проверкой того, что (\ref{F}) действительно дает функцию Грина
оператора $4 \pi \beta (\partial_{\tau}^2 - 2
\partial_{\tau} \frac{1}{\tau})\tau (\partial_{\tau}^2 + \frac{2}{\tau})$
присутствующего в (\ref{Z2_}).

\section{Мера в двумерной дискретизованной квантовой гравитации}

\subsection{Постановка задачи}

Выше нами рассматривалась двумерная гравитация в пределе слабой связи, когда
конкретный вид меры интегрирования по полям не важен по сравнению с действием,
поскольку оно содержит фактор $\beta \rightarrow \infty$.  Теперь рассмотрим
подробно вопрос о мере интегрирования по геометриям и то, как это
интергирование должно осуществляться в дискретизованной гравитации. Мы
предлагаем версию двумерного исчисления Редже, в котором площади симплексов
равны друг другу. В этой дискретизации мера Лунда - Редже по линковым
переменным существенно упрощается. Наша мера локальна в противоположность
обычному исчислению Редже с мерой Лунда - Редже, где мера оказывается
нелокальной, а реальные симуляции существенно осложнены.

Как уже отмечалось выше, одна из основных трудностей квантовой гравитации - это
широта ее калибровочной группы - группы общих координатных преобразований.
Из-за этой широты почти невозможно положить теорию на решетку без потери
калибровочной инвариантности. В настоящий момент единственно возможной
калибровочно - инвариантной дискретизацией является исчисление Редже. Однако,
ценой калибровочной инвариантности является проблема с определением меры
интегрирования. Здесь вместо метрики, которая была в непрерывной теории
локальным полем, фундаментальными калибровочно - инвариантными величинами
являются длины линков. А метрика выражается через них нелокально. Поэтому и
мера оказывается нелокальной.

   Не смотря на то, что в  \cite{W} и связанных публикациях утверждается,
    что определенный выбор
   локальной меры интегрирования по длинам линков может оказаться соответствующим непрерывной
   теории, точка зрения автора настоящей диссертации иная. Подробно мы аргументируем нашу
   точку зрения в Главе 8, где обсуждается мера, предлагаемая в  \cite{W} и обсуждаются
   недостатки этого подхода. В настоящей главе мы принимаем в качестве
   отправной точки для построения меры интегрирования в исчислении Редже так
   называемую меру Лунда - Редже (см., например, \cite{A1},
   где она также называется мерой Де Витта). В отличии от меры, предлагаемой в
   \cite{W} и являющейся эвристической, эта мера последовательным путем
   получена из  соответствия
   с нормой на пространстве римановых геометрий. Как отмечалось выше,
   она в общем случае нелокальна. В действительности ее форма столь сложна, что
   представляется невозможным использовать ее в реальных численных симуляциях.

Однако, в настояшей главе мы показываем, как эта трудность можно преодолеть в
двумерном случае \footnote{К сожалению, предлагаемый способ не удается
перенести на многомерный случай. Поэтому в следующей главе, посвященной
четырехмерным моделям мы предлагаем иное решение для построения локальной меры
интегрирования в дискретизованных моделях гравитации, основанное на расширении
Римановой геометрии  до геометрии Римана - Картана.}. Для того, чтобы сделать
это мы принимаем компромиссное решение. А именно, мы начинаем дискретизацию
гравитации, стартуя с непрерывной теории, в которой калибровка частично
фиксирована фиксацией конформной моды. В результате мы приходим к версии
исчисления Редже, в котором площади всех симплексов равны друг другу. В этой
дискретизации мера Лунда - Редже становится локальной, что позволяет
осуществлять численные симуляции.

\subsection{Мера Лунда - Редже}

Прежде всего напомним основные факты о мере интегрирования в квантовой
гравитации. Пространством, по которому производится интегрирование, является
множество всех компактных  $D$ - мерных Римановых многообразий. На этом
пространстве определена следующая метрика
\begin{equation}
\|\delta g\|^2 = \frac{1}{2}\int d^D x \sqrt{|g|} (g^{\mu \nu}g^{\rho \eta} +
g^{\rho \nu}g^{\mu \eta} + C g^{\mu \rho}g^{\nu \eta})\delta g_{\mu \rho}
\delta g_{\nu \eta}\label{7NORM}
\end{equation}

Хорошо известно, что в конечномерном случае любая метрика (на компактном
пространстве) продуцирует единственную меру  $\mu$. В двух словах процедура
построения метрики работает следующим образом. Прежде всего выберем некоторое
$\epsilon$. Далее, наполним шарами радиуса $\epsilon$ все пространство. Шары
должны быть расположены плотно, то есть их количество - максимально возможное
для данного $\epsilon$. Затем для любого данного множества $\Omega \subset
\Lambda$ (все пространство обозначено $\Lambda$) вычисляем количество шаров,
расположенных внутри $\Omega$. Обозначим это число $n_{\epsilon}(\Omega)$.
Окончательно
\begin{equation}
\mu(\Omega) = {\rm lim}_{\epsilon \rightarrow 0}
\frac{n_{\epsilon}(\Omega)}{n_{\epsilon}(\Lambda)}
\end{equation}

В принципе эта процедура может быть обобщена и на бесконечномерный случай.
Пространство может быть представлено как предел последовательности
конечномерных компактных пространств. А именно, каждое Риманово пространство
может быть рассмотрено как предел последовательности кусочно - линейных
многообразий, построенных из все большего количества симплексов (симплекс - $D$
- мерное обобщение треугольника и пирамиды). Эта процедура и называется Редже -
дискретизация.  Тогда можно попытаться построить непрерывную меру на
пространстве всех Римановых геометрий как предел мер, определенных на указанных
выше кусочно  - линейных пространствах. Мера на Редже - дискретизованном
Римановом пространстве с фиксированной триангуляцией (то есть взаимном
расположением симплексов), но различными размерами симплексов, соответствующая
норме (\ref{7NORM}), - и называется ниже мерой Лунда - Редже \cite{A1}.

 Для того, чтобы вычислить ее в явном виде, снабдим каждый симплекс  $[\Gamma_0 \Gamma_1 ...
\Gamma_D]$ базисом  $\{ \frac{\Gamma_0 \Gamma_1}{|\Gamma_0 \Gamma_1|}$ ,
$\frac{\Gamma_0 \Gamma_2}{|\Gamma_0 \Gamma_2|}$ , ... , $\frac{\Gamma_0
\Gamma_D}{|\Gamma_0 \Gamma_D|} \}$. В этом базисе метрика равна
\begin{equation}
g_{i j} = \frac{1}{2}(|\Gamma_0 \Gamma_i|^2 + |\Gamma_0 \Gamma_j|^2 - |\Gamma_i
\Gamma_j|^2). \label{7G}.
\end{equation}
Здесь $|\Gamma_i \Gamma_j|$ - это расстояние между точками $\Gamma_i$ и
$\Gamma_j$. Подставим (\ref{7G}) в (\ref{7NORM}) и получим форму, квадратичную
по вариациям длины линков (мы обозначили длину линков $a_i$).

Перед этим перепишем (\ref{7NORM}) в следующем виде:
\begin{eqnarray}
\|\delta g\|^2 & = & \frac{1}{2}\int d^D x \sqrt{|g|} (g^{\mu \nu}g^{\rho \eta}
+ g^{\rho \nu}g^{\mu \eta} + C g^{\mu \rho}g^{\nu \eta})\delta g_{\mu \rho}
\delta g_{\nu \eta}\nonumber\\
& = & \int d^D x \sqrt{|g|} ( - \delta g^{\mu \rho} \delta g_{\mu \rho} +
\frac{C}{2} [\delta {\rm log} |g| ]^2)\label{N1}
\end{eqnarray}
Мы использовали здесь, что $\delta |g| = |g| g^{ik} \delta g_{ik}$ и
$g_{ji}\delta g^{ik} = - g^{ik} \delta g_{ji}$.

Представим $g^{ij}$ как
\begin{equation}
g^{kl} = \frac{1}{(D-1)!g}\epsilon^{kab...d}\epsilon^{lvg...s} g_{av} g_{bg}
... g_{ds}\label{g}
\end{equation}

Тогда, подставив (\ref{g}) в (\ref{N1}), получим:
\begin{equation}
\|\delta g\|^2  = \int d^D x  [-
\frac{\epsilon^{kab...d}\epsilon^{lvg...s}\delta g_{kl}\delta g_{av} g_{bg} ...
g_{ds}}{\sqrt{|g|}(D-2)!}
 + \frac{C+2}{2} (\delta {\rm log} |g|)^2 \sqrt{|g|}]\label{N2}
\end{equation}

Теперь введем новую переменную ${\hat g}_{ij} = |g|^{-\frac{1}{D}} g_{ij}$.
Детерминант $|{\hat g}|$ равен единице по построению. Исходная полевая
переменная $g_{ij}$ тогда представляется как произведение этой новой переменной
и конформного фактора $|g|^{\frac{1}{D}}$. В результате (\ref{N2})
переписывается как
\begin{equation}
\|\delta g\|^2  = \int d^D x  [-
\frac{\epsilon^{kab...d}\epsilon^{lvg...s}\delta {\hat g}_{kl}\delta {\hat
g}_{av} {\hat g}_{bg} ... {\hat g}_{ds}}{(D-2)!}
 + \frac{C+\frac{2}{D}}{2} (\delta {\rm log} |g|)^2] \sqrt{|g|}\label{N2a}
\end{equation}
Таким образом, мы видим, что конформная мода $|g|$ ортогональна полю  ${\hat
g}_{ij}$.

 На симплициальном многообразии выражение (\ref{N2}) имеет вид:
\begin{equation}
\|\delta a\|^2 = \sum_{\rm simplices} [-
\frac{\epsilon^{kab...d}\epsilon^{lvg...s}\delta g_{kl}\delta g_{av} g_{bg} ...
g_{ds}}{\sqrt{|g|}(D-2)!}
 + \frac{C+2}{2} (\delta {\rm log} |g|)^2 \sqrt{|g|}]
 \label{NORMD2}
\end{equation}
Здесь \footnote{Следует отметить, что аналогичным образом выражение (\ref{N2a})
может быть перенесено на симплициальное многообразие. Однако, тогда основное
свойство (\ref{N2a}) теряется. А именно, на решетке мы снова можем ввести новые
переменные (вместо длин линков). Эти новые переменные - это площади
треугольников и остающиеся углы. Наивно можно ожидать, что по аналогии с ${\hat
g}_{ij}$ и $|g|$ подпространства, натянутые на эти два множества переменных
ортогональны. Однако, это не так, потому, что теперь индуцированная  ${\hat
g}_{ij}$ на симплексах функционально зависит и от упомянутых угловых
переменных, и от площадей треугольников. } на каждом симплексе $g_{ij}$
выражается через длины линков $a_m$.

 Собирая все члены, квадратичные по $\delta
a_m$, можно получить окончательный результат
\begin{equation}
\| \delta a \|^2 = \sum_{i,j \in C_1} O_{i j}[a] \delta (a_i^2) \delta
(a_j^2),\label{na}
\end{equation}
с матрицей $O$, которая (в общем) зависит от длин линков довольно сложным
образом. Далее мы покажем, однако, что в двух измерениях это выражение
существенно упрощается если мы налагаем некое дополнительное ограничение на
Редже - скелетон.

Из (\ref{na}) следует, что результирующая мера дается \footnote{Мы обозначаем
множество линков решетки как $C_1$, а множество треугольников как $C_2$. Число
точек обозначается как $N_0$, число треугольников обозначается как $N_2$, число
линков обозначается как $N_1$. } \cite{W,A1}:
\begin{equation}
D a =  {\rm Det}^{\frac{1}{2}} O[a] \Pi_{i\in C_1}  d a^2_i,\label{DA}
\end{equation}

Выражение (\ref{DA}) было названо в \cite{A1} мерой Де - Витта. Однако, мы
полагаем более правильным называть ее мерой Лунда - Редже, поскольку выведена
она была из выражения для метрики, впервые данного Лундом и Редже в их
неопубликованном препринте. Как уже отмечалось, в общем случае  (\ref{DA}) не
подходит для практических вычислений из - за сложной и нелокальной формы. Ниже
будет показано, что в двух измерениях можно достигнуть существенного упрощения,
позволяющего использовать меру Лунда - Редже в практических численных
исследованиях.

\subsection{Двумерный случай}

Итак, обратимся к двумерному случаю. Прежде всего, возвратимся к непрерывной
формулировке. Прежде, чем дискретизовать модель, частично фиксируем калибровку,
требуя, чтобы конформная мода $|g|$ имела значение, не зависящее от положения
на поверхности:
\begin{equation}
\sqrt{|g(x)|} = \sqrt{|g(x_0)|},
\end{equation}
где $g(x)$ - метрический тензор, а $x_0$ - фиксированная точка на поверхности.
Процедура Фаддеева - Попова (будучи применена к статсумме модели с действием
$S[g]$) дает
\begin{eqnarray}
Z & = & \int Dg \exp( - S[g])\int D f \delta({\rm log}\,\sqrt{|g^f|} - {\rm log}\, \sqrt{|g^f(x_0)|})\Delta_{FP}[g]\nonumber\\
& = & \int Dg^{f^{-1}} D f \exp( - S[g^{f^{-1}}])  \delta({\rm log}\,\sqrt{|g|} - {\rm log}\,\sqrt{|g(x_0)|}) \Delta_{FP}[g^{f^{-1}}]\nonumber\\
& = & \int Dg \exp( - S[g]) \delta({\rm log}\,\sqrt{|g|} - {\rm log}\,
\sqrt{|g(x_0)|}) \Delta_{FP}[g]
\end{eqnarray}
Здесь\footnote{Важно отметить, что определение дельта - функции на Римановом
многообразии требует задания фоновой метрики $g_0$. А именно, эта функция может
быть представлена как  $\delta(h(x)) = {\rm lim}_{t\rightarrow \infty} \exp(-t
\int h^2(x) \sqrt{g_0(x)} d^2 x)$. В наших формулах мы предполагаем, что некая
фоновая метрики задана. И эта метрика не имеет ничего общего с динамической
переменной $g$. (Разумеется, таким же образом мера по репараметризациям $D f$
определяется по отношению к этой фоновой метрике.)} репараметризации обозначены
как $f$, а их действие на метрику обозначены $g^f$. Учитываем, что действие
$S[g]$, мера  $Dg$,  и детерминант Фаддеева - Попова $\Delta_{FP}[g]$
репараметризационно инвариантны.

Детерминант Фаддеева - Попова выражается как (репараметризации записаны здесь
как $ x \rightarrow x - y(x)$) :
\begin{eqnarray}
&& \Delta_{FP}^{-1}[g]|_{\sqrt{|g(x)|} = \sqrt{|g(x_0)|}}  =   \int D y
\delta({\rm log} \, \frac{|g(x - y(x))|^{1/2}(1 +
\frac{1}{2}\partial_{\mu}y^{\mu}(x))}{|g(x_0 - y(x_0))|^{1/2}(1 +
\frac{1}{2}\partial_{\mu}y^{\mu}(x_0))})\nonumber\\
& = & \int D y \delta({\rm log} \, \frac{1 +
\frac{1}{2}\partial_{\mu}y^{\mu}(x)}{1 +
\frac{1}{2}\partial_{\mu}y^{\mu}(x_0)})
 = {\rm  const}
\end{eqnarray}

  Таким образом,
$\Delta_{FP}(g)$ не зависит от метрики.

Итак, мы начинаем дискретизацию квантовой гравитации с модели с статсуммой
\begin{equation}
Z =  \int Dg \exp( - S[g]) \delta({\rm log}\,\sqrt{|g(x)|} - {\rm log}\,
\sqrt{|g(x_0)|}) \label{CGF}
\end{equation}

В Редже дискретизации (\ref{CGF}) мы используем меру Лунда - Редже (\ref{DA})
вместо $Dg$. Вместо дельта - функции $\delta({\rm log}\,\sqrt{|g(x)|} - {\rm
log}\, \sqrt{|g(x_0)|}) $ мы используем ее дискретизованную версию:
\begin{equation}
\Theta[a] = \Pi_{J\in C_2, J \ne J_0} \delta({\rm log} \, A^J - {\rm log}\,
A^{J_0} ),
\end{equation}
где произведение - по треугольникам (всем, кроме данного треугольника $J_0$)
симплициального многообразия, а  $A^J$ - площадь треугольника. Мы также можем
представить последнее выражение в следующем виде:
\begin{equation}
\Theta[a] = \int \Pi_{J\in C_2} \delta({\rm log} \, A^J - {\rm log}\, (V/N_2) )
\frac{d V}{V},
\end{equation}
где $V$ - общий инвариантный объем.

Таким образом, статсумма дискретизованной модели имеет вид
\begin{equation}
 Z  =   \int Da \exp(-S[a]) \label{CGFD}
\end{equation}
где $S[a]$ - дискретизованное действие, а $Da$ - мера:
\begin{equation}
 Da = \{{\rm Det}^{\frac{1}{2}} O[a]\,  \frac{d
V}{V}\}\, \{\Pi_{i\in C_1} d a^2_i\}  \,\{\Pi_{J\in C_2} \delta({\rm log} \,
A^J - {\rm log}\, (V/N_2) )\} \label{CGFDM}
\end{equation}

В (\ref{CGFDM}) появилось дополнительное ограничение на линковые переменные. Мы
удерживаем площади всех треугольников равными друг другу. Исходно имелось $N_1$
степеней свободы. Есть $N_2$ дополнительных условий. В каждой точке имеется
также $2$ лишних локальных степеней свободы. Проще всего понять их природу,
рассматривая плоскую поверхность. Тогда локальные смещения каждого узла решетки
внутри поверхности оставляет аппроксимируемую поверхность неизменной. Итак,
полное число лишних степеней свободы равно   $2 N_0 = N_2 + 4(1-h)$, где $h$ -
род поверхности. Для тора число дополнительных условий равно числу лишних
степеней свободы. В случае сферы число дополнительных условий меньше, чем число
лишних степеней свободы. Для $h > 1$, однако, наше дополнительное условие может
стать слишком ограничительным. Поэтому для поверхности рода $h
> 1$ число дополнительных условий должно быть уменьшено.
Далее мы ограничиваем себя случаями $h = 0,1$.

Основное упрощение достигается в выражении для детерминанта  ${\rm Det} \,
O[a]$  в специальном случае $C = -2$ когда площади всех треугольников равны
друг другу. В этом случае (\ref{NORMD2}) имеет вид:
\begin{equation}
\|\delta a\|^2 = - \frac{N_2}{2V} \sum_{J\in C_2}
\epsilon^{ka}\epsilon^{lv}\delta g^J_{kl}\delta g^J_{av}, \label{NORMD2L}
\end{equation}
Это выражение может быть легко переписано в терминах длин линков $a_i$:
\begin{equation}
\|\delta a\|^2 = \frac{1}{V}\sum_{ij} U^{ij} \delta (a^2_{i}) \delta (a^2_{j})
\end{equation}
где матрица $U$ зависит от триангуляции и не зависит от длин линков. Таким
образом, в этом случае ${\rm Det} \, O[a] \sim \frac{1}{V^{N_1}}$.

Поэтому мы приходим к следующему {\bf выводу:} В двух измерениях (для $h = 0,
1; C = -2$) мера по дискретизованным геометриям имеет вид\footnote{Здесь и
далее мы рассматриваем меру с точностью до фактора, зависящего только от
триангуляции.}:
\begin{equation}
D a = \{V^{- 1 + N_1/2} d V\} \, \{ \Pi_{i\in C_1} d [\frac{a^2_i}{V/N_2}]\}\,
\{\Pi_{J\in C_2} \delta({\rm log} \, A^J - {\rm log}\, (V/N_2) )\},\label{DAC2}
\end{equation}

Выражение (\ref{DAC2}) и есть основной результат настоящего раздела.

\subsection{Модель с квадратом кривизны в действии}

Теперь рассмотрим пример конкретной $2D$ модели. А именно, рассмотрим
дискретизованную $R^2$ модель. Статсумма имеет вид:
\begin{equation}
Z = \int D a \exp( - \beta \, \sum_{i\in C_0} \frac{\theta_i^2}{B_i[a]} )
\label{DGF}
\end{equation}
Полный угол поворота (deficit angle) в точке $i$ обозначен как $\theta_i$.
Здесь мы используем традиционное определение для дискретизованной скалярной
кривизны $\frac{\theta_i}{ B_i[a]}$ (где $B_i[a] = \sum_{i \in J} \frac{1}{3}
A^J$ - сумма по треугольникам, имеющим данную точку $i$ в качестве одной из
своих вершин).

Теперь рассмотрим статсумму для фиксированного инвариантного объема (мы
перескалируем $a$ как $ a \rightarrow \sqrt{2 V/N_2} a$):
\begin{equation}
Z(V) =  V^{N_1/2-1} \hat{Z}(\hat{V}),
\end{equation}
где
\begin{equation}
\hat{Z}(\hat{V}) =   \int  \Pi_{i\in C_1} d a^2_i \exp( - \frac{1}{\hat{V}}\,
\frac{N_2}{2} \sum_{i\in C_0} \frac{\theta_i^2}{B_i[a]} )\Pi_{J\in
C_2}\delta(2A^J[a]-1), \label{RDGF}
\end{equation}
и $\hat{V} = \frac{V}{\beta}$.

Для того, чтобы вычислить $Z(V)$ и извлечь струнную восприимчивость мы должны
исследовать коррелятор
\begin{equation}
F(V,\beta)  =   \frac{\partial \, {\rm log}\, { Z}}{\partial \, {\rm log} { V}}
- (\frac{N_1}{2}-1) = \frac{N_2}{2 {\hat Z} {\hat V}}  < \sum_{i\in C_0}
\frac{\theta_i^2}{B_i[a]}
> \label{CRDGF}
\end{equation}
где усреднение производится в модели, определенной статсуммой (\ref{RDGF}).

При ${\hat V} >> 1$ структура ${ Z}({ V})$, как ожидается, должна
соответствовать следующему выражению
\begin{equation}
{ Z}({ V}) \sim { V}^{ - 3 + \gamma} \exp( m  V)), \label{GM}
\end{equation}
где $\gamma$ - это так называемая струнная восприимчивость (string
succeptibility), а $m$ - перенормированная космологическая постоянная
\cite{KPZ}.

С другой стороны, мы ожидаем, что в пределе $N_2
>> \hat{V}
>> 1$ система в лидирующем порядке ведет себя как собрание
$N_d$ осцилляторов, где $N_d$ - эффективное число степеней свободы. Итак,
ведущая часть (\ref{CRDGF}):
\begin{equation}
F({\hat V}) = \frac{N_d}{2} + o(N_2)
\end{equation}

Наивно можно предположить, что $N_d = N_1-N_2$.  Тогда мы можем представить
$\gamma$ как (мы здесь использовали, что $N_1 = \frac{3}{2}N_2$):
\begin{equation}
\gamma =  N_2 + \gamma_{fin} + O(\frac{1}{N_2}), \label{GAMMA}
\end{equation}
где $ \gamma_{fin}$ не зависит от $N_2$.

 С другой стороны, формула КПЗ\cite{KPZ} дает конечный результат:
\begin{equation}
\gamma = 2 - \frac{5}{2} (1 - h)
\end{equation}

Сравнение двух указанных выражений для струнной восприимчивости показывает их
несовпадение. Однако, вблизи к непрерывному пределу число степеней свободы
может быть уменьшено благодаря восстанавливающейся симметрии (которая была
потеряна в дискретизованной модели). При этом может получиться иное выражение
для струнной восприимчивости в дискретизованной модели. Впрочем, исследование
этого вопроса выходит за рамки нашего анализа.

\subsection{Модификация алгоритма Метрополиса}

Важно отметить, что обычный алгоритм Метрополиса, будучи применен к системе
(\ref{RDGF}) должен быть видоизменен с тем, чтобы реализовать дополнительные
ограничения на площади треугольников. При изменении длин линков условие $A_i =
0.5$ разрешается следующим образом. Случайным образом выбирается линк,
предлагается новое значение его длины,  и рассматриваются два треугольника,
имеющие его в качестве общей стороны. В каждом из этих треугольников мы
выбираем вершину, не принадлежащую данному линку. Далее рассматриваются все
треугольники, имеющие одну из этих точек в качестве вершины. Результирующая
фигура состоит из двух звезд с двумя отмеченными выше точками в качестве
центров.

Если есть $n_1$ внутренних линков этой фигуры (то есть не принадлежащих границе
этой фигуры), тогда есть либо $n_1 - 1$, либо $n_1 - 2$ треугольников,
принадлежащих фигуре. Таким образом, оставляя прежними длины граничных линков и
исходного линка, чья длина уже была обновлена, мы можем в обоих случаях
разрешить условие  $A^J = 0.5$ внутри фигуры. Однако, неравенства треугольника
могут запретить разрешение нашего дополнительного условия для некоторых
значений длины исходного линка. Если так, предлагаем иное значение длины
исходного линка и вновь пытаемся разрешить дополнительные условия.

На практике мы используем доп. условие
\begin{equation}
|A^J - 0.5| < \delta \label{CON}
\end{equation}
на каждом треугольнике. Здесь $\delta$ выбрана достаточно малой.

Чтобы разрешить дополнительные условия мы делаем несколько шагов вдоль всех
линков фигуры, разрешенных к изменению. Подстраивая каждый линк, минимизируем
величину  $(A_1^2 - \frac{1}{4})^2 + (A_2^2 - \frac{1}{4})^2$, где  $A_1$ и
$A_2$ - площади треугольников, имеющих данный линк в качестве общей стороны.
Минимизация достигается решением кубического уравнения. Процедура повторяется
итеративно пока условие (\ref{CON}) не достигается на каждом треугольнике.

После выполнения данной процедуры формируется предложение новых длин внутренних
линков фигуры. Затем это предложение принимается с обычной вероятностью ( $ p =
1$ если $\Delta S < 0$, иначе $ p = e^{-\Delta S}$, где $\Delta S$ -
соответствующее изменение действия).

\section{Публикации}

Результаты настоящей главы опубликованы в работах

"2D R**2 gravity in weak coupling limit", M.A. Zubkov.
Phys.Lett.B594:375-380,2004

"Measure in the 2D Regge quantum gravity", M.A. Zubkov.
Phys.Lett.B616:221-227,2005

\chapter{Четырехмерная квантовая гравитация. Дискретизация Телепараллелизма и
Пуанкаре - гравитации} \label{ch8}

\section{Гравитация с действием, включающим члены, квадратичные по кривизне}

Квантовая гравитация является одной из важных проблем современной теоретической
физики. Данные о структуре гравитации на Планковской шкале энергий приходят к
нам от единственного реального эксперимента, которым является история
Вселенной. Эти данные могут быть получены из астрофизических наблюдений. Эти
энергии не могут быть достигнуты в экспериментах, созданных вручную, что
существенно усложняет ситуацию.

Математическая структура квантовой гравитации должна быть связана с Римановой
геометрией, которая должна появиться, как минимум, в классическом пределе
квантовой теории. Разумеется, есть множество геометрических и алгебраических
структур, которые могут служить основой квантовой теории гравитации. Однако, в
настоящий момент у нас нет оснований для того, чтобы выбрать ту или иную
математическую основу для будущей теории.  Поэтому естественным является
"минимальный" выбор, возвращающий нас к Римановой геометрии. Даже если такая
теория и не будет исчерпывающей, она может появиться как хорошее
низкоэнегритическое приближение.

Как известно, перенормируемые асимптотически свободные теории хорошо определены
на малых расстояниях. Мы не можем сказать то же про перенормируемые теории,
которые не являются асимптотически свободными и, тем более, о неперенормируемых
теориях. Поэтому естественно было бы начинать попытки построения квантовой
теории гравитации с требования перенормируемости и асимптотической свободы.

Также известно, что ультрафиолетовые расходимости в квантовой гравитации с
действием квадратичным по кривизне могут быть поглощены перенормировкой
констант связи \cite{renormalizable}. Модель, рассмотренная в
\cite{renormalizable} и связанных публикациях имеет следующее действие (будучи
преобразована к Евклидовой сигнатуре пространства - времени):
\begin{equation}
S = \int \{ \alpha (R_{AB} R_{AB} - \frac{1}{3}R^2) + \beta R^2 - \gamma m_p^2
R + \lambda m_P^4\} |E| d^4 x, \label{S1}
\end{equation}
где $|E| = {\rm det} E^A_{\mu}$, $E^A_{\mu}$ - поле тетрады, тетрадные
компоненты тензора Риччи обозначены $R_{AB}$, а $R$ - скалярная кривизна.
Константы связи $\alpha, \beta, \gamma$, и $\lambda$ - безразмерные, в то
время, как $m_p$ - размерный параметр. Линеаризованная (возле плоского фона)
теория содержит гравитон и дополнительные тензорные и скалярные возбуждения.
Пропагатор ведет себя как $\frac{1}{q^4}$ в ультрафиолете при $\alpha, \beta
\ne 0$. Тензорное возбуждение является духом, что ведет к потере унитарности.
Полное непертурбативное рассмотрение асимптотических состояний, однако, может
привести к решению этой проблемы  \cite{unitarity}. Вероятно, традиционный
подход здесь не работает поскольку взаимодействие между квантовыми
возбуждениями  (а также их формирование) - более сложны, чем предполагается при
использовании обычных методов теории возмущений. Тем не менее, теория
возмущений для функций Грина может содержать важную информацию. Мы ожидаем, что
благодаря перенормируемости теории это разложение является самосогласованным и
может, в принципе, рассматриваться как приближенная схема вычислений.

Однако, проблема унитарности - не единственная проблема, встречающаяся при
рассмотрении теории с действием  (\ref{S1}). А именно, требование того, чтобы
действие (\ref{S1}) было ограничено снизу ведет к появлению тахиона. Это
означает, что плоское пространство не является истинным вакуумом модели. Тахион
исчезает, если рассматривается фоновая метрика, минимизирующая  (\ref{S1})
\cite{a_free}. В дополнение к ультрафиолетовым расходимостям теория возмущений
может также содержать и инфракрасные расходимости. Для того, чтобы отделить их
рассмотрение от рассмотрения ультрафиолета, следует использовать дополнительную
регуляризацию. Например, мы можем зафиксировать полный инвариантный объем
системы. Тогда в обычной (скажем, размерной, или решеточной) регуляризации
ультрафиолетовые расходимости поглощаются переопределением констант связи, и
каждый член ряда теории возмущений становится конечным. В теории с
фиксированным инвариантным объемом космологическая постоянная не влияет на
динамику, а действие ограничено снизу если ($\alpha \ge 0$, $\beta > 0$,
$\gamma\ne 0$) или ($\alpha \ge 0$, $\beta \ge 0$, $\gamma = 0$).
Ренормгрупповой анализ показывает, что \cite{a_free} при $\alpha, \beta
>0$ существует область констант связи, такая, что
теория асимптотически свободна относительно  $\alpha$ и $\beta$, в то время,
как $\gamma$ - может быть выбрана постоянной (в однопетлевом приближении). Мы
не обсуждаем  ниже возможные расходимости, которые могут возникнуть в пределе
$V \rightarrow \infty$. Отметим, однако, что аналогичные расходимости,
появляющиеся в КЭД, компенируются излучением мягких фотонов. Вероятно, подобный
механизм компенсации может реализоваться и в квантовой гравитации.

К сожалению, классический Ньютоновский предел нельзя получить из действия
 (\ref{S1}) пока оно ограничено снизу. Однако, если мы стартуем с чистой гравитации
 с действием  (\ref{S1}) (с $\lambda = 0$) и поворачиваем теорию обратно к
 сигнатуре пространства Минковского, решения уравнений Эйнштейна в пустоте
 будут удовлетворять классические уравнения движения  \footnote{Это не единственные решения уравнений движения.
 Однако, при $\gamma
= \lambda = 0$ пространства Эйнштейна минимизируют (Евклидово) действие.
Поэтому представляет интерес рассмотрение теории с действием  (\ref{S1}) при
условии, что на некоторой шкале энергий {\it перенормированные} константы
$\gamma$ и $\lambda$ зануляются.}. Тогда массивные точечные объекты могут быть,
в принципе, рассмотрены как сингулярности пространства - времени
\cite{mass_singularities}, а Ньютоновский предел появляется как асимптотика
решений вида черных дыр. Предположим, сингулярность вдоль некоторой линии
внедрена в пространство - время. Тогда уравнения Эйнштейна в пустом
пространстве ведут к уравнениям Эйнштейна в присутствии частицы, чья мировая
линия совпадает с линией сингулярности. Масса частицы не фиксируется
уравнениями поля, но показано, что будучи задана, она не изменяется со временем
\cite{mass_singularities}. Это указывает на то, что материя может быть введена
в квантовую теорию гравитации с действием  (\ref{S1}) таким образом, что
воспроизводятся результаты общей теории относительности при  $\alpha > 0,\,
\beta > 0, \lambda = 0$.

\section{Дискретизации}

Далее будут рассматриваться дискретизации  как на гиперкубической, так и на
симплициальной решетке модели с действием  (\ref{S1}) при $\alpha > 0,\,\beta
> 0$.  Традиционный подход к дискретизации базируется на геометрической интерпретации
в терминах Римановой геометрии. Поэтому основные переменные - это метрика и
 $SO(3,1)$ (ил $SO(4)$) связность. В некоторых схемах используются расширения
 связности
 (скажем, связности группы Пуанкаре или группы Де Ситтера).
 Тетрада вместе с $SO(3,1)$ калибровочным полем появляются как часть соответствующей
 связности, а метрика компонуется из тетрады \cite{Loll,Menotti}.
Требование того, чтобы кручение занулялось приводит к появлению Римановой
геометрии.  Ниже нами развивается соответствующий подход (см.п. 8.4.).

Альтернативный подход к дискретизации Римановой геометрии связан с исчислением
Редже  \cite{Regge, Regge_R2, Regge_measure}, в котором пространство - время
приближается множеством симплексов (четырехмерный аналог треугольника и
пирамиды) склеенных вместе. Каждый симплекс предполагается плоским. Геометрия
определяется размерами симплексов. Численные исследования соответствующих
моделей  производились на протяжении последних $20$ лет. В частности, теория с
действием квадратичным по кривизне была исследована в этом подходе (см.
\cite{Regge_R2}). К сожалению подход Редже дискретизации связан с проблемой в
определении меры интегрирования (см. ниже п. 8.4.1, а также главу 7)
\cite{Regge_measure}.

В подходе динамических триангуляций (ДТ) реализуется вариант исчисления Редже
\cite{DT} с фиксированными размерами симплексов. В этом подходе различные
способы склеивания симплексов и являются динамическими переменными (см.
следующую главу). К сожалению, за исключением случая  $D = 2$ симплициальное
многообразие, использующееся в подходе ДТ не в состоянии воспроизвести плоское
пространство. Обычно предполагается, что плоское пространство появляется
приближенно. Но в этом случае локальное действие квадратичное по кривизне не
переходит в действие (\ref{S1}) с $\gamma = \lambda = 0$ в непрерывном пределе:
всегда появляется член линейный по $R$ \cite{DTR2}.

Ниже мы изложим, два подхода к дискретизации Римановой геометрии \footnote{Иные
подходы к дискретизации римановой геометрии обсуждаются в обзоре
 \cite{Loll}. Некоторые из конструкций упомянутых в  \cite{Loll} содержат
 теории с действием (\ref{S1}) в качестве частного случая или могут быть изменены соответствующим образом.
 Исчерпывающие численные исследования проводились только в Редже и ДТ подходах
\cite{Regge_R2,DT}, и в модели Де Ситтера \cite{Smolin, dS_numerical}. Модель,
соответствующая действию (\ref{S1}) численно исследовалась только в рамках
исчисления Редже  \cite{Regge_R2} на очень маленьких решетках. Однако,
изучалась упрощенная модель с $\alpha = 0$.}. Один из них основан на
дискретизации теории гравитации в телепараллельной формулировке
\cite{teleparallel}, а другой основан на дискретизации Пуанкаре гравитации, то
есть гравитации с кручением.

\section{Телепараллелизм}

\subsection{Пространство Вейценбока}

 Геометрическая конструкция, на которой основан телепараллелизм, -
это пространство Вейценбока, которое является частным случаем пространства
Римана - Картана. Пространство Римана - Картана - это касательное расслоение,
снабженное связностью группы Пуанкаре. Группа Пуанкаре состоит из
преобразований Лоренца и трансляций. Трансляционная часть связности
идентифицируется с полем тетрады и определяет метрику. Соответствующая часть
кривизны переходит в кручение. Риманова геометрия возникает когда зануляется
кручение. Геометрия Вейценбока - это противоположный случай: Лоренцева часть
кривизны зануляется, а кручение отлично от нуля. Телепараллелизм - это теория
геометрии Вейценбока, или трансляционная калибровочная теория.

Если пространственно - временное многообразие параллелизуемо  (как это
предполагается ниже), Лоренцева связность может быть выбрана равной нулю при
нулевой кривизне. Поэтому единственной динамической переменной является
тетрада, рассматриваемая как трансляционная связность. Обычно в
телепараллельной гравитации действие выражается через трансляционную кривизну
(кручение). Пространственно - временные и внутренние индексы могут спариваться
с использованием поля тетрады. Эквивалентность между непрерывными теориями
Римановой и Вейценбоковой геометрий задается, если в Римановой геометрии все
выражается через метрику, а последняя строится из тетрады, которая
идентифицируется с трансляционной связностью геометрии Вейценбока. Например,
существует специальный выбор действия телепараллелизма, квадратичного по
трансляционной кривизне, которое переходит при указанной выше идентификации в
действие Эйнштейна - Гильберта. Таким образом, существует телепараллельная
версия ОТО.

Следует отметить, что  в рамках дискретизованной Пуанкаре - гравитации кручение
может появиться как дислокации симплициальной решетки \cite{Poincare_lat}.
Решеточная дискретизация телепараллелизма может, в принципе, быть построена,
стартуя с \cite{Poincare_lat}, если симплициальное многообразие удерживается
соответствующим плоскому пространству. В соответствующей модели следует
определить поле тетрады, функциональную меру и решеточную производную кручения.
Иное представление решеточной телепараллельной гравитации рассматривалось в
\cite{teleparallel_lat}, где базовой конструкцией является обычное
симплициальное многообразие, использующееся в исчислении Редже.   Тензор Римана
на симплициальном многообразии может быть выражен через вторые производные от
поля тетрады. В результате в \cite{teleparallel_lat} можно определить
компоненты решеточного кручения, соответствующие телепараллельному эквиваленту
теории Редже с действием Эйнштейна. Этот подход позволяет трактовать
симплициальное Риманово многообразие как приближение решеточного пространства
Вейценбока и выразить классические уравнения Эйнштейна в Редже дискретизаци в
терминах кручения. К сожалению, расширения этого формализма на гравитацию с
высшими производными не производилось.

Ниже нами предлагается альтернативный способ дискретизации телепараллельной
гравитации, позволяющий работать с действием  (\ref{S1}). В нашем подходе
геометрия Вейценбока переносится на решетку непосредственно, без использования
дополнительных структур (таких, как симплициальное Риманово многообразие,
используемое в отмеченных выше подходах \cite{Poincare_lat,teleparallel_lat}).
Наша основная переменная - это решеточная связность Абелевой калибровочной
группы трансляций и все величины выражаются непосредственно через нее.

\subsection{Дискретизуемая модель}

Прежде, чем перейти непосредственно к дискретизации, кратко напомним,
определение непрерывной модели. Поле тетрады обозначается как   $E_{\mu}^A$
(все пространственно - временные индексы обозначены греческими буквами) и
рассматривается как трансляционная связность. Тензор Римана выражается через
трансляционную связность следующим образом ($T_{ABC} = E^A_{[\mu,\nu]}E^{\mu}_B
E^{\nu}_C$):
\begin{equation}
R_{ABCD} = \gamma_{AB[C,D]} + \gamma_{ABF}\gamma_{F[CD]} +
\gamma_{AFC}\gamma_{FBD} - \gamma_{AFD}\gamma_{FBC},\label{R}
\end{equation}
где
$\gamma_{ABC} = \frac{1}{2}(T_{ABC} + T_{BCA} - T_{CAB})$.
Квантовая теория определяется функциональным интегралом статсуммы:
\begin{equation}
Z = \int D E exp( - S[E]) \label{Z}
\end{equation}
Мера интегрирования по $E$ определяется как мера по трансляционной
калибровочной группе.

Рассматриваемая теория инвариантна относительно диффеоморфизмов  $(x_{\mu}
\rightarrow \tilde{x}_{\mu}; E^A_{\mu} \rightarrow \frac{\partial
\tilde{x}_{\mu} }{\partial x_{\nu} } E^A_{\nu}$). Подставляя единицу Фаддеева -
Попова мы фиксируем калибровку, в которой  $|E| = {\rm const}$. Таким образом
замораживается общий инвариантный объем. Ниже мы пренебрегаем его флуктуациями.
Соответствующий детерминант Фаддеева - Попова не зависит от  $E$. Таким
образом, имеем
\begin{equation}
Z = \int D E exp( - S[E]) \delta(|E| - {\rm const}),
\end{equation}

Как уже было сказано выше, непрерывная версия модели перенормируема и
асимптотически свободна при соответствующем выборе констант связи
\footnote{Выбор меры $D E$ не влияет на этот результат. Однако, для получения
перенормируемой теории может оказаться необходимым включить в рассмотрение
флуктуации общего объема. Технически это не вызывает никаких трудностей.
Однако, в нашем изложении мы этого не делаем.}. Для этих значений констант
связи действие ограничено снизу (если общий объем удерживается постоянным).
Таким образом, мы надеемся, что решеточная модель должна быть хорошо
определена.

\subsection{Дискретизация телепараллелизма}


Рассматривается либо симплициальная, либо гиперкубическая решетка. Далее мы
будем говорить как о симплексах, так и о гиперкубах как об элементах решетки.
Пространство внутри каждого элемента решетки предполагается плоским. Форма
элементов решетки фиксируется векторами ${\bf e}_{\mu}$, соединяющими центр
элемента с центрами сторон его границы. Выражение ${\bf e}_{\mu}$ через
ортогональный базис общий для всей решетки  ${\bf f}_A$ ($A = 1,2,3,4$)
является основной переменной нашей конструкции. Имеем
\begin{equation}
{\bf e}_{\mu}(x) = \sum_A E^A_{\mu}(x) {\bf f}_A
\end{equation}
Не все вектора ${\bf e}_{\mu}$ независимы. А именно, на гиперкубической решетке
${\bf e}_{\mu} + {\bf e}_{-\mu} = 0$ ($\mu = \pm 4, \pm 3, \pm 2, \pm1$),
где ${\bf e}_{\mu}$ и ${\bf e}_{-\mu}$ связывают центр гиперкуба с
противоположными сторонами его границы. В случае симплициальной решетки имеем
 ($\mu = 1,2,3,4,5$)
$\sum_{\mu} {\bf e}_{\mu} = 0$,
Также мы предполагаем, что все стороны границы элемента упорядочены.
Независимые переменные в обоих случаях обозначены  $E^A_{\mu}$ с $\mu =
1,2,3,4$.

Трансляционная кривизна (кручение) локализована на двумерных симплексах
симплициальной решетки и плакетах гиперкубической решетки, называемых далее
двумерными элементами (bones). Рассмотрим замкнутый путь, соединяющий центры
решеточных элементов, содержащие данный двумерный элемент. Часть пути,
соединяющая центры $x$ и $y$ соседних элементов решетки состоит из двух
векторов  ${\bf e}_{M_{xy}}(x)$ и ${\bf e}_{M_{yx}}(y)$, соединяющих  $x$ и $y$
с центром общей стороны. Целочисленная функция $M_{xy}$ определена на парах
центров соседних элементов решетки. Мы обозначим ${\bf e}_{xy} = {\bf
e}_{M_{xy}}(x) - {\bf e}_{M_{yx}}(y)$  и ${\bf e}_{yx} = {\bf e}_{M_{yx}}(y) -
{\bf e}_{M_{xy}}(x)$. Для замкнутого пути $x \rightarrow y \rightarrow z
\rightarrow ... \rightarrow w \rightarrow x$ вокруг данного двумерного элемента
решеточное кручение равно
\begin{equation}
T_{xyz...w} = \frac{1}{s}({\bf e}_{xy} + {\bf e}_{yz} + ... + {\bf e}_{wx})
\end{equation}
где $s$ - площадь фигуры, имеющий данный путь своей границей. Если расстояние
между центрами соседних элементов равно единице, тогда $s = 1$ в
гиперкубическом случае, и $s = O\sqrt{\frac{D+1}{D-1}}$ с $D = 4$ в
симплициальном случае.
 Мы также обозначим
$T_{xyz...w} = T^A_{M_{xy} M_{xw}}(x) {\bf f}_A$,
где предполагается суммирование по $A$.

Теперь выразим базисные вектора ${\bf f}$ через $\bf e$. Обозначим
\begin{equation}
{\bf f}_A = \sum_{\mu} F^{\mu}_A(x) {\bf e}_{\mu}(x)
\end{equation}
Здесь сумма - по всем векторам $\bf e$ (а не только по четырем независимым).
Для гиперкубического случая обозначим $\bf E$ матрицы $4\times 4$, состоящие из
$E^A_{\mu}$ с положительным $\mu$. Тогда $F$ может быть выбрано в виде
$F^{\mu}_A = \frac{{\rm sign}(\mu)}{2}({\bf E}^{-1})^{|\mu|}_A$.
В симплициальном случае ситуация сложнее. Для того, чтобы получить симметричное
выражение обозначим $\bar{E}_{\mu} = [E^1_{\mu} ... E^4_{\mu}]^T$ and
${\bf E}_{\mu} = [\bar{E}_1 ... \bar{E}_{\mu - 1} \bar{E}_{\mu+1} ...
\bar{E}_{5}]$.
Тогда $F$ можно выбрать в виде:
$F^{\mu}_A = \frac{1}{5}\sum_{\nu \ne \mu}({\bf E}_{\nu}^{-1})^{\rho(\nu,
\mu)}_A$,
где $\rho(\nu,\mu) = \mu$ если $\nu > \mu$ и $\rho(\nu,\mu) = \mu - 1$ если
$\nu < \mu$ . Теперь тетрадные компоненты кручения легко вычисляются
\begin{equation}
T_{ABC}(x) = \sum_{\mu,\nu} F^{\mu}_B F^{\nu}_C T^A_{\mu \nu}(x)
\end{equation}

Мы также определяем тетрадные компоненты решеточной производной от  $T$:
$T_{ABC,D}(x) = \sum_{\mu} F^{\mu}_D T_{ABC,\mu}(x)$,
где
$T_{ABC,\,\mu}(x) = T_{ABC}(y) - T_{ABC}(x)$
для $\mu = M_{xy}$.

Формально применяя выражение (\ref{R}) к определенным выше решеточным
переменным мы получаем определение решеточного тензора Римана. Тогда тензор
Риччи и скалярная кривизна определены как обычно. Решеточное действие тогда
строится следующим образом:

\begin{equation}
S = \sum_x \{ \alpha (R_{AB}(x) R_{AB}(x) - \frac{1}{3}R(x)^2) + \beta R(x)^2 -
\gamma m_p^2 R \} \label{SL}
\end{equation}

Мы опустили здесь фактор $|E|$ и член, пропорциональный $\lambda$ поскольку
наложен решеточный аналог условия  $|E| = {\rm const}$. Именно в этой
калибровке следует рассматривать модель если мы хотим удерживать постоянным
общий объем системы. Решеточный аналог непрерывного калибровочного условия
 $|E| = {\rm const}$ - это условие постоянства инвариантного объема каждого элемента решетки.
 Если мы обозначим
${\bf E} = {\bf E}_5$ для симплициального случая, соответствующее условие может
быть выбрано в виде
${\rm det}\, {\bf E} - v = 0$
(для обеих решеток), где $v$ - это константа, пропорциональная общему объему
решетки.

Тогда ${\bf E}$ - это динамическая переменная, все $E^A_{\mu}$ выражены через
нее, и статсумма модели имеет вид
\begin{equation}
Z = \int \Pi_x  D{\bf E}(x) {\rm exp} ( - S[{\bf E}]) ,
\end{equation}
где
\begin{equation}
D {\bf E} = (\Pi_{A,B} d {\bf E}^A_B) \delta({\rm det}{\bf E} - v)
\end{equation}

\subsection{Глобальные свойства многообразия}

Для того, чтобы закончить конструкцию, следует задать глобальные свойства
многообразия  $\cal M$. Прежде всего, следует отметить, что рассматривать
открытое пространственно - временное многообразие невозможно поскольку мы не
определили граничные члены в действии. Вместо того, чтобы строить эти члены, мы
ограничиваем себя случаем замкнутых многообразий. Естественное ограничение на
топологию - это то, что многообразия должны быть параллелизуемыми. Иначе наши
переменные не описывают геометрию Вейценбока.

Гиперкубическая решетка удобна для рассмотрения случая четырехмерного тора
$T^4$, соответствующего обычным периодическим граничным условиям.

Если мы хотим рассматривать случай более сложной топологии, необходимо
использовать симплициальную дискретизацию. При этом неприемлемо требование
параллелизуемости. Поэтому в этом случае мы должны дополнить указанную выше
конструкцию рассмотрением $SO(4)$ связности с нулевой кривизной. Эта связность
локализована на 3-мерных симплексах и позволяет соединять разные части
симплициального многообразия (каждая из частей имеет фиксированный базис).
Соответственно, производные и их дискретные аналоги следует заменить
ковариантными с данной $SO(4)$ связностью.

Следует отметить, что в общем случае топологии $4$ - мерных многообразий
невозможно классифицировать. Для того чтобы иметь дело с топологией,
поддающейся классификации, следует наложить сильное ограничение на  $\cal M$.
Обычно предполагают существование спинорной структуры \cite{Hawking}. В этом
случае сигнатура $\tau$ и эйлерова характеристика $\chi$ характеризуют
многообразие с точностью до гомотопии (если $|\tau| \ne \chi - 2$). Чтобы
исследовать процессы с изменением топологии симплициальное многообразие должно
быть динамическим (подобно моделям ДТ). В дополнение к сохраняюшим топологию
изменениям триангуляции следует определить изменяющие топологию  (см.,
например, \cite{topology_DT}). Альтернативным способом исследовать динамическую
топологию могло бы также послужить исследование матричной модели, эквивалентной
симплициальной теории \cite{de_Bakker}.

\section{Пуанкаре - гравитация}

\subsection{Еще раз о проблемах с мерой интегрирования в Редже гравитации}

 Как уже говорилось в главе 7, для того, чтобы определить меру интегрирования
 на множестве всех Римановых геометрий, пользуются инвариантной метрикой,
 заданной на этом множестве:
\begin{equation}
\|\delta g\|^2 = \frac{1}{2}\int d^D x \sqrt{|g|} (g^{\mu \nu}g^{\rho \eta} +
g^{\rho \nu}g^{\mu \eta} + C g^{\mu \rho}g^{\nu \eta})\delta g_{\mu \rho}
\delta g_{\nu \eta}\label{NORM}
\end{equation}
где $C$ - константа, такая, что $C\ne-2/D$. Можно постулировать, что правильная
мера на пространстве Римановых геометрий соответствует этой метрике в том же
смысле, в каком мера  $dx$ на $R^1$ соответствует метрике  $\|\delta x\|^2 =
(\delta x)^2$. К сожалению, переход к бесконечномерному случаю представляется
нетривиальным. В главе 7 мы обсуждали меру Лунда - Редже в исчислении Редже,
основанную на естественном способе перехода от континуума Римановых геометрий к
конечномерному случаю Редже дискретизации. В двумерном случае удается сделать
меру Лунда - Редже локальной и тем самым подходящей для применения численных
методов. В четырехмерном случае этого сделать не удалось.

В ряде работ, использующих численные симуляции Редже гравитации использовалась
эвристически выбранная мера интегрирования по длинам линков. Ход рассуждений,
приводящих к этой мере следующий (см., например,  \cite{Regge_measure}).
Бесконечномерное обобщение конечномерного выражения для меры выбирается в виде
\begin{equation}
Dg  =  \Pi_x (\sqrt{|g(x)|})^{\sigma} \Pi_{\mu \ge \nu}
dg_{\mu\nu}(x)\label{Measure}
\end{equation}
Здесь выбор $\sigma = (D-4)(D+1)/4$ соответствует суперметрическому тензору в
форме предложенной в \cite{DeWitt,Fijikawa}, а выбор $\sigma=-(D+1)$
рассматривался в \cite{Misner}. Очевидно, эта неоднозначность связана с тем,
что произведение по точкам  $\Pi_x$ неопределено, если  $x$ принадлежит $R^D$.
На самом деле конкретное выражение может быть придано (\ref{Measure}) только
когда выбрана какая - то дискретизация и $x$ не должен пробегать континуум
значений.

В практических вычислениях использовалась мера  \cite{Regge_measure}:
\begin{equation}
Dl  =  [\Pi_x V_x^{\sigma}] [\Pi_{ij} dl^2_{ij}]
\Theta(l_{ij})\label{MeasureL},
\end{equation}
где произведение  $\Pi_x$ - по симплексам, а произведение $\Pi_{ij}$ - по
вершинам симплициального многообразия. $\Theta(l_{ij})$ - функция,
обеспечивающая неравенства треугольника.  $V_x$ - объем симплекса $x$, а
$l_{ij}$ - длина линка, соединяющего вершины $i$ и $j$. Оправдывается выбор
меры (\ref{MeasureL}) следующим образом. На каждом симплексе $V_x^{\sigma}
\Pi_{ij} dl^2_{ij}(x)$  - это непосредственная дискретизация выражения
$(\sqrt{|g(x)|})^{\sigma} \Pi_{\mu \ge \nu} dg_{\mu\nu}(x)$ (где $i$ и $j$ -
вершины симплекса, а постоянный фактор опущен). В системе симплексов, склеенных
друг с другом длины линков не независимы. Следует наложить соответствующее
ограничение. В \cite{Regge_measure} оно выбрано в виде: $\Pi_{xy} \Pi_{ij}
\delta(l_{ij}^2(x)-l_{ij}^2(y))$, где произведение $\Pi_{xy}$ - по парам
соседних симплексов в то время, как произведение  $\Pi_{ij}$ - по парам вершин
сторон, общих для $x$ и $y$. Тогда и получается решеточная мера в форме
(\ref{MeasureL}).

С нашей точки зрения вывод непрерывной формулы (\ref{Measure}), а тем более -
ее дискретизованной версии не является математически строгим. На это указывает
и наличие неоднозначности в выборе постоянной  $\sigma$. Присутствует также
неоднозначность в выборе ограничения, накладываемого на длины линков соседних
симплексов. А именно, такое ограничение можно выбрать и в виде
$\Omega[l_{ij}]\Pi_{xy} \Pi_{ij} \delta(l_{ij}^2(x)-l_{ij}^2(y))$, где
$\Omega[l_{ij}]$ зависит от длин линков. Априори неясно, как выбрать этот
функционал.

Возможным решением указанной проблемы было бы наличие симметрии непрерывной
теории, которая будучи перенесена на решетку, зафиксировала бы форму решеточной
меры.  К сожалению, такой симметрии мы не находим в Редже гравитации. В то же
время, в телепараллельном представлении гравитации, рассмотренном выше такая
симметрия есть - это калибровочная симметрия группы трансляций. Она однозначно
фиксирует меру интегрирования по трансляционной связности. Однако, в этой
дискретизации теряется значительная часть симметрии действия, связанная с
преобразованиями Лоренца во внутреннем пространстве. Ниже нами будет
рассматриваться случай дискретизации пространства Римана - Картана, где
калибровочная симметрия группы Пуанкаре также фиксирует меру интегрирования. В
то же время, группа Лоренца остается группой симметрии дискретизованной модели.

\subsection{Дискретизация пространства Римана - Картана}

Напомним, что пространство Римана - Картана - это обобщение Риманова
пространства на случай ненулевого кручения. Связность состоит из двух частей -
трансляционной связности и связности группы Лоренца. В случае, если
трансляционная кривизна (кручение) зануляется, мы приходим к Риманову
пространству. В случае, если зануляется кривизна группы Лоренца, мы получаем
пространство Вейценбока.

Следует отметить, что дискретизация Пуанкаре гравитации рассматривалась в ряде
работ. Так, в  \cite{Magnea,Drummond} независимая $SO(4)$ связность была
введена в исчисление Редже. Связность сингулярна и локализована на сторонах
симплексов. В принципе, эта конструкция схожа с предлагаемой ниже. Однако, мы
используем гиперкубическую решетку, которая гораздо более удобна для
практических вычислений. Кроме того, нами рассматривается инвариантная мера по
динамическим переменным, чего не было сделано в  \cite{Magnea,Drummond}.
Отметим также, что в \cite{Magnea,Drummond} не рассматривались члены в
действии, квадратичные по кривизне.

Пуанкаре гравитация на гиперкубической решетке рассматривалась, скажем, в
\cite{unitarity, Magnea, Smolin, Manion, Kaku, Menotti}. В этих работах
решеточная дискретизация рассматривалась посредством наивной дискретизации
методом конечных разностей. Поэтому соответствующие конструкции теряют
симметрию непрерывной теории. В этих работах некоторые из рассматриваемых
моделей инвариантны относительно решеточных калибровочных преобразований,
принадлежащих группе Пуанкаре. Однако, существование этой симметрии не
означает, что теория инвариантна относительно общекоординатных преобразований.
Более подробно мы описываем конструкции работ \cite{unitarity, Magnea, Smolin,
Manion, Kaku, Menotti} ниже в п. 8.4.5.

 Мы обобщаем конструкцию Редже на пространство Римана - Картана. В нашем подходе
 мы аппроксимируем любое заданное многообразие кусочно - линейным многообразием Римана - Картана (с гиперкубическими элементами).
 Ключевое отличие нашей конструкции от подхода  \cite{unitarity, Magnea, Smolin,
Manion, Kaku, Menotti} заключается в том, что мы используем геометрическую
конструкцию. То есть мы дискретизуем исходное многообразие инвариантными
объектами (по отношению к группе общекоординатных преобразований). В результате
наша решеточная модель калибровочно инвариантна по построению.

Итак, мы рассматриваем пространство Римана - Картана, в котором  $SO(4)$
связность \footnote{Предполагается, что выполнен поворот Вика к Евклидовой
сигнатуре пространства - времени. }
 и поле тетрады (трансляционная связность) являются динамическими переменными.
Дискретизованное пространство само является пространством Римана - Картана. Оно
состоит из плоских гиперкубических объектов, склеенных друг с другом. Ниже мы
называем гиперкубы элементами решетки. В принципе мы можем строить таким
образом приближение к пространствам различной топологии. Однако, для простоты в
дальнейшем мы ограничиваемся топологией четырехмерного тора, что соответствует
периодическим граничным условиям.

Форма элемента решетки фиксируется множеством векторов  ${\bf e}_{\mu}$,
соединяющих центр гиперкуба с его вершинами. Заметим, что здесь используется
определение векторов ${\bf e}_{\mu}$ иное, чем выше, когда мы рассматривали
дискретизацию телепараллелизма (там обозначенные так вектора соединяли центры
элементов решетки с центрами сторон их границ). Выражение ${\bf e}_{\mu}$ через
элементы ортонормированного базиса ${\bf f}_A$ ($A = 1,2,3,4$) (общего для всех
решеточных элементов) - это одна из основных базовых переменных нашей
конструкции. Имеем
\begin{equation}
{\bf e}_{\mu} = \sum_A E^A_{\mu} {\bf f}_A
\end{equation}
(Везде пространственно - временные индексы обозначены греческими буквами, а
тетрадные - латинскими.) В общем случае все Вектора ${\bf e}_{\mu} (\mu = 0,
1,2,..., 15)$ предполагаются независимыми. Это означает, что противоположные
стороны гиперкубов в общем случае непараллельны. Однако, условие параллельности
может быть наложено, и это существенно упрощает решеточные выражения  (см.
обсуждение в п. 8.4.4.). Положение начальной вершины каждого гиперкуба
определяется вектором ${\bf e}_0$. Вектора ${\bf e}_{\mu} (\mu = 1,2,3,4)$
указывают на ее соседей. Мы выбираем начальную вершину в каждом гиперкубе так,
что если бы пространство было плоским, все вектора ${\bf e}_{\mu} - {\bf e}_0
(\mu = 1,2,3,4)$ были направлены в положительном направлении.

Переменные $E^A_{\mu}$ представляют трансляции из центров решеточных элементов
к вершинам. Метрика (или поле тетрады) предполагается постоянной внутри каждого
решеточного элемента. Смещение центра решеточного элемента на вектор  $v^A$
вызывает трансформацию переменной $E^A_{\mu} \rightarrow E^A_{\mu} + v^A$,
которое может рассматриваться как калибровочное преобразование по отношению к
трансляционной калибровочной группе. Оно соответствует трансляции решеточного
элемента внутри соответствующей карты.

В дополнение к трансляционной связности, определенной как множество переменных
 $E^A_{\mu}$, каждое смещение от одного решеточного элемента к другому
 сопровождается поворотом четырехмерного касательного пространства. Другими словами, есть
 $SO(4)$ связность, которая сингулярна и локализована на сторонах гиперкубов.
 Мы обозначим посредством $U_{IJ}$  $SO(4)$ матрицу, которая прикреплена к
 стороне, общей для гиперкубов $I$ и $J$.

Данное пространство Римана - Картана имеет сингулярную связность. В этом случае
определение кривизны и кручения становится двусмысленным. Поэтому мы должны
выбрать одно из возможных определений. Ниже этот вопрос будет освещен подробно.
Следует отметить, что в случае ненулевого кручения длина одного и того же
линка, наблюдаемая из различных гиперкубов, оказывается различной. Или, что то
же, два линка разной длины оказываются на бесконечно малом расстоянии друг от
друга. Эта ситуация соответствует наличию дислокаций решетки и специфична для
ненулевого кручения. Требование того, чтобы кручение занулялось, приводит к
отсутствию дислокаций.

Калибровочные преобразования по отношению к полной группе Пуанкаре
представляется трансляциями и $SO(4)$ вращениями, которые приводят к следующим
преобразованиям базовых переменных:
\begin{equation}
E^A_{\mu} \rightarrow \Theta^A_B E^B_{\mu} + v^A,\label{trans}
\end{equation}
где $\Theta \in SO(4)$, а $v$ представляет трансляции.

\subsection{Кривизна и кручение кусочно - линейного пространства Римана - Картана}

 $SO(4)$ связность сингулярна и локализована на трехмерных кубах, являющихся сторонами
 гиперкубов.
 Кривизна сконцентрирована на плакетах. Следующее интегральное уравнение
 выбирается в качестве определения  $SO(4)$ кривизны.
\begin{eqnarray}
&& 2 \int_{y\in{\Sigma}}\Omega(z,y) R_{\mu\nu}(y)\Omega^+(z,y)dy^{\mu}\wedge
dy^{\nu} = P {\rm exp} (\int^z_{z\in \partial \Sigma} \omega_{\mu} dx^{\mu})
\nonumber\\&& - [P {\rm exp} (\int^z_{z\in \partial \Sigma} \omega_{\mu}
dx^{\mu})]^+ \,  \, , |\Sigma|\rightarrow 0
\end{eqnarray}

Здесь $\omega_{\mu}$ - это $SO(4)$ связность \footnote{$U_{IJ}=P {\rm exp}
(\int \omega_{\mu} dx^{\mu})$, где интеграл по пути минимальной длины,
соединяющем центры гиперкубов $I$ и $J$.}. $\Sigma$ - малая поверхность,
пересекающая данный плакет (в четырехмерии они пересекаются в одной точке), $
|\Sigma|$ ее площадь. $\partial \Sigma$ - граница $\Sigma$. Ее ориентация
соответствует ориентации $\Sigma$. $\Omega(z,y) = P {\rm exp} (\int^y_z
\omega_{\mu} dx^{\mu})$ - оператор параллельного переноса вдоль пути,
соединяющего фиксированную точку на $\partial \Sigma$ с точкой $y$. Мы выбираем
этот путь таким образом, что он оборачивается вокруг данного плакета в том же
направлении, что и  $\partial \Sigma$ и имеет минимальную длину. Интеграл в
правой стороне указанного выше выражения - по пути  $\partial \Sigma$, который
начинается и заканчивается в точке $z$.

Следует отметить, что данное определение не противоречит общепринятому в случае
гладкой связности.

Зафиксируем решеточный гиперкуб. Внутри него кривизна равна
\begin{equation}
R_{\mu\nu B}^A(y) = \sum_{b\in bones}\int_{x\in{b}}\epsilon_{\mu \nu \rho
\sigma}dx^{\rho}\wedge dx^{\sigma}\delta^{(4)}(y-x)
\frac{[\Pi_{i}U^b_{I_iI_{i+1}}-
(\Pi_{i}U^b_{I_iI_{i+1}})^+]^A_B}{2D!},\label{RB1}
\end{equation}
Здесь сумма - по плакетам (bones), принадлежащим данному гиперкубу. Интеграл -
по поверхности плакетов. Произведение $\Pi_{i}U^b_{I_iI_{i+1}}$ - вдоль
замкнутого пути вокруг данного плакета $b$, этот путь состоит из линков,
соединяющих центры гиперкубов. Предполагается, что этот путь начинается внутри
данного гиперкуба и имеет минимальную решеточную длину.

Вычислим кручение, сконцентрированное на трехмерных сторонах решеточных
гиперкубов. Поле кручения $T_{\mu \nu}^A$ определяется интегральным  уравнением
\begin{equation}
 \int_{y\in{\Sigma}}\Omega^A_B(z,y)
T^B_{\mu\nu}(y)dy^{\mu}\wedge dy^{\nu} = \int_{\partial \Sigma}\Omega^A_B(z,y)
b^B_{\mu}(y)dy^{\mu}
\end{equation}
Здесь $b^A_{\mu}(x)$ - поле тетрады, выраженное через переменные  $E^A_{\mu}$
внутри каждого гиперкуба если задана некая ее параметризация.

Это уравнение удовлетворяется следующим выражением  (справедливым внутри
гиперкуба $\bf I$):
\begin{eqnarray}
T^A_{\mu\nu}(y) & = & \sum_{s\in sides} [\int^{\bf J^s}_{x\in{s}}\frac{[U_{\bf
 IJ^s}]^A_B b^B_{[\mu}(x)\epsilon_{\nu] \tau \rho \sigma}}{D!}dx^{\tau}\wedge dx^{\rho}\wedge
dx^{\sigma}\delta^{(4)}(y-x)\nonumber\\
&& - \int^{\bf I}_{x\in{s}}\frac{b^A_{[\mu}(x)\epsilon_{\nu ]\tau \rho
\sigma}}{D!} dx^{\tau}\wedge dx^{\rho}\wedge
dx^{\sigma}\delta^{(4)}(y-x)]\label{TT}
\end{eqnarray}
Здесь первый интеграл в сумме - по данной стороне  $s$, наблюдаемой из
соседнего решеточного гиперкуба $\bf J^s$ (сторона $s$ - общая для $\bf I$ и
$\bf J^s$). (Мы предполагаем, что в (\ref{TT}) данный гиперкуб и все его соседи
имеют общую параметризацию.)

Определим внутри каждого гиперкуба следующие переменные: ${\cal E}^A_{\mu} =
E^A_{\mu} - E^A_0, \mu = 1,2,3,4$. Эти переменные могут рассматриваться как
поле тетрады внутри заданного гиперкуба. Также обозначим ${\cal E}^{\mu}_A$
элементы обратной матрицы ${\cal E}^{-1}$. В тетрадных компонентах имеем:
\begin{equation}
R_{C F B}^A(y) = {\cal E}^{\mu}_{C} {\cal E}^{\nu}_{F}\sum_{b\in
bones}\int_{x\in{b}}\epsilon_{\mu \nu \rho \sigma}dx^{\rho}\wedge
dx^{\sigma}\delta^{(4)}(y-x) \frac{[\Pi_{i}U^b_{I_iI_{i+1}}-
(\Pi_{i}U^b_{I_iI_{i+1}})^+]^A_B}{2D!},\label{RB}
\end{equation}

Кручение выражается как
\begin{eqnarray}
T^A_{C F}(y) & = & \frac{{\cal E}^{\mu}_{{\bf I}C} {\cal E}^{\nu}_{{\bf
I}F}}{D!} \sum_{s\in sides} \int_{x\in{s}} dx^{\tau}\wedge dx^{\rho}\wedge
dx^{\sigma}\delta^{(4)}(y-x) \nonumber\\&&([U_{\bf
 IJ^s}]^A_B {\cal E}^B_{{\bf J^s}[\mu}\epsilon_{\nu] \tau \rho \sigma} - {\cal E}^A_{{\bf I}
[\mu}\epsilon_{\nu ]\tau \rho \sigma})\label{TSING}
\end{eqnarray}
Здесь $ {\cal E}^B_{{\bf I}\mu}$ вычисляется внутри заданного решеточного
гиперкуба $\bf I$ в то время, как $ {\cal E}^B_{{\bf J^s}\mu}$ - вычисляется
внутри его соседа $\bf J^s$.

\subsection{Действие}

Мы рассматриваем следующее действие:

\begin{eqnarray}
S &=& \int \{ \alpha (R_{AB} R_{AB} - \frac{1}{3}R^2) + \beta R^2 - \gamma
m_p^2 R + \lambda m_P^4\} |{\cal E}| d^4 x \nonumber\\&& +\,  \delta m_P^2 \int
T^A_{BC} T^A_{BC} |{\cal E}| d^4 x, \label{S2}
\end{eqnarray}
где $|{\cal E}| = {\rm det} {\cal E}^A_{\mu}$, ${\cal E}^A_{\mu}$ - поле
тетрады, тетрадные компоненты тензора Риччи обозначены $R_{AB}$, а $R$ -
скалярная кривизна. Константы связи $\alpha, \beta, \gamma$ и $\lambda$ -
безразмерные, в то время, как $m_p$ - размерный параметр.

Если потребовать, чтобы кручение занулялось, условие $T^A_{BC} = 0$ будет
означать, что длина каждого линка, вычисленная в любом из гиперкубов, к
которому он относится, одна и та же. Это означает отсутствие дислокаций. Если
при этом потребовать, чтобы противоположные стороны каждого гиперкуба были
параллельны друг другу, то можно зафиксировать длины всех линков решетки
равными друг другу. При этом единственными динамическими переменными окажутся
углы между линками каждого гиперкуба. При приближении к непрерывному пределу,
когда длина линков стремится к нулю, такое кусочно линейное пространство
является приближением к Риманову пространству, что становится очевидным, если
рассмотреть Риманову геометрию в калибровке, в которой $g_{ii} = 1$
(суммирование по $i$ не предполагается). То, что такая калибровка может быть
зафиксирована при евклидовой сигнатуре пространства - времени очевидно,
поскольку она означает наличие 4 - х условий в каждой точке при 4 - х
неизвестных функциях $\tilde{x}^i(x)$, описывающих репараметризацию.

  Если противоположные стороны гиперкубов предполагаются параллельными,
  решеточная модель существенно упрощается (см. ниже) и становится гораздо более приспособлена к
  численным симуляциям. Ниже мы приводим выражение для решеточного действия в
   соответствующей модели. Аналогичное рассмотрение может быть выполнено и для
   общего случая, но при этом полученное выражение оказывается существенно
   более сложным.

  Таким образом, ниже в  данном пункте вектора ${\bf e}_{\mu} (\mu = 0, 1,2,3,4)$ предполагаются независимыми.
Остальные ${\bf e}_{\mu} (\mu = 5, ...,15)$ определяются так, что
противоположные стороны гиперкубов параллельны друг другу. Гиперкубическая
решетка периодична.

Вычислим действие (\ref{S2}) на кусочно - линейном пространстве Римана -
Картана, состоящим из гиперкубов.

Чтобы переписать выражения (\ref{RB}) и (\ref{TSING}) для решеточной кривизны и
кручения в удобном виде мы переходим к дуальной решетке. Тогда матрицы вращений
прикреплены к линкам, а поле тетрады определено в точках. Обозначим
$U_{\mu}(x)$ матрицы, соответствующие линкам, которые начинаются в точке  $x$ и
указывают в направлении $\mu$ ($\mu = \pm 4, \pm 3, \pm 2, \pm1$). Мы обозначим
посредством $\Omega_{\mu \nu}(x) = U_{\mu}(x) ... $ произведение линковых
матриц вдоль границы плакета, помещенного в плоскости $(\mu \nu)$. Тетрада,
прикрепленная к точке $x$, обозначается ${\cal E}^A_{\mu} = E^A_{\mu} - E^A_0,
\mu = 1,2,3,4$. Для отрицательных значений $\mu$ определяем ${\cal E}^A_{-\mu}
= - {\cal E}^A_{\mu}$. Обратная матрица для положительных значений индексов
обозначается ${\cal E}^{\mu}_A(x) = \{{\cal E}(x)^{-1}\}^{\mu}_A$. Мы также
распространяем это определение на отрицательные значения индексов: ${\cal
E}^{\mu}_A(x) = {\rm sign}(\mu)\{[{\cal E}(x)]^{-1}\}^{|\mu|}_A$. Обозначим
посредством $\Delta x_{\mu}$ смещение на один шаг решетки в  $\mu$ - м
направлении ($\Delta x_{-\mu} = - \Delta x_{\mu} $). Тогда, $x+\Delta x_{\mu}$
это точка, полученная смещением из точки $x$ на один шаг решетки в направлении
 $\mu$ в то время, как $x-\Delta x_{\mu}$ получается смещением в противоположном направлении.
 Тогда $U_{-\mu}(x) = U^{-1}_{\mu}(x-\Delta x_{\mu})$. Далее, определим
 $\Delta_{\mu} {\cal E}_{\nu}(x) = U_{\mu}(x){\cal E}_{\nu}(x+\Delta
x_{\mu}) - {\cal E}_{\nu}(x)$. Везде предполагается суммирование по
повторяющимся индексам  $\mu,\nu,...$ - по $\pm 4, \pm 3, \pm 2, \pm1$.

Вводим решеточные тетрадные компоненты кручения и кривизны, прикрепленные к
узлам решетки и положительным и отрицательным направлениям  (в этом параграфе и
в следующем суммирование по греческим индексам не подразумевается):
\begin{eqnarray}
&& {\bf [R^{\mu\nu}]}_{C F B}^A(x) = \frac{1}{2}{\cal E}^{\mu}_C {\cal E}^{\nu}_F [\Omega_{\mu \nu}-\Omega_{\nu \mu}]^A_B \nonumber\\
&& {\bf [R^{\mu\nu}]}_{F B}(x) = {\bf [R^{\mu\nu}]}_{A F B}^A (x)\nonumber\\
&& {\bf [R^{\mu\nu}]}(x) = {\bf [R^{\mu\nu}]}_{A A} (x)\nonumber\\
&& {\bf R}(x) = \sum_{\mu,\nu} {\bf [R^{\mu\nu}]} (x)\nonumber\\
&& {\bf [T^{\mu}]}^A_{C F}(x) = \sum_{\nu}\Delta_{\mu} {\cal E}^A_{\nu} [{\cal
E}^{\mu}_C {\cal E}^{\nu}_F-{\cal E}^{\nu}_C {\cal E}^{\mu}_F]
\end{eqnarray}

Теперь мы готовы выразить действие (\ref{S2}) на кусочно - линейном
многообразии через $E^A_{\mu}(x)$ и $U_{\mu}(x)$. Для этого мы используем
выражения для кручения и кривизны (\ref{RB}) и (\ref{TSING}). Мы опускаем
промежуточные шаги и представляем окончательный результат, используя
обозначения для решеточного кручения и кривизны.
\begin{eqnarray}
S &=& \sum_{x\in sites} |{\cal E}(x)| {\large \bf ( } \sum_{\mu,\nu} \{
\bar{\alpha} {\bf [R^{\mu\nu}]}_{FB}(x){\bf [R^{\mu\nu}]}_{FB}(x) +
(\bar{\beta}-\frac{1}{3}\bar{\alpha}){\bf [R^{\mu\nu}]^2}(x)\nonumber\\&&
 - \bar{\gamma} m_p^2 {\bf [R^{\mu\nu}]}(x) \}  +\, \bar{\delta} m_P^2 \sum_{\mu}{\bf
[T^{\mu}]}^A_{BC}(x){\bf [T^{\mu}]}^A_{BC}(x)+ \bar{\lambda} m_P^4 {\large \bf
)} , \label{S31}
\end{eqnarray}
где $|{\cal E}(x)| = |{\rm det}{\cal E}(x)| $ - это объем гиперкуба, дуального
узлу  $x$ дуальной решетки.

Здесь мы ввели решеточные константы связи $\bar{\alpha}, \bar{\beta},
\bar{\delta}, \bar{\lambda}, \bar{\gamma}$, которые отличаются от исходных
факторами, которые формально бесконечны, поскольку происходят от дельта -
функций из выражений  (\ref{RB}) и (\ref{TSING}) для кручения и кривизны.
Разумеется, здесь предполагается, что использована некоторая регуляризация,
которая делает эти факторы конечными. Мы полагаем, что после перенормировки
каждая физическая величина может быть выражена через физические константы
связи. Бесконечности, указанные выше, также должны поглотиться
перенормированными постоянными.

В нашей решеточной модели (на дуальной решетке) трансляционная связность
$E^A_{\mu}(x)$ прикреплена к узлам, а $SO(4)$ матрицы $U_{\mu}(x)$ прикреплены
к линкам. Действие модели компактным образом выражается через эти переменные.
Легко понять, что (\ref{trans}) - симметрия действия. Итак, мы имеем решеточную
модель с группой Пуанкаре в качестве калибровочной группы.

\subsection{Сравнение с предыдущими работами по дискретизации Пуанкаре гравитации на гиперкубической решетке}

Как уже отмечалось, решеточная гиперкубическая реализация Пуанкаре гравитации
рассматривалась в ряде работ (см., например, \cite{unitarity, Magnea, Smolin,
Manion, Kaku, Menotti}). После того как нами изложен наш подход мы готовы
сравнить его с подходом упомянутых работ. Прежде всего, во всех этих работах
непрерывная модель была перенесена на решетку посредством применения простой
наивной дискретизации методом конечных разностей. Это означает, что
рассматривались определенные переменные, прикрепленные к узлам и линкам
гиперкубической решетки. При этом задавалось соответствие между ними и
связностью группы Пуанкаре. В  \cite{Magnea,
 Kaku, Menotti} и тетрада, и  $SO(4)$ связность на решетке прикреплены к линкам. В
  \cite{Smolin} группа Пуанкаре рассматривалась как предельный случай
  группы де Ситтера, а линковые переменные принадлежат $SO(5)$. В
\cite{unitarity, Manion}  $SO(4)$ связность прикреплена к линкам, в то время,
как тетрада прикреплена к узлам решетки. Определение решеточной модели в
\cite{Manion}, в принципе, близко к нашей модели, сформулированной на дуальной
решетке. Однако, модели, тем не менее отличаются. Прежде всего, в \cite{Manion}
действие - в форме Эйнштейна - Гильберта в формулировке Палатини, то есть
содержит только первую степень решеточной кривизны. Член $-\bar{\gamma} m_p^2
\sum_{x\in sites}  {\bf R}(x)|{\cal E}(x)|$ из (\ref{S31}) мог бы совпадать с
действием \cite{Manion}, если бы в последнем было выполнена симметризация по
различным направлениям. В \cite{unitarity} были рассмотрены члены квадратичные
по кривизне общего вида. И линковая $O(4)$ связность, и тетрада были
рассмотрены в $4\times 4$ спинорном представлении. Окончательное выражение для
решеточного действия довольно сложное и не совпадает с (\ref{S31}).

Ключевое отличие отмеченных подходов и нашего подхода - в том, что мы
используем регулярную процедуру и аппроксимируем пространство Римана - Картана
кусочно - линейными пространствами Римана - Картана. Эта процедура дает нам
дискретизацию исходного многообразия, которая инвариантна относительно
общекоординатных преобразований по построению. В противоположность этому
конструкции, рассмотренные в \cite{unitarity, Magnea, Smolin, Manion, Kaku,
Menotti} нарушают инвариантность относительно диффеоморфизмов  и не дают
калибровочно инвариантной дискретизации Пуанкаре гравитации.

\subsection{Мера интегрирования}

В нашем случае есть два поля: $SO(4)$ связность и трансляционная связность.
Таким образом, естественно использовать меру, инвариантную относительно
решеточной реализации калибровочных преобразований. Наш выбор - это мера,
инвариантная относительно калибровочных преобразований и локальная. Такая мера
оказывается единственной, что существенно отличает нашу дискретизацию от Редже
дискретизации, где локальная эвристическая мера не является однозначно
определенной (см. п. 8.4.1.), а мера Лунда - Редже, соответствующая решеточной
реализации метрики на множестве кусочно - линейных римановых пространств, не
является локальной (см. главу 7).

Обозначим заданную дискретизацию (с переменными $E$ и $U$) как $\cal M$.
Рассмотрим множество соответствующих независимых переменных $\{E^A_{\mu}({\bf
I}); U_{\bf IJ}\}$. (В случае гиперкубов, у которых противоположные стороны
параллельны мы имеем 5 независимых переменных $E^A_{\mu}({\bf I})$ в каждом
гиперкубе. В случае, если такое условие не накладывается, их - 16).
Калибровочные преобразования соответствуют сдвигу каждого гиперкуба решетки на
вектор $v^A ({\bf I})$ и вращению $\Theta_{\bf I} \in SO(4)$. Это
преобразование действует как $\{E^A_{\mu}({\bf I}); U_{\bf IJ}\}\rightarrow
\{\Theta_{\bf I}E_{\mu}({\bf I})+v({\bf I}); \Theta_{\bf I}U_{\bf
IJ}\Theta^T_{\bf J}\}$.

Локальность меры означает следующее. Полная мера может быть представлена как
\begin{equation}
D_{\cal M} (E;U) = \Pi_{\bf I} \Pi_{\mu}D E^A_{\mu}({\bf I}) \Pi_{\bf I, J}
DU_{\bf IJ},
\end{equation}
Здесь произведение - по сторонам гиперкубов и по линкам, соединяющим центры
гиперкубов с вершинами. Мера по линковым матрицам $U_{\bf IJ}$ обозначается
$DU_{\bf IJ}$. Мера по векторам $E^A_{\mu}({\bf I})$ обозначается $D
E^A_{\mu}({\bf
 I})$. Мы называем решеточную меру локальной, если внутри
 каждого гиперкуба $DE^A_{\mu}$ для данного $\mu$ зависит только от $E^A_{\mu}$, а
  $DU_{\bf IJ}$ для данных $\bf I, J$ зависит только от $U_{\bf I J}$. Очевидно,
  что это требование фиксирует
единственный выбор $DE^A_{\mu}$ и $DU_{\nu}$: $D E^A_{\mu} = \Pi_{A,\mu} d
E^A_{\mu}$, в то время, как $DU$ - инвариантная мера на $SO(4)$.

\subsection{Алгоритм Метрополиса}

Следует отметить, что в реальных численных симуляциях удобно выразить каждую
$SO(4)$ линковую матрицу (на дуальной решетке) как функцию $SL(2,C)$ матрицы.
Соответствие задается привычным спинорным представлением $SO(4)$ вращений.
Тогда инвариантная мера на $SL(2,C)$ генерирует инвариантную меру на  $SO(4)$.
Технически гораздо проще симулировать $SL(2,C)$ поле, чем собственно $SO(4)$
поле.

Алгоритм Метрополиса для симуляции нашей модели (когда противоположные стороны
каждого гиперкуба предполагаются параллельны друг другу) описывается следующим
образом. На каждом шагу алгоритма выбирается один из линков и один из концов
этого линка. Формируется предложение линковой $SL(2,C)$ матрицы и $4\times4$
матрицы, прикрепленной к выбранному узлу, являющемуся выбранным концом данного
линка \footnote{Можно зафиксировать калибровку и потребовать $E^A_0 = 0 $.
Соответствующий детерминант Фаддеева - Попова - константа.}. Затем вычисляются
члены в действии (\ref{S31}), которые содержат кручение и соответствуют соседям
данного узла. Затем, вычисляются члены в действии, которые содержат кривизну и
соответствуют точкам "бабочки", соответствующей данному линку. Термин "бабочка"
("butterfly") означает фигуру, состоящую из всех плакетов, содержащих данный
линк.  Обозначим сумму указанных членов как $S^{\rm new}$. Ту же сумму,
вычисленную при старых значениях предлагаемых переменных обозначим $S^{\rm
old}$. Предложение принимается, если $S^{\rm new}<S^{\rm old}$. Иначе оно
принимается с вероятностью ${\rm exp}\,(S^{\rm old}-S^{\rm new})$. Затем
выбирается другой линк и один из концов этого линка, и процедура повторяется.

Для того, чтобы ускорить симуляцию следует сохранять в компьютерной памяти
значения всех плакетных переменных $\Omega_{\mu \nu}$, значения производных
$\Delta_{\mu} {\cal E}_{\nu}$, и знгачения обратных матриц $\{{\cal
E}^A_{\mu}\}^{-1}$.

\section{Выводы}

В настоящей главе нами рассмотрены два способа дискретизации квантовой
гравитации. Оба они инспирированы трудностями предыдущих дискретизаций. А
именно, из предложенных ранее конструкций только Редже дискретизация
калибровочно инвариантна по построению. В этой дискретизации, однако,
присутствуют проблемы с мерой интегрирования по полям, которые невозможно
устранить в четырех измерениях. В предыдущей главе была предложена конструкция,
которая делает последовательно построенную меру Лунда - Редже в двух измерениях
локальной и приемлемой для практических вычислений. В четырех измерениях
добиться того же не удается. Эвристическая мера, использованная в практических
вычислениях в рамках исчисления Редже, с нашей точки зрения, не выдерживает
критики.

Для преодоления указанной проблемы с мерой мы рассмотрели прежде всего
дискретизацию телепараллельного представления теории гравитации.
Телепараллелизм - это способ смотреть на Риманову геометрию как на геометрию
Вейценбока, а на квантовую гравитацию - как на калибровочную теорию группы
трансляций. Дискретизация строится как кусочно - линейное пространство
Вейценбока и, таким образом, является калибровочно инвариантной по построению.
Инвариантная локальная мера на группе трансляций фиксирует единственную меру
интегрирования по полям. К трудностям этой дискретизации следует отнести то,
что дискретизованное действие теряет $SO(4)$ инвариантность. Кроме того, оно
достаточно громоздко, если мы хотим включить в него члены, квадратичные по
Римановой кривизне (последняя, разумеется, выражается соответствующим образом
через трансляционную связность пространства Вейценбока, являющегося
телепараллельным эквивалентом Риманова пространства).

Второй способ дискретизации, который мы предлагаем, разрешает трудности первого
за счет включения в рассмотрение расширения Риманова пространства -
пространства Римана Картана, в котором в общем случае кручение не зануляется.
Нами построена дискретизация, основывающаяся на кусочно - линейных
пространствах Римана - Картана. Причем ячейки этих пространств имеют
гиперкубическую форму, что сильно упрощает практические применения в численных
расчетах. Снова инвариантная мера на группе Пуанкаре фиксирует нам единственную
локальную меру интегрирования. В то же время решеточная модель имеет группу
Пуанкаре в качестве калибровочной. Переход к генерации Римановой геометрии в
этом подходе сожет быть выполнен динамически, когда при увеличении значения
$\delta$ в (\ref{S2}) подавляются флуктуации кручения.

\section{Публикации}

Результаты настоящей главы опубликованы в работах

"Teleparallel gravity on the lattice", M.A. Zubkov, Phys.Lett.
B582:243-248,2004, [hep-lat/0311036]

"Gauge invariant discretization of Poincare quantum gravity",  M.A. Zubkov.
Phys.Lett.B638:503-508,2006, Erratum-ibid.B655:307,2007, [hep-lat/0604011]

\chapter{Многомерные динамические триангуляции}

\label{ch9}
В предыдущих двух главах мы рассматривали дискретизацию квантовой гравитации, и
разрешили определенные технические трудности предыдущих подходов к этому
вопросу. В главе 7 нами построен вариант Редже дискретизации в двух измерениях,
в котором мера Лунда - Редже локальна. В главе 8 нами построены две различные
дискретизации четырехмерной гравитации с локальной мерой интегрирования,
которая фиксируется калибровочной инвариантностью. В настоящей главе нами
рассматривается еще один подход к дискретизации квантовой гравитации,
основанный на динамических триангуляциях. Этот подход существенно отличается от
рассмотренных в главах 7 и 8 подходов тем, что в нем определяющую роль играет
энтропийный фактор. Поэтому подход динамических триангуляций близок по духу к
подходу, основанному на представлении о гравитации как об энтропийной силе
\cite{Entropy_force_}.

 Свойства динамических триангуляций в двух, трех, и четырех
измерениях хорошо известны. Также проводились численные исследования и
пятимерной модели. В настоящей главе мы решаем техническую задачу о численном
исследовании многомерных динамических триангуляций. А именно, нами исследуется
$10$ - мерная модель. Следует отметить, что затравочная размерность в моделях
динамических триангуляций вовсе не обязана совпадать с физической размерностью
симулируемого пространства. Последняя определяется как фрактальная размерность
и оказывается близка к двум, трем, и четырем в зависимости от того, в какой
области констант связи рассматривается модель и каков общий объем системы.
Таким образом, исследуемая модель дает динамическое определение размерности
пространства - времени и эффективно в состоянии описывать теорию меньшей (чем
10) размерности.

В \cite{Triangulation} (и следуя ссылкам внутри этой работы)  можно найти
подробное описание результатов, полученных  в
 $2$, $3$ и $4$ мерных моделях динамических триангуляций (ДТ). Результаты исследований $5$ -
  мерной модели содержатся в  \cite{5D}. При $D = 2$ ДТ модель имеет хорошо определенный непрерывный предел,
  соответствующий предсказаниям непрерывной теории
  \cite{2D}. При $D = 3,4$ Евклидовы модели имеют две фазы: хаотическая фаза (crumpled phase)
  с бесконечной фрактальной размерностью, и фаза ветвящихся полимеров  (branched polymer) с
  фрактальной размерностью, близкой к $2$. Введение в модель каузальной структуры
  (что соответствует переходу от Евклидовой к Лоренцевой квантовой гравитации) или
  введение взаимодействия с материей изменяет поведение модели и делает его более
  реалистическим \cite{Lorentzian,Gauge_Triang}. Тем не менее,чистая Евклидова
  модель динамических триангуляций
остается структурным блоком более сложных моделей и заслуживает исследования
сама по себе.

Нами рассматривается десятимерная модель ДТ сферической топологии. Ожидается,
что поведение такой модели аналогично поведению $3,4$, и  $5$ - мерных моделей.
Частично это предположение подтверждается. Однако, оказывается, что есть и
новые свойства, которые отсутствуют в моделях меньших размерностей. Например,
$10D$ модель также имеет две фазы аналогичные подобным фазам в  $3$, $4$ и $5$
- мерных моделях. Однако, в нашем случае фазовый переход (по крайней мере для
наблюдаемых объемов $V = 8000$ и $V = 32000$) имеет место при отрицательном
значении затравочной гравитационной постоянной в то время как в низших
размерностях он происходит при положителном значении константы связи.  Это
означает, что вакуум, определяющийся одной лишь энтропией в $10$ D системе
(когда решеточное действие выключено) имеет структуру ветвящихся полимеров в то
время, как в низкоразмерных системах это была бы хаотическая фаза.

Следует также заметить, что рассматриваемые нами объемы не могут
соответствовать системе "физической" размерности $D = 10$. Это следует из того,
что уже при линейном размере $3$ гиперкубическая $10D$ решетка имеет объем $V =
3^{10} = 59049$. Однако, наблюдаемая эффективная (фрактальная) размерность
триангулируемого пространства, скажем, при $V = 32000$ равна $\sim 4$ в
хаотической фазе, и $\sim 2$ в фазе ветвящихся полимеров. Таким образом,
линейный размер системы ожидается порядка $\sim (32000)^{\frac{1}{4}}\sim 10$ и
$\sim (32000)^{\frac{1}{2}} \sim 100 $ соответственно. Это и подтверждается
прямыми численными расчетами.

\section{Динамические триангуляции}

\begin{figure}
\begin{center}
\epsfig{figure=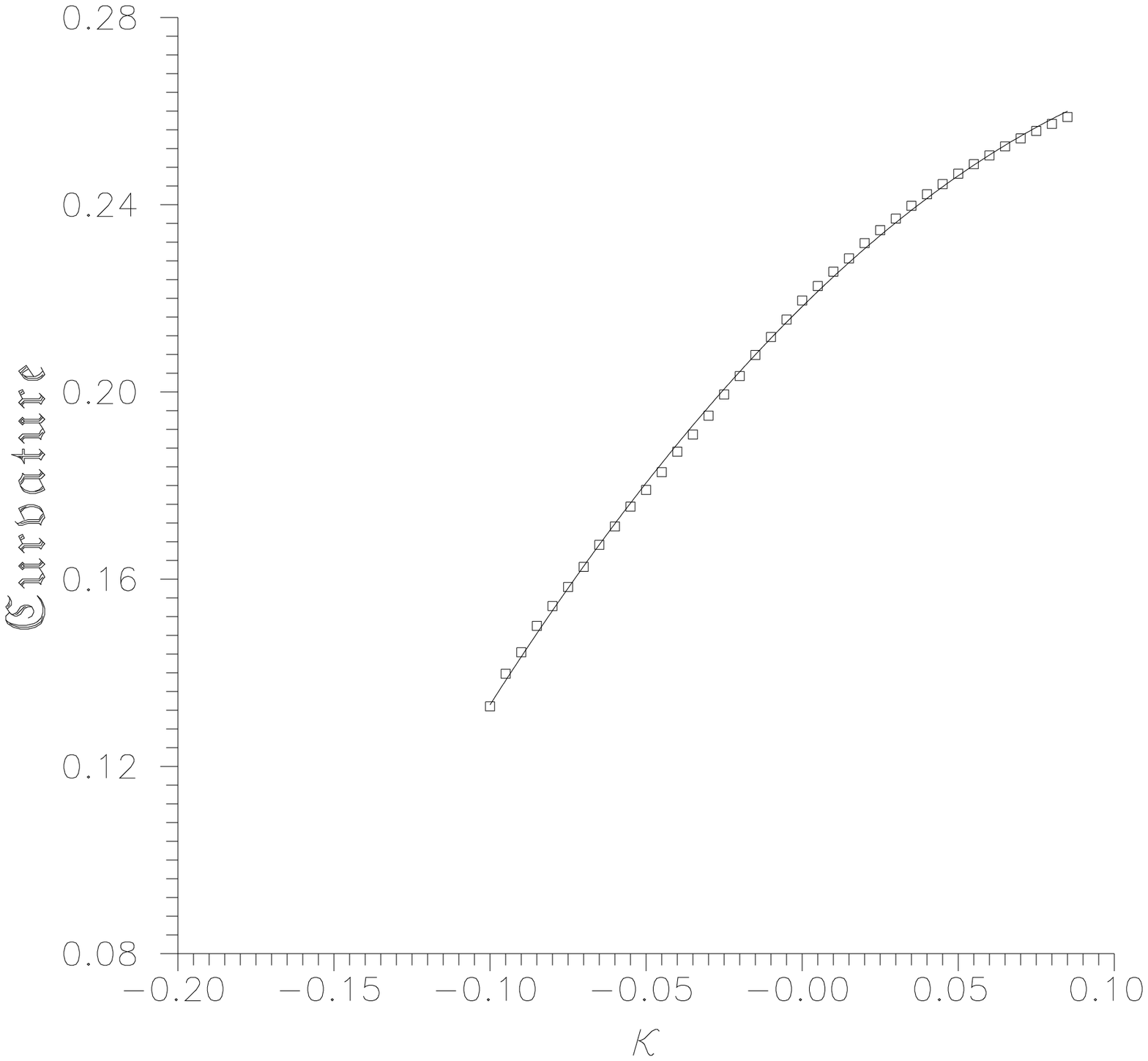,height=100mm,width=100mm,angle=0}
\caption{\label{fig.02} Кривизна.
 \label{fig.2}}
\end{center}
\end{figure}

В этом разделе мы кратко опишем модель ДТ и численный алгоритм. Обзор ДТ можно
найти в \cite{Triangulation}. Подробное описание численного алгоритма в
произвольной размерности дано в \cite{Catterall}.

В подходе динамических триангуляций Риманово пространство (Евклидовой
сигнатуры) аппроксимируется симплициальным комплексом, состоящим из склеенных
друг с другом $D$ - мерных симплексов (пирамид). Каждый симплекс имеет $D+1$
вершину. Все линки имеют одинаковую длину $a$. Метрика внутри каждого симплекса
- плоская. Поэтому Риманова связность сконцентрирована на границах симплексов.
Скалярная кривизна $R$ зануляется везде, кроме $D-2$ - мерных симплексов.
Традиционно в качестве действия для динамических триангуляций используют
комбинацию общего числа $D-2$ - мерных (bones) и $D$ - мерных симплексов
(simplices), которая в наивном непрерывном пределе переходит в действие
Эйнштейна - гильберта на рассматриваемом кусочно - линейном Римановом
пространстве:
\begin{eqnarray}
S & = & - \frac{1}{16 \pi G} \int R(x) \sqrt{g} d^D x =  - \frac{{\rm
Vol}_{D-2}}{16 \pi G}\sum_{\rm bones} (2\pi - O({\rm bone}) cos^{-1}
(\frac{1}{D}))
\nonumber\\
 & = & - \frac{{\rm Vol}_{D-2}}{8 G} ( N_{\rm bones} -
 \frac{D(D+1)}{4\pi} N_{\rm simplices} cos^{-1} (\frac{1}{D})),
\end{eqnarray}
где $O({\rm bone})$ - число $D$ - симплексов, которые содержат данный $D-2$ -
мерный симплекс. ${\rm Vol}_{j} = \frac{a^j \sqrt{j+1}}{j!\sqrt{2^j}}$ - это
объем $j$ - мерного симплекса с ребрами длины $a$, $N_{\rm bones}$ - общее
число $D-2$ мерных симплексов, $N_{\rm simplices}$ - общее число $D$ - мерных
симплексов.

Метрика триангулируемого многообразия полностью определяется способом склеивать
$D$ - мерные симплексы друг другом. Поэтому функциональный интеграл в этом
подходе вырождается в суммирование по различным триангуляциям  (мы ограничиваем
себя случаем сферической топологии):
\begin{equation}
\int D g \rightarrow \sum_{T}\frac{1}{C_T}
\end{equation}
где сумма - по триангуляциям $T$, аппроксимирующим разные Римановы пространства
\footnote{Две формально различных триангуляции могут аппроксимировать одно и то
же Риманово пространство. Поэтому следует ввести соответствующий фактор.} а
$C_T$ - симметрийный фактор самой триангуляции (порядок группы автоморфизмов).

Мы рассматриваем модель, в которой флуктуации глобального объема подавлены.
Статсумма имеет вид:
\begin{equation}
Z_V = \sum_T \frac{1}{C_T}{\rm exp} ( - S(T)) = \sum_T \frac{1}{C_T}{\rm
exp}(\kappa_{D - 2} N_{D - 2} - \kappa_{D} N_D - \gamma (N_D - V)^2
)\label{ZVD}
\end{equation}
где мы обозначили $N_D = N_{\rm simplices}$ , $N_{D-2} = N_{\rm bones}$ , и
$\kappa_{D-2} = \frac{{\rm Vol}_{D-2}}{8 G}$. К сожалению, невозможно построить
алгоритм, который генерирует сумму по триангуляциям одного и того же объема.
Поэтому постоянная $\gamma$ остается конечной. $\kappa_D(V,\kappa_{D-2})$
выбирается так, что
\begin{equation}
< N_D > = \sum_T \frac{1}{C_T}{\rm exp}(\kappa_{D - 2} N_{D - 2} - \kappa_{D}
N_D - \gamma (N_D - V)^2 ) N_D = V \label{kappa}
\end{equation}
Это обеспечивает то, что объем флуктуирует около требуемого значения $V$.
Флуктуации - порядка $\delta V \sim \frac{1}{\sqrt{\gamma}}$. Для того, чтобы
приблизиться к модели с постоянным $N_D$ мы должны удерживать $\frac{\delta
V}{V} << 1$. На практике мы используем $\gamma = 0.005$ для $V = 16000 ,
32000$. Таким образом, $\frac{\delta V}{V} \sim 10^{-3}$.

Для численных исследований мы используем алгоритм Метрополиса в его форме,
описанной в \cite{Catterall}. Он основан на следующем Марковском процессе.
Каждый шаг цепочки - это предложение деформации $T_i \rightarrow T_{f}$ данной
триангуляции $T_i$, которое принимается или отвергается с вероятностью  ${\cal
P}(T_i \rightarrow T_f)$, удовлетворяющей условию
\begin{equation}
{\rm exp}(-S(T_i)){\cal P}(T_i \rightarrow T_f) = {\rm exp}(-S(T_f)){\cal
P}(T_f \rightarrow T_i)
\end{equation}

Определение предлагаемой деформации базируется на следующей идее. Рассмотрим
некоторое замкнутое   $D$ - мерное симплициальное многообразие топологии $D$ -
мерной сферы. Тогда, если связная часть нашей исходной триангуляции равна части
этого многообразия, мы можем заменить ее остающейся частью данного
многообразия. Таким образом мы получаем деформированную триангуляцию той же
топологии, что и исходная. Далее, выберем границу $\partial s_{D+1}$ симплекса
$D+1$ - размерности $s_{D+1}$ как отмеченное выше многообразие. Есть $D+1$
возможностей выделить часть $\partial s_{D+1}$ соответствующую $p$ - мерным
подсимплексам $s_{D+1}$ ($ p = 0, ..., D$). Результирующая деформация
называется $(p, D - p)$ движение \cite{Catterall}.

Показано, что посредством таких движений возможно, стартуя с любой
триангуляции, достичь триангуляцию, которая {\it комбинаторно эквивалентна}
\footnote{Две триангуляции комбинаторно эквивалентны, если они имеют разбиения,
которые равны с точностью до переобозначения подсимплексов. Разбиение $T_s$
данной триангуляции $T$ - это другая триангуляция, такая, что множество ее
вершин содержит все вершины $T$, а любой симплекс $s \in T_s$ либо принадлежит
$T$, либо принадлежит какому - то симплексу $T$.} произвольно выбранной
триангуляции \cite{ergodicity}. Это свойство называется эргодичностью.
Благодаря этому свойству начиная с триангуляции, аппроксимирующей
\cite{Geometry} данное Риманово пространство, возможно достичь триангуляции,
которая аппроксимирует почти любое Риманово многообразие. Исключительные
случаи, в которых это невозможно, как считается обычно, не могут повлиять на
физические результаты.

На практике мы выбираем случайным образом движение ($p \in \{ 0, ... , D \}$),
симплекс триангуляции и его $p$ - мерный подсимплекс. После этого проверяем,
есть ли окрестность этого $p$ - мерного симплекса, эквивалентная требуемой
части  $\partial s_{D+1}$. Если так, предлагаемое движение (и соответствующий
подсимплекс) называется допустимым и мы проверяем возможность его выполнить
\footnote{Движение геометрически неосуществимо, если результирующий новый
симплекс уже существует в данной триангуляции.}. Если движение разрешено, мы
принимаем или отвергаем его с вероятностью:
\begin{equation}
p(T_i \rightarrow T_f) = \frac{1}{1+(1+\frac{N_D(T_f) -
N_D(T_i)}{N_D(T_i)}){\rm exp}(S(T_f) - S(T_i))} \label{P}
\end{equation}

В наших вычислениях мы начинаем с триангуляции минимального размера, то есть с
$\partial s_{D+1}$. Тогда мы позволяем ему расти случайным образом до того
момента, пока объем не достигнет данного значения $V$. После этого мы начинаем
нормальный процесс алгоритма Метрополиса. Во время этого процесса постоянная
$\kappa_D$ подстраивается автоматически с тем, чтобы удовлетворялось
(\ref{kappa}). Это достигается переопределением $\kappa \rightarrow \kappa + 2
\gamma (<N_D>-V)$ после каждых $10$ шагов \footnote{Один шаг - это $V$
предложений {\it допустимых} движений.}.

\section{Численные результаты}

\begin{figure}
\begin{center}
 \epsfig{figure=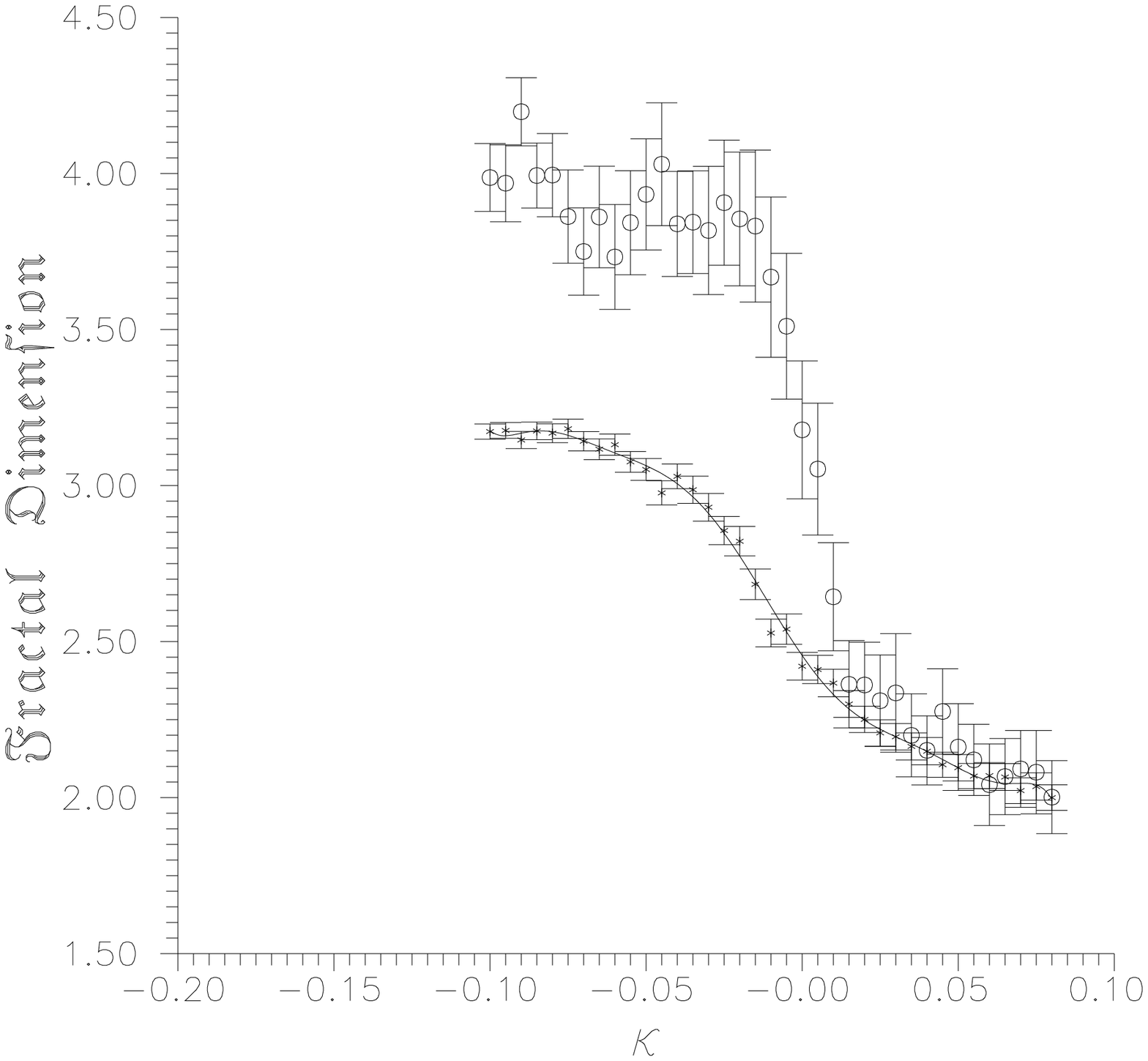,height=100mm,width=100mm,angle=0}
 \caption{\label{fig.04} Фрактальная размерность. Данные для $V=8000$ представлены точками и непрерывной линией.
 Данные для $V=32000$ представлены кругами.
 \label{fig.4}}
\end{center}
\end{figure}

Мы исследуем поведение  $10D$ модели для $V = 8000$ и $V = 32000$.
Рассматриваются значения $\kappa_D$, варьирующиеся в пределах от $-0.1$ до
$0.1$, в которых система обнаруживает существование двух фаз. Фазовый переход
имеет место при $\kappa_D \sim -0.03 $ для  $V = 8000$ и про $\kappa_D \sim
-0.01 $ для $V = 32000$.

Самоподстраивающееся значение $\kappa_{10}$ оказывается независящим от $V$. В
соответствии с асимптотикой большого объема, полученной в \cite{Geometry}, ее
зависимость от $\kappa_8$ - линейна с хорошей точностью:
\begin{equation}
\kappa_{10} = 15.57(1) \kappa_{8} + 0.42(1)
\end{equation}

Мы исследуем следующие переменные, которые отражают свойства триангулируемого
пространства.

1. Средняя кривизна, которую несет симплекс. Нормировка выбрана так, что она
определяется как
\begin{equation}
R = \frac{ 4\pi N_{\rm bones}}{D(D+1) N_{\rm simplices} cos^{-1} (\frac{1}{D})}
- 1
\end{equation}

Результаты для $V = 8000$ и $V = 32000$ совпадают с хорошей точностью. Они
представлены на рис.  ~\ref{fig.02}. В противоположность четырехмерному случаю
 (при наблюдаемых значениях $\kappa_8$) $R
> 0$ (в четырехмерном случае кривизна становится отрицательной при $\kappa_{D-2}$ близком к $0$).
Хорошим фитом для данных рис. ~\ref{fig.02} является:
\begin{equation}
R = 0.218(1) + 0.654(2) \kappa_8 - 1.97(3) \kappa_8^2
\end{equation}

2. Геодезическое расстояние между двумя симплексами - длина кратчайшего пути
между ними. Точки этого пути соответствуют симплексам. Линки соответствуют
парам соседних симплексов. Мы обозначаем геодезическое расстояние между
симплексами $u$ и $v$ посредством $\rho(u,v)$. Зафиксируем симплекс $s$. Тогда
шар $B_R(s)$ радиуса $R \in Z$ состоит из симплексов $u$ таких, что $\rho(u,s)$
меньше, или равно $R$. Объем шара определяется как число симплексов,
содержащихся внутри него.

Одна из наиболее информативных характеристик триангулируемого многообразия -
это средний объем ${\cal V}(R)$ шаров (радиуса $R$) как функция этого радиуса.
На практике при измерениях мы вычисляем $V_R(s)$ для произвольно выбранного
симплекса $s$. Тогда мы выполняем усреднение по измерениям (отдельно для
каждого $R$). ${\cal V}(R)$ становится постоянным, начиная с некоего значения
$R$. Это значение - это усредненное наибольшее значение между двумя симплексами
многообразия, называемое также диаметром $d$. Мы находим, что для $V = 8000$
при $1 < R < \frac{d}{2}$ зависимость ${\rm log} \,{\cal V}$ от ${\rm log} \,R$
линейна. Для $V = 32000$ то же имеет место при $4 < R < \frac{d}{2}$. Угол
наклона кривой дает нам определение фрактальной (Хаусдорфовой) размерности
многообразия: ${\rm log} \,{\cal V} = {\rm const} + {\cal D}{\rm log} \,R$.

3. Средняя фрактальная размерность ${\cal D}(V)$  многообразия для наблюдаемых
объемов $V$ как функция $\kappa_{D-2}$ представлена на рис. ~\ref{fig.04}.
Данная зависимость указывает на то, что имеет место фазовый переход при
критическом $\kappa_{D-2} = \kappa_c(V)$. Для наблюдаемых объемов $\kappa_c$
оказывается малым и отрицательным ($\kappa_c(8000) \sim -0.3$ и
$\kappa_c(32000) \sim -0.1$). При $\kappa > \kappa_c$ фрактальная размерность
близка к двум в соответствии с предположением о том, что аналогично
четырехмерному случаю мы имеем дело с фазой ветвящихся полимеров. Для  $\kappa
< \kappa_c$ ${\cal D}(8000) \sim 3.2$ и ${\cal D}(32000) \sim 4$. Это находится
в соответствии с ожиданиями, что эта фаза имеет сингулярную природу и
соответствует  ${\cal D}(\infty) = \infty$. Далее мы будем называть указанные
фазы соответственно хаотической фазой и фазой ветвящихся полимеров.

4. Линейный размер системы можно оценить как $V^{\frac{1}{\rm D}}$. Таким
образом, мы ожидаем, что в хаотической фазе диаметр $d \sim 10$ в то время, как
в фазе ветвящихся полимеров $d \sim 100$. Эти ожидания находятся в соответствии
с прямым измерением диаметра, а также иного параметра, называемого линейное
расширение (linear extent, см. \cite{Catterall}). Линейное расширение  ${\cal
L}$ определяется как среднее расстояние между двумя симплексами триангуляции:
\begin{equation}
{\cal L} = \frac{1}{V^2}\sum_{u,v}<\rho(u,v)>
\end{equation}
По построению  $\cal L$ должно быть близко к половине диаметра. В наших
измерениях мы вычисляем линейное расширение, используя несколько иное
определение (которое, тем не менее, должно приводить к тому же результату после
усреднения по измерениям). А именно, мы вычисляем
\begin{equation}
{\cal L}(s) = \frac{1}{V}\sum_{v}<\rho(s,v)> = \frac{1}{V}\sum_R R
(V_R(s)-V_{R-1}(s)).
\end{equation}
Выполняя усреднение по измерениям, мы получаем требуемое значение среднего
линейного расширения. Наш результат для $\cal L$ представлен на рис.
~\ref{fig.01}. Можно увидеть, что линейный размер многообразия быстро
увеличивается в фазе ветвящихся полимеров, в то время, как в хаотической фазе
он остается почти постоянным (и близким к значению, полученному  в $3$, $4$, и
$5$ - мерных моделях \cite{Triangulation}). Мы нашли, что флуктуации линейного
расширения почти отсутствуют в хаотической фазе и порядка среднего размера
многообразия - в фазе ветвящихся полимеров.

\section{Обсуждение}

\begin{figure}
\begin{center}
 \epsfig{figure=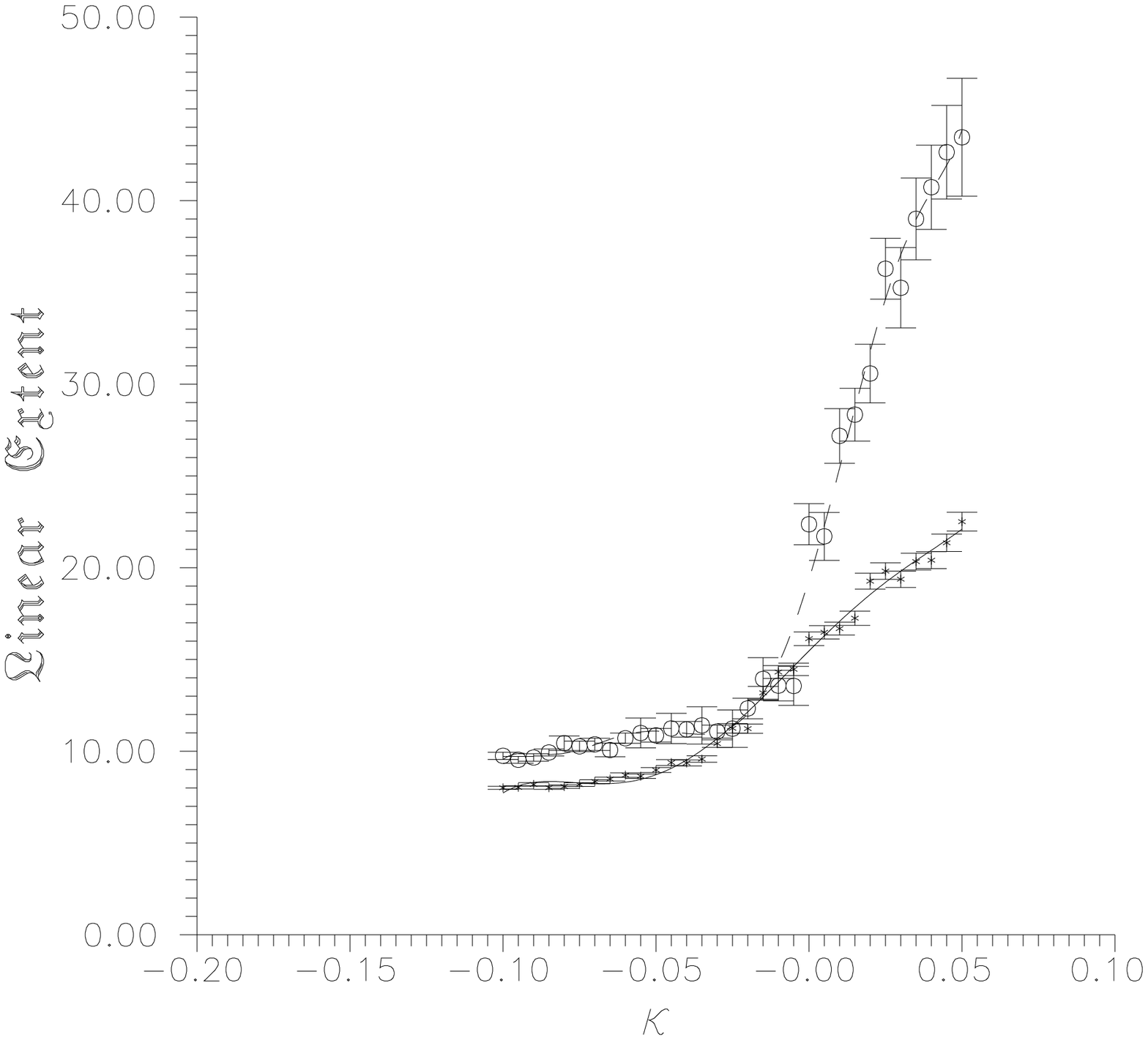,height=100mm,width=100mm,angle=0}
 \caption{\label{fig.01} Линейное расширение. Данные для $V=8000$ представлены точками и непрерывной
 линией. Данные для  $V=32000$ представлены кругами и пунктирной линией.
 \label{fig.1}}
\end{center}
\end{figure}

Исследование $5$ - мерной модели (см. \cite{5D}) указывает на то, что она имеет
усложненную фазовую структуру. А именно, при $\kappa_{D-2} < -5$ появляются
несколько различных вакуумов. И система может тунеллировать между ними. В нашем
исследовании $10$ - мерной модели мы ограничиваемся областью $\kappa_8 \in
(-0.1, 0.1)$, где наблюдается фазовый переход между хаотической фазой и фазой
ветвящихся полимеров.  Однако, имея в виду указанное свойство $5D$ мерной
модели, мы ожидаем, что фазовая структура рассматриваемой модели, возможно, не
ограничена наблюдаемыми нами двумя фазами.

Также остался неисследованным нами вопрос о роде наблюдаемого фазового
перехода. Хотя мы имеем ряд указаний на то, что это переход первого рода,
окончательного вывода на основании полученных данных мы не можем сделать.

Отдельным и важным вопросом является роль действия в модели динамических
триангуляций. Как уже отмечалось в предыдущей главе, наиболее реалистичным
вариантом действия является действие квадратичное по кривизне. Однако, в модели
ДТ оно не может в принципе применяться. Дело в том, что используемое кусочно -
линейное пространство всегда имеет сингулярную кривизну. В отличие от модели
Редже дискретизации не существует способа представить плоское пространство в
виде используемого кусочно - линейного пространства (исключение составляет
двумерный случай). Поэтому вычисляемая в модели ДТ кривизна, строго говоря, не
имеет физического смысла. Кривизна приближаемого посредством ДТ Риманова
многообразия может быть получена только посредством некоторого усреднения по
достаточно большому количеству симплексов. Впрочем, интеграл от кривизны,
представляющий действие Эйнштейна, по видимому, имеет смысл, поскольку
оказывается пропорциональным глобальной характеристике многообразия (полное
число симплексов разной размерности) и потому не зависит от указанной процедуры
усреднения. Члены, квадратичные по кривизне, таким образом трактоваться не
могут. Действие Эйнштейна неограничено снизу. Поэтому в зависимости от знака
эффективной гравитационной постоянной будут доминировать структуры с большой
отрицательной или большой положительной кривизной, что мы и наблюдаем.
Поскольку локальное действие, делающее модель реалистичной, не может быть в
принципе введено в модель ДТ, особую роль в ней играет энтропийный фактор. Это
и позволяет нам проводить аналогию между подходом ДТ и подходом, трактующим
гравитацию как энтропийную силу \cite{Entropy_force_}.

\section{Публикации}

Результаты настоящей главы опубликованы в работах

 "10-D Euclidean quantum gravity on the lattice", A.I. Veselov, M.A. Zubkov,
Nucl.Phys.Proc.Suppl.129:797-799,2004, [hep-lat/0308025]

"10D Euclidean dynamical triangulations", A.I. Veselov, M.A. Zubkov.
Phys.Lett.B591:311-317,2004.

\part{Возможные пути построения моделей новой физики на масштабе ТэВ}

В этой части мы рассматриваем подходы к построению моделей новой физике на
масштабе ТэВ. В десятой главе рассматриваются модели Малого объединения.
Исследуются появляющиеся монополи и связанная с ними дополнительная симметрия
Стандартной Модели ($Z_6$, $Z_3$, или $Z_2$). В одиннадцатой главе делается
попытка продолжить $Z_6$ симметрию Стандартной Модели на модели техницвета.
Показано, что лишь немногие из рассмотренных моделей допускают такое
продолжение. В двенадцатой главе предлагается заменить группу Техницвета
группой Лоренца. Исследуется возможность динамического нарушения Электрослабой
симметрии благодаря нарушению киральной инвариантности в калибровочной теории
группы Лоренца. При этом оказывается возможным с легкостью придать
реалистические значения массам фермионов Стандартной Модели, в отличие от
моделей техницвета, где приходится вводить дополнительное калибровочное поле
Расширенного техницвета.

\chapter{Малое объединение на масштабе ТэВ и $Z_6$ симметрия Стандартной
Модели}

\label{ch10}

В этой главе мы обсуждаем возможность наблюдения $Z_6$ симметрии Стандартной
Модели. Под $Z_6$ симметрией Стандартной Модели мы понимаем тот факт, что ее
фермионный сектор устроен так, что калибровочная группа может быть определена в
виде $SU(3)\times SU(2) \times U(1)/{\cal Z}$ (${\cal Z} = Z_6$, $Z_3$ or
$Z_2$) вместо обычной $SU(3)\times SU(2) \times U(1)$. При этом теории с
различными факторами $\cal Z$ идентичны на уровне теории возмущений. Нами
высказывалось предположение, что непертурбативные эффекты могут привести к
различной динамике этих моделей в некоторой области энергий. Однако, численные
решеточные исследования показали, что это не так (см. часть II). В области
физических значений констант связи модели с разными факторами $\cal Z$
неразличимы при энергиях много меньших ультрафиолетового обрезания. Ситуация,
однако, существенно меняется при повышении шкалы энергии. А именно, если
предположить, что объединение взаимодействий происходит на шкале ТэВ (так
называемое малое объединение \cite{PUT,PUT1}), то нарушение симметрии $G$
объединенной модели происходит по схеме $G \rightarrow SU(3)\times SU(2) \times
U(1)/{\cal Z}$. При этом разные модели объединения приводят к разным факторам
$\cal Z$. В свою очередь, эти факторы существенно влияют на свойства
монопольных состояний объединенных моделей. Эти монопольные состояния имеют
массы порядка $40$ ТэВ и несут различные магнитные заряды. Ниже мы выясняем
какие схемы нарушения симметрии имеют место для различных моделей малого
объединения (МО, Petite Unification) и каковы в этих моделях свойства
монопольных решений. При этом рассматриваются три различных модели МО.

\section{Дополнительная $Z_6$ симметрия в стандартной модели}

Известно, что спонтанное нарушение $SU(5)$ симметрии в моделях Большого
объединения ведет к возникновению калибровочной группы Стандартной Модели
$SU(3)\times SU(2) \times U(1)/Z_6$ вместо привычной $SU(3)\times SU(2) \times
U(1)$ (см., например, \cite{Z6}). Однако, указанный $Z_6$ фактор не есть
свойство одного лишь $SU(5)$ объединения. Ниже мы покажем, что дополнительная
$Z_6$ симметрия присутствует в Стандартной Модели вне всякой связи с какой -
либо моделью объединения. И сама структура $SU(5)$ объединения возможна только
благодаря наличию этой дополнительной симметрии.  При этом, вообще говоря,
модели объединения могут как содержать так и не содержать эту симметрию, что
будет обсуждаться далее. Ниже приводится объяснение того, что такое эта
дополнительная симметрия.

Для любого (незамкнутого) пути $\cal C$ мы можем вычислить элементарные
операторы параллельного переноса
\begin{eqnarray}
\Gamma &=& {\rm P} \, {\rm exp} (i\int_{\cal C} C^{\mu} dx^{\mu}) \nonumber\\
U &=& {\rm P} \, {\rm exp} (i\int_{\cal C} A^{\mu} dx^{\mu}) \nonumber\\
e^{i\theta} &=& {\rm exp} (i\int_{\cal C} B^{\mu} dx^{\mu}) ,\label{Sing_}
\end{eqnarray}
здесь $C$, $A$, и $B$ - это соответственно $SU(3)$, $SU(2)$, и $U(1)$
калибровочные поля  Стандартной Модели.

Операторы параллельного переноса, соответствующие каждому из фермионов
Стандартной Модели (и полю Хиггса) это некие произведения указанных выше
элементарных операторов параллельного переноса. Можно легко установить, что
операторы параллельного переноса в Стандартной Модели встречаются только в
комбинациях указанных в  таблице \ref{10tab.01_}.

\begin{table}
\begin{center}
\begin{tabular}{|c|l|}
\hline
$U\, e^{-i\theta}$ & {\rm левые лептоны } \\
\hline
$e^{-2 i \theta}$ & {\rm правые лептоны} \\
\hline
$ \Gamma \, U \, e^{ \frac{i}{3} \theta}$ & {\rm левые кварки}\\
\hline
$ \Gamma \, e^{ -\frac{2i}{3} \theta}$ &{\rm правые $d$, $s$, и $b$ - кварки} \\
\hline
$ \Gamma \, e^{ \frac{4i}{3} \theta}$ &{\rm правые $u$, $c$, и $t$ - кварки} \\
\hline
$  U \, e^{  i \theta}$ &{\rm Хиггс}\\
\hline
\end{tabular}
\end{center}
\caption{Комбинации операторов параллельного переноса, встречающиеся в
Стандартной Модели.} \label{10tab.01_}
\end{table}
Ясно, что {\it все} перечисленные комбинации инвариантны относительно следующих
 $Z_6$ преобразований:
\begin{eqnarray}
 U & \rightarrow & U e^{i\pi N}, \nonumber\\
 \theta & \rightarrow & \theta +  \pi N, \nonumber\\
 \Gamma & \rightarrow & \Gamma e^{(2\pi i/3)N},
\label{symlat}
\end{eqnarray}
где $N$ - произвольное целое число. Эта симметрия позволяет определить
Стандартную Модель с калибровочной группой $SU(3)\times SU(2) \times U(1)/{\cal
Z}$ (${\cal Z} = Z_6$, $Z_3$ или $Z_2$) вместо обычной $SU(3)\times SU(2)
\times U(1)$.

То, что Стандартная Модель может иметь калибровочную группу $SU(3)\times SU(2)
\times U(1)/Z_6$ видно из структуры $SU(5)$ модели объединения \cite{Z6}.
Оператор параллельного переноса в $SU(5)$ теории при низких энергиях имеет вид
\begin{equation}
\Omega = \left( \begin{array}{c c}

\Gamma^+ e^{\frac{2i\theta}{3}} & 0  \\
0 & Ue^{-i\theta}

\end{array}\right)\in SU(5), \label{SU(5)}
\end{equation}
где $\Gamma, U$, и $e^{i\theta}$ - элементарные операторы параллельного
переноса Стандартной Модели, введенные выше.  (\ref{SU(5)}) очевидным образом
инвариантно относительно (\ref{symlat}), что означает то, что катрина
спонтанного нарушения симметрии  $SU(5) \rightarrow SU(3)\times SU(2) \times
U(1)/Z_6$, а не $SU(5) \rightarrow SU(3)\times SU(2) \times U(1)$. Легко
понять, что именно наличие в Стандартной Модели симметрии (\ref{symlat}) и
позволяет расположить фермионы Стандартной Модели в двух мультиплетах $SU(5)$
объединения (фундаментальном и антисимметричном двухиндексном).

 Ясно, что благодаря наличию указанной  выше дополнительной симметрии
 калибровочной группой Стандартной Модели может быть также и
  $SU(3)\times SU(2) \times U(1)/{\cal Z}$ с ${\cal Z} = Z_6$,
$Z_3$ или $Z_2$.

\section{Возможность объединения взаимодействий на шкале ТэВ. Монополи в моделях объединения.}

Научное сообщество ожидает появления новой физики на шкале ТэВ, доступной
Большому Адронному Коллайдеру (БАК). То есть предполагают, что на этой шкале
энергий Стандартная Модель более не в состоянии описывать действительность. И
на ее место должна придти некая другая модель, содержащая Стандартную Модель в
качестве низкоэнергитического приближения. Основной причиной, побуждающей
думать таким образом, является проблема иерархий в Стандартной Модели
\cite{M_H,Extention}, заключающаяся в том, что массовый член поля Хиггса
получает положительную квадратично расходящуюся поправку в теории возмущений.
Поэтому для того, чтобы получить перенормированное значение $m^2$ отрицательным
и конечным необходимо, чтобы затравочное значение $m^2$ было отрицательным и по
порядку величины равным квадрату ультрафиолетового обрезания $\Lambda^2$. При
этом должна иметь место тонкая подстройка (fine tuning) этих двух параметров.
Такая тонкая подстройка считается неестественной. Поэтому так называемое
требование естественности (naturalness) приводит к тому, что нужно потребовать,
чтобы поправка к $m^2$, пропорциональная $\Lambda^2$, не превышала
перенормированное значение $m^2$ более, чем в 10 раз. Из этого требования
выводим $\Lambda \sim 1$ ТэВ.

Таким образом, ниже мы предполагаем, что Стандартная Модель работает до шкалы в
$1$ ТэВ, а выше работает (неизвестная пока) новая теория. Это означает, что в
пространстве могут существовать области размерами  порядка $1 \, {\rm
Tev}^{-1}$, где поля Стандартной Модели не определены.  Эти области могут
представлять точечные или струноподобные объекты. В результате мы вынуждены
рассматривать топологию Стандартной Модели в пространственно - временном
многообразии $\cal M$ с нетривиальной $\pi_2({\cal M})$ и (или) $\pi_1({\cal
M})$. Ниже объясняется, что такие объекты могут появиться с массами порядка
$40$ ТэВ.

К настоящему моменту исследовалось несколько картин объединения взаимодействий
на шкале ТэВ. Среди них отметим так называемые Модели Малого Хиггса (Little
Higgs models) \cite{little_higgs}, в которых $SU(2)\times U(1)$ подгруппа
внедрена в большую группу,  часть которой становится калибровочной.
Соответствующая симметрия нарушена при нескольких ТэВ. Некоторые из
Голдстоуновских бозонов становятся массивными благодаря радиационным поправкам
и играют роль поля Хиггса в Стандартной Модели. Топологические объекты,
появляющиеся в этих моделях были рассмотрены в
 \cite{TopLittleHiggs}.

Второй пример - это так называемое Малое Объединение (МО) \cite{PUT,PUT1}. В
соответствующих моделях калибровочная симметрия Стандартной Модели расширяется
до большей группы на масштабе ТэВ. Результирующие модели имеют две различные
константы связи, соответствующие сильным и Электрослабым взаимодействиям в
отличие от Моделей Большого Объединения (Grand Unified models), в которых есть
только одна константа связи и объединение достигается на шкале $10^{15}$ ГэВ.

Нам сейчас не важно, какая конкретно модель описывает физику ТэВ. Единственное,
что мы от нее потребуем - это то, что ее калибровочная группа достаточно велика
(что это значит, вскоре станет понятно).

Зафиксируем замкнутую поверхность $\Sigma$ в $4$ - мерном пространстве $R^4$.
Для любой замкнутой петли $\cal C$, которая оборачивается вокруг этой
поверхности, мы можем вычислить петли Вильсона $\Gamma = {\rm P} \, {\rm exp}
(i\int_{\cal C} C^{\mu} dx^{\mu})$, $U = {\rm P} \, {\rm exp} (i\int_{\cal C}
A^{\mu} dx^{\mu})$, и $e^{i\theta} = {\rm exp} (i\int_{\cal C} B^{\mu}
dx^{\mu})$, где $C$, $A$, и $B$ - соответственно $SU(3)$, $SU(2)$ и $U(1)$
калибровочные поля Стандартной Модели. В обычной реализации Стандартной Модели,
когда ее калибровочная группа - $SU(3)\times SU(2) \times U(1)$, такие петли
Вильсона должны стремиться к единице, когда длина $\cal C$ стремится к нулю
($|{\cal C}| \rightarrow 0$). Однако, в $SU(3)\times SU(2) \times U(1)/{\cal
Z}$ калибровочной теории разрешены следующие значения петель Вильсона при
$|{\cal C}| \rightarrow 0$:
\begin{eqnarray}
\Gamma &=& {\rm P} \, {\rm exp} (i\int_{\cal C} C^{\mu} dx^{\mu}) = e^{N
\frac{2\pi
i}{3}}\nonumber\\
U &=& {\rm P} \, {\rm exp} (i\int_{\cal C} A^{\mu} dx^{\mu}) =
e^{-N \pi i}\nonumber\\
e^{i\theta} &=& {\rm exp} (i\int_{\cal C} B^{\mu} dx^{\mu}) = e^{N \pi
i},\label{Sing}
\end{eqnarray}
где $N = 0,1,2,3,4,5$ для ${\cal Z}=Z_6$, $N = 0,2,4$ для ${\cal Z}=Z_3$, и $N
= 0,3$ для ${\cal Z}=Z_2$. Тогда поверхность $\Sigma$ может нести $Z_2$ поток
$\pi [N\, {\rm mod}\,2]$ для ${\cal Z} = Z_2, Z_6$. Она также может нести $Z_3$
поток
 $\frac{2\pi [N\, {\rm mod}\,3]}{3}$ для ${\cal Z} = Z_3, Z_6$.

Любая конфигурация с сингулярностью типа (\ref{Sing}) может быть устранена
сингулярным калибровочным преобразованием. Поэтому ее появление в теории с
калибровочной группой $SU(3)\times SU(2) \times U(1)/{\cal Z}$ не может
повлиять на динамику. Однако, если поверхность имеет границу, конфигурация
становится нетривиальной.

Рассмотрим незамкнутую поверхность $\Sigma$. Малая окрестность ее границы
$U(\partial \Sigma)$ представляет точечное солитонное состояние объединенной
модели. Это означает, что поля Стандартной Модели определены везде, кроме
$U(\partial \Sigma)$. Рассмотрим такие конфигурации, что на бесконечно малом
контуре $\cal C$ (наматывающемся вокруг $\Sigma$) указанные выше петли Вильсона
выражаются как в (\ref{Sing}). Для  $N \ne 0$ невозможно распространить
определение таких полей на $U(\partial \Sigma)$. Однако, это может быть
возможно внутри объединенной модели если калибровочная группа Стандартной
Модели $SU(3)\times SU(2) \times U(1)/{\cal Z}$ погружена в односвязную группу
 $\cal H$. Это следует немедленно из того, что любая замкнутая
 петля в такой $\cal H$ может быть непрерывным образом деформирована в единицу.
 В действиетльности, в такой $\cal H$ имеем $\pi_2({\cal H}/[SU(3)\times SU(2)
\times U(1)/{\cal Z}]) = \pi_1(SU(3)\times SU(2) \times U(1)/{\cal Z})$. Это
означает, что в такой объединенной модели монопольные состояния допустимы.
Конфигурации с (\ref{Sing}) и $N\ne 0$ представляют фундаментальные монополи в
объединенной модели. Такие конфигурации рассматривались \cite{Z6}, например, в
модели $SU(5)$ объединения. Однако, в \cite{Z6} предполагалось, что такие
солитонные состояния могут появиться с массами порядка шкалы Большого
Объединения ($10^{15}$ ГэВ). В нашем случае появление таких объектов ожидается
уже при энергиях порядка $40$ ТэВ, как это будет объяснено ниже. Другие
монополи могут быть построены из фундаментальных как из элементарных блоков. В
объединенной модели, которая нарушается до Стандартной Модели с калибровочной
группой $SU(3)\times SU(2) \times U(1)$ такие конфигурации с $N < 6$ появиться
не могут.

Объединенная модель, которая нарушается до Стандартной Модели с группой
$SU(3)\times SU(2) \times U(1)$ также содержит монополи поскольку $\pi_2({\cal
H}/[SU(3)\times SU(2) \times U(1)]) = \pi_1(SU(3)\times SU(2) \times U(1)) =
Z$. Они соответствуют Дираковской струне с $ \int_{\cal C} B^{\mu} dx^{\mu} = 6
\pi K, K\in Z$ и отличаются от монополей Стандартной Модели с дополнительной
дискретной симметрией магнитным потоком $U(1)$ поля гиперзаряда.

Следует отметить, что в Стандартной Модели также существуют и другие
монополеподобные состояния. Это так называемые монополи Намбу, подробно
изучавшиеся нами в части II. Они существенно отличаются от монополей,
рассматриваемых в настоящей главе. Прежде всего, они тпологически нестабильны.
Поэтому могут появиться только в виде пары монополь - антимонополь, связанной Z
- струной. Далее, они имеют нецелый магнитный заряд. Электромагнитное поле
выражается через $A$ и $B$ следующим образом (в соответствующей калибровке):
\begin{equation}
 A_{\rm em}  =  2 B - 2 \,{\rm sin}^2\, \theta_W (A_3+B).
\label{A_em}
\end{equation}
Полный гиперзарядовый магнитный поток монополя Намбу равен нулю. Поэтому
обычный магнитный поток пропорционален $4 \pi \,{\rm sin}^2\, \theta_W$.
Монополи, рассмотренные выше, имеют нетривиальный гиперзарядовый поток и
обычный магнитный поток, пропорциональный $2\pi$.

Другой тип монополей, рассматривавшийся в Стандартной Модели - это монополь Хо
- Мейсона \cite{Cho}. Монополи этого вида несут гиперзарядовый поток $2\pi$ и
электромагнитный поток $4\pi$. Они имели бы бесконечную собственную энергию,
если бы Стандартная Модель имела бесконечное Ультрафиолетовое обрезание. Их
свойства позволяют идентифицировать их как монополи объединенной модели.

Используя аналогию с монополями т'Хофта - Полякова \cite{HooftPolyakov}, мы
можем оценить массы фундаментальных монополей, несущих нетривиальный $U(1)$
поток как  $\frac{e \Lambda N}{\alpha} \sim 40 N$ ТэВ, где $\Lambda \sim 1 \,
{\rm Tev}$ - шкала нарушения симметрии, а $\alpha = \frac{e^2}{4\pi}$ -
постоянная тонкой структуры ($\alpha(M_Z)\sim \frac{1}{128}$) (см., например,
\cite{Weinb}, где рассматривались монополи в для случая произвольной компактной
калибровочной группы в пределе Богомольного - Прасада - Зоммерфельда (BPS
limit). В соответствии с (\ref{A_em}) обычный магнитный поток фундаментальных
монополей равен $2\pi N$. Если бы Стандартная Модель имела общепринятую форму
калибровочной группы, тогда эти монополи могли бы появиться только с $N$
пропорциональным $6$, а их магнитный поток был бы пропорционален $12\pi$.
Теперь становится ясно, что магнитный поток монополей оказывается связан с
различием между факторами $\cal Z$, введенными выше.

\section{Модели Малого Объединения}

Для иллюстрации того, как может возникнуть дополнительная $Z_3$, или $Z_2$
симметрия Стандартной Модели рассмотрим Малое объединение сильных и
электрослабых взаимодействия, обсуждаемое, например, в \cite{PUT, PUT1}. В этих
работах три возможности сконструировать объединенную теорию на масштабе ТэВ
были выделены из всего множества возможностей. А именно, рассмотрим в качестве
объединенной группы произведение $SU(4)_{PS}$ и $SU(N)^k$, где $SU(4)_{PS}$
объединяет лептонное число с цветом как в моделях Пати - Салама \cite{PATI}. В
теории есть две независимые константы связи $\alpha_s$ и $\alpha_W$,
соответствующие двум указанным группам. Тогда, если мы потребуем, что
спонтанное нарушение $SU(4)_{PS}\otimes SU(N)^k$ происходит на шкале ТэВ, мы
приходим к трем возможностям: ${\rm PUT}_0(N=2,k=4);\,{\rm PUT}_1 (N=2,k=3);\,
{\rm PUT}_2(N=3,k=2)$. Другой выбор $N$ и $k$ не может обеспечить приемлемых
значений констант связи на Электрослабой шкале.

Нам будет удобно представить картину нарушения симметрии в моделях $\rm PUT_0,
PUT_1, PUT_2$ в терминах петлевых переменных $\Gamma, U$, и $\theta$,
вычисленных вдоль произвольного замкнутого контура  $\cal C$.

В модели $\rm PUT_2$ на электрослабой шкале оператор параллельного переноса
$SU(4)_{PS}\otimes SU(3)^2$,  обозначаемый $\Omega$,  вдоль контура $\cal C$
выражается через $\Gamma, U$, и $\theta$ следующим образом:

\begin{equation}
\Omega = \left( \begin{array}{c c}

\Gamma^+ e^{\frac{2i\theta}{3}} & 0  \\
0 & e^{-2i\theta}

\end{array}\right) \otimes\left( \begin{array}{c c c}

e^{\frac{-4i\theta}{3}} & 0 & 0 \\
0 & e^{\frac{2i\theta}{3}} & 0 \\
0 & 0 & e^{\frac{2i\theta}{3}}

\end{array}\right)\otimes\left( \begin{array}{c  c}

U e^{-\frac{i\theta}{3}} & 0  \\
0 & e^{\frac{2i\theta}{3}}

\end{array}\right)\label{PUT2}
\end{equation}

Из (\ref{PUT2}) следует, что значения (\ref{Sing}) петель Вильсона  $\Gamma$,
$U$, и $e^{i\theta}$ с $N = 0, 3 \in Z_2$ ведут к $\Omega = {\bf 1}$.
Напряженность $SU(4)_{PS}\otimes SU(3)^2$ калибровочного поля выражается через
 $\Omega$, вычисленного вдоль бесконечно малого контура. Поэтому чистое калибровочное поле
 в низкоэнергитическом приближении (на Электрослабой шкале) инвариантно относительно
 дополнительной $Z_2$ симметрии. Это значит, что в $\rm PUT_2$ картина нарушения симметрии
 - $SU(4)_{PS}\otimes SU(3)^2 \rightarrow SU(3)\times SU(2)
\times U(1)/Z_2$, а не $SU(4)_{PS}\otimes SU(3)^2 \rightarrow SU(3)\times SU(2)
\times U(1)$. Поэтому мы ожидаем возникновения в этой объединенной модели $Z_2$
монополей с массами порядка $40*3 = 120$ ТэВ.

Здесь мы использовали значения Электрослабых зарядов, вычисленное в \cite{PUT}
для того, чтобы представить картину нарушения симметрии в форме, удобной для
нашего рассмотрения. Можно непосредственно проверить, что элемент калибровочной
группы в форме (\ref{PUT2}) действует соответствующим образом на фермионы
Стандартной Модели, размещенные ы представлениях, перечисленных в \cite{PUT}.
Та же проверка может быть выполнена также для моделей $\rm PUT_1$ and $\rm
PUT_0$, рассматриваемых ниже.

В Модели $\rm PUT_1$ на Электрослабой шкале оператор параллельного переноса для
поля $SU(4)_{PS}\otimes SU(2)^3$ (обозначен $\Omega$) вдоль контура $\cal C$
выражается следующим образом:

\begin{equation}
\Omega = \left( \begin{array}{c c}

\Gamma^+ e^{\frac{2i\theta}{3}} & 0  \\
0 & e^{-2i\theta}

\end{array}\right)\otimes U \otimes \left( \begin{array}{c  c}

e^{-i\theta} & 0  \\
0 & e^{i\theta}

\end{array}\right) \otimes\left( \begin{array}{c  c}

e^{i\theta} & 0  \\
0 & e^{-i\theta}

\end{array}\right)\label{PUT1}
\end{equation}

Ясно, что значения (\ref{Sing}) петель Вильсона $\Gamma$, $U$, и $e^{i\theta}$
с $N = 0, 2, 4 \in Z_3$ ведут к $\Omega = {\bf 1}$. Это значит, что в $\rm
PUT_1$ схема нарушения - $SU(4)_{PS}\otimes SU(2)^3 \rightarrow SU(3)\times
SU(2) \times U(1)/Z_3$, а не $SU(4)_{PS}\otimes SU(2)^3 \rightarrow SU(3)\times
SU(2) \times U(1)$. Таким образом, должны существовать  $Z_3$ монополи с
массами порядка $40*2 = 80$ ТэВ.

В $\rm PUT_0$ на Электрослабой шкале оператор параллельного переноса поля
$SU(4)_{PS}\otimes SU(2)^4$ (также обозначен $\Omega$) вдоль контура
 $\cal C$ выражается через $\Gamma, U$, и $\theta$ следующим образом:

\begin{equation}
\Omega = \left( \begin{array}{c c}

\Gamma e^{\frac{i\theta}{3}} & 0  \\
0 & e^{-i\theta}

\end{array}\right)\otimes U \otimes\left( \begin{array}{c  c}

e^{i\theta} & 0  \\
0 & e^{-i\theta}

\end{array}\right)\otimes U \otimes\left( \begin{array}{c  c}

e^{i\theta} & 0  \\
0 & e^{-i\theta}

\end{array}\right)\label{PUT0}
\end{equation}

Ясно, что в $\rm PUT_0$ картина нарушения - $SU(4)_{PS}\otimes SU(2)^4
\rightarrow SU(3)\times SU(2) \times U(1)/Z_3$, а не $SU(4)_{PS}\otimes SU(2)^4
\rightarrow SU(3)\times SU(2) \times U(1)$. Таким образом, в этой модели также
должны существовать $Z_3$ монополи с массами порядка $80$ Tэв. Здесь следует
отметить, что, по всей вероятности,  $\rm PUT_0$ исключена по вероятности
процесса $K_L \rightarrow \mu e$.

\section{Выводы}

Таким образом, мы рассмотрели объединенные модели, нарушающиеся до Стандартной
Модели на масштабе ТэВ. Из-за нарушения возникают монополи, имеющие массы
порядка $40$ ТэВ. Эти объекты могут оказаться легчайшими топологически
стабильными магнитными монополями. В принципе, во время высокоэнергитических
столкновений могут появляться монополь - антимонопольные пары. Размер монополей
должен быть порядка $1$ $\rm Tev^{-1}$ в то время, как  типичный размер
объектов, создаваемых во время столкновений должен быть в районе $80$ $\rm
Tev^{-1}$. Поэтому появление монополь - антимонопольной пары должно быть
подавлено отношением этих двух масштабов. Теоретическое рассмотрение рождения
пары монополь - антимонополь является проблемой, до сих пор не получившей
решения даже для монополей т'Хофта - Полякова (см., например, \cite{MONOPOLE} и
ссылки внутри этой работы). Поэтому мы не можем здесь представить оценку
соответствующего сечения. Следует также отметить, что порог рождения таких
объектов недостижим на Большом Адронном Коллайдере и может быть достигнут
только на коллайдерах следующего поколения.

Появление указанных монополей в ранней вселенной может иметь определенные
следствия для космологии. В частности, мы ожидаем, что симметрия объединенной
модели может восстановиться при высоких температурах. Мы ожидаем, что в ранней
Вселенной при температурах, близких к температуре $T_c$ соответствующего
перехода, монополи, рассматриваемые в настоящей главе могут появляться в
элементарных процессах с большой вероятностью. При $T>T_c$ они могут оказаться
сконденсированы подобно монополям Намбу при температуре выше температуры
Электрослабого фазового перехода (см. главу 5). Следует также отметить, что
существует ряд космологических моделей, в которых исключено появление магнитных
монополей с массами, вычисленными выше. В этих моделях, соответственно,
исключены рассмотренные нами модели Малого Объединения.

\section{Публикации}

Результаты настоящей главы опубликованы в работах

"The Observability of Z(6) symmetry in the standard model", M.A. Zubkov.
Phys.Lett.B649:91-94,2007,Erratum-ibid.B655: 91,2007, [hep-ph/0609029]

"Monopoles, topology of the standard model, and unification of interactions at
TeV scale", Mikhail A. Zubkov, PoS LAT2007:285,2007.

\chapter{Продолжение $Z_6$ симметрии Стандартной Модели на теории
техницвета}

\label{ch11}

В настоящей главе мы рассматриваем возможность продолжить $Z_6$ симметрию
Стандартной Модели, обсуждавшуюся в главе 10, на Модели техницвета. Выясняется,
что среди $SU(N)$ моделей Вайнберга - Сасскинда и $SU(N)$ моделей Фари -
Сасскинда для $N>2$ только $SU(4)$ модель Фари - Сасскинда может содержать эту
симметрию. Мы также рассматриваем минимальную почти конформную модель (Minimal
Walking) с групой $SU(2)$, в которой можно выбрать гиперзаряды таким образом,
что имеет место указанная дополнительная симметрия.

\section{Техницвет}

Как уже отмечалось в главе 10, есть основания полагать, что при энергиях,
превышающих  $1$ ТэВ, Стандартная Модель перестает работать. В свою очередь,
КХД имеет хорошо определенный непрерывный предел, что явилось причиной
возникновения ряда моделей, схожих с КХД, и претендующих на описание физики
ТэВ. Эти модели получили название моделей техницвета \cite{Technicolor, WS,FS}.
В них добавляется новое неабелево взаимодействие со шкалой $\Lambda_{TC} \sim
1$ ТэВ, где $\Lambda_{TC}$ - это аналог $\Lambda_{QCD}$. Это новое
взаимодействие получило название техницвет (ТЦ). Соответствующие новые фермионы
названы технифермионами. Электрослабая калибровочная группа действует на
технифермионы. Поэтому нарушение киральной симметрии в модели техницвета ведет
к нарушению Электрослабой симметрии, что дает массы трем из калибровочных
бозонов. Однако, чистая модель ТЦ не в состоянии обеспечить объяснение
появления масс фермионов.

Для того, чтобы дать массы фермионам обычно добавляется новое калибровочное
поле, называемое расширенным техницветом (РТЦ)
\cite{Technicolor,ExtendedTechnicolor}. В этой новой калибровочной теории
фермионы Стандартной Модели и технифермионы входят в одно представление группы
РТЦ. Это приводит к возникновению перехода между ними и возникновению
взаимодействия фермионов Стандартной Модели с конденсатом технифермионов. Это
взаимодействие, в свою очередь, обеспечивает появление масс у фермионов
Стандартной Модели. К сожалению, первые модели РТЦ страдали от проблем с
переходами между фермионами различного аромата, а также  от недопустимо больших
вкладов технифермионов в Электрослабые поляризационные операторы. Возможный
способ преодоления этих трудностей связан с поведением киральных калибровочных
теорий при большом числе фермионов или для высших представлений калибровочной
группы, в которых они расположены. А именно, почти конформное поведение модели
Техницвета позволяет подавить опасные нейтральные токи, меняющие аромат (FCNC)
и уменьшить вклад в S - параметр \cite{Appelquist,minimal_walking}. (Следует
отметить, однако, что генерация массы $t$ - кварка в этих моделях все еще
вызывает серьезные трудности.)

Существует огромное количество моделей Техницвета и моделей РТЦ. Поэтому важно
иметь некоторый принцип, который позволил бы сделать выбор. Ниже мы предлагаем
в качестве такового то, что модель техницвета должна содержать в себе некоторое
естественное обобщение  $Z_6$ симметрии Стандартной Модели. А именно, мы
предлагаем, как $Z_6$ симметрия может быть обобщена на модели техницвета. Затем
мы требуем, чтобы модели техницвета удовлетворяли этой симметрии и находим, что
это требование приводит к существенному ограничению на выбор модели. Кроме
того, мы показываем, что модель РТЦ содержит монополи, чьи свойства в случае,
если дополнительная симметрия имеет место отличаются от свойств монополей в
РТЦ, когда дополнительной дискретной симметрии нет.

\section{Продолжение $Z_6$ симметрии на модели Техницвета}

Определение $Z_6$ симметрии Стандартной Модели было дано в предыдущей главе.
Здесь мы отметим лишь, что она является достаточно ограничительной. Например,
требование того, что операторы параллельного переноса, соответствующие
фермионам $Z_6$ - симметричны, приводит к запрету появления, скажем, левых
$SU(2)$ дублетов с нулевым гиперзарядом.

Природа этой симметрии связана с центрами $Z_3$ и $Z_2$ групп $SU(3)$ и
$SU(2)$. Эта симметрия связывает центры $SU(2)$ и  $SU(3)$ подгрупп
калибровочной группы. Основываясь на этом, мы предлагаем следующий способ
продолжить эту симметрию на модели ТЦ.

Мы связываем центр группы ТЦ с центрами $SU(3)$ и $SU(2)$. Пусть $SU(N_{TC})$ -
группа ТЦ. Тогда преобразование (\ref{symlat}) обобщается до
\begin{eqnarray}
 U & \rightarrow & U e^{i\pi N}, \nonumber\\
 \theta & \rightarrow & \theta +  \pi N, \nonumber\\
 \Gamma & \rightarrow & \Gamma e^{(2\pi i/3)N},\nonumber\\
 \Theta & \rightarrow & \Theta e^{(2\pi i/N_{TC})N}.
\label{symlatWS}
\end{eqnarray}
Здесь $\Theta$ - это  оператор параллельного переноса группы $SU(N_{TC})$.
Остальные переменные определены в 10.1. Операторы параллельного переноса,
соответствующие технифермионам, должны быть инвариантны относительно
(\ref{symlatWS}). Следует отметить, что результирующая группа симметрии не есть
произведение $Z_6$ и $Z_{N_{TC}}$.


\section{Минимальная Модель Техницвета Вайнберга и Сасскинда}

Рассмотрим простейшую модель ТЦ, предложенную в \cite{WS} (см. также
\cite{Technicolor}). Модель содержит технифермионы
\begin{equation}
 \left(
 \begin{array}{c}
  T^a \\
  B^a
 \end{array}
 \right)_L , \quad
 \left(
 \begin{array}{c}
 T^a \\
  B^a
 \end{array}
 \right)_R .
\end{equation}
Гиперзаряд $Y=0$ для левых технифермионов и $Y = \pm 1$ для правых
технифермионов. Индекс $a$ соответствует группе ТЦ $SU(N_{TC})$. Модель имеет
локальную  $SU(2)_L$ калибровочную симметрию и глобальную $SU(2)_R$ симметрию.
Нарушение киральной симметрии обеспечивает нарушение Электрослабой симметрии и
образование массивных $W$ и $Z$ бозонов. Можно также рассмотреть $N_D \ne 1$
копий технифермионов.

Потребуем, чтобы операторы параллельного переноса, соответствующие
технифермионам были инвариантны относительно (\ref{symlatWS}). Это приводит к
следующему условию:

\begin{equation}
\frac{2\pi N}{N_{TC}} + \pi N = 2 \pi k(N), \, k(N)\in Z
\end{equation}

Единственное решение этого уравнения - это $N_{TC} = 2, k(N) = N$. Таким
образом мы заключаем, что данная модель инвариантна относительно дополнительной
дискретной симметрии только для группы ТЦ $SU(2)$. Соответствующая
дополнительная симметрия - это $Z_{6}$.

Следует отметить, что выравнивание вакуума (vacuum alignment) в $SU(2)$ модели
ТЦ не приводит к правильному нарушению Электрослабой симметрии \cite{Align}.
Напомним, почему это происходит. Определим поле
\begin{equation}
 Q^{\dot{\alpha},a} = \left(
 \begin{array}{c}
  T_L^{\dot{\alpha},a} \\
  B_L^{\dot{\alpha},a}\\
 \epsilon^{a a^\prime} [T_R^{\alpha,a^\prime}]^* \\
  \epsilon^{a a^\prime} [B_R^{\alpha,a^\prime}]^* \\
   \end{array}
 \right)
\end{equation}
(Здесь $\alpha, \beta, \dot{\alpha}, \dot{\beta}$ - спинорные индексы.
Четырехкомпонентный спинор $T$ состоит из двухкомпонентных спиноров: $T =
(T_R^{1}, T_R^{2}, \epsilon_{1 \alpha} T_L^{\dot{\alpha}}, \epsilon_{2 \alpha}
T_L^{\dot{\alpha}})^T$.)

 $Q^{\dot{\alpha},a}$ преобразуется как левополяризованный пунктирный спинор при действии
  $SL(2,C)$ и как элемент фундаментального представления под действием группы ТЦ $SU(2)$.
  Если Электрослабые взаимодействия выключаются, логранжиан ТЦ инвариантен относительно
  глобальной  $SU(4)$ симметрии.
$SU(2)$ и $SL(2,C)$ инвариантные билинейные комбинации $Q$:
\begin{equation}
\Phi_{AB} = \epsilon_{ab}\epsilon_{\alpha \beta} Q^{\dot{\alpha},a}_A Q^{
\dot{\beta},b}_B,
\end{equation}
где $A,B$ - это $SU(4)$ индекс.

Низкоэнергитический эффективный потенциал $V(\Phi)$ инвариантен относительно
действия $SU(4)$ на $\Phi$. Вакуумное значение $\Phi$ выбирается, когда
учитываются возмущения, нарушающие $SU(4)$. Этот процесс называется
выравниванием вакуума (vacuum alignment) \cite{Align}.

Обычный вакуум Электрослабой теории получается, если $\Phi$ пропорционально
\begin{equation}
 \Phi  = \left(
 \begin{array}{c c c c}
  0 & 0 & 1 & 0 \\
  0 & 0 & 0 & 1 \\
 -1 & 0 & 0 & 0  \\
  0 & -1 & 0 & 0  \\
   \end{array}
 \right)
\end{equation}
Соответствующий киральный конденсат: $\langle \delta_{ab} \epsilon_{\alpha
\beta} [T^{\beta b}_R]^* T^{\dot{\alpha}a}_L  + \delta_{ab}\epsilon_{\alpha
\beta}[B^{ \beta b}_R]^* B^{\dot{\alpha} a}_L \rangle$.

Однако, допустимо также и следующее значение
\begin{equation}
 \Phi  = \left(
 \begin{array}{c c c c}
  0 & 1 & 0 & 0 \\
  -1 & 0 & 0 & 0 \\
 0 & 0 & 0 & 1  \\
  0 & 0 & -1 & 0  \\
   \end{array}
 \right)
\end{equation}
В этом случае конденсат имеет вид $\langle \epsilon_{ab}\epsilon_{\alpha \beta}
T^{\dot{\alpha}a}_L B^{ \dot{\beta}b}_L + \epsilon_{ab}\epsilon_{\alpha \beta}
[T^{\alpha a}_R]^+ [B^{ \beta b}_R]^+\rangle $, и Электрослабая $SU(2)$
остается ненарушенной, в то время, как фотон становится массивным.

В общем случае нарушение киральной симметрии - довольно сложное явление и его
физика не изучена детально. Поэтому неясно до конца, какая из двух указанных
выше возможностей должна реализоваться. Анализ \cite{Align} показывает, что
малые возмущения, вызванные Электрослабыми взаимодействиями выбирают вакуум, в
котором Электрослабая симметрия нарушена минимальным образом, то есть сумма
квадратов масс калибровочных бозонов минимизируется. Это правило приводит к
тому, что в $SU(2)$ модели Вайнберга - Сасскинда реализуется вторая из
указанных выше возможностей и Электрослабая симметрия нарушается неправильным
образом.

\section{Модель Фари - Сасскинда}

Эта модель \cite{FS,Technicolor} содержит четыре дублета
\begin{eqnarray}
 \left(
 \begin{array}{c}
  U_i^a \\
  D_i^a
 \end{array}
 \right)_L , \quad
 \left(
 \begin{array}{c}
 U_i^a \\
  D_i^a
 \end{array}
 \right)_R \nonumber\\
 \left(
 \begin{array}{c}
  N^a \\
  E^a
 \end{array}
 \right)_L , \quad
 \left(
 \begin{array}{c}
 N^a \\
  E^a
 \end{array}
 \right)_R .
\end{eqnarray}
Здесь  $a$ - это $SU(N_{TC})$ индекс, а индекс $i$ соответствует цветной группе
$SU(3)$. Цветные фермионы называются техникварками, остальные называются
технилептонами. Гиперзаряд левых технилептонов обозначен $Y_L$. Гиперзаряды
правых технилептонов обозначены  $Y^{1,2}_R$. Гиперзаряд левых техникварков
обозначен $Y^c_L$. Гиперзаряд правых техникварков обозначен $(Y^c)^{1,2}_R$.
Киральные конденсаты инвариантны относительно электромагнитных $U(1)$
преобразований, если $Y^{1,2}_R = Y_L \pm 1$, $(Y^c)^{1,2}_R = Y^c_L \pm 1$.

Теория не содержит аномалий, если $Y_L + 3 Y^c_L = 0$. Потребуем, что модель
инвариантна относительно дополнительной симметрии (\ref{symlatWS}). Получим

\begin{eqnarray}
&&[\frac{2}{N_{TC}} + \frac{2}{3} + 1 + Y^c_L ] \, {\rm mod}\, 2 = 0 \nonumber\\
&&[\frac{2}{N_{TC}} +  1 + Y_L ] \, {\rm mod}\, 2 = 0\nonumber\\
&&   Y_L + 3 Y^c_L = 0
\end{eqnarray}

Данная система уравнений имеет два решения:

1) $N_{TC} = 2, Y^c_L = -\frac{Y_L}{3}, Y_L = 2(1-3k), k \in Z$;

2) $N_{TC} = 4, Y^c_L = -\frac{Y_L}{3}, Y_L = \frac{1}{2} - 6k , k\in Z$.
Только группы  $SU(2)$ и $SU(4)$ могут быть группами ТЦ, если мы требуем, что
теория инвариантна относительно (\ref{symlatWS}). Следующие группы могут быть
калибровочными группами модели, включающей ТЦ и Стандартную Модель:
\begin{equation}
SU(2)\otimes SU(3)\times SU(2) \times U(1)/Z_6
\end{equation}
или
\begin{equation}
SU(4)\otimes SU(3)\times SU(2) \times U(1)/Z_{12}
\end{equation}

Как уже говорилось, в $SU(2)$ модели ТЦ Вайнберга - Сасскинда Электрослабая
симметрия нарушается неправильно. Точно так же в $SU(2)$ модели Фари -
Сасскинда появляются проблемы в технилептонном секторе \cite{Align}. Поэтому из
двух групп $SU(2)$ и $SU(4)$ только группа $SU(4)$ может быть приемлемой
группой ТЦ.

\section{Объединение ТЦ и калибровочной группы СМ}

Рассмотрим, как, в принципе, ТЦ и калибровочная группа Стандартной Модели могут
быть объединены в общую калибровочную группу. Мы не будем обсуждать здесь
детали механизма нарушения симметрии и то, как обеспечивается сокращение
киральной аномалии в модели РТЦ. Наша цель заключается лишь в том, чтобы
продемонстрировать, как дополнительная дискретная симметрия (\ref{symlatWS})
может появиться при спонтанном нарушении симметрии объединенной модели.

Для определенности рассмотрим $N_{TC} = 4$. Пусть $U(10)$ - это группа
объединения. Схема нарушения симметрии следующая: $U(10)\rightarrow
SU(4)\otimes SU(3)\times SU(2) \times U(1)/Z_{12}$. Мы предполагаем, что при
низких энергиях оператор параллельного переноса  $U(10)$ поля принимает вид:

\begin{equation}
\Omega = \left( \begin{array}{c c c c }
e^{-2i\theta} & 0 & 0 & 0\\
0 & Ue^{-i\theta} & 0 & 0 \\
0 & 0 & \Gamma e^{-\frac{2i\theta}{3}} & 0  \\
0 & 0 & 0 & \Theta e^{-\frac{2i\theta}{4}} \\
\end{array}\right)\in U(N_{ETC})\label{U(10)}
\end{equation}

Форма этого оператора демонстрирует, что симметрия (\ref{symlatWS})
действительно имеет место. Фермионы каждого поколения $\Psi^{i_1 ... i_N}_{j_1
... j_K}$ несут индексы $i_k$ фундаментального представления $U(N_{ETC})$ и
индексы $j_k$ сопряженного представления. Они могут быть идентифицированы с
фермионами Стандартной Модели и технифермионами модели Фари - Сасскинда
следующим образом (мы рассматриваем здесь лишь первое поколение):

\begin{eqnarray}
&& \Psi^{1} = e_R; \, \Psi_{1}^{1} = \nu_R; \, \Psi^{i_2} =
\left(\begin{array}{c} \nu_L
\\ e^-_L\end{array}\right) ;\nonumber\\
&&\Psi^{i_3} = d_{i_3,R} ; \, \Psi^{i_3}_{1} = u_{i_3,R} ;\, \Psi^{i_2 i_3}_1 =
\left(\begin{array}{c} u^{i_3}_L \\ d^{i_3}_L \end{array}\right) ;\nonumber\\
&&\Psi^{i_4} = E_{i_4,R} ; \, \Psi_{1}^{i_4} = N_{i_4,R} ;\, \Psi^{i_2 i_4}_1 =
\left(\begin{array}{c} N^{i_4}_L \\ E^{i_4}_L \end{array}\right) ;\nonumber\\
&&\Psi^{i_3 i_4} = D_{i_3 i_4,R} ; \, \Psi_1^{i_3 i_4} = U_{i_3
i_4,R} ;\, \Psi^{i_2 i_3 i_4}_{1} = \left(\begin{array}{c} U^{i_3 i_4}_L \\
D^{i_3 i_4}_L
\end{array}\right) \nonumber\\&& (i_2 = 1,3; \, i_3 =
4,5,6;\, i_4 = 7,8,9,10);\label{ferm}
\end{eqnarray}

 Из (\ref{ferm}) следует, что все фермионы Стандартной Модели могут переходить в
 технифермионы с излучением $U(N_{ETC})$ калибровочных бозонов.
 Это позволяет им приобрести массы, а данную модель объединения рассматривать как вариант
 модели РТЦ. Разумеется, данная схема РТЦ не описывает появление реалистических масс известных частиц.
 Однако, она дает пример того, как это может быть сделано в принципе. Ниже мы опускаем детали
 данной модели РТЦ, не описываем, благодаря чему происходит нарушение симметрии, как исчезают из низкоэнергитической физики
 лишние фермионы и как сокращаются аномалии.
Предположим, однако, что все в порядке, и РТЦ модель работает, приводя при
низких энергиях к оператору $U(10)$ параллельного переноса в форме
(\ref{U(10)}). Тогда все операторы параллельного переноса в теории инвариантны
относительно (\ref{symlatWS}). Калибровочная группа $U(10)$ - односвязна.
Поэтому  модель РТЦ содержит монополи. Похожая ситуация была нами рассмотрена в
предыдущей главе. В частности, должны присутствовать монопольные конфигурации с
гиперзарядовым потоком $\pi$.

Предположим теперь, что модель РТЦ построена так, что схема нарушения - $G
\rightarrow SU(4)\otimes SU(3)\times SU(2) \times U(1)$. Также предположим, что
гиперзаряды фермионов - рациональные числа $\frac{P}{Q}$ с целыми $P$ и $Q$, и
максимальное значение $Q$ равно $3$. Тогда монополи в теории будут совершенно
другими. А именно, минимальный гиперзарядовый поток будет равен $6 \pi$ (если
максимальное значение $Q$ равно $Q_{max}
> 3$, то гиперзарядовый поток монополя будет $ 2 Q_{max} \pi  $).


Таким образом, есть существенная разница между монопольным содержанием модели
РТЦ в двух рассматриваемых случаях. А именно, в присутствии симметрии
(\ref{symlatWS}), появляются монополи с потоком, равным $\pi$, в противном
случае их нет.


\section{Почти конформный Техницвет}

Модели техницвета с их нарушением киральной симметрии способны обеспечить
нарушение Электрослабой симметрии. Однако, одни эти модели не в состоянии
обеспечить фермионы реалистическими массами. Фермионы Стандартной Модели
становятся массивными, если они могут превращаться в технифермионы, скажем, с
излучением новых массивных калибровочных бозонов. Тогда кварковые и лептонные
массы оцениваются как  $m_{q,l}\sim
\frac{N_{TC}\Lambda_{TC}^3}{\Lambda_{ETC}^2}$, где  $\Lambda_{TC}$ - шкала ТЦ,
а $\Lambda_{ETC}$ - шкала нового сильного взаимодействия (РТЦ). (Спонтанное
нарушение РТЦ приводит к возникновению масс новых калибровочных бозонов порядка
$\Lambda_{ETC}$.)

Число фермионов (размещенных в фундаментальном представлении группы ТЦ), для
которого поведение модели становится близким к конформному оценивается как
\cite{Appelquist}  $N_f \sim 4 N_{TC}$. В этом случае эффективный заряд
становится "идущим" (walking) вместо бегущего \cite{walking}. В соответствующей
теории РТЦ нейтральные токи с изменением аромата могут быть подавлены, что
позволяет приблизиться к реалистическому описанию генерации фермионных масс.
Следует отметить, что реалистическая масса t кварка не может быть сгенерирована
таким образом. Поэтому возникновение массы t кварка является отдельной
проблемой в построении моделей РТЦ (см., например, \cite{Sannino_t}). В модели
Фари - Сасскинда конформный режим достигается если число поколений
технифермионов $N_D$ равно $N_D = 2$ при $N_{TC} = 4$.

Если технифермионы размещаются в фундаментальном представлении модели почти
конформного техницвета (walking Technicolor model), то пертурбативный вклад в S
параметр все еще остается опасно большим \cite{minimal_walking}. Одним из путей
преодоления этой проблемы является рассмотрение высших представлений группы ТЦ.
Минимальный выбор здесь $N_{TC} = 2$ с одним поколением технифермионов,
принадлежащих двухиндексному симметричному представлению $SU(2)$. Эта
минимальная модель содержит технифермионы симметричные по ТЦ $SU(2)$ индексам
$a$ и $b$:
\begin{eqnarray}
 L^{a,b} = \left(
 \begin{array}{c}
  U^{a,b} \\
  D^{a,b}
 \end{array}
 \right)_L , \quad
 R^{a,b} =\left(
 \begin{array}{c}
 U^{a,b} \\
  D^{a,b}
 \end{array}
 \right)_R \nonumber\\
 \left(
 \begin{array}{c}
  N \\
  E
 \end{array}
 \right)_L , \quad
 \left(
 \begin{array}{c}
 N \\
  E
 \end{array}
 \right)_R .
\end{eqnarray}
Здесь добавлено дополнительное поколение лептонов Стандартной Модели для того,
чтобы сократить киральную аномалию. В этой модели вклад в S - параметр
существенно меньше, чем в модели с технифермионами из фундаментального
представления \cite{minimal_walking}.

Аномалия отсутствует, если $3 Y^c_L + Y_L = 0$, где  $Y^c_L$ - гиперзаряды
левых технилептонов, а $Y_L$ - гиперзаряд новых левых лептонов, являющихся
синглетами относительно ТЦ. Важно, что данное двухиндексное представление ТЦ
группы $SU(2)$ не чувствует центр группы $SU(2)$. Поэтому оператор
параллельного переноса соответствующий новым фермионам инвариантен относительно
(\ref{symlatWS}) с $N_{TC} = 2$ если
\begin{eqnarray}
 &&[1 + Y^c_L ]  {\rm mod} 2 = 0 \nonumber\\
 &&[1 + Y_L ]  {\rm mod} 2 = 0 \nonumber\\
&& Y_L + 3 Y^c_L = 0
\end{eqnarray}

Решение этих уравнений имеет вид $Y^c_L = -\frac{Y_L}{3}, Y_L = 3(1-2k), k \in
Z$. Таким образом, минимальная модель почти конформного техницвета инвариантна
относительно расширения $Z_6$ симметрии Стандартной Модели при данном выборе
гиперзарядов.

Отметим также, что в данной модели вакуумное среднее
\begin{equation}
\langle \epsilon_{cd}\epsilon_{ab}\epsilon_{\alpha \beta}
U^{a,c,\dot{\alpha}}_L D^{b,d, \dot{\beta}}_L +
\epsilon_{cd}\epsilon_{ab}\epsilon_{\alpha \beta} [U_R^{a,c,\alpha}]^+
[D_R^{b,d, \beta}]^+\rangle
\end{equation}
 может появиться вместо привычного
\begin{equation}
\langle \epsilon_{cd}\epsilon_{ab}\epsilon_{\alpha \beta}
[U^{a,c,\dot{\alpha}}_L]^+ U^{b,d,
\beta}_R+\epsilon_{cd}\epsilon_{ab}\epsilon_{\alpha \beta}
[D^{a,c,\dot{\alpha}}_L]^+ D^{b,d, \beta}_R \rangle
\end{equation}
 Если бы вакуум
соответствовал этому конденсату, Электрослабая симметрия была бы нарушена
неправильно. Однако, может быть показано  \cite{Align}, что в данном случае
сумма квадратов масс калибровочных бозонов больше, чем при обычном нарушении
симметрии. Таким образом, ориентация вакуума в данном случае соответствует
правильному нарушению Электрослабой симметрии.

\section{Надстройка над моделью Фари - Сасскинда}

Выше мы показали, что среди различных моделей ТЦ лишь некоторые симметричны
относительно дополнительной дискретной симметрии (\ref{symlatWS}). В частности
в модели Фари - Сасскинда, привлекающей наше внимание своим сходством со
Стандартной Моделью (ее технифермионы - копия фермионов Стандартной Модели (СМ)
с дополнительным ТЦ индексом) допустима лишь группа $SU(4)$. Эта модель (вместе
с СМ имеет калибровочную группу $SU(4) \otimes SU(3) \otimes SU(2) \otimes
U(1)$. Гиперзаряды всех фермионов фиксируются дополнительной симметрией с
точностью до целого числа.

Структура калибровочной группы  $SU(4) \otimes SU(3) \otimes SU(2) \otimes
U(1)$ подсказывает нам возможное продолжение как последовательности $SU(N)$
подгрупп:
\begin{equation}
G = ... \otimes SU(6)\otimes SU(5)\otimes SU(4) \otimes SU(3) \otimes SU(2)
\otimes U(1)/{\cal Z}, \label{G}
\end{equation}
где $\cal Z$ - дискретная группа, определяемая ниже.

Эта возможность кажется нам интересной. Ниже она анализируется с той же точки
зрения, что и исходная модель Фари - Сасскинда. Мы размещаем фермионы в
фундаментальных представлениях подгрупп $SU(N)$ группы (\ref{G}). В общем
случае модель с группой (\ref{G}) содержит аномалии различных типов. Априори
неочевидно, что существует выбор гиперзарядов такой, что отсутствует киральная
аномалия в то время, как удовлетворяется дополнительная дискретная симметрия.
Ниже мы показываем, однако, что такой выбор существует.

Ниже мы не конкретизируем, какой механизм приводит к возникновению масс
фермионов. Это может быть механизм РТЦ или какой то вариант механизма,
предлагаемого в следующей главе, или, какой - либо иной. Любой из таких
механизмов определяет соответствие между левыми и правыми фермионами,
идентифицируемый с сопряжением четности спиноров.  Наша модель также, как и
модель Фари - Сасскинда, содержит левые дублеты и правые синглеты. Поэтому
четность связывает пары правых синглетов с левыми дублетами. В частности,
компонента, соответствующая правому электрону, называется левым электроном.
Обозначим левый дублет первого поколения лептонов СМ  $\Theta$. Данное
соответствие может быть записано как $\Omega^1_{i}\Theta^{i} = \nu_L;\,
\Omega^2_{i}\Theta^{i} = e^-_L$, где мы ввели вспомогательное поле  $\Omega \in
SU(2)$. То же поле $\Omega$ примененное к другим левым дублетам дает партнеров
по четности оставшихся правых синглетов. Физические величины не зависят от
$\Omega$ и это поле всегда может быть выбрано единичным. Этот выбор
соответствует унитарной калибровке. Массовый член дает амплитуду перехода между
правыми синглетами и их партнерами по четности. В то же время динамические
фермионные члены содержат смешивание. Требование того, что $\Omega$ - одна и та
же для всех левых дублетов необходимо для того, чтобы реализовалось правильное
нарушение Электрослабой симметрии. Следует отметить, что $SU(4)$ модель Фари -
Сасскинда, в которой РТЦ взаимодействия рассматриваются как возмущения,
выравнивает вакуум правильным образом \cite{Align}. Также отметим, что поле
$\Omega$ - не динамическое, то есть нет интегрирования по $\Omega$ в
функциональном интеграле. Фиксация  $\Omega$ означает, что выбран один из
эквивалентных вакуумов при спонтанном нарушении Электрослабой симметрии.

Дополнительные $SU(N)$ ($N>4$) присутствующие в (\ref{G}) могут наблюдаться при
энергиях выше шкалы ТЦ. Ниже мы называем $SU(N)$ подгруппы для $N>4$ группами
гиперцвета (ГЦ). Кроме того, вся последовательность (\ref{G}) называется нами
башней гиперцвета.

\subsection{Предлагаемая модель}

Модель содержит  $U(1)$ калибровочные поля и калибровочные $SU(N)$ поля с любым
 $N$. Калибровочная группа модели - (\ref{G}). Присутствуют фермионы, относящиеся к фундаментальным
 представлениям подгрупп $SU(N)$ группы $G$. Таким образом, возможные фермионы - это правые
 $\Psi_{A, Y}^{\alpha i_{k_N} ... i_{k_3} i_{k_2}}$ и левые
 $\Theta_{\dot{\beta} A, Y}^{ i_{k_N} ... i_{k_3} i_{k_2}}$, где
$\alpha$ и $\dot{\beta}$ - спинорные индексы, $A$ нумерует поколения, а индекс
 $i_k$ относится к подгруппе $SU(k)$. Здесь $Y$ - $U(1)$ заряд данного фермиона.
 В частности, присутствуют фермионы $\Psi_{A; Y}$, не имеющие индексов, а единственная группа,
 действующая на $\Psi_{A; Y}$ - это $U(1)$.
 Более того, мы предполагаем, что присутствуют фермионы, на которые $G$ не действует совсем.
 Мы обозначаем их $\Psi_{A;0}$. Все фермионы - двхкомпонентные спиноры. Мы также предполагаем с самого
 начала, что группа $SU(2)$ действует только на левые спиноры.
Для простоты мы опускаем ниже спинорные индексы и индексы, нумерующие
поколения. Таким образом мы имеем следующие фермионы

\begin{eqnarray}
U(1): && \Psi_0, \Psi_{Y_{1}}, \Psi_{Y^{\prime}_{1}},...;\nonumber\\
U(1), SU(2): && \Theta^{i_2}_{Y_2}, \Theta^{i_2}_{Y^{\prime}_2}, ...; \nonumber\\
U(1), SU(3): && \Psi^{i_3}_{Y_3}; \Psi^{i_3}_{Y^{\prime}_3}, ...; \nonumber\\
U(1), SU(2), SU(3): && \Theta^{i_3 i_2}_{Y_{32}},\Theta^{i_3 i_2}_{Y^{\prime}_{32}},... ;\nonumber\\
U(1), SU(4): && \Psi^{i_4}_{Y_4},\Psi^{i_4}_{Y^{\prime}_4},...;\nonumber\\
U(1), SU(2), SU(4): && \Theta^{i_4 i_2}_{Y_{42}},\Theta^{i_4 i_2}_{Y^{\prime}_{42}},...;\nonumber\\
U(1), SU(3), SU(4): && \Psi^{i_4 i_3}_{Y_{43}}, \Psi^{i_4 i_3}_{Y^{\prime}_{43}},...;\nonumber\\
U(1), SU(2),SU(3),SU(4): &&  \Theta^{i_4 i_3 i_2}_{Y_{432}},\Theta^{i_4 i_3
i_2}_{Y^{\prime}_{432}},...; \nonumber\\
&&...\label{F}
\end{eqnarray}
Здесь в каждой строке перечисляются подгрупы $G$, действующие на фермионы этой
строки. В каждой строке значения $U(1)$ заряда обозначены $Y, Y^{\prime}$, и
т.д.

Рассмотрим первую строку. Здесь для того, чтобы воспроизвести СМ мы
ограничиваемся значениями $Y$, равными $0$  и $-2$. Далее, вторая строка
содержит единственный элемент с $Y = -1$. Третья строка содержит два элемента с
$Y = \frac{4}{3}$ и $Y = -\frac{2}{3}$. В четвертой строке есть единственный
элемент с $Y = \frac{1}{3}$. Эта строка завершает СМ и мы переходим к ее
Ультрафиолетовому дополнению.

Следующие 4 строки в (\ref{F}) выбираются так, чтобы представить $SU(4)$ модель
Фари - Сасскинда \cite{FS}. Выше были получены гиперзаряды такие, что
отсутствует киральная инвариантность, а дополнительная симметрия сохраняется. В
результате в $5$ - й строке есть два элемента с $Y_4 = \frac{1}{2}- 6K + 1$ и
$Y^\prime_4 = \frac{1}{2}- 6K -1$ (где $K$ - произвольное целое число). В $6$ -
й строке присутствует один элемент с $Y_{42} = \frac{1}{2}- 6K$, где $K$ - то
же, что и выше. В $7$ - й строке есть два элемента с $Y_{43} =
-\frac{\frac{1}{2}- 6K}{3} + 1$ и $Y^\prime_{43} = -\frac{\frac{1}{2}- 6K}{3}
-1$. В $8$ - й строке есть один элемент с $Y_{432} = -\frac{\frac{1}{2}-
6K}{3}$. Снова, в этих двух строках $K$ - то же, что и выше.

Преобразование четности $\cal P$ действует на фермионы следующим образом. Если
бы присутствовали только два элемента  $\chi^{\alpha}$ и $\eta_{\dot{\alpha}}$
то было бы ${\cal P}\chi^{\alpha}(t,\bar{r}) = i
\eta_{\dot{\alpha}}(t,-\bar{r}); {\cal P} \eta_{\dot{\alpha}}(t,\bar{r})= i
\chi^{\alpha}(t,-\bar{r})$. В нашем случае мы потребуем, чтобы для любой
последовательности $SU(N)$ ($N
> 2$) индексов существовали два правых спинора и один $SU(2)$ дублет.
Четность связывает каждый из правых спиноров с компонентой $SU(2)$ дублета.
Тогда
\begin{eqnarray}
&& {\cal P}\Psi_0(t,\bar{r}) = i \Omega^1_{i_2}(t,-\bar{r})
\Theta^{i_2}_{-1}(t,-\bar{r}); {\cal P}\Psi_{-2} = i
\Omega^2_{i_2}\Theta^{i_2}_{-1};\nonumber\\
&& {\cal P}\Psi^{i_3}_{\frac{4}{3}} = i \Omega^1_{i_2} \Theta^{i_3
i_2}_{\frac{1}{3}}; {\cal P}\Psi^{i_3}_{-\frac{2}{3}} = i
\Omega^2_{i_2}\Theta^{i_3 i_2}_{\frac{1}{3}};\nonumber\\
&& {\cal P}\Psi^{i_4}_{Y_{4}} = i \Omega^1_{i_2} \Theta^{i_4 i_2}_{Y_{42}};
{\cal P}\Psi^{i_4}_{Y^\prime_{4}} = i
\Omega^2_{i_2}\Theta^{i_4 i_2}_{Y_{42}};\nonumber\\
&& {\cal P}\Psi^{i_4 i_3}_{Y_{43}} = i \Omega^1_{i_2} \Theta^{i_4
i_2}_{Y_{432}}; {\cal P}\Psi^{i_4 i_3}_{Y^\prime_{43}} = i
\Omega^2_{i_2}\Theta^{i_4 i_3 i_2}_{Y_{432}};\nonumber\\
&& ... \label{P}
\end{eqnarray}

Здесь $\Omega$  - это вспомогательное $SU(2)$ поле, отмеченное в Введении.
Физическое значение поля $\Omega$ - то, что оно выбирает партнера по четности
для каждого правого спинора.

Соответствие между нашими обозначениями и общеупотребительными - следующее (мы
здесь приводим соответствие для случая $K = 0$ только для первого поколения):
\begin{eqnarray}
 && \Psi_0 = \nu_R; \Psi_{-2} = e^-_R; {\cal P}\Psi_0(t,\bar{r}) = i \nu_L(t,-\bar{r}); {\cal P}\Psi_{-2} = i e^-_L ;\nonumber\\
 && \Psi^{i_3}_{\frac{4}{3}} = u_R; \Psi^{i_3}_{-\frac{2}{3}} = d_R; {\cal P}\Psi^{i_3}_{\frac{4}{3}} = i
 u_L;  {\cal P}\Psi^{i_3}_{-\frac{2}{3}} = i d_L;\nonumber\\
 && \Psi^{i_4}_{\frac{3}{2}} = N_R; \Psi^{i_4}_{-\frac{1}{2}}=E_R; {\cal P}\Psi^{i_4}_{\frac{3}{2}} = i
 N_L;  {\cal P}\Psi^{i_4}_{-\frac{1}{2}} = i E_L ;\nonumber\\
 && \Psi^{i_4 i_3}_{\frac{5}{6}} = U_R; \Psi^{i_4 i_3}_{-\frac{7}{6}}= D_R;
 {\cal P}\Psi^{i_4 i_3}_{\frac{5}{6}} = i U_L; {\cal P}\Psi^{i_4 i_3}_{-\frac{7}{6}} =
 i D_L.\label{i}
\end{eqnarray}

 Следует отметить, что фермионы первого поколения перечисленные здесь не
 диагонализируют массовую матрицу. А диагонализируют ее линейные комбинации с
 учетом смешивания.

Прежде, чем перейти к следующим строкам рассмотрим, как $Z_6$ симметрия СМ
может быть продолжена на калибровочные поля гиперцвета.

\subsection{$\cal Z$ симметрия}

Выше мы предложили способ продолжить $Z_6$ симметрию СМ на модели техницвета.
Здесь мы продолжим эту симметрию на все подгруппы группы G:
\begin{eqnarray}
 U & \rightarrow & U e^{i\pi N}, \nonumber\\
 \theta & \rightarrow & \theta +  \pi N, \nonumber\\
 \Gamma & \rightarrow & \Gamma e^{(2\pi i/3)N},\nonumber\\
 \Pi_4 & \rightarrow & \Pi_4 e^{(2\pi i/4)N},\nonumber\\
 \Pi_5 & \rightarrow & \Pi_5 e^{(2\pi i/5)N}, \nonumber\\
 \Pi_6 & \rightarrow & \Pi_6 e^{(2\pi i/6)N}, \nonumber\\
 ...
\label{symlatWS1}
\end{eqnarray}
Здесь $\Pi_K$ - это $SU(K)$ оператор параллельного переноса. Мы строим нашу
модель так, что операторы параллельного переноса инвариантны относительно
(\ref{symlatWS1}). Результирующая симметрия обозначается $\cal Z$ и входит в
определение (\ref{G}).

\subsection{$SU(N)$ группы с $N>4$}

Проанализируем последовательность (\ref{F}) в форме (\ref{i}). Заметим, что
вторая пара строк - это копия первых двух с дополнительным $SU(3)$ индексом.
Далее, вторая четверка строк - копия первой четверки с дополнительным  $SU(4)$
индексом. Предположим, что этот процесс продолжается бесконечно. Тогда
последовательность фермионов имеет вид:
\begin{eqnarray}
... \nonumber\\
U(1), SU(5): && \Psi^{i_5}_{Y_5}, \Psi^{i_5}_{Y_5^\prime};\nonumber\\
U(1), SU(2), SU(5): && \Theta^{i_5 i_2}_{Y_{52}}; \nonumber\\
U(1), SU(3), SU(5): && \Psi^{i_5 i_3}_{Y_{53}}; \Psi^{i_5 i_3}_{Y_{53}^{\prime}}; \nonumber\\
U(1), SU(2), SU(3), SU(5): && \Theta^{i_5 i_3 i_2}_{Y_{532}};\nonumber\\
U(1), SU(4), SU(5): && \Psi^{i_5 i_4}_{Y_{54}},\Psi^{i_5 i_4}_{Y_{54}^{\prime}};\nonumber\\
U(1), SU(2), SU(4), SU(5): && \Theta^{i_5 i_4 i_2}_{Y_{542}};\nonumber\\
U(1), SU(3), SU(4), SU(5): && \Psi^{i_5 i_4 i_3}_{Y_{543}}, \Psi^{i_5 i_4 i_3}_{Y_{543}^\prime};\nonumber\\
U(1), SU(2),SU(3),SU(4), SU(5): &&  \Theta^{i_5 i_4 i_3 i_2}_{Y_{5432}}; \nonumber\\
&&... \nonumber\\
U(1), ... , SU(K): && \Psi^{i_K ... }_{Y_{K...}}, \Psi^{i_K ...}_{Y_{K ...}^\prime};\nonumber\\
U(1), SU(2), ... , SU(K): &&  \Theta^{i_K ... i_2}_{Y_{K...2}}; \nonumber\\
... \label{SMFS100}
\end{eqnarray}

Ниже мы выводим гиперзаряды для всех фермионов, требуя, что сохраняется
симметрия $\cal Z$ в то время, как киральные аномалии отсутствуют.

Сокрашение аномалий необходимо для того, чтобы модель была хорошо определена.
Что касается требования, что сохраняется $\cal Z$, это - наше предположение.

{\bf Предложение}

 Ниже мы покажем, что указанные требования приводят к следующему выражению для
гиперзарядов
\begin{eqnarray}
&& Y_2  =  -1 \nonumber\\
&&Y_{i_1 i_2 i_3 ... i_{M-1} i_M 2} = -1 + 2(1 - \frac{1}{i_M})  +  2 \sum_{k =
1}^{M-1}[\theta(i_k - i_{k+1} - 1) - \frac{1}{i_k}] + 2 N_{i_1 i_2 i_3 ...
i_{M-1} i_M 2}
\,\nonumber\\
&&Y_{ i j ... l} = Y_{ i j ... l 2} + 1; \, Y^\prime_{ i j ... l }  = Y_{ i j
... l 2} - 1\label{Y0}
\end{eqnarray}
где $\theta(x) = 1 \,  \,(x>0); \, \theta(x) = 0 \,  \,(x\le 0)$. Во второй
строке  $\, M \ge 1$. Для любых $K$ целые числа $N_{i_1 i_2 i_3 ... i_{M-1} i_M
2}$, которые входят в (\ref{Y0}), удовлетворяют уравнению
\begin{equation}
\sum_{K > i > j > ... > l > 2} i j ... l \, N_{K ij...l2} = 0 \label{Y11}
\end{equation}
Здесь сумма - по всем (неупорядоченным) наборам различных целых чисел
$i,j,...,l$ таких, что  $2<i,j,...,l <K$.

{\bf Доказательство}

Прежде всего, если (\ref{symlatWS}) - это симметрия теории, имеет место
следующее рекурсивное соотношение:
\begin{equation}
Y_{K i j ... l 2} = Y_{i j ... l 2} - \frac{2}{K} + 2 M_{K ij...l2}; Y_{K i j
... l} = Y_{K i j ... l 2} + 1; Y^\prime_{K i j ... l }  = Y_{K i j ... l 2} -
1,
\end{equation}
где $M_{K ij...l2}$ - целые числа.

Потребуем, чтобы для любых  $K$
\begin{equation}
\sum_{K > i > j > ... > l > 2} i j ... l \, Y_{K i j ... l 2} = 0,\label{Y}
\end{equation}
Это означает, что киральная аномалия отсутствует даже если последовательность
(\ref{G}) заканчивается фактором  $SU(K)$ для некоторого $K$.

А именно, могут появиться аномалии следующих типов \cite{Weinberg}:
\begin{eqnarray}
&&1) SU(N) - SU(N) - SU(N),\, N > 2 \nonumber\\
&&2) SU(N) - SU(N) - U(1),\, N>2 \nonumber\\
&&3) SU(2) - SU(2) - U(1) \nonumber\\
&&4) U(1) - U(1) - U(1)
\end{eqnarray}

Аномалия первого типа зануляется поскольку число левых фермионов равно числу
правых в то время, как и те и другие находятся в фундаментальном представлении
 $SU(N)$. Аномалии второго типа сокращаются, поскольку $Y_{ i j ... l} =
Y_{ i j ... l 2} + 1; \, Y^\prime_{ i j ... l }  = Y_{ i j ... l 2} - 1$.
Аномалии третьего типа и четвертого типа зануляются если сумма гиперзарядлв по
левым дублетам равна нулю. Это приводит к (\ref{Y}).

Покажем, что для любого  $K$ целые числа $M_{K ij...l2}$ могут быть выбраны
так, что удовлетворяется (\ref{Y}). Пусть $\sum_{K^\prime
> i > j
> ... > l > 2} i j ... l Y_{K^\prime i j ... l 2} = 0$ for $K^\prime < K$ (Это было уже продемонстрировано
для  $K^\prime = 4$.). Тогда
\begin{equation}
\sum_{K > i > j > ... > l > 2} i j ... l \, Y_{K i j ... l 2} =  - 2
\frac{K!}{3!K} + 2\sum_{K > i > j
> ... > l > 2} i j ... l \, M_{K ij...l2} \label{KM}
\end{equation}
Здесь использовано равенство $\sum_{K > i > j > ... > l
> 2} i j ... l = \frac{K!}{3!}$.

Введем следующие обозначения:
\begin{eqnarray}
M_{K ij...l2} = M_{K ij...l2}^\prime + 1,\, {\rm for}\, \, K-1 > i > j > ... >
l > 2;\nonumber\\
M_{K ij...l2} = M_{K ij...l2}^\prime,\, {\rm for}\, \, K-1 = i > j > ... > l >
2
\end{eqnarray}

Тогда
\begin{equation}
-\frac{K!}{3!K} + \sum_{K > i > j > ... > l > 2} i j ... l \, M_{K ij...l2}  =
\sum_{K > i > j > ... > l > 2} i j ... l \, M^\prime_{K ij...l2}
\end{equation}

Соотношения, определяющие фермионные гиперзаряды, могут быть переписаны
следующим образом:
\begin{eqnarray}
&&Y_{K i j ... l} = Y_{K i j ... l 2} + 1; \, Y^\prime_{K i j ... l }  = Y_{K i
j ... l 2} - 1; Y_{K i j ... l 2} = Y_{i j ... l 2} - \frac{2}{K} + 2 + 2
M^\prime_{K ij...l2} \,\nonumber\\ && ( {\rm for}\, \, K-1
> i > j > ... > l > 2, \, {\rm or} \, K = 3);\nonumber\\
&&Y_{K i j ... l 2} = Y_{i j ... l 2} - \frac{2}{K}  + 2 M^\prime_{K ij...l2}
\, ({\rm for}\, \, K-1 = i > j > ... > l > 2)
\end{eqnarray}
Здесь целые числа $M^\prime_{K ij...l2}$ выбраны так, что  $\sum_{K
> i > j > ... > l > 2} i j ... l \, M^\prime_{K ij...l2} = 0$.

Наконец, мы приходим к решению  (\ref{Y}) в форме (\ref{Y0}). В частности,
выбор  $N_{i_1 i_2 i_3 ... i_{M-1} i_M 2}=0$ приводит к  $Y_{i_1 i_2 i_3 ...
i_{M-1} i_M 2} = -1 + 2(1 - \frac{1}{i_M}) + 2 \sum_{k = 1}^{M-1}[\theta(i_k -
i_{k+1} - 1) - \frac{1}{i_k}]$.
 Таким образом, симметрия  (\ref{symlatWS}) фиксирует гиперзаряды с точностью до
 целых чисел  $N_{i_1 i_2 i_3 ... i_{M-1} i_M
2}$ таких, что удовлетворяется  (\ref{Y11}).

Отметим, что если $\cal Z$ симметрия не накладывается, то гиперзаряды
определяются только условием сокращения аномалий. При этом гиперзаряды
получаются соответствующими (\ref{Y0}), $N_{i_1 i_2 i_3 ... i_{M-1} i_M 2}$ -
это уже действительные числа.

\subsection{Свойства техницвета и гиперцвета}

Динамика техницвета связана обычным образом с числом фермионов $N_f$. А именно,
бета - функция в однопетлевом приближении имеет вид $\beta_{SU(K)}(\alpha) = -
\frac{11 K - 2 N_f}{6 \pi} \alpha^2$ где $\alpha = \frac{g^2_{SU(K)}}{4\pi}$.
Если $N_f < \frac{11}{2} K$, однопетлевое вычисление указывает на появление
асимптотической свободы. Двухпетлевое вычисление \cite{Appelquist,Appelquist_}
указывает, что нарушение киральной симметрии происходит при $N_f < N_c \sim K
\frac{100 K^2 - 66}{25 K^2 - 15} \sim 4 K $. Это требуется для появления массы
калибровочного бозона.

У нас есть три поколения технифермионов. Поэтому их число $24
> 4 N_{TC} = 16$. Однако, вносят вклад только те, чьи массы меньше, чем $\Lambda_{TC}$
($\Lambda_{TC}$ - это $SU(4)$ аналог $\Lambda_{QCD}$). Тогда предположим, что
массы третьего поколения технифермионов и, возможно, массы некоторых из
технифермионов второго поколения существенно выше шкалы ТЦ. Мы также
предполагаем, что массы фермионов, несущих индексы высших групп гиперцвета
существенно выше шкалы ТЦ. Поэтому они не оказывают влияния на динамику ТЦ. Как
результат, $SU(4)$ взаимодействия ведут к нарушению киральной симметрии и
возникновению масс $W$ и $Z$ бозонов.

Если число технифермионов приближается к $N_c \sim 4 N_{TC}$, то поведение
техницвета становится близким к конформному \cite{walking}. Таким образом, в
нашем случае (два поколения, и $N_{TC} = 4$) поведение ТЦ может быть почти
конформным.

Что касается высших групп гиперцвета, уже для $SU(5)$ взаимодействий число
гиперфермионов (фермионов несущих $SU(5)$ индекс) равно  $2(1 + 3 + 4 + 12) =
40
> \frac{55}{2} = 27.5$. Предположим, их массы близки друг к другу.
Тогда при $K
> 4$ взаимодействия не асимптотически свободны.
В эффективных зарядах может присутствовать полюс Ландау. Это значит, что наша
модель не имеет непрерывного предела, и должна подобно СМ рассматриваться как
модель с конечным обрезанием (finite cutoff model). При энергиях порядка
ультрафиолетового обрезания должна появиться новая теория, включающая описанную
выше в качестве низкоэнергитического приближения. В принципе, эта шкала может
быть очень высокой, даже порядка массы Планка в зависимости от значения
$g^2_{SU(K)}$ при низких энергиях. Грубо эту шкалу, например, для эффективного
 $SU(5)$ заряда можно оценить как $\Lambda_h = e^{\frac{6
\pi}{(2N_f - 55) \alpha_{SU(5)}(1 \, {\rm Tev})}} \, $ ТэВ. Если, скажем, три
поколения включаются  в рассмотрение, и $\alpha_{SU(5)}(1 \, {\rm Tev}) =
\frac{1}{300}$, то полюс Ландау в $SU(5)$ заряде возникает при $\Lambda_h \sim
10^{13}\, {\rm Tev}\sim 10^{-3} M_{\rm Plank}$. При энергиях много меньше
$\Lambda_h$  $SU(5)$ взаимодействия могут быть учтены пертурбативно также, как
и в КЭД.

\section{Публикации}

Результаты настоящей главы были опубликованы в работах

 "Z(6) symmetry of the Standard Model and technicolor theory", M.A. Zubkov.
Phys.Lett.B674:325-329,2009, [arXiv:0707.0731]

 "A Superstructure over the Farhi-Susskind Technicolor model", M.A. Zubkov.
Mod. Phys. Lett. A25:679-689,2010, [arXiv:0910.4771]

\chapter{Калибровочная теория группы Лоренца как возможный источник динамического нарушения Электрослабой симметрии}

\label{ch12}

В настоящей главе мы рассматриваем возможность нарушения Электрослабой
симметрии и возникновения масс фермионов Стандартной Модели благодаря
существованию массивных мод кручения пространства - времени с массами порядка
ТэВ. Предлагаемая конструкция заменяет группу Техницвета группой Лоренца.
Используемая модель по сути дела является калибровочной теорией группы Лоренца
в пространстве Минковского \cite{Minkowsky}. При этом динамическая квантовая
теория метрического поля, то есть, собственно, гравитация, может существовать,
или не существовать, что не оказывает на наше рассмотрение никакого влияния.
Ниже мы будем предполагать для определенности, что квантовая теория поля
метрики существует и имеет шкалу массы Планка.

Хорошо известно, что квантовая гравитация в формализме первого порядка с
действием Палатини или действием Хольста, ведет к четырехфермионному
взаимодействию между спинорными полями \cite{Rovelli}. Это четырехфермионное
взаимодействие может привести к конденсации фермионов, которое в свою очередь,
рассматривалось в некоторых космологических моделях, например, в качестве
источника темной энергии \cite{Xue,Alexander1,Alexander2}.

Мы рассматриваем фермионы, взаимодействующие неминимальным образом с кручением
пространства - времени \cite{Shapiro} в качестве источника Динамического
Нарушения Электрослабой Симметрии (ДНЭС). При этом предполагается, что массовый
параметр поля кручения близок к $1$ ТэВ. Как уже говорилось выше,
предполагается, что либо квантовой теории метрического поля нет (и классическая
гравитация имеет иной источник), либо ее масштаб - масса Планка.

Ясно, что кручение, связанное с 4 - спинорными полями фермионов Стандартной
Модели, не может привести к требуемому ДНЭС, поскольку оно взаимодействует со
всеми фермионами одинаково. Поэтому если  сконденсированы дополнительные
фермионы (называемые далее технифермионами), то сконденсированы и фермионы
Стандартной Модели (СМ). Однако, если разместить все фермионы Стандартной
Модели в левополяризованных двухкомпонентных спинорах, а технифермионы - в
правополяризованных, и при этом предположить, что действие не сохраняет
четность, то после интегрирования по кручению может возникнуть асимметрия между
правыми и левыми фермионами. В результате правые фермионы могут оказаться
сконденсированы, а левые - нет. Так мы и получим конденсат технифермионов,
оставляя фермионы СМ несконденсированными.

Действие Хольста \cite{Hehl,Holst:1995pc,imir,Rovelli,Khriplovich:2005jh}
нарушает четность. Поэтому, когда Пуанкаре гравитация с действием Хольста
связана с фермионами, она может обеспечить появление нарушающих четность
четырехфермионных взаимодействий
\cite{Freidel:2005sn,Randono:2005up,Mercuri:2006um,Alexandrov}. Мы
рассматриваем действие Хольста в качестве низкоэнергитической части действия
для кручения. При этом также рассматривается неминимальное взаимодействие
фермионов с Пуанкаре гравитацией, которое само нарушает четность
\cite{Alexandrov}. Для того, чтобы обеспечить фермионы СМ массами нами
рассматривается массовый член для спиноров, чья левая компонента содержит поля
СМ, а правая - технифермионы. В рассматриваемом подходе ряд проблем моделей РТЦ
получает разрешение.

\section{Фермионы в пространстве Римана - Картана}

Рассмотрим действие безмассового Дираковского спинора в пространстве Римана -
Картана \cite{Alexandrov}:

\begin{eqnarray}
S_f & = & \frac{i}{2}\int E \{ \bar{\psi} \gamma^{\mu} (\zeta - i\chi\gamma^5)
D_{\mu} \psi - [D_{\mu}\bar{\psi}](\bar{\zeta} - i\bar{\chi}\gamma^5)
\gamma^{\mu}\psi \} d^4 x \label{Sf}
\end{eqnarray}

Здесь  $\zeta=\eta + i\theta$ и $\chi = \rho+i\tau$ - это константы связи, $E =
{\rm det} E^a_{\mu}$, $E^a_{\mu}$ - поле тетрады, $\gamma^{\mu} = E^{\mu}_a
\gamma^a$, ковариантная производная $D_{\mu} =
\partial_{\mu} + \frac{1}{4}(\omega_{\mu}^{ab}+C_{\mu}^{ab})\gamma_{[a}\gamma_{b]}$;
$\gamma_{[a}\gamma_{b]} =
\frac{1}{2}(\gamma_{a}\gamma_{b}-\gamma_{b}\gamma_{a})$. Спиновая связность без
кручнения обозначается $\omega_{\mu} $, в то время, как $C_{\mu}$ -
конторсионный тензор (contorsion). Они связаны с $E^a_{\mu}$, Афинной
связностью $\Gamma^{i}_{jk}$, и кручением $T^a_{.\mu \nu}=
T^{\rho}_{.\mu\nu}E^a_{\rho}$ следующим образом:
\begin{eqnarray}
\nabla_{\nu} E_{\mu}^{a} &=& \partial_{\nu}E^a_{\mu} - \Gamma^{\rho}_{\mu
\nu}E^a_{\rho} + \omega^{a}_{ . b\nu}E^b_{\mu}+ C^{a}_{ . b\nu}E^b_{\mu}=0\nonumber\\
\tilde{D}_{[\nu} E_{\mu]}^{a} &=& \partial_{[\nu}E^a_{\mu]} +
\omega^{a}_{.b[\nu }E^b_{\mu]}=0\nonumber\\
T^a_{.\mu \nu} & = & {D}_{[\nu} E_{\mu]}^{a} = \partial_{[\nu}E^a_{\mu]} +
\omega^{a}_{.b[\nu }E^b_{\mu]}+ C^{a}_{ . b[\nu}E^b_{\mu]}=C^{a}_{ .
b[\nu}E^b_{\mu]}
\end{eqnarray}

Это приводит к:
\begin{eqnarray}
\Gamma^{\rho}_{\mu \nu}&=& \{^{\rho}_{\mu \nu}\} + C^{\rho}_{.\mu \nu}
\nonumber\\
C^{\rho}_{.\mu \nu} & = & \frac{1}{2}(T^{\rho}_{.\mu \nu}-T^{.\rho}_{\nu
.\mu}+T^{..\rho}_{\mu \nu})\nonumber\\
\{^{\alpha}_{\beta \gamma}\} & = &
\frac{1}{2}g^{\alpha\lambda}(\partial_{\beta}g_{\lambda
\gamma}+\partial_{\gamma}g_{\lambda \beta}-\partial_{\lambda}g_{\beta \gamma})\nonumber\\
\omega_{ab\mu} & = &\frac{1}{2}( c_{abc}-c_{cab}+c_{bca})E^{c}_{\mu}\nonumber\\
C_{ab\mu} & = &\frac{1}{2}( T_{abc}-T_{cab}+T_{bca})E^{c}_{\mu}
\end{eqnarray}
Здесь $c_{abc} = \eta_{ad}E^{\mu}_b E_c^{\nu}\partial_{[\nu}E^d_{\mu]}$;
$T_{abc} = \eta_{ad}E^{\mu}_b E_c^{\nu}T_{.\mu\nu}^d$; $g_{\mu \nu} =
E^a_{\mu}E^b_{\nu}\eta_{ab}$; $\Gamma^{\rho}_{\mu \nu}-\Gamma^{\rho}_{\nu \mu}
= T^{\rho}_{.\mu \nu}$; индексы поднимаются и опускаются с помощью $g$ и $E$
как обычно.

(\ref{Sf}) переписывается в виде:

\begin{eqnarray}
S_f & = & \frac{1}{2}\int E\{i\bar{\psi} \gamma^{\mu}(\zeta - i\chi\gamma^5)
\tilde{D}_{\mu} \psi - i[\tilde{D}_{\mu}\bar{\psi}](\bar{\zeta} -
i\bar{\chi}\gamma^5) \gamma^{\mu}\psi\nonumber\\&& + \frac{i}{4}
C_{abc}\bar{\psi}[\{ \gamma^{[a}\gamma^{b]},\gamma^c\}(\eta+\tau\gamma^5) - i[
\gamma^{[a}\gamma^{b]},\gamma^c](\theta-\rho\gamma^5)] \psi\} d^4 x \label{Sf1}
\end{eqnarray}
Здесь $\tilde{D}$ - ковариантная производная ОТО. Получаем:
\begin{eqnarray}
S_f & = & \frac{1}{2}\int E\{i\bar{\psi} \gamma^{\mu}({\zeta} -
i{\chi}\gamma^5) \tilde{D}_{\mu} \psi -
i[\tilde{D}_{\mu}\bar{\psi}](\bar{\zeta} - i\bar{\chi}\gamma^5)
\gamma^{\mu}\psi\nonumber\\&& - \frac{1}{4}
C_{abc}\bar{\psi}[-2\epsilon^{abcd}\gamma^5 \gamma_d(\eta+\tau\gamma^5) +4
\eta^{c[a} \gamma^{b]}(\theta-\rho\gamma^5)] \psi\} d^4 x \label{Sf2}
\end{eqnarray}

Введем неприводимые компоненты кручения:
\begin{eqnarray}
S^i& =& \epsilon^{jkli}T_{jkl}\nonumber\\
T_i& =& T^j_{.ij}\nonumber\\
T_{ijk}& =& \frac{1}{3}(T_j \eta_{ik}-T_k\eta_{ij}) -
\frac{1}{6}\epsilon_{ijkl}S^l + q_{ijk}
\end{eqnarray}

В терминах $S$ и $T$ (\ref{Sf2}) может быть представлено в виде:
\begin{eqnarray}
S_f & = & \frac{1}{2}\int E\{i\bar{\psi} \gamma^{\mu}(\zeta - i\chi\gamma^5)
\tilde{D}_{\mu} \psi - i[\tilde{D}_{\mu}\bar{\psi}](\bar{\zeta} -
i\bar{\chi}\gamma^5) \gamma^{\mu}\psi\nonumber\\&& +
\frac{1}{4}\bar{\psi}[\gamma^5 \gamma_d (\eta S^d - 4\rho T^d) -(\tau
S^b+4\theta T^b) \gamma_{b}] \psi\} d^4 x \label{Sf22}
\end{eqnarray}

\section{Действие Хольста и Дираковские фермионы}

Рассмотрим действие Хольста:
\begin{equation}
S_T = -M_T^2 \int E E_a^{\mu}E_b^{\nu}G^{ab}_{\mu\nu} d^4x -
\frac{M_T^2}{\gamma}\int E E_a^{\mu}E_b^{\nu} {^*}G^{ab}_{\mu\nu}d^4x\label{ST}
\end{equation}

Здесь $G^{ab}_{\mu\nu} = [D_{\mu},D_{\nu}]$ - это $SO(3,1)$ кривизна
пространства Римана - Картана, а  ${^*}G^{ab}_{\mu\nu} = \frac{1}{2}
\epsilon^{ab}_{..cd}G^{cd}_{\mu\nu}$ - ее дуальный тензор. В отсутствии
кручения второй член - это интеграл от полной производной и, поэтому, исчезает
из классического рассмотрения. Однако, в присутствии кручения, он дает
нетривиальный вклад в фермионные взаимодействия.

Представим действия Хольста в терминах кручения и Римановой кривизны
 \cite{Mercuri:2006um}:

\begin{eqnarray}
S_T = M_T^2 \int E\{- R + \frac{2}{3}T^2 - \frac{1}{24}S^2  +
\frac{1}{3\gamma}TS \}d^4x + \tilde{S}
\end{eqnarray}
Здесь $R$ - Риманова скалярная кривизна, $\tilde{S}$ содержит члены, зависящие
от $q$, а также так называемый инвариант Ние - Яна (Nieh - Yan invariant).

Предположим на некоторое время, что в полном действии нет других членов,
зависящих от кручения.  Тогда интегрирование по кручению может быть выполнено.
Результат интегрирования:

\begin{eqnarray}
S_f& = & \frac{1}{2}\int E\{i\bar{\psi} \gamma^{\mu}(\zeta - i\chi\gamma^5)
\tilde{D}_{\mu} \psi - i[\tilde{D}_{\mu}\bar{\psi}](\bar{\zeta} -
i\bar{\chi}\gamma^5) \gamma^{\mu}\psi\}d^4x\nonumber\\&& -
\frac{3\gamma^2}{(1+\gamma^2)32M_T^2} \int E
\{V^2[\theta^2-\tau^2+\frac{2\theta\tau}{\gamma}] +
A^2[\rho^2-\eta^2-\frac{2\eta\rho}{\gamma}]\nonumber\\&&+2AV[\theta\rho+\tau\eta+\frac{\rho\tau-\theta\eta}{\gamma}]
\} d^4x + S_{eff}[E]\label{F42}
\end{eqnarray}
Здесь мы определили:
\begin{eqnarray}
V_{\mu} & = & \bar{\psi} \gamma_{\mu}  \psi \nonumber\\A_{\mu} & = &
\bar{\psi}\gamma^5 \gamma_{\mu}  \psi
\end{eqnarray}
$S_{eff}$ - эффективное действие, зависящее только от поля метрики. Оно
возникает из функционального интеграла по кручению. Если отсутствуют члены,
содержащие производные кручения, $S_{eff}$ сводится к перенормировке
космологической постоянной. По этой причине ниже мы его опускаем.
Четырехфермионный член в (\ref{F4}) отличается от полученного в
\cite{Alexandrov} общим знаком и знаком параметра Иммирци  $\gamma$ из - за
соответствующей разницы в определении действия (\ref{ST}).

Введем правые и левые токи:
\begin{eqnarray}
J_+^{\mu} & = & \bar{\psi}_+ \gamma^{\mu}  \psi_+ \nonumber\\J_-^{\mu} & = &
\bar{\psi}_- \gamma^{\mu}  \psi_-
\end{eqnarray}
Здесь $\psi_+$ - правая компонента $\psi$, а $\psi_-$ - левая компонента. В
дальнейшем рассмотрении мы имеем дело с $E^a_{\mu}=\delta^a_{\mu}$, и
$\omega_{\mu}=0$. Это соответствует тому, что динамическая теория этих двух
полей (то есть, собственно, квантовая гравитация) либо имеет шкалу массы Планка
и потому приводит к тривиальному вакууму на рассматриваемых нами энергиях ниже
$1$ ТэВ, либо такая теория отсутствует вовсе, и $E^a_{\mu}=\delta^a_{\mu}$,
$\omega_{\mu}=0$, - по построению. Оба случая по сути соответствуют реализации
при данных энергиях рассматриваемой теории как калибровочной теории группы
$SO(3,1)$ в пространстве Минковского \cite{Minkowsky}.

Мы перескалируем левые и правые компоненты $\psi$:
\begin{equation}
\psi_-\rightarrow \frac{1}{\sqrt{\eta+\tau}}\psi_-\, ; \,\psi_+\rightarrow
\frac{1}{\sqrt{\eta-\tau}}\psi_+
\end{equation}

Теперь (\ref{F42}) может быть переписано как:
\begin{eqnarray}
S_f& = & \frac{1}{2}\int \{i\bar{\psi} \gamma^{\mu} \partial_{\mu} \psi -
i[\partial_{\mu}\bar{\psi}] \gamma^{\mu}\psi\} d^4x\nonumber\\&& -
\frac{3\gamma^2}{(1+\gamma^2)32M_T^2} \int \{J_+^2 [-1 +
\frac{(\theta+\rho)^2}{(\eta-\tau)^2}
-\frac{2(\theta+\rho)}{(\eta-\tau)\gamma}] \nonumber\\&&+ J_-^2 [-1 +
\frac{(\theta-\rho)^2}{(\eta+\tau)^2}
+\frac{2(\theta-\rho)}{(\eta+\tau)\gamma}]\nonumber\\&&+ 2J_+J_-[1 +
\frac{\theta^2-\rho^2}{\eta^2-\tau^2}
+\frac{\theta+\rho}{(\eta-\tau)\gamma}-\frac{\theta-\rho}{(\eta+\tau)\gamma}]
\} d^4x \label{F4}
\end{eqnarray}

  Указанная $SO(3,1)$ калибровочная теория нам неизвестна. Ее действие может содержать члены квадратичные
  по кривизне и, возможно, взаимодействие с иными неизвестными нам полями. Ниже мы будем предполагать, что
  такая теория может быть хорошо определена, то есть является перенормируемой. Соответствующую ей
  шкалу энергий обозначим $\Lambda_{\chi}$.
  Мы полагаем $\Lambda_{\chi}\sim$ 1 ТэВ или $\Lambda_{\chi}\sim$ 10 ТэВ. Эффективный заряд, входящий в
  действие перед производными кручения зависит от отношения  $\epsilon/\Lambda_{\chi}$, где
   $\epsilon$ - характерная энергия рассматриваемого процесса.
Массовый параметр $M_T$ из  (\ref{ST}) также предполагается в районе ТэВ.

  На шкале $\Lambda_{\chi}$ в дополнение к (\ref{ST}) часть действия, соответствующая
  кручению, содержит члены, зависящие от производных от $T$ и $S$.
Это означает, в частности, что могут появиться следующие члены в действии:
\begin{eqnarray}
S_T &=&  \beta_1 \int E G^{abcd}G_{abcd}d^4x+\beta_2 \int
EG^{abcd}G_{cdab}d^4x\nonumber\\&& +\beta_3 \int EG^{ab}G_{ab}d^4x+\beta_4 \int
EG^{ab}G_{ba}d^4x\nonumber\\&& +\beta_5 \int EG^2d^4x+\beta_6 \int
EA^{abcd}A_{abcd}d^4x\label{ST2}
\end{eqnarray}
с константами связи $\beta_{1,2,3,4,5,6}$. Здесь
$G^{abcd}=E^c_{\mu}E^d_{\nu}G^{ab}_{\mu\nu}$, $G^{ac}=G^{abc}_{...b}$, $G =
G^a_a$, $A_{abcd} =
\frac{1}{6}(G_{abcd}+G_{acdb}+G_{adbc}+G_{bcad}+G_{bdca}+G_{cdab})$. В
действительности, действие (\ref{ST2}) - наиболее общий вид действия
квадратичного по кривизне и не содержащего нарушения четности.

Теперь интегрирование по кручению приводит к (\ref{F4}) в низкоэнергитическом
приближении $\epsilon << \Lambda_{\chi}$. Поэтому полученная модель с
четырехфермионным взаимодействием - только эффективное низкоэнергитическое
приближение, работающее при энергиях, много меньших шкалы $\Lambda_{\chi}$.

\section{Предельные случаи}

Рассмотрим различные предельные случаи (\ref{F4}). Наша цель заключается в том,
чтобы найти возможность существования притягивающего взаимодействия между
правыми фермионами в то время, как взаимодействие между левыми фермионами -
либо отталкивающее, либо пренебрежимо мало. При этом также необходимо, чтобы
взаимодействие между левыми и правыми фермионами было мало по сравнению с
взаимодействием между правыми фермионами. Все это нам необходимо, чтобы
обеспечить конденсацию правых фермионов, благодаря которой Электрослабая
симметрия нарушается спонтанно. При этом мы также получим отсутствие
конденсации левых фермионов, которые используются для размещения фермионов
Стандартной Модели.

\subsection{Действие Палатини}

Этот случай соответствует бесконечному значению параметра Иммирци $\gamma$.
Имеем:
\begin{eqnarray}
S_f& = & \frac{1}{2}\int \{i\bar{\psi} \gamma^{\mu} \partial_{\mu} \psi -
i[\partial_{\mu}\bar{\psi}] \gamma^{\mu}\psi\} d^4x\nonumber\\&& -
\frac{3}{64M_T^2} \int \{J_+^2 [-1 + \frac{(\theta+\rho)^2}{(\eta-\tau)^2} ]
\nonumber\\&&+ J_-^2 [-1 + \frac{(\theta-\rho)^2}{(\eta+\tau)^2}
]\nonumber\\&&+ 2J_+J_-[1 + \frac{\theta^2-\rho^2}{\eta^2-\tau^2} ] \} d^4x
\label{F4E}
\end{eqnarray}

Перекрестные члены исчезают, если $1 + \frac{\theta^2-\rho^2}{\eta^2-\tau^2}
=0$ то есть $|\zeta|=|\chi|$. Отталкивание между $J_-$ и притяжение между $J_+$
возникает при
\begin{eqnarray}
&&|\zeta|=|\chi|\nonumber\\&& |{\rm Re} \zeta + {\rm Im} \chi|<|{\rm Im} \zeta
- {\rm Re} \chi|\nonumber\\&& |{\rm Re} \zeta - {\rm Im} \chi|>|{\rm Im} \zeta
+ {\rm Re} \chi|\label{C1}
\end{eqnarray}

Взаимодействия между  $J_-$ исчезают, если $|{\rm Re} \zeta + {\rm Im}
\chi|=|{\rm Im} \zeta - {\rm Re} \chi|$.

\subsection{Второй член в действии Хольста}
Рассмотрим ситуацию, когда действие Палатини отсутствует, и присутствует только
второй член в действии Хольста:

\begin{eqnarray}
S_f& = & \frac{1}{2}\int \{i\bar{\psi} \gamma^{\mu} \partial_{\mu} \psi -
i[\partial_{\mu}\bar{\psi}] \gamma^{\mu}\psi\} d^4x\nonumber\\&& -
\frac{3\gamma}{32M_T^2} \int \{J_+^2 [-\frac{2(\theta+\rho)}{(\eta-\tau)}] +
J_-^2 [\frac{2(\theta-\rho)}{(\eta+\tau)}]\nonumber\\&&+
2J_+J_-[\frac{\theta+\rho}{(\eta-\tau)}-\frac{\theta-\rho}{(\eta+\tau)}] \}
d^4x \label{F4H}
\end{eqnarray}

Для того, чтобы исчез перекрестный член необходимо, чтобы
\begin{equation}
\gamma\frac{\theta+\rho}{(\eta-\tau)}=\gamma\frac{\theta-\rho}{(\eta+\tau)}=\alpha
\label{C2}
\end{equation}
где $\alpha$ - новая константа связи.

 Имеем тогда:
\begin{eqnarray}
S_f& = & \frac{1}{2}\int \{i\bar{\psi} \gamma^{\mu} \partial_{\mu} \psi -
i[\partial_{\mu}\bar{\psi}] \gamma^{\mu}\psi\} d^4x\nonumber\\&& -
\frac{3\alpha}{16M_T^2} \int \{J_-^2 - J_+^2
 \}
d^4x \label{F4H}
\end{eqnarray}

Отталкивание между $J_-$ и притяжение между $J_+$ возникает при
\begin{eqnarray}
\gamma\frac{\theta+\rho}{(\eta-\tau)}=\gamma\frac{\theta-\rho}{(\eta+\tau)}=\alpha>0
\end{eqnarray}

\subsection{Общий случай}

В общем случае для того, чтобы получить притяжение между технифермионами и
отталкивание между фермионами СМ нам необходимо:
\begin{eqnarray}
&&\frac{\gamma^2}{(1+\gamma^2)2}\{1 - \frac{(\theta+\rho)^2}{(\eta-\tau)^2}
+\frac{2(\theta+\rho)}{(\eta-\tau)\gamma}\}=\alpha_+>0\nonumber\\&&\frac{\gamma^2}{(1+\gamma^2)2}
\{-1 + \frac{(\theta-\rho)^2}{(\eta+\tau)^2}
+\frac{2(\theta-\rho)}{(\eta+\tau)\gamma}\}=\alpha_->0\label{C3}
\end{eqnarray}

Для того, чтобы исключить перекрестный член $J_+J_-$ нам необходимо
\begin{equation}
|\frac{\gamma^2}{(1+\gamma^2)2}\{1 + \frac{\theta^2-\rho^2}{\eta^2-\tau^2}
+\frac{\theta+\rho}{(\eta-\tau)\gamma}-\frac{\theta-\rho}{(\eta+\tau)\gamma}\}|=|\alpha_{+-}|<<
\alpha_+\label{req}
\end{equation}

 Рассмотрим следующую область констант связи:
\begin{eqnarray}
&&|\eta-\tau|<<1\nonumber\\&&|\eta+\tau|\sim|\theta-\rho|\sim|\theta+\rho|\sim
1\nonumber\\&&\gamma\sim 1
\end{eqnarray}
В этой области мы можем пренебречь членами с $J_-^2$ и $J_-J_+$, и действие
принимает вид:
\begin{eqnarray}
S_f& = & \frac{1}{2}\int \{i\bar{\psi} \gamma^{\mu} \partial_{\mu} \psi -
i[\partial_{\mu}\bar{\psi}] \gamma^{\mu}\psi\} d^4x\nonumber\\&& -
\frac{3\gamma^2}{(1+\gamma^2)32M_T^2}\frac{(\theta+\rho)^2}{(\eta-\tau)^2} \int
J_+^2
 d^4x \label{F41}
\end{eqnarray}

Это отталкивание между правыми фермионами. Для того, чтобы получить притяжение
следует поменять общий знак в (\ref{ST}). Формально это эквивалентно замене
$M_T \rightarrow i M_T$. Следует отметить, что данная ситуация соответствует
знаку действия Палатини, противоположному общепринятому.

\subsection{Обсуждение}

Если действие (\ref{ST}) присутствует с конечным $\gamma$, в то время, как
фермионное действие содержит ненулевые затравочные константы $\zeta$ и $\chi$,
то все эффективные заряды $\theta,\eta,\rho,\tau$ получают вклады от петлевых
поправок благодаря динамическому кручению. Разумеется, это может разрушить
условия (\ref{C1}), (\ref{C2}), (\ref{req}), (\ref{C3}) и необходима будет
некоторая тонкая настройка для того, чтобы удержать точное (или, почти точное)
требование (\ref{req}).

Интересен в этой связи случай, когда действие Палатини и константа $\chi$
отсутствуют. Тогда мы имеем единственную безразмерную константу $\alpha =
\alpha_+ = \alpha_-= \frac{\theta\gamma}{\eta}$. Необходимо наложить условие на
члены в действии, зависящие от производных кручения, что они не приведут
эффективно к появлению действия Палатини и ненулевой эффективной константы
$\chi$. Если это окажется возможным, мы приходим к теории, в которой
реализуется требуемая область констант связи (при $\alpha
> 0$).

\section{Два Дираковских спинора}

Ниже мы считаем, что трансляционная связность $E^a_{\mu}$ равна
$\delta^a_{\mu}$, и обычные символы кристоффеля зануляются. Рассмотрим два
спинора $\psi$ и $\phi$.  Фермионное действие имеет вид:
\begin{eqnarray}
S_f & = & \frac{1}{2}\int \{i\bar{\psi} \gamma^{\mu} ({\zeta} -
i{\chi}\gamma^5)D_{\mu} \psi - i[D_{\mu}\bar{\psi}](\bar{\zeta} -
i\bar{\chi}\gamma^5) \gamma^{\mu}\psi \} d^4 x \nonumber\\&& +\frac{1}{2}\int
\{i\bar{\phi^c} \gamma^{\mu} ({\zeta} - i{\chi}\gamma^5)D_{\mu} \phi^c -
i[D_{\mu}\bar{\phi^c}](\bar{\zeta} - i\bar{\chi}\gamma^5) \gamma^{\mu}\phi^c \}
d^4 x \label{Sf2}
\end{eqnarray}

Здесь $\phi^c = i \gamma^2  \left(\begin{array}{c}\phi_-\\
\phi_+\end{array}\right)^*=\left(\begin{array}{c}i\sigma^2 \phi^*_+\\
-i\sigma^2 \phi^*_-\end{array}\right)$. Ниже используется следующее
представление $\gamma$ матриц: $\gamma^{\mu} = \left(\begin{array}{cc}0&\sigma^{\mu}\\
\bar{\sigma}^{\mu}&0\end{array}\right)$, where $\bar{\sigma}^0 = \sigma^0 = 1;
\bar{\sigma}^i = -\sigma^i \, (i=1,2,3)$; $\gamma^5 = -i\gamma^0\gamma^1\gamma^2\gamma^3=\left(\begin{array}{cc}1&0\\
0&-1\end{array}\right)$.

В соответствии со сказанным выше интегрирование по кручению после
соответствующего перескалирования фермионных полей приводит к
\begin{eqnarray}
&&S_f  =  \int \{i\psi_+^+ {\sigma}^{\mu} \partial_{\mu} \psi_+  +i\psi_-^+
\bar{\sigma}^{\mu}
\partial_{\mu} \psi_-
+ i\phi_+^+ {\sigma}^{\mu}
\partial_{\mu} \phi_+  +
i\phi_-^+ \bar{\sigma}^{\mu} \partial_{\mu} \phi_-  \nonumber\\&&
+\frac{3\alpha_+}{16M_T^2}({\psi}^+_+{\sigma}^i \psi_+-{\phi}^+_-\bar{\sigma}^i
\phi_-)^2-\frac{3\alpha_-}{16M_T^2}({\phi}^+_+{\sigma}^i
\phi_+-{\psi}^+_-\bar{\sigma}^i \psi_-)^2
 \} d^4 x
\label{Sf_2}
\end{eqnarray}
(Здесь мы предполагаем, что константы связи выбраны так, чтобы отсутствовал
перекрестный член с взаимодействием между правыми и левыми токами.)

Образуем новые спиноры $\psi_t = \left(\begin{array}{c}\phi_-\\
\psi_+\end{array}\right)$ и $\psi_s = \gamma^5\left(\begin{array}{c}\psi_-\\
\phi_+\end{array}\right)$. Тогда приходим к следующему выражению для
эффективного действия:

\begin{eqnarray}
S_f & = & \int \{i\bar{\psi}_s \gamma^{\mu} \partial_{\mu} \psi_s -
\frac{3\alpha_-}{16M_T^2}(\bar{\psi}_s\gamma^i \gamma^5
\psi_s)(\bar{\psi}_s\gamma_i \gamma^5  \psi_s)\} d^4 x \nonumber\\&&+ \int
\{i\bar{\psi}_t \gamma^{\mu}
\partial_{\mu} \psi_t  +
\frac{3\alpha_+}{16M_T^2}(\bar{\psi}_t\gamma^i \gamma^5
\psi_t)(\bar{\psi}_t\gamma_i \gamma^5  \psi_t)\} d^4 x \label{Sf22}
\end{eqnarray}

Как и выше, мы полагаем $\alpha_-,\alpha_+>0, \alpha_{+-} = 0$. Тогда возникает
притяжение между $\psi_t$ и отталкивание между $\psi_s$. Между собой $\psi_t$ и
 $\psi_s$ не взаимодействуют. Это открывает возможность того, что $\psi_t$ сконденсировано, а
 $\psi_s$ - нет.

\section{Нарушение Электрослабой симметрии}

Разместим все фермионы Стандартной Модели в левых частях спиноров.
Соответственно, технифермионы размещаются в правых частях. Эффективное
низкоэнергитическое действие имеет вид:
\begin{eqnarray}
&&S_f  =  \int \{i\bar{\psi}^a_s \gamma^{\mu} D_{\mu} \psi^a_s -
\frac{3\alpha_-}{16M_T^2}(\bar{\psi}^a_s\gamma^i \gamma^5
\psi^a_s)(\bar{\psi}^b_s\gamma_i \gamma^5  \psi^b_s)\} d^4 x \nonumber\\&&
+\int \{i\bar{\psi}^a_t \gamma^{\mu} D_{\mu} \psi^a_t  +
\frac{3\alpha_+}{16M_T^2}(\bar{\psi}^a_t\gamma^i \gamma^5
\psi^a_t)(\bar{\psi}^b_t\gamma_i \gamma^5  \psi^b_t)\} d^4 x\label{Sf22s}
\end{eqnarray}
Здесь индексы  $a,b$ нумеруют дираковские спиноры, а производная $D$ содержит
все поля СМ. Применим преобразование Фирца к четырехфермионному члену в
(\ref{Sf22s}):
\begin{eqnarray}
S_{4}  &=&  \int \{ -\frac{3\alpha_-}{16M_T^2}(\bar{\psi}^a_s\gamma^i \gamma^5
\psi^a_s)(\bar{\psi}^b_s\gamma_i \gamma^5  \psi^b_s)\} d^4 x \nonumber\\&&
+\int \{ \frac{3\alpha_+}{16M_T^2}(\bar{\psi}^a_t\gamma^i \gamma^5
\psi^a_t)(\bar{\psi}^b_t\gamma_i \gamma^5  \psi^b_t)\} d^4 x\nonumber\\
&=& \frac{3\alpha_+}{16M_T^2}\int\{4
(\bar{\psi}^a_{t,L}\psi^b_{t,R})(\bar{\psi}^b_{t,R} \psi^a_{t,L})\nonumber\\&&
+[(\bar{\psi}^a_{t,L}\gamma_i\psi^b_{t,L})(\bar{\psi}^b_{t,L}\gamma^i
\psi^a_{t,L})+(L\leftarrow \rightarrow R)]\} d^4 x\nonumber\\&&
-\frac{3\alpha_-}{16M_T^2}\int\{4(\bar{\psi}^a_{s,L}\psi^b_{s,R})(\bar{\psi}^b_{s,R}
\psi^a_{s,L})\nonumber\\&&
+[(\bar{\psi}^a_{s,L}\gamma_i\psi^b_{s,L})(\bar{\psi}^b_{s,L}\gamma^i
\psi^a_{s,L})+(L\leftarrow \rightarrow R)]\} d^4 x
\end{eqnarray}
В этом виде действие имеет форму аналогичную расширенной модели Намбу - Йона -
Лазинио (РНЙЛ) для $\psi_t$ (см (4),  (5),  (6) в \cite{ENJL}) (впрочем, с
отрицательным $G_V$). Также мы имеем отталкивание между $\psi_s$.

 Для последующего рассмотрения мы обозначаем ${\cal N} = N_t = 24$ число технифермионов;
  $G_S = \frac{3\alpha_+
N_t\Lambda^2_{\chi}}{16M_T^2\pi^2}$; $G_V =-\frac{1}{4}G_S$. Здесь
$\Lambda_{\chi}$ - ультрафиолетовое обрезание, которое в данном случае является
физическим параметром модели РНЙЛ. Его значение зависит от свойств модели,
описывающей появление четырехфермионных взаимодействий. В нашем случае
$\Lambda_{\chi}$ должно вычисляться в модели упомянутой выше $SO(3,1)$
калибровочной теории. Соответственно, эта величина по порядку величины
совпадает с шкалой этой калибровочной теории, что и ведет к использованному
нами тождеству обозначений шкалы $SO(3,1)$ модели и обрезания в РНЙЛ модели. Мы
также обозначаем $g_s =
\frac{4\pi^2G_S}{N_t\Lambda_{\chi}^2}=\frac{3\alpha_+}{4M_T^2}$.

Далее, вводятся вспомогательные поля $H$, $L_i$, и $R_i$, и новое действие для
$\psi_t$ имеет вид:
\begin{eqnarray}
S_{4,t}  &=& \int\{ -(\bar{\psi}^a_{t, L}H^+_{ab} \psi^b_{R} + (h.c.)) -
\frac{4 M_T^2}{3\alpha_+} \, H_{ab}^+H_{ab}\}d^4x \nonumber\\&&
+\int\{(\bar{\psi}^a_{t,L}\gamma^i L^{ab}_i\psi^b_{t,L})
-\frac{4M_T^2}{3\alpha_+} {\rm Tr}\,L^i L_{i} +(L\leftarrow \rightarrow R)\}
d^4 x\label{eff}
\end{eqnarray}

Интегрируя по фермионам, мы приходим к эффективному действию для указанных
вспомогательных полей (и источников для фермионных билинейных комбинаций).
Результирующее эффективное действие имеет минимум при $H = m_t {\bf 1}$, где
$m_t$ играет роль массы технифермионов (равной для всех технифермионов).

Применим следующую регуляризацию:
\begin{equation}
\frac{1}{p^2+m^2} \rightarrow \int_{\frac{1}{\Lambda_{\chi}^2}}^{\infty} d\tau
e^{-\tau (p^2+m^2)}
\end{equation}

С этой регуляризацией выражение для конденсата $\psi_t$ равно (после Виковского
поворота):
\begin{eqnarray}
<\bar{\psi}_t\psi_t> &=& N_t \int\frac{d^4p}{(2\pi)^4}\frac{1}{p\gamma + m}=
-N_t m_t\int\frac{d^4p}{(2\pi)^4}\frac{1}{p^2 + m^2_t}\nonumber\\&=&-N_t
m_t\int_{\frac{1}{\Lambda_{\chi}^2}}^{\infty} d\tau
\int\frac{d^4p}{(2\pi)^4}e^{-\tau(p^2 + m^2_t)}\nonumber\\&=&
 -\frac{N_t}{16\pi^2}4m_t^3
\Gamma(-1,\frac{m_t^2}{\Lambda_{\chi}^2})\label{GM}
\end{eqnarray}

Здесь $\Gamma(n,x) = \int_x^{\infty}\frac{dz}{z}e^{-z}z^{n}$. Приближение
среднего поля дает:
\begin{equation}
m_t = -g_s<\bar{\psi}_t\psi_t>
\end{equation}

То есть
\begin{equation}
m_t = G_S m_t
\{\exp(-\frac{m_t^2}{\Lambda_{\chi}^2})-\frac{m_t^2}{\Lambda_{\chi}^2}
 \Gamma(0,\frac{m_t^2}{\Lambda_{\chi}^2})\}
\end{equation}
Это уравнение не зависит от $G_V$. Очевидно, существует критическое значение
$G_S$: при $G_S > 1$ есть ненулевое решение для $m_t$, а для $G_S < 1$ - нет.
Это означает, что в данном приближении конденсация технифермионов происходит
при

\begin{equation}
M_T < M_T^{\rm critical} = \sqrt{3\alpha_+ N_t}\frac{\Lambda_{\chi}}{4\pi}\sim
\sqrt{\alpha_+} \Lambda_{\chi}\label{cond}
\end{equation}
Например, при  $\Lambda_{\chi}\sim 10$ ТэВ и $\alpha_+\sim 1/100$ мы можем
получить $M_T^{\rm critical} \sim 1 $ ТэВ.

Технипионная константа  $F_T$ в данном приближении равна
\begin{equation}
F_t = \frac{N_t m_t^2}{4\pi^2}
 \Gamma(0,\frac{m_t^2}{\Lambda_{\chi}^2})
\end{equation}
Тггда,
\begin{equation}
F^2_t =
\frac{N_t\Lambda_{\chi}^2}{4\pi^2}e^{-\frac{m^2_t(M_T,\Lambda_{\chi})}{\Lambda_{\chi}^2}}
- \frac{4 M_T^2}{3\alpha_+}
\end{equation}
Чтобы иметь правильные значения масс $W$ и $Z$ - бозонов, нам нужно иметь
$F_T\sim 250$ ГэВ. При $M_T = M_T^{\rm critical}$ имеем $m_t = 0$ и $F_T = 0$.
Когда $M_T$ уменьшается, $m_t$ увеличивается и достигает значения порядка
$\Lambda_{\chi}$ где - то в районе  $M_T = M_T^{\rm critical}/2$. При этом
$F_T\sim \frac{\sqrt{N_t}\Lambda_{\chi}}{4\pi}$. Поскольку  $\Lambda_{\chi}
> 1$ ТэВ, нам нужно иметь $\frac{ M_T^{\rm critical}-M_T }{M_T^{\rm critical}} << 1$.
Обычные аргументы естественности (naturalness) приведут к  $x = \frac{
[M_T^{\rm critical}]^2-M^2_T }{[M_T^{\rm critical}]^2} \sim 0.1$. Меньшие
значения этого отношения означали бы, что необходима тонкая подстройка,
считающаяся неестественной. При малых $x$ имеем: $F_T \sim \sqrt{N_t/2}
\frac{\Lambda_{\chi}}{2\pi}x \sim 0.25$ ТэВ. Поэтому естественность приводит к
тому, что слишком большие значения $\Lambda_{\chi}$  не должны реализоваться и
следует ограничиться   $\Lambda_{\chi}$ в пределах между $1$ ТэВ и $10$ ТэВ.

Отрицательные $G_V$ ведут к появлению члена в действии с $(\rho_L^2 +
\rho_R^2)$, где $\rho^{ab}_L = (\bar{\psi}^a_{t,L}\gamma^0\psi^b_{t,L})$ и
$\rho^{ab}_R = (\bar{\psi}^a_{R,L}\gamma^0\psi^b_{R,L})$ - плотности правых и
левых технифермионов. Это взаимодействие эффективно соответствует
положительному сдвигу химического потенциала $\mu$. Поэтому отрицательные $G_V$
сдвигают восстановление симметрии к меньшим значениям $\mu$. Однако, мы
ожидаем, что это изменение при $G_V =-\frac{1}{4}G_S$ не влияет на физику при
$\mu = 0$, хотя это должно быть предметом дополнительного исследования.

В отсутствии калибровочных полей СМ симметрия $SU({\cal N})_L\otimes SU({\cal
N})_R$ выражения (\ref{Sf22s}) нарушена до $SU({\cal N})_V$ (здесь ${\cal N}$ -
полное число фермионов СМ). Относительная ориентация калибровочной группы СМ
$G_W = SU(3)\otimes SU(2)\otimes U(1)$ и $SU({\cal N})_V$ из $SU({\cal
N})_L\otimes SU({\cal N})_R \rightarrow SU({\cal N})_V$ не имеет значения.
Однако, когда включаются поля СМ, эффективный потенциал, возникающий из-за
обмена калибровочными бозонами СМ, зависит от этой ориентации. Минимум
потенциала достигается в вакуумном состоянии и определяет картину нарушения
$G_W$. Этот процесс известен как вакуумное выравнивание (см., например,
\cite{Align, Align1}). Эффективный потенциал имеет вид \cite{Align}:
\begin{eqnarray}
V(U) &=& 4 \sum_{\alpha = SU(3), SU(2), U(1); \, k} e_{\alpha}^2 \, {\rm Tr} \,
(\theta^{\alpha, k}_L U \theta^{\alpha, k}_R U^+) \,\nonumber\\&&
(-\frac{i}{2}) \int d^4 x \Delta^{\mu \nu} (x) <0|T[ J^A_{\mu L} J^A_{\nu R}
|0>\nonumber\\&& = -\frac{3}{32\pi^2} (F^2 \Delta^2)\sum_{\alpha = SU(3),
SU(2), U(1); \, k}e_{\alpha}^2 \, {\rm Tr} \, (\theta^{\alpha, k}_L U
\theta^{\alpha, k}_R U^+)
\end{eqnarray}
Здесь нет суммирования по $A$. $\theta^{\alpha, k}_{L,R}$ - генераторы  $G_W$,
$\Delta^{\mu \nu} (x)$ - пропагатор калибровочного бозона, $J^A_{\mu
L;R}=(\bar{\psi}^a_{t,L;R}\lambda_{ab}^A\gamma_i\psi^b_{t,L;R})$ -
технифермионные токи; матрицы $\lambda_{ab}^A$ - генераторы $ SU({\cal N})$. $U
\in SU({\cal N})$ определяет относительную ориентацию $ SU({\cal N})_V$ и
$G_W$. $F$ - технипионная константа. В общем случае $\Delta^2$ может быть
отрицательной.  Однако, в \cite{Align} приводятся аргументы в пользу того, что
$\Delta^2>0$. А именно, было показано, что если взаимодействие ТЦ
перенормируемо и асимптотически свободно, то имеют место правила сумм для
спектральных функций. Предполагая, что в спектральных функциях, соответствующих
векторному и аксиальному векторному каналам $<0|T[ J^A_{\mu L} J^A_{\nu R} |0>$
доминируют единственные состояния, находим, что $\Delta^2>0$. В нашем случае
динамическое кручение играет роль техницвета. Пользуясь аналогией с ТЦ мы
предполагаем, что динамическая теория кручения приводит к $\Delta^2>0$. При
этом предположении также, как в \cite{Align}, мы приходим к выводу, что $G_W$
нарушено минимальным образом. Это означает то, что подгруппы  $G_W$ ненарушены
до тех пор, пока этого нельзя избежать. Форма конденсата (\ref{GM}) требует,
что $SU(2)$ и $U(1)$ - нарушены. Поэтому, также, как и в $SU(N_{TC})$ модели
Фари - Сасскинда в нашем случае $SU(3)$ группа ненарушена, а Электрослабая
группа нарушена правильным образом.

\section{Массовый член}

Рассмотрим действие с дополнительным массовым членом для $\psi$ и $\phi$:

\begin{eqnarray}
S_f & = & \int \{\frac{i}{2}\bar{\psi}_a \gamma^{\mu}({\zeta} -
i{\chi}\gamma^5) D_{\mu} \psi_a + (c.c.) \} d^4 x \nonumber\\&& +\int
\{\frac{i}{2}\bar{\phi^c}_b \gamma^{\mu}({\zeta} - i{\chi}\gamma^5)
\bar{D}_{\mu} \phi_b^c + (c.c.)\} d^4 x \nonumber\\&& - \int(\delta_{a
a^{\prime}}\bar{\psi}_{a} \psi_{a^{\prime}} + {\bf f}_{b
b^{\prime}}\bar{\phi}_{b} \phi_{b^{\prime}}) m_0 d^4 x \label{Sf2SM__}
\end{eqnarray}
Здесь $m_0$ - константа размерности массы, а ${\bf f}$ - эрмитова матрица
констант связи. Интегрируя по кручению, получаем:
\begin{eqnarray}
&&S_f  =  \int \{i\bar{\psi}^a_s \gamma^{\mu} D_{\mu} \psi^a_s -
\frac{3\alpha_-}{16M_T^2}(\bar{\psi}^a_s\gamma^i \gamma^5
\psi^a_s)(\bar{\psi}^b_s\gamma_i \gamma^5  \psi^b_s)\} d^4 x \nonumber\\&&+
\int \{i\bar{\psi}^a_t \gamma^{\mu} D_{\mu} \psi^a_t  +
\frac{3\alpha_+}{16M_T^2}(\bar{\psi}^a_t\gamma^i \gamma^5
\psi^a_t)(\bar{\psi}^b_t\gamma_i \gamma^5  \psi^b_t)\} d^4 x \nonumber\\&& -
\frac{m_0}{\sqrt{[{\rm Re}\zeta]^2-[{\rm Im} \chi]^2}}\int(\bar{\psi}_{s,a}
[\frac{\delta_{a a^{\prime}}-{\bf f}_{a a^{\prime}}}{2}-\gamma^5
\frac{\delta_{a a^{\prime}}+{\bf f}_{a a^{\prime}}}{2}]\psi_{t,a^{\prime}} +
(c.c.))  d^4 x\nonumber\\ \label{Sf22SM___}
\end{eqnarray}

Мы составили новые спиноры
$\psi^a_t = \sqrt{{[{\rm Re}\zeta]-[{\rm Im} \chi]}}\left(\begin{array}{c}\phi^a_-\\
\psi^a_+\end{array}\right)$ и $\psi^a_s = {\gamma^5}\sqrt{{[{\rm Re}\zeta]+[{\rm Im} \chi]}}\left(\begin{array}{c}\psi^a_-\\
\phi^a_+\end{array}\right)$.

Далее, пренебрегаем калибровочными полями СМ, которые должны рассматриваться
как возмущения. Мы также вводим вспомогательные поля, как в подходе РНЙЛ:
\begin{eqnarray}
S_f & = & \int \{i\bar{\psi}^a_s \gamma^{\mu} D_{\mu} \psi^a_s -
\frac{3\alpha_-}{16M_T^2}(\bar{\psi}^a_s\gamma^i \gamma^5
\psi^a_s)(\bar{\psi}^b_s\gamma_i \gamma^5  \psi^b_s)\} d^4 x \nonumber\\&&+
\int \{i\bar{\psi}^a_t \gamma^{\mu} D_{\mu} \psi^a_t \} d^4 x
\nonumber\\&&+\int\{ -(\bar{\psi}^a_{t,L}H^+_{ab} \psi^b_{t,R} + (h.c.)) -
\frac{4M_T^2}{3\alpha_+} {\rm Tr}\, H^+H\}d^4x \nonumber\\&&
+\int\{(\bar{\psi}^a_{t,L}\gamma^i L^{ab}_i\psi^b_{t,L}) -
\frac{4M_T^2}{3\alpha_+}{\rm Tr}\,L^iL_i +(L\leftarrow \rightarrow R)\} d^4
x\nonumber\\&& - \frac{m_0}{\sqrt{[{\rm Re}\zeta]^2-[{\rm Im}
\chi]^2}}\int(\bar{\psi}_{s,a} [\frac{\delta_{a a^{\prime}}-{\bf f}_{a
a^{\prime}}}{2}-\gamma^5 \frac{\delta_{a a^{\prime}}+{\bf f}_{a
a^{\prime}}}{2}]\psi_{t,a^{\prime}} + (c.c.))  d^4 x\nonumber\\
\label{Sf22SMNJL}
\end{eqnarray}

Интегрирование по технифермионам ведет к появлению эффективного потенциала для
$H$ имеющего минимум при  $H = m_t {\bf 1}$. So, $H = m_t {\bf 1} + h$, где $h$
- нуль. Имеем:
 \begin{eqnarray}
&&S_f  =  \int  \{i\bar{\psi}^a_s \gamma^{\mu} D_{\mu} \psi^a_s -
\frac{3\alpha_-}{16M_T^2}(\bar{\psi}^a_s\gamma^i \gamma^5
\psi^a_s)(\bar{\psi}^b_s\gamma_i \gamma^5  \psi^b_s)\} d^4 x+ S_{eff}[L,R,H]
\nonumber\\&&-\frac{m^2_0}{{[{\rm Re}\zeta]^2-[{\rm Im}
\chi]^2}}\int\{ {\left(\begin{array}{c}\psi_{s,L}\\
-{\bf f} \psi_{s,R}\end{array}\right)}^+\gamma^0   [i \gamma^{\mu} D_{\mu}- m_t
{\bf 1}-h] ^{-1}{\left(\begin{array}{c}\psi_{s,L}\\
-{\bf f} \psi_{s,R}\end{array}\right)} \}d^4x \nonumber\\\label{Sf22SMNJL_}
\end{eqnarray}
где $D_{\mu} = (\partial_{\mu} -i \frac{1+\gamma_5}{2}L_{\mu}-i
\frac{1-\gamma_5}{2}R_{\mu})$. Здесь мы обозначили $S_{eff}= -i{\rm Sp}\,{\rm
Log}[i \gamma^{\mu} D_{\mu}- m_t {\bf 1}-h] $.

 Предположим, что $m_t >> m_0$. Далее, при энергиях много меньших, чем $M_T$ мы можем
 опустить фермионные члены для  $\psi_s$.
 Мы также пренебрегаем флуктуациями $h$, $L$, и $R$ возле их нулевого вакуумного значения:

\begin{eqnarray}
&&S_f  =  \int \bar{\psi}_s (i\gamma^{\mu} \partial_{\mu}-\frac{m^2_0}{{[{\rm
Re}\zeta]^2-[{\rm Im} \chi]^2}} {\bf f} [ m_t ]^{-1})\psi_sd^4x
\label{Sf22SMNJLF}
\end{eqnarray}

 В результате массовый член для
$\psi_s$ появляется с массовой матрицей

\begin{equation}
m_s = \frac{m^2_0}{{[{\rm Re}\zeta]^2-[{\rm Im} \chi]^2}}\frac{{\bf f}}{m_t}
\end{equation}

Следует отметить, что для того, чтобы иметь положительно определенное  $m_s$
необходимо, чтобы было $[{\rm Re}\zeta]^2>[{\rm Im} \chi]^2$, а матрица $\bf f$
положительно определена. Когда $[{\rm Re}\zeta]^2<[{\rm Im} \chi]^2$, мы можем
скомпоновать $\psi_s$ следующим образом: $\psi^a_s = \sqrt{{[{\rm Re}\zeta]+[{\rm Im} \chi]}}\left(\begin{array}{c}\psi^a_-\\
\phi^a_+\end{array}\right)$ и получить $m_s = \frac{m^2_0}{{[{\rm Im}
\chi]^2-[{\rm Re}\zeta]^2}}\frac{{\bf f}}{m_t}$.

\section{Выводы}

В этой главе мы рассмотрели фермионы, взаимодействующие неминимальным образом с
Пуанкаре гравитацией. Последняя содержит динамическую теорию метрического поля
на шкале массы Планка и калибровочную теорию $SO(3,1)$ группы на шкале
$\Lambda_{\chi}$  от $1$ ТэВ до $10$ ТэВ. Наша модель содержит действие Хольста
с массовым параметром на шкале ТэВ. После интегрирования по фермионам
появляются эффективные четырехфермионные взаимодействия. Мы размещаем все
фермионы СМ в левополяризованных частях Дираковских спиноров, а дополнительные
технифермионы - в правополяризованных. Определенный выбор констант связи
приводит к тому, что указанные четырехфермионные взаимодействия между
технифермионами имеют характер притяжения, а между фермионами СМ - отталкивания
(или пренебрежимо малы), при этом перекрестные члены между токами фермионов СМ
и технифермионов  пренебрежимо малы по сравнению с взаимодействием между
технифермионами. При определенных условиях технифермионы оказываются
сконденсированы, что приводит к нарушению Электрослабой симметрии и появлению
масс $W$ и $Z$ - бозонов.

Для того, чтобы обеспечить появление масс фермионов СМ, мы добавляем массовый
член для Дираковских спиноров, содержащих в своих левополяризованных
компонентах фермионы СМ, а в правополяризованных компонентах - технифермионы. В
результате появляется массовый член для фермионов Стандартной Модели.

\section{Публикации}

Результаты настоящей главы опубликованы в работе

"Torsion instead of Technicolor",  M.A. Zubkov, Mod. Phys. Lett.
A25:2885-2898,2010


\clearpage

{\Large \bf Заключение}

 \label{ch13}
\vspace{0.5cm}

Итак, кратко перечислю основные результаты диссертации.

\vspace{0.5cm}

Первая часть. В первой главе были изучены фрактальные и перколяционные свойства
центральных вихрей в глюодинамике в максимальной центральной проекции. Было
предложено понятие центрального монополя. Во второй главе предложена простая
центральная проекция, в рамках которой изучены свойства центральных вихрей и
рассмотрена их связь с конфайнментом. В третьей главе получено точное Абелево
представление для неабелевой петли Вильсона на решетке.

Вторая часть. В четвертой главе была рассмотрена решеточная модель Вайнберга -
Салама, изучены ее свойства при нефизически больших значениях постоянной тонкой
структуры.  В пятой главе было показано, что монополи Намбу сконденсированы при
темературах выше температуры Электрослабого перехода. В шестой главе было
продемонстрировано существование флуктуационной области   в окрестности
фазового перехода в решеточной модели Вайнберга-Салама при нулевой температуре.

Третья часть. В седьмой главе рассматривалась двумерная квантовая гравитация с
действием, квадратичным по кривизне. Получено решение в пределе слабой связи.
Показано, что на дискретизацию Редже можно наложить дополнительное условие, при
котором мера Лунда - Редже становится локальной. В восьмой главе предложена
калибровочно - инвариантная дискретизация телепераллелизма и калибровочно -
инвариантная дискретизация Пуанкаре - гравитации. В девятой главе численно
исследована модель динамических триангуляций при затравочной размерности 10.

Четвертая часть. В десятой главе изучены монополи в моделях малого объединения.
В одиннадцатой главе предложено продолжение $Z_6$ симметрии Стандартной Модели
на модели Техницвета. Показано, что лишь немногие из моделей Техницвета
обладают этой дополнительной дискретной симметрией. В двенадцатой главе
предложено использовать калибровочную теорию группы Лоренца для обеспечения
динамического нарушения Электрослабой симметрии.

В заключении я хотел бы выразить благодарность тем людям, которые так или иначе
помогали мне и без которых написание этой диссертации было бы невозможно.
Прежде всего, это мои родители, Татьяна Михайловна Зубкова и Александр
Михайлович Зубков. Особую роль в моей научной судьбе сыграл проф. Михаил
Игоревич Поликарпов. Он был научным руководителем моего диплома и кандидатской
диссертации.  В 1997 году я фактически оставил науку. И вернулся в нее лишь В
2002 году. Без поддержки М.И. Поликарпова это возвращение было бы невозможно.
Также я сердечно благодарен ныне покойному проф. К.А. Тер - Мартиросяну, к
которому я пришел сдавать теоретический минимум, будучи студентом МФТИ.
Вероятно, я был одним из последних учеников этого уникального человека. Также я
благодарен и проф. Ю.А.Симонову за многочисленные обсуждения, поддержку и
постоянный интерес к моей научной судьбе.

Я благодарен моим коллегам из ГНЦ РФ ИТЭФ за научные обсуждения и за
великолепную научную атмосферу, которую они создавали и создают. Среди этих
людей я особенно хотел бы отметить В.И.Захарова, Э.Т.Ахмедова, Ф.В.Губарева,
М.Н.Чернодуба, В.И.Шевченко, Ю.М.Макеенко, М.И.Высоцкого, В.А.Новикова,
О.В.Канчелли, Б.Л.Йоффе, В.Г.Ксензова, В.М.Вайнберга, А.У.Дубина,
Б.В.Мартемьянова, С.И.Блинникова, Е.В.Лущевскую, В.Г.Борнякова,
П.В.Буйвидовича, В.К.Митрюшкина. Э.Т.Ахмедову я особенно благодарен за
постоянную дружескую поддержку. Я благодарен соавтору многих моих работ и
старшему товарищу, проф. Б.Баккеру из свободного университета г. Амстердама. За
обсуждения я хотел бы поблагодарить Д.Гринсайта, В.А. Рубакова, Т.Томбулиса, Я.
Смита, Я.Амбьорна, Ф.Де Форкрана, А.В.Смилгу.

Особенно я хотел бы поблагодарить Александра Ивановича Веселова, прекрасного
мастера своего дела, соавтора многих из работ, вошедших в диссертацию, без
которого ее написание было бы невозможно.

\eject

\vspace{0.5cm}
 Я хотел бы
поблагодарить весь коллектив кафедры теоретической физики МФТИ, где я
проработал в 2008 - 2010 г.г. и особенно ее руководителя, Ю.М. Белоусова. Также
я благодарен педагогическому коллективу математической школы № 444 г. Москвы,
среди которых хотел бы отметить классного руководителя и преподавателя физики
М.Д.Гельфанд и преподавателя математики А.Г.Александрова, привившего мне, как и
многим другим, любовь к математике. Также я хочу поблагодарить своих товарищей
по МФТИ, с которыми мы вместе начинали входить в науку - С.Тополя и А.Комеча.

Кроме того, я хотел бы поблагодарить семью Трофимовых за четверть века дружбы.
Именно эта семья пробудила во  мне тягу к знанию.

Наконец, я хочу поблагодарить свою жену, С.Н. Кофман, по чьему настоянию я
вернулся в науку в 2002 г. и провел ту работу, результаты которой изложены в
диссертации.

\end{rm}


\begin{thebibliography}{99}



\bibitem{CenterGaugeFirst} L.Del~Debbio {\it et al.},
Phys.Rev. D55 (1997) 2298.

\bibitem{NumerousCenter} L.Del~Debbio {\it et al.}, hep-lat/9802003.

\bibitem{MaA} A.S.~Kronfeld, M.L. Laursen, G. Schierholz, U.J.
Wiese, Phys.Lett. 198B (1987) 516.

\bibitem{Alford} M.G.~Alford and F.Wilczek, {Phys.Rev.Lett.}, {62}
(1989) 1071; M.G.~Alford, J.~March--Russel and F.Wilczek, { Nucl.Phys.}, {
B337} (1990) 695; J.~Preskill and L.M.~Krauss, { Nucl.Phys.}, { B341} (1990)
50.
\bibitem{LinkingConf} L.Del~Debbio {\it et~al.},
Nucl.Phys.Proc.Suppl. 63 (1998) 552.
\bibitem{Percolation} A.V.~Pochinsky, M.I.~Polikarpov and
B.N. Yur\-chen\-ko, Phys.Lett.~A154 (1991) 194;

\bibitem{Abrikosov} A.A. Abrikosov, Sov.Phys. JETP 32 (1957) 1442;
H.B. Nielsen and P.Olesen, Nucl.Phys. B61 (1973) 45.



\bibitem{Suzuki}
Y. Matsubara, S. Ejiri, T. Suzuki, Nucl. Phys. Proc. Suppl. 34 (1994) 176

\bibitem{Tomboulis_}
T. G. Kovacs, E.T. Tomboulis,  hep-lat/9808046

\bibitem{Wiese_}
A.S. Kronfeld, M.L. Laursen, G. Schierholz, U.J. Wiese Phys. Lett. B 198 (1987)
516

\bibitem{forms}
M.I. Polikarpov, U.J. Wiese, M.A. Zubkov, Phys. Lett. B 309 (1993) 133

\bibitem{engels} G. Boyb, et. al. Nucl. Phys. B469 (1996) 419



\bibitem{Gr2000}
J. Greensite, Talk at {\em Confinement 2000}, Osaka, Japan, March 7-10, 2000;
hep-lat/0005001.

\bibitem{DFGGO1998}
L. Del Debbio, M. Faber, J. Giedt, J. Greensite, and S. Olejnik, Phys. Rev.
{\bf D 58} (1998) 094501.


\bibitem{BKPV2000}
V.G. Bornyakov, D.A. Komarov, M.I. Polikarpov, and A.I. Veselov, Talk at {\em
Confinement 2000}, Osaka, Japan, March 7-10, 2000; hep-lat/0002017.

\bibitem{FE1999}
Ph. de Forcrand and M. D'Elia, Phys. Rev. Lett. {\bf 82} (1999) 4582; C.
Alexandrou, M. M. D'Elia, and Ph. de Forcrand, hep-lat/9907028.

\bibitem{FGO1999a}
M. Faber, J. Greensite, S. Olejnik, and D. Yamada, hep-lat/9912002.


\bibitem{Vo1999}
G.E. Volovik, Pisma Zh. Eksp. Teor. Phys. {\bf 70} (1999) 776;
cond-mat/9911374.



\bibitem{Og1999}
M.C. Ogilvie, Nucl. Phys. Proc. Suppl., {\bf 73} (1999), 542: hep-lat/9809167

\bibitem{CPV1997}
M.N.Chernodub, M.I. Polikarpov, and A.I.Veselov, Phys. Lett. {\bf B 399} (1997)
267; hep-lat/9610007.



\bibitem{BBMS1996}
G.S. Bali, V.G. Bornyakov, M. M\"uller-Preussker, and K. Schilling, Phys. Rev.
{\bf D 54} (1996) 2863; hep-lat/9603012

\bibitem{FGO1999b}
M. Faber, J. Greensite, and S. Olejnik, HEP, 9901:008 (1999); hep-lat/9810008

\bibitem{Tomboulis}
T. G. Kovacs and E.T. Tomboulis, Nucl. Phys. Proc. Suppl. {\bf 73} (1999) 566;
hep-lat/9808046

\bibitem{Wiese}
A.S. Kronfeld, M.L. Laursen, G. Schierholz, and  U.J. Wiese, Phys. Lett. {\bf B
198} (1987) 516

\bibitem{IPP1993}
T.L. Ivanenko, A.V. Pochinsky, and M.I. Polikarpov, Phys. Lett. {\bf B 302}
(1993) 458

\bibitem{AGG1999}
J. Ambj{\o}rn, J. Geidt, and J. Greensite, hep-lat/9907021

\bibitem{FGO2000}
M. Faber, J. Greensite, and S. Olejnik, hep-lat/0005017


\bibitem{Streit}
 L. Streit, {\em Functional Integrals for Quantum Theory}, in
H. Latal and W. Schweiger (Eds.), ``Methods of Quantization'', Lecture Notes in
Physics LNP 572, Springer-Verlag, (Berlin, Heidelberg 2001)

\bibitem{C}
M.N. Chernodub, F.V. Gubarev, M.I. Polikarpov, and V.I. Zakharov,\\
Nucl. Phys. {\bf B 592} (2001) 107; hep-lat/0003138

\bibitem{D}
V. Dzhunushaliev, D. Singleton, hep-th/9912194

\bibitem{F}
F.Lenz, S.Woerlen, hep-th/0010099



\bibitem{Greensite__}
M. Faber, J. Greensite, and S. Olejnik, in ``Confinement, Topology, and other
Non-Perturbative Aspects of QCD'', NATO Advanced Research Workshop, Stara
Lesna, Slovakia, 2002





\bibitem{PvortSing}
F. V. Gubarev, A. V. Kovalenko, M. I. Polikarpov, S. N. Syritsyn and V. I.
Zakharov, Phys. Lett. B574, 136 (2003).

\bibitem{Triang}
J. Ambjorn, J. Jurkiewicz, Y. Watabiki, Nucl.Phys. B454 (1995) 313-342


\bibitem{PvortSing2}
A. V. Kovalenko, M. I. Polikarpov, S. N. Syritsyn, V. I. Zakharov, Phys.Rev.
D71 (2005) 054511

\bibitem{Creutz}
M. Creutz, {\em Quarks, gluons and lattices}, Cambridge University Press,
(Cambridge, 1985).

\bibitem{Scaling}
J.Fingberg, U.Heller, F.Karsch,\\
hep-lat/9208012

\bibitem{fractal1}
B.B. Mandelbrot, ``Fractals - Form, Chance and Dimension'' (Freeman, San
Francisco, 1977)\\ D.A. Russel, J.D. Hanson, and E. Ott, Phys. Rev. Lett. 45,
(1980), 1175\\ P. Grassberger and I. Procaccia Phys. Rev. Lett. 50 (1983), 346

\bibitem{fractal}
M.I. Polikarpov Phys.Lett. {\bf B 236} (1990), 61;\\ T.L. Ivanenko, A.V.
Pochinsky, and M.I.Polikarpov, Phys. Lett. {\bf B252} (1990), 631

\bibitem{BCGPSVZ}
V.G. Bornyakov, M.N. Chernodub, F.V. Gubarev, M.I. Polikarpov, T. Suzuki, A.I.
Veselov, and V.I Zakharov, {\em Anatomy of the lattice magnetic monopoles},
hep-lat/0103032 and Phys. Lett. B.



\bibitem{tH}
't Hooft {Nucl.\ Phys.}{B 190},
  {455} ({1981}).

\bibitem{P}
{~M.I.} {Polikarpov}
  {Nucl.\ Phys. Proc. Suppl.} {53},
  {134} ({1997}).





\bibitem{DP}
{~D.} {Diakonov}, {~V.} {Petrov},
  {Phys.\ Lett.} {B 224},
{131} ({1989}).

  \bibitem{DP1}
{~D.} {Diakonov},{~V.} {Petrov} {hep-th/0008004}.

\bibitem{KT}
{~K.-I.} {Kondo}, {~Y.} {Taira},{hep-th/9911242}.




\bibitem{Iv} M.{Faber}, {A.~N.} {Ivanov}, {N.~I.} {Troitskaya} {M.} {Zach},
{Phys. \ Rev.} {D 62}, {025019} ({2000}).






\end{thebibliography}

\begin{thebibliography}{99}



\bibitem{TEV}
J.A.~Casas, J.R.~Espinosa, and I.~Hidalgo, hep-ph/0607279.



\bibitem{UV}
K.Holland, arXiv:hep-lat/0409112

Zoltan Fodor, Kieran Holland, Julius Kuti, Daniel Nogradi, Chris Schroeder, PoS
(LATTICE 2007) 056, arXiv:0710.3151



\bibitem{SU2U1}
R.~Shrock, Phys. Lett. B {\bf 162}, 165 (1985); Nucl. Phys. B {\bf 267}, 301
(1986).


\bibitem{1}
F. Csikor, Z. Fodor, J. Heitger Phys.Rev.Lett. 82 (1999) 21-24 Phys.Rev. D58
(1998) 094504 Nucl.Phys.Proc.Suppl. 63 (1998) 569-571

\bibitem{2}
F. Csikor, Z. Fodor, J. Heitger Phys.Lett. B441 (1998) 354-362

\bibitem{3}
F. Csikor, Z. Fodor, J. Hein, A. Jaster, I. Montvay Nucl.Phys. B474 (1996)
421-445

\bibitem{4}
Joachim Hein (DESY),   Jochen Heitger, Phys.Lett. B385 (1996) 242-248

\bibitem{5}
F. Csikor, Z. Fodor, J. Hein, J. Heitger, A. Jaster, I. Montvay
Nucl.Phys.Proc.Suppl. 53 (1997) 612-614

\bibitem{6}
Z. Fodor, J. Hein, K. Jansen, A. Jaster, I. Montvay Nucl.Phys. B439 (1995)
147-186

\bibitem{7}
F. Csikor, Z. Fodor, J. Hein, J. Heitger,
 Phys.Lett. B357 (1995) 156-162

\bibitem{8}
F. Csikor, Z. Fodor, J. Hein, K.Jansen, A. Jaster, I. Montvay
 Nucl.Phys.Proc.Suppl. 42 (1995) 569-574

\bibitem{9}
F. Csikor, Z. Fodor, J. Hein, K.Jansen, A. Jaster, I. Montvay
 Phys.Lett. B334 (1994) 405-411

\bibitem{10}
Y. Aoki, F. Csikor, Z. Fodor, A. Ukawa Phys.Rev. D60 (1999) 013001
Nucl.Phys.Proc.Suppl. 73 (1999) 656-658

\bibitem{11}
Y. Aoki Phys.Rev. D56 (1997) 3860-3865



\bibitem{12}
W.Langguth, I.Montvay, P.Weisz Nucl.Phys.B277:11,1986.

\bibitem{13}
W. Langguth, I. Montvay (DESY) Z.Phys.C36:725,1987

\bibitem{14}
Anna Hasenfratz, Thomas Neuhaus, Nucl.Phys.B297:205,1988

\bibitem{Montvay}
I.~Montvay, Nucl. Phys. B {\bf 269}, 170 (1986).




\bibitem{T}
 Bohdan Grzadkowski, Jacek Pliszka, Jose Wudka
 Phys.Rev.
D69 (2004) 033001

\bibitem{Kertesz}
M.N. Chernodub, Phys.Rev.Lett. 95 (2005) 252002


\bibitem{Nambu}
Y.~Nambu, Nucl.Phys. B {\bf 130}, 505 (1977);\\
Ana~Achucarro and Tanmay~Vachaspati, Phys. Rept. {\bf 327}, 347 (2000); Phys.
Rept. {\bf 327}, 427 (2000).




\bibitem{Jersak}
J.~Jersak, C.B.~Lang, T.~Neuhaus, G.~Vones, Phys.Rev. D {\bf 32},2761 (1985).\\
%
H.G.~Evertz, J.~Jersak, C.B.~Lang, T.~Neuhaus, Phys. Lett. B {\bf 171}, 271
(1986).
%
H.G.~Evertz, V.~Grosch, J.~Jersak, H.A.~Kastrup, T.~Neuhaus, D.P.~Landau,
J.L.~Xu, Phys. Lett. B {\bf 175}, 335 (1986).
%



\bibitem{phi4}
I.~Montvay, BNL Gauge Theor.Symp. (1986) 235; Nucl. Phys. B 293 (1987) 479.


\bibitem{EW_T}
M.~Gurtler, E.-M.~Ilgenfritz, and A.~Schiller,
Phys. Rev. D {\bf 56}, 3888 (1997).\\
K.~Rummukainen, M.~Tsypin, K.~Kajantie, M.~Laine, and M.~Shaposhnikov,
Nucl. Phys. B {\bf 532}, 283 (1998);\\
Yasumichi Aoki, Phys. Rev. D {\bf 56}, 3860 (1997);\\
N.~Tetradis, Nucl. Phys. B {\bf 488}, 92 (1997);\\
B.~Bunk, Ernst-Michael~Ilgenfritz, J.~Kripfganz, and A.~Schiller
(BI-TP-92-46), Nucl. Phys. B {\bf 403}, 453 (1993);\\
B.~Bunk, Ernst-Michael~Ilgenfritz, J.~Kripfganz, and A.~Schiller (BI-TP-92-12),
Phys. Lett. B {\bf 284}, 371 (1992).

\bibitem{U1monopole}
Thomas~A.~DeGrand, Doug~Toussaint, Phys. Rev. D {\bf 22}, 2478 (1980).

\bibitem{Ranft}
J.~Ranft, J.~Kripfganz, G.~Ranft, Phys.Rev.D {\bf 28}, 360 (1983).

\bibitem{Mperc}
Wolfgang~Franzki, John B.~Kogut, M.P.~Lombardo, Phys. Rev. D {\bf 57}, 6625
(1998).

\bibitem{Chernodub}
M.N.~Chernodub, F.V.~Gubarev, E.M.~Ilgenfritz, and A.~Schiller,
Phys. Lett. B {\bf 434}, 83 (1998);\\
M.N.~Chernodub,~F.V. Gubarev, E.M.~Ilgenfritz, and A.~Schiller, Phys. Lett. B
{\bf 443}, 244 (1998).

\bibitem{Chernodub_Nambu} M.N.~Chernodub, JETP Lett. {\bf 66}, 605 (1997)







\bibitem{Kertesz3}
D.~Stauffer, A.~Aharony, ``Introduction to percolation theory", (Taylor \&
Francis, London, 1994).



\bibitem{propagator}
O.~Oliveira, P. J.~Silva, E.-M.~Ilgenfritz, A.~Sternbeck, arXiv:0710.1424




\bibitem{SM_present_status}
U.M.~Heller, Nucl. Phys. Proc. Suppl. {\bf 34}, 101 (1994);

I.~Montvay, hep-lat/9703001;

I.~Montvay, W.~Langguth, and P.~Weisz, Nucl. Phys. B {\bf 277}, 11 (1986);

A.~Hasenfratz and T.~Neuhaus, Nucl. Phys. B {\bf 297}, 205 (1988);

U.M.~Heller, M.~Klomfass, H. ~Neuberger, and P.~Vranas, Nucl. Phys. B {\bf
405}, 555 (1993).

\bibitem{triviality}
M.~L{\"{u}}scher and P.~Weisz, Nucl. Phys. B {\bf 318}, 705 (1989);

I.~Montvay, Nucl. Phys. B {\bf 293}, 479 (1987);

M.~Klomfass,  Nucl. Phys. B {\bf 412}, 621 (1994).


\bibitem{path_integral}
R.A.~Brandt, F.~Neri, and D.~Zwanziger, Phys. Rev. D {\bf 19}, 1153 (1979).

\bibitem{lattice_paths}
F.V.~Gubarev and V.I.~Zakharov, hep-lat/0211033.



\bibitem{SO3}
P.~de Forcrand, O.~Jahn, Nucl.Phys. B {\bf 651}, 125 (2003).









\bibitem{Greensite}
J.~Greensite, Prog. Part. Nucl. Phys. {\bf 51} (2003) 1



\bibitem{GZP}
F.V.~Gubarev, A.V.~Kovalenko, M.I.~Polikarpov, S.N.~Syritsyn, and
V.I.~Zakharov, Phys. Lett. B {\bf 574} 136 (2003)




\bibitem{lattice_fermions}
N.B.~Nielsen and M.~Ninomiya,
Nucl. Phys. B {\bf 185}, 20 (1981); {\it ibid}, 173;\\
M.~L{\"{u}}scher, Phys. Lett. B {\bf 428}, 342 (1998);\\
H.~Neuberger, Phys. Lett. B {\bf 417}, 141 (1998)\\

\bibitem{noncompact}
A.~Patrascioiu, E.~Seiler, and I.O.~Stamatescu,
Phys. Lett. B {\bf 107}, 364 (1981); \\
E.~Seiler, I.O.~Stamatescu, and D.~Zwanziger,
Nucl. Phys. B {\bf 239}, 177 (1984); \\
Y.~Yotsuyanagi, Phys. Lett. B {\bf 135}, 141 (1984); \\
K.~Cahill, S.~Prasad, R.~Reeder, and B.~Richter,  \\
Phys. Lett. B {\bf 181}, 333 (1986); \\
K.~Cahill, Phys. Lett. B {\bf 231}, 294 (1989)





\bibitem{M_H}
Bohdan Grzadkowski, Jose Wudka, IFT-31/2001, UCRHEP-T321, Acta  Phys. Polon. B
{\bf 32} (2001) 3769-3782

\bibitem{M_W_T}
Peter Arnold and Olivier Espinosa,  Phys. Rev. D {\bf 47} (1993) 3546.\\
Z. Fodor and A. Hebecker,  Nucl. Phys. B {\bf 432} (1994) 127.\\
W. Buchmuller, Z. Fodor, and A. Hebecker, Nucl. Phys. B {\bf 447} (1995) 317.\\







\bibitem{Rum}
K. Kajantie, M. Laine, K. Rummukainen, and M. Shaposhnikov, Phys. Rev. Lett.
{\bf 77} (1996) 2887-2890.


\bibitem{Rubakov}
V.A.Kuzmin, V.A.Rubakov, and M.E.Shaposhnikov, Phys. Lett. B {\bf 155}, 36
(1985)

\bibitem{Janke_T}
W. Janke, D.A. Johnston, R. Kenna, XXIIIrd Int. Symp. Latt. Field Theory,
Proceedings of Science (LAT2005) 244, hep-lat/0512022

\bibitem{Janke}
Sandro Wenzel, Elmar Bittner, Wolfhard Janke, Adriaan M.J. Schakel, A.
Schiller, Phys. Rev. Lett. {\bf 95}, 051601 (2005)













\end{thebibliography}

\begin{thebibliography}{99}




\bibitem{Loll}
R. Loll, gr-qc/9805049; Living Reviews in Relativity, www.livingreviews.org

\bibitem{nrenorm}
M.H.Goroff, A.Sagnotti, Nucl. Phys. {\bf B 266}, 709 (1986)

\bibitem{R2}
I. G. Avramidi, Soviet Journal of Nuclear Physics, 44 (1986) 160-164,  "Heat
Kernel and Quantum Gravity", Lecture Notes in Physics, Series Monographs, LNP:
m64 (Berlin: Springer-Verlag,  2000), hep-th/9510140


\bibitem{mass_singularities}
L.Infeld, Rev. Mod. Phys. {\bf 29}, 398 (1957)




\bibitem{Odintsov}
E.Elizalde, S. Naftulin and S.D. Odintsov, Phys. Lett. {\bf B 323}, 124 (1994)

S.Naftulin and S.D. Odintsov, Mod.Phys.Lett. {\bf A 10}, 2071 (1995)


\bibitem{2DR2perturb}
J.Nishimura, S.Tamura, A.Tsuchiya, Mod.Phys.Lett. {\bf A9} 3565 (1994)


\bibitem{R2CFT}
H.Kawai, R.Nakayama, Phys. Lett. {\bf B306}, 224 (1993)

\bibitem{R2synchronic}
H.Kawai, N.Kawamoto, T.Mogami, Y.Watabiki, Phys. Lett. {\bf B306}, 19 (1993)


\bibitem{W}
H. W. Hamber, R. M. Williams, Phys.Rev. D59 (1999) 064014,  Nucl.Phys. B487
(1997) 345

\bibitem{A1}
J. Ambjorn, J. Nielsen, J. Rolf, G. Savvidy,  Class.Quant.Grav. 14 (1997)
3225-3241

\bibitem{KPZ}
V.G.Knizhnik, A.M.Polyakov, A.B. Zamolodchikov, Mod. Phys. Lett. A3, 819 (1988)

F.David, Mod. Phys. Lett. A3 (1988) 1651

J.Distler and H.Kawai, Nucl.Phys. B321 (1989)509



\bibitem{renormalizable}
K.S.Stelle, Phys. Rev. {\bf D 16} , 953 (1977)

\bibitem{unitarity}
E.T.Tomboulis, Phys. Rev. Lett. {\bf 52}, 1173 (1984)

\bibitem{Hawking}
S.W. Hawking, Nucl. Phys. {\bf B114}, 349 (1978)


\bibitem{a_free}
I. G. Avramidi, Soviet Journal of Nuclear Physics, 44 (1986) 160-164,  "Heat
Kernel and Quantum Gravity",  Lecture Notes in Physics, Series Monographs, LNP:
m64 (Berlin:  Springer-Verlag,  2000), hep-th/9510140

I. G. Avramidi and A. O. Barvinsky,  Physics Letters B, 159 (1985) 269-274




\bibitem{Menotti}
P.Menotti, A.Pelisseto, Phys. Rev. {\bf D35}, 1194 (1987)

\bibitem{Regge}
B.A. Berg, W. Beirl, B. Krishnan, H. Markum, J. Riedler, Phys.Rev. D54 (1996)
7421-7425

E. Bittner, W. Janke, H. Markum,  Phys.Rev. D66 (2002) 024008

W.Beirl, B.A. Berg, hep-lat/0309002


\bibitem{Regge_R2}
Herbert W. Hamber, Ruth M. Williams, Nucl.Phys. B248 (1984) 392, Nucl.Phys.
B269 (1986) 712, Nucl.Phys. B435 (1995) 361

\bibitem{Regge_measure}
H.W. Hamber, R.M Williams, Phys.Rev. D59 (1999) 064014, Nucl.Phys. B487 (1997)
345-408



\bibitem{DT}
J.Ambjorn, J.Jurkiewicz, Y.Watabiki, J.Math.Phys. {\bf 36} (1995) 6299-6339

\bibitem{DTR2}
J.Ambjorn, J.Jurkiewicz, C.F.Kristjansen, Nucl. Phys. {\bf B393}, 601 (1993)


\bibitem{dS_numerical}
S.Caracciolo, A.Pelissetto, Nucl. Phys. {\bf B299}, 693 (1988)

\bibitem{teleparallel}
K.Hayashi, T.Shirafuji, Phys. Rev. {\bf D19}, 3524 (1979)

Yu. N. Obukhov, J. G. Pereira, Phys.Rev. D67 (2003) 044016

V. C. de Andrade, J. G. Pereira, Phys.Rev. D56 (1997) 4689-4695

\bibitem{Poincare_lat}
Juergen Schmidt, Christopher Kohler, Gen.Rel.Grav. 33 (2001) 1799-1808

\bibitem{teleparallel_lat}
J.G. Pereira, T. Vargas,  Class.Quant.Grav. 19 (2002) 4807-4816

\bibitem{topology_DT}
H.Hagura, hep-lat/0304011

\bibitem{de_Bakker}
B.V. de Bakker,  Nucl.Phys.Proc.Suppl. {\bf 42} (1995) 716-718

\bibitem{topological_solutions}
T.Eguchi, B.Gilkey, A.J.Hanson, Phys. Rep. {\bf 66}, 213 (1980)


\bibitem{DeWitt}
Bryce S. DeWitt, Phys.Rev.160:1113-1148,1967, Phys.Rev.162:1239-1256,1967,
Phys.Rev.162:1195-1239,1967


\bibitem{Misner}
C.W.Misner, Rev.Mod.Phys. 29 (1957), 497

\bibitem{Popov}
L.Fadeev,V.Popov, Sov.Phys.Usp. 16 (1974) 777

\bibitem{Fijikawa}
Kazuo Fujikawa, INS-461, Nucl.Phys.B226:437,1983






\bibitem{Magnea}
M. Caselle, A. D'Adda  , Lorenzo Magnea . DFTT-18/89, ITP-SB-89-68,
Phys.Lett.B232:457,1989

\bibitem{Drummond}
 I.T. Drummond DAMTP-85/28,  Nucl.Phys.B273:125,1986





\bibitem{Smolin}
Lee Smolin,  HUTP-78/A035, Nucl.Phys.B148:333,1979

\bibitem{Manion}
C.L.T. Mannion, J.G. Taylor  Phys.Lett.B100:261,1981

\bibitem{Kaku}
A. Das, M. Kaku, P.K. Townsend Phys.Lett.B81:11,1979





\bibitem{NONABELIAN}
A.~Patrascioiu, E.~Seiler, and I.O.~Stamatescu,
Phys. Lett. B {\bf 107}, 364 (1981); \\
E.~Seiler, I.O.~Stamatescu, and D.~Zwanziger,
Nucl. Phys. B {\bf 239}, 177 (1984); \\
Y.~Yotsuyanagi, Phys. Lett. B {\bf 135}, 141 (1984); \\
K.~Cahill, S.~Prasad, R.~Reeder, and B.~Richter,  \\
Phys. Lett. B {\bf 181}, 333 (1986); \\
K.~Cahill, Phys. Lett. B {\bf 231}, 294 (1989)













\bibitem{Entropy_force_}
 Erik P.
Verlinde, arXiv:1001.0785

\bibitem{Strings}
 O. Aharony, S.S. Gubser, J. Maldacena, H. Ooguri, Y. Oz,
 Phys.Rept. 323 (2000) 183-386



\bibitem{Triangulation}
J.Ambjorn, hep-th/9411179


\bibitem{2D}
J. Ambjorn, K. N. Anagnostopoulos, U. Magnea, G. Thorleifsson, Phys.Lett. B388
(1996) 713-719



\bibitem{5D}
Alun George, hep-lat/9909033


\bibitem{Lorentzian}
J. Ambjorn, J. Jurkiewicz, R. Loll,  Nucl.Phys. B610 (2001) 347-382


\bibitem{Gauge_Triang}
H.S.Egawa, S.Horata, T.Yukawa,  Nucl.Phys.Proc.Suppl. 106 (2002) 971-973

\bibitem{Catterall}
S. Catterall, Comput.Phys.Commun. 87 (1995) 409-415

\bibitem{Geometry}
J. Ambjorn, M. Carfora, A. Marzuoli, hep-th/9612069

\bibitem{ergodicity}
U.Pachner, Europe J. Combinatorics, 12 (1991) 129 - 145

\bibitem{deBakker}
B.V. de Bakker, hep-lat/9508006

\bibitem{Kaluza}
J. M. Overduin, P. S. Wesson, Phys.Rept. 283 (1997) 303


\end{thebibliography}

\begin{thebibliography}{99}


\bibitem{Z6f}
K.S.~Babu, I.~Gogoladze, and K.~Wang, Phys. Lett. B {\bf 570}, 32 (2003);\\
K.S.~Babu, I.~Gogoladze, and K.~Wang, Nucl. Phys. B {\bf 660}, 322 (2003);



\bibitem{M_H}
Bohdan Grzadkowski, Jose Wudka, IFT-31/2001, UCRHEP-T321, Acta  Phys. Polon. B
{\bf 32} (2001) 3769-3782



\bibitem{Extention}
F. del Aguila, R. Pittau,  Acta Phys.Polon. B35 (2004) 2767-2780







\bibitem{little_higgs}
Martin Schmaltz, David Tucker-Smith, Ann.Rev.Nucl.Part.Sci. 55 (2005) 229-270

\bibitem{TopLittleHiggs}
Mark Trodden, Tanmay Vachaspati, Phys.Rev. D70 (2004) 065008


\bibitem{PUT}
Andrzej J. Buras, P.Q. Hung, Phys.Rev. D68 (2003) 035015

\bibitem{PUT1}
Andrzej J. Buras, P.Q. Hung, Ngoc-Khanh Tran, Anton Poschenrieder, Elmar
Wyszomirski, Nucl.Phys. B699 (2004) 253





\bibitem{Cho}
Y. M. Cho, D. Maison,  Phys.Lett. B {\bf 391}, 360 (1997)

\bibitem{HooftPolyakov}
Gerard 't Hooft, Nucl.Phys.B79:276-284,1974

Alexander M. Polyakov, JETP Lett.20:194-195,1974, Pisma
Zh.Eksp.Teor.Fiz.20:430-433,1974

\bibitem{Weinb}
Erick J. Weinberg, Nucl.Phys.B167:500,1980

\bibitem{PATI}
J.C.Pati, S.Radjpoot, A.Salam, Phys. Rev.  {\bf D 17}, 131 (1978)





\bibitem{MONOPOLE}
 K. A. Milton,  Rept.Prog.Phys. 69
(2006) 1637-1712


\bibitem{Rubakov_}
V.A. Rubakov,  JETP Lett.33:644-646,1981, Pisma
Zh.Eksp.Teor.Fiz.33:658-660,1981

V.a. Rubakov, M.s. Serebryakov, Nucl.Phys.B218:240-268,1983.





\bibitem{WS}
S.Weinberg, Phys.Rev.D 13, 974, 1976

L.Susskind, Phys.Rev.D 20, 2619, 1979

\bibitem{FS}
E.Farhi, L.Susskind, Phys.Rev.D 20, 3404, 1979











\bibitem{PUT3}

 Mehrdad~Adibzadeh and
P.Q.~Hung, hep-ph/0705.1154.


\bibitem{walking}
Thomas Appelquist, John Terning, L.C.R. Wijewardhana,   Phys.Rev.Lett. 77
(1996) 1214-1217

Thomas Appelquist, Anuradha Ratnaweera, John Terning, L. C. R. Wijewardhana,
Phys.Rev. D58 (1998) 105017

\bibitem{Sannino_t}
F. Sannino, arXiv:0804.0182

N. Evans and F. Sannino, arXiv:hep-ph/0512080.

\bibitem{Shrock}
 Thomas Appelquist, Maurizio Piai, Robert Shrock,
 Phys.Rev. D69
(2004) 015002

 Thomas Appelquist, Robert Shrock, Phys.Lett. B548 (2002)
204-214






\bibitem{Z6}
C.Gardner, J.Harvey, Phys. Rev. Lett. {\bf 52} (1984) 879

\bibitem{Z6_}
Tanmay Vachaspati, Phys.Rev.Lett. 76 (1996) 188-191

\bibitem{Z6__}
Hong Liu, Tanmay Vachaspati, Phys.Rev. D56 (1997) 1300-1312



\bibitem{Technicolor}
 Christopher T. Hill, Elizabeth H. Simmons,
 Phys.Rept. 381 (2003) 235-402;
Erratum-ibid. 390 (2004) 553-554

\bibitem{Technicolor_}
Kenneth Lane, hep-ph/0202255

\bibitem{Technicolor__}
R. Sekhar Chivukula, hep-ph/0011264

\bibitem{ExtendedTechnicolor}
 Thomas Appelquist, Neil Christensen, Maurizio Piai, Robert Shrock, Phys.Rev. D70 (2004) 093010

\bibitem{ExtendedTechnicolor_}
 Adam Martin, Kenneth Lane, Phys.Rev. D71 (2005) 015011

\bibitem{ExtendedTechnicolor__}
 Thomas Appelquist, Maurizio Piai, Robert Shrock, Phys.Rev. D69
(2004) 015002

\bibitem{ExtendedTechnicolor___}
Robert Shrock, hep-ph/0703050

\bibitem{ExtendedTechnicolor____}
 Adam Martin, Kenneth Lane, Phys.Rev. D71 (2005) 015011



\bibitem{minimal_walking}
R. Foadi, M.T. Frandsen, T. A. Ryttov, F. Sannino, arXiv:0706.1696

\bibitem{minimal_walking_}
Sven Bjarke Gudnason, Chris Kouvaris, Francesco Sannino,   Phys.Rev. D73 (2006)
115003

\bibitem{minimal_walking__}
D.D. Dietrich (NBI), F. Sannino (NBI), K. Tuominen, Phys.Rev. D72 (2005) 055001











\bibitem{Weinberg}
S.Weinberg, "The Quantum Theory of Fields", Cambridge, University press, 2001













\bibitem{Appelquist}
 Thomas Appelquist, John Terning, L.C.R. Wijewardhana, Phys.Rev.Lett.
79 (1997) 2767-2770

\bibitem{Appelquist_}
 Neil D. Christensen, Robert Shrock, Phys.Rev. D72 (2005) 035013




\bibitem{Hehl}
  R.~Hojman, C.~Mukku and W.~A.~Sayed,
    Phys.\ Rev.\  D {\bf 22} (1980) 1915--1921.

 P.~C.~Nelson,
   Phys.\ Lett.\  A {\bf 79} (1980) 285--287.

 H.~T.~Nieh and M.~L.~Yan,  J.\ Math.\ Phys.\ {\bf 23}
    (1982) 373.

 F.~W.~Hehl, J.~D.~McCrea,  Found.\
   Phys.\ {\bf 16} (1986) 267--293.

J.~D.~McCrea, F.~W.~Hehl and E.~W.~Mielke,
    Int.\ J.\ Theor.\ Phys.\ {\bf 29} (1990)
    1185--1206.





\bibitem{Holst:1995pc}
S.~Holst,  Phys.\ Rev.\ D {\bf 53}, 5966 (1996) [gr-qc/9511026].


\bibitem{imir}
G.~Immirzi, Nucl.\ Phys.\ Proc.\ Suppl.\ {\bf 57}, 65 (1997) [gr-qc/9701052].

\bibitem{Rovelli}
A.Perez, C.Rovelli, Phys.Rev. D73 (2006) 044013,  ArXiv:gr-qc/0505081


\bibitem{Khriplovich:2005jh}
I.~B.~Khriplovich and A.~A.~Pomeransky,  Phys.\ Rev.\  D {\bf 73} (2006) 107502
[arXiv:hep-th/0508136].

\bibitem{Freidel:2005sn}
L.~Freidel, D.~Minic and T.~Takeuchi,  Phys.\ Rev.\  D {\bf 72} (2005) 104002
[arXiv:hep-th/0507253].




\bibitem{Randono:2005up}
A.~Randono, arXiv:hep-th/0510001.


\bibitem{Mercuri:2006um}
S.~Mercuri,  Phys.\ Rev.\  D {\bf 73} (2006) 084016 [arXiv:gr-qc/0601013].

\bibitem{Alexandrov}
 Sergei Alexandrov,
Class.Quant.Grav.25:145012,2008

\bibitem{Xue}
She-Sheng Xue, Phys.Lett.B665:54-57,2008, ArXiv:0804.4619

\bibitem{Alexander1}
S.Alexander, T.Biswas, G.Calcagni, Phys. Rev. D 81, 043511 (2010),
ArXiv:0906.5161

\bibitem{Alexander2}
S.Alexander, D.Vaid, ArXiv:hep-th/0609066





\bibitem{Shapiro}
A.S.Belyaev, I.L.Shapiro, Nucl.Phys. B543 (1999) 20-46, ArXiv:hep-ph/9806313





\bibitem{Kagan}
Alex Kagan,  CCNY-HEP-91-12,  Proc. of 15th Johns Hopkins Workshop on Current
Problems in Particle Theory, Baltimore, MD, Aug 26-28, 1991,
 Johns Hopkins Wrkshp 1991:217-242
(QCD161:J55:1991)

\bibitem{Dobrescu_Kagan}
Bogdan A. Dobrescu, Nucl.Phys.B449:462-482,1995

D.Atwood, A.Kagan and T.G.Rizzo,
Phys.\ Rev.\ D {\bf 52}, 6264 (1995) [arXiv:hep-ph/9407408].

A.L.Kagan,
Phys.\ Rev.\ D {\bf 51}, 6196 (1995) [arXiv:hep-ph/9409215].


B.A.~Dobrescu and E.H.Simmons,
Phys.\ Rev.\ D {\bf 59}, 015014 (1999) [arXiv:hep-ph/9807469].


B.A.Dobrescu and J.Terning,
Phys.\ Lett.\ B {\bf 416}, 129 (1998) [arXiv:hep-ph/9709297].


\bibitem{ENJL}
J.Bijnens, C.Bruno, E. de Rafael, Nucl.Phys. B390 (1993) 501-541, hep-ph/920623

\bibitem{ConformalGrav}
 Philip D. Mannheim, Prog.Part.Nucl.Phys. 56 (2006) 340-445

V.V. Zhytnikov, Int.J.Mod.Phys.A8:5141-5152,1993.


\bibitem{Sesgin}
E. Sezgin, P. van Nieuwenhuizen, Phys.Rev.D22:301,1980.

\bibitem{Elizalde}
E. Elizalde, S.D.Odintsov, Int.J.Mod.Phys.D2:51-58,1993


\bibitem{Align}
J. Preskill, Nucl. Phys. B177, 21 (1981)

\bibitem{Align1}
 M. E. Peskin, Nucl. Phys. B175, 197
(1980).


\bibitem{Minkowsky}
Nakia Carlevaro, Orchidea Maria Lecian, Giovanni Montani, Int. J. Mod. Phys. A
23, 1282-1285 (2008)

Nakia Carlevaro, Orchidea Maria Lecian, Giovanni Montani,
Mod.Phys.Lett.A24:415-427,2009











\end{thebibliography}

\begin{thebibliography}{99}
\begin{rm}

\bibitem{CPVZ1999} "Aharonov-Bohm effect, center monopoles and center vortices in SU(2) lattice
gluodynamics", M.N. Chernodub, M.I. Polikarpov, A.I. Veselov, M.A. Zubkov.
Nucl.Phys.Proc.Suppl.73:575-577,1999, hep-lat/9809158

\bibitem{BVZ1999} "Central dominance and the confinement mechanism in gluodynamics", B.L.G.
Bakker, A.I. Veselov, M.A. Zubkov, Phys.Lett.B471:214-219,1999, hep-lat/9902010

\bibitem{BVZ2000} "Central dominance and the confinement mechanism", B.L.G. Bakker, A.I.
Veselov, M.A. Zubkov, Nucl.Phys.Proc.Suppl.83:565-567,2000.

\bibitem{BVZ2001} "The Simple center projection of SU(2) gauge theory", B.L.G Bakker, A.I.
Veselov, M.A. Zubkov, Phys.Lett.B497:159-164,2001, hep-lat/0007022

\bibitem{BVZ2001_2} "The simple center projection of SU(2) gauge theory",  B.L.G. Bakker, A.I.
Veselov, M.A. Zubkov, Nucl.Phys.Proc.Suppl.94:478-481,2001.



\bibitem{BVZ2002} "Evidence for the reality of singular configurations in SU(2) gauge
theory",  B.L.G. Bakker, A.I. Veselov, M.A. Zubkov,
Phys.Lett.B544:374-379,2002, hep-lat/0205027



\bibitem{Z2002_2} "Abelian representation of nonAbelian Wilson loop and nonAbelian Stokes
theorem on the lattice",  M.A. Zubkov, Phys.Rev.D68:054503,2003,
hep-lat/0212001



\bibitem{BVZ2003} "A hidden symmetry in the Standard Model",   B.L.G. Bakker, A.I. Veselov,
M.A. Zubkov. Phys.Lett.B583:379-382,2004, hep-lat/0301011

\bibitem{BVZ2004} "An Additional symmetry in the Weinberg: Salam model", B.L.G. Bakker, A.I.
Veselov, M.A. Zubkov. Yad.Fiz.68:1045-1053,2005, Phys.Atom.Nucl.68:
1007-1015,2005, hep-lat/0402004

\bibitem{BVZ2005}  "Standard model with the additional Z(6) symmetry on the lattice", B.L.G.
Bakker, A.I. Veselov, M.A. Zubkov, Phys.Lett.B620:156-163,2005, hep-lat/0502006

\bibitem{BVZ2006} "Z(6) symmetry, electroweak transition, and magnetic monopoles at high
temperature", B.L.G. Bakker, A.I. Veselov, M.A. Zubkov,
Phys.Lett.B642:147-152,2006, hep-lat/0606010

\bibitem{BVZ2007} "Nambu monopoles in lattice Electroweak theory",  B.L.G. Bakker, A.I.
Veselov, M.A. Zubkov. J.Phys.G36:075008,2009, arXiv:0707.1017

\bibitem{BVZ2007_2} "Lattice study of monopoles in the Electroweak theory",  A.I. Veselov,
B.L.G. Bakker, M.A. Zubkov, PoS LAT2007:337,2007, arXiv:0708.2864

\bibitem{VZ2008} "Upper bound on the cutoff in lattice Electroweak theory", M.A.
Zubkov, A.I. Veselov. JHEP 0812:109,2008, arXiv:0804.0140

\bibitem{VZ2009L} "Upper bound on the cutoff in the Standard Model", M.A. Zubkov, A.I. Veselov,
proceedings of 27th International Symposium on Lattice Field Theory (Lattice
2009), Beijing, China, 25-31 Jul 2009, arXiv:0909.2840


\bibitem{BVZ2008} "Monopoles in lattice Electroweak theory", B.L.G. Bakker, A.I. Veselov,
M.A. Zubkov, proceedings of SPMTP08, arXiv:0809.1757

\bibitem{Z2009} "The Fluctuational region on the phase diagram of lattice Weinberg-Salam
model",  M.A. Zubkov. Phys.Lett.B684:141-146,2010, arXiv:0909.4106

\bibitem{Z2010Q} "The vicinity of the phase transition in the lattice Weinberg - Salam
Model",  M.A. Zubkov, proceedings of QUARKS2010, arXiv:1007.4885

\bibitem{Z2010PRD} "How to approach continuum physics in lattice Weinberg - Salam model", M.A.
Zubkov, Phys.Rev.D82:093010,2010

\bibitem{PZ2011} "Effective constraint potential in lattice Weinberg - Salam model",
M.I.Polikarpov, M.A.Zubkov, Phys. Lett. B 700 (2011) pp. 336-342,
[arXiv:1104.1319]

\bibitem{Z2004} "2D R**2 gravity in weak coupling limit", M.A. Zubkov,
Phys.Lett.B594:375-380,2004, gr-qc/0404087

\bibitem{Z2005} "Measure in the 2D Regge quantum gravity", M.A. Zubkov,
Phys.Lett.B616:221-227,2005, hep-lat/0501017

\bibitem{Z2004_2} "10-D Euclidean quantum gravity on the lattice", A.I. Veselov, M.A.
Zubkov, Nucl.Phys.Proc.Suppl.129:797-799,2004, hep-lat/0308025

\bibitem{Z2004_3} "Teleparallel gravity on the lattice", M.A. Zubkov. Phys.Lett.
B582:243-248,2004, hep-lat/0311036

\bibitem{Z2004_4} "10D Euclidean dynamical triangulations", A.I. Veselov, M.A. Zubkov.
Phys.Lett.B591:311-317,2004.

\bibitem{Z2006} "Gauge invariant discretization of Poincare quantum gravity", M.A. Zubkov.
Phys.Lett.B638:503-508,2006, Erratum-ibid.B655:307,2007, hep-lat/0604011


\bibitem{Z2006_2} "The Observability of Z(6) symmetry in the standard model", M.A.
Zubkov, Phys.Lett.B649:91-94,2007,Erratum-ibid.B655: 91,2007, hep-ph/0609029

\bibitem{Z2007} "Monopoles, topology of the standard model, and unification of interactions
at TeV scale", M. A. Zubkov, PoS LAT2007:285,2007.

\bibitem{Z2007_2} "Z(6) symmetry of the Standard Model and technicolor theory", M.A. Zubkov.
Phys.Lett.B674:325-329,2009, arXiv:0707.0731

\bibitem{Z2009_} "A Superstructure over the Farhi-Susskind Technicolor model", M.A. Zubkov.
Mod. Phys. Lett. A25:679-689,2010, arXiv:0910.4771



\bibitem{Z2010_2} "Torsion instead of Technicolor", M.A. Zubkov,
ITEP-LAT-2010-03, Mod. Phys. Lett. A25:2885-2898,2010, arXiv:1003.5473




\end{rm}
\end{thebibliography}
\end{document}